\pdfoutput=1 
\documentclass[11pt]{article}

\usepackage[utf8]{inputenc}
\usepackage[margin=1in]{geometry}
\usepackage[round]{natbib}
\usepackage{amsmath,amsthm,amssymb}
\IfFileExists{bbm.sty}{\usepackage{bbm}}{}
\providecommand{\mathbbm}[1]{\mathbf{#1}}
\newcommand{\notnotleftrightarrow}{\mathrel{\not\mkern-2mu\not\leftrightarrow}}
\usepackage{graphicx}
\usepackage{float}
\usepackage[section]{placeins}

\usepackage{microtype}
\usepackage{setspace}
\usepackage{xcolor}
\usepackage{hyperref}
\usepackage[english]{isodate}
\usepackage{booktabs}
\usepackage{adjustbox}
\usepackage{longtable,array}
\usepackage{nicefrac}
\usepackage[T1]{fontenc}
\usepackage{lmodern}

\usepackage{euscript}

\DeclareMathOperator{\diag}{diag}
\DeclareMathOperator{\spr}{spr}

\theoremstyle{plain}

\newtheorem{proposition}{Proposition}
\newtheorem{lemma}{Lemma}
\newtheorem{corollary}{Corollary}
\newtheorem{assumption}{Assumption}

\newtheorem{definition}{Definition}

\title{\textbf{Economic Power in International Trade}}

\author{Ashwin Bhattathiripad\thanks{Economics Area, Indian Institute of Management Kozhikode, Kerala 673 570, India.} \and  Vipin P. Veetil\footnotemark[1]}

\date{\small\printdayoff\today}

\begin{document}
\maketitle
\begin{abstract}
\setstretch{1.3}
Economic power is a country's capacity to harm another more than itself by
withdrawing from a trading relationship. This paper develops a general equilibrium model of the world trade network
to measure that power through counterfactual sanctions.  Using a world input--output table of eighty economies and fifty industries, we solve for $3{,}160$ such
sanction equilibria, one for each bilateral severance of a trading relationship. We
read the asymmetry between the two sides' losses as power. Our results show that mutual trade dependence is anything but mutual. The distribution of power in world trade is highly skewed. In the median trade relationship, one country inflicts on its trading partner more than four times the loss it bears. Only one pair in seven is close to balanced. The United
States is more powerful than all seventy-nine of its partners and
China against all but one. Economic power is not determined by the bilateral trade balance. 
\end{abstract}

\newpage
\setstretch{1.3}
\section{Introduction}

The classical case for international trade is that it lets
countries consume beyond what they could make alone. The gains are mutual in every country pair. But
trade also creates dependence, which tends not to be mutual to the same degree. A country that can be starved of an input it cannot replace, or denied a market it cannot do without, has in effect placed a part of its fortunes in the hands of whoever sits at the other end of the trading relationship. In this paper, we study not how
much trade enriches but how unevenly it empowers. More specifically, we ask how the structure of
who buys what from whom shapes the capacity of
one nation to injure another by the bare act of withdrawal.

We build a short-run network model of world trade to compute power by imposing counterfactual sanctions. We take each country-sector in the world input--output matrix as a production unit. A sanction is then a pattern of attenuated entries in the matrix.  Buyers \emph{and} sellers adapt to a sanction by finding alternative arrangements. The re-routing of trade, which is always imperfect, generates inefficiencies in production by altering the proportions in which intermediate inputs are combined. The analytical problem is then one of characterizing the new trading pattern in which all these re-routing and production changes are consistent with one another. We characterize such a sanction general equilibrium assuming the losses from production inefficiency take a CES form with inputs as complements. Prices are fixed in the short run, so world production reconfigures through rationing, and what we measure is the threat point of a trading relationship, the loss a country bears if the relationship breaks off today.

We use this model of trade disruption to measure the consequences of sanctions using counterfactual experiments. 
 More specifically, we take our model to the world input--output table with eighty economies and fifty industries, together making for four thousand country-sectors. For every pair of countries, we solve for the post-sanction fixed point after bilateral severances of trade links. This gives us $3{,}160$ counterfactual equilibria. Each is a world economy in which a single trade relationship has been cut and everything else allowed to adjust. These equilibria return what the trade severance costs each side as a share of its own economic activity, that side's vulnerability to the cut. The asymmetry between the two losses is what we read as power. The United States is the more powerful side in every one of its seventy-nine relationships, China in all but one. In nearly three in ten pairs of trading nations the larger of the two losses is ten times the smaller or more. These power asymmetries are stable across years and realistic parametric specifications.  The skewness of power is not confined to the binary measure of who is more powerful but extends to the distribution of losses. For instance, a trade severance with Russia would cost Belarus over a tenth of its gross economic activity, and one with China would cost Vietnam a tenth, against a fraction of a percent for the more powerful partner. Canada and Mexico would each lose about a tenth of their economic activity to a rupture with the United States, while the United States would lose about one percent or less to either. Across all country pairs these losses span nearly seven orders of magnitude, from the negligible to the catastrophic.
 
The ability to disrupt economic life through international trade sanctions is therefore very asymmetric. There is such a thing as `power' in international trade. From a proximate point of view, an economy's size is closely related to its power, and nearly the whole of that relation is that the same lost dollars are a smaller fraction of a larger economy. Trade deficits, by contrast, appear to have little to do with it. Prima facie there is little ground to claim that the nations most capable of hurting a country are the ones with which it runs the largest trade deficits.\footnote{Erasing the trade imbalance of a pair from the trade table hardly moves the tilt in power between its two countries, and in a wide sweep of regressions the trade balance adds a median of one thousandth to the explained variance of power.} The sharpest illustration of how little trade deficits have to do with power comes from the case of the United States and China. Our measures show that the United States is the more powerful party despite running large deficits in its trade with China. What decides the tilt in power between two nations is not who sells more but how much of each nation's production rests on the trade that cannot be replaced.\footnote{An interactive simulator built on the
model is available at \href{https://sanction-simulator.streamlit.app/}{sanction-simulator.streamlit.app}.}

\subsection{Related literature}
\label{subsec:literature}

This paper draws on two streams of work on trade and sanctions: the political economy literature on trade dependence and the empirical literature that estimates the economic consequences of particular sanctions. The political economy literature on trade dependence is as old as the science of economics itself. Hamilton, Fichte, and List all developed theories in which commercial interdependence is itself a form of subordination. Within the confines of the narrow problem considered in this paper, however, it is \citeauthor{hirschman1945} who is our closest predecessor. He argued that a country's dependence is decided by how well it could substitute, as buyer and as seller, if a trade relation were cut. His own empirical work, however, and that of nearly everyone who followed, took observed trade flows as the proxy for that dependence \citep{keohanenye1977,baldwin1985,farrell2019}. Neither the substitution nor its consequences were ever computed. In this sense, we offer a faithful examination of Hirschman's hypothesis.

Work that speaks most directly to our paper is the one which measures the impact of particular sanction episodes. That literature uses a wide variety of models to structure and discipline these estimates, including dynamic general equilibrium models \citep{ghironi2025sanctions,bachmann2022whatif}, more reduced-form econometric approaches \citep{felbermayr2020gsdb,crozethinz2020}, and models tailored for the study of particular sanction designs \citep{hausmannschetter2024}.\footnote{\citet{liuyang2026power} too study the question of power, but production in their model uses labor alone, so that there is no production network through which a disruption of trade can spread. Their model therefore assumes away many of the pronounced difficulties that arise from the discrete and sizeable disruption of trade.} The two papers closest to our own are \citet{imbspauwels2024} and \citet{joshi2024networks}. What connects us to them is the premise that the cost of a sanction is a network object, borne on both sides of the severed trade relation. We generalize their approach by considering a production economy with substitution among the suppliers of each input on sufficiently general trade networks. 

We must add that our paper can be read as a complement to \citet{clayton2026geoeconomics}. They develop a general equilibrium of rent extraction by a hegemon, and examine the optimal policy for the hegemon and its welfare consequences. The hegemon is a `hegemon' because of its ability to inflict asymmetric damage through sanctions.  \citeauthor{clayton2026geoeconomics} do not concern themselves with the network origins of this `power' to inflict asymmetric damage. Those origins are the subject of this paper. An analysis of how structural power influences the international terms of trade goes back to \citet{prebisch1950} and \citet{singer1950}, who argued that the center-periphery structure of world trade turns the terms of trade against the periphery. What \citeauthor{prebisch1950}  and \citeauthor{singer1950} read off price series, and what \citeauthor{clayton2026geoeconomics} derive from an optimizing hegemon, we locate in the world trade network.

\subsection{Organization of the paper}

Section~\ref{sec:model} develops the fixed-price general equilibrium model of sanctions. Section~\ref{sec:statistics} estimates country vulnerabilities and with them the bilateral power between nations of the world. Section~\ref{sec:power_size} asks how much of that power is the size of an economy. Section~\ref{sec:power_balance} asks how much of it is the balance of trade between two nations. Section~\ref{sec:validation} takes the measure of power to the historical record of economic coercion. We ask whether the state that initiates a sanction is the one our estimates place on the more powerful side. Section~\ref{sec:conclusion} offers concluding remarks. Appendix~\ref{app:m-coreproofs} proves the formal claims and states the auxiliary results they rest on. The Online Appendix carries the supplementary constructions, the calibration, and the robustness exercises, and it solves the same trade severances in an economy in which prices clear markets (Online Appendix~\ref{app:flex}).

\section{The Short-Run Sanction Economy}
\label{sec:model}

\subsection{The benchmark economy and the trade severance}
\label{subsec:model_benchmark}

The full world input--output network contains $CS$ country-sectors, since each of the $C$ countries operates $S$ sectors.\footnote{Throughout the paper matrices are uppercase bold and vectors lowercase bold, with the household matrix $\mathbb A$ and the indicator vector of a country double-struck. Greek letters denote parameters and derived quantities, calligraphic letters denote sets, mappings, and the statistics of a pair or of a country, and inequalities between vectors and matrices are entrywise. The indices $i,j$ range over the $N$ active country-sectors, the indices $c,c'$ range over the $C$ countries, and the indices $r,r'$ range over the $S$ sectors. The index $b$ ranges over the buyers, country-sector producers and country-households alike. The symbol $\circ$ denotes the entrywise product of two vectors. A caron $\check x$ denotes a disrupted object taken before the rationed exchange, and a subscript or parenthetical superscript $\omega$ marks a quantity evaluated under the trade sanction $\omega$. An up arrow $x^{\uparrow}$ marks a quantity measured on a country's sourcing side, upstream toward its suppliers, and a down arrow $x^{\downarrow}$ marks a quantity measured on its sales side, downstream toward its customers. Online Appendix~\ref{app:notation} summarizes the principal notation of the paper and the appendices.} We drop the country-sectors with zero benchmark activity and relabel the retained ones as $\{1,\dots,N\}$, where $N\le CS$. Every vector and matrix in the model is defined on this active set. Two projections return a country-sector's coordinates, its country $\mathcal C(i)\in\{1,\dots,C\}$ and its sector $\mathcal S(i)\in\{1,\dots,S\}$, and the set $\mathcal N_r:=\{\,i:\mathcal S(i)=r\,\}$ collects the country-sectors of sector $r$.

Throughout, rows are sellers and columns are buyers, and a buyer is either a producer, one of the $N$ country-sectors, or a country-household, one per country. Let $\mathbf A=[a_{ij}]$ be an $N \times N$ nonnegative matrix with unit column sums, $\sum_{i=1}^Na_{ij}=1$ for each country-sector producer $j$. The entry $a_{ij}$ is the benchmark share of producer $j$'s intermediate input bundle sourced from seller $i$. The share of input sector $r$ in that bundle, whatever country the sector's output comes from, is
\[
\alpha_{rj}:=\sum_{i\in\mathcal N_r}a_{ij}.
\]
The set $\mathcal R_j:=\{r:\alpha_{rj}>0\}$ collects the input sectors producer $j$ uses.

A leakage parameter $\beta_j\in(0,1)$ is the fraction of country-sector $j$'s gross economic activity that leaves the intermediate goods network, the counterpart of the value-added share in input--output accounts. On the production side, this fraction corresponds to the non-intermediate block, including labor. We collect these parameters in $\boldsymbol\beta:=\diag(\beta_1,\dots,\beta_N)$, and the benchmark propagation matrix is $\mathbf A(\mathbf I-\boldsymbol\beta)$, with entries $(1-\beta_j)a_{ij}$. 

Country $c$'s household, which aggregates the country's final consumers, its government, and its investors, buys $f^{c}_i\ge0$ of country-sector $i$'s output and spends $F^{c}:=\sum_if^{c}_i$ in all. Its sourcing shares $a_{ic}:=f^{c}_i/F^{c}$ form the country's column of the $N\times C$ household matrix $\mathbb A=[a_{ic}]$. The household matrix has unit column sums, and the share of sector $r$ in country $c$'s column is $\alpha_{rc}:=\sum_{i\in\mathcal N_r}a_{ic}$. Indexing the household's column by its country lets one symbol serve both kinds of buyer, so that $a_{ib}$ is the share of buyer $b$'s spending sourced from seller $i$ whether buyer $b$ is a producer or a household, and $\alpha_{rb}$ is the share of sector $r$ in that spending. What remains of the demand for country-sector $i$'s output, $f^{o}_i\ge0$, we call its residual final demand. It is the part that no sanction between two countries of the trade table reaches, the purchases of the `rest of the world sector' together with inventory changes and statistical discrepancies.  Final demand is the sum of the two parts,
\begin{equation}
\mathbf f=\mathbb A\mathbf F+\mathbf f^{o}.
\label{eq:m-final-split}
\end{equation}
Here $\mathbf F:=(F^{1},\dots,F^{C})^{\top}$ is the vector of household budgets. The leakage $\beta_j$ measures what leaves the intermediate network at producer $j$, and the final demand $\mathbf f$ measures the orders that reach the network from outside it.

The benchmark activity vector $\mathbf s$ satisfies
\begin{equation}
\mathbf s=\underbrace{\mathbf f}_{\text{final demand}}+\underbrace{\mathbf A(\mathbf I-\boldsymbol\beta)\,\mathbf s}_{\text{intermediate orders}}.
\label{eq:m-benchmark}
\end{equation}
Country-sector $i$'s activity is its final demand plus the intermediate input demand directed to it by all producers. We write $s_i$ for the activity, or size, of country-sector $i$, its gross output in the language of input--output accounts, and $Q(c):=\sum_{i:\mathcal C(i)=c}s_i$ for the size of country $c$. 

The benchmark transaction flow from seller $i$ to buyer $b$ is the sourcing share $a_{ib}$, the entry of $\mathbf A$ for a producer and of $\mathbb A$ for a household, scaled by the buyer's benchmark spending,
\begin{equation}
\begin{aligned}
w_b&:=
\begin{cases}
w_j:=(1-\beta_j)s_j, & \text{$b=j$, a producer},\\
w_c:=F^{c}, & \text{$b=c$, a country-household}
\end{cases}\\
Z_{ib}&:=a_{ib}\,w_b.
\end{aligned}
\label{eq:m-flows}
\end{equation}
The matrix $\mathbf Z:=[Z_{ib}]$ collects the flows over producers and households alike. We assume that the network of benchmark trading links, which joins a seller and a buyer when the flow between them is positive, is connected. The 2022 world table on which we compute our estimates satisfies the assumption.

 The sanction $c'\notnotleftrightarrow c$ bars every sector of country $c'$ from supplying any sector and the household of country $c$, and every sector of country $c$ from supplying any sector and the household of country $c'$. A link from seller $i$ to buyer $b$ is thus barred whenever the two sit in different countries of the pair. The pair's households lose their suppliers in the partner exactly as the pair's producers do, and a producer loses its customers in the partner whether they are producers or households. The admissible sanctions, collected in the set $\Omega$, are the bilateral sanctions over the country pairs of interest, the group trade severances that bar every link between a set of acting countries and a set of target countries in the same way, and the sector-level refinements of either. A sanction enters the model only through the entries of $\mathbf A$ and $\mathbb A$ it bars.

Since no sanction is watertight, and some trade evades it, a sanction $\omega\in\Omega$ attenuates rather than fully zeroes the entries it bars. The barred-link attenuation $\delta$ is the weight each barred link carries in the disrupted matrix before buyers and sellers adjust, and it satisfies
\begin{equation}
\delta\in(0,\bar\delta],\qquad 0<\bar\delta<1.
\label{eq:m-leak}
\end{equation}
Because the attenuation is positive, it keeps every benchmark relationship alive and creates no trading link that the benchmark lacks.

For a sanction $\omega\in\Omega$ and an attenuation $\delta$ as in \eqref{eq:m-leak}, the disrupted matrices $\check{\mathbf A}=[\check a_{ij}]$ and $\check{\mathbb A}=[\check a_{ic}]$ carry the entries
\begin{equation}
\check a_{ib}
:=
\begin{cases}
a_{ib}, & \text{$\omega$ leaves the link $i\to b$ alone},\\
\delta\,a_{ib}, & \text{$\omega$ severs it}
\end{cases}.
\label{eq:m-disrupted}
\end{equation}
The disrupted flows are
\[
\check Z_{ib}:=\check a_{ib}\,w_b.
\]
A severed link keeps $\delta Z_{ib}$ of its benchmark flow and displaces the remaining $(1-\delta)Z_{ib}$.

\subsection{Rationed exchange at fixed prices}
\label{subsec:model_rerouting}

Buyers and sellers redirect part of their lost trade to other partners. Prices are fixed in the short run, so the redirection cannot run through price adjustment and takes the form of rationing. The rationing is proportional to incumbent relationships. A buyer that loses a supplier seeks the same input from its remaining suppliers in proportion to what it already bought from each, and a seller that loses a customer seeks buyers among its remaining customers in proportion to what it already sold to each.\footnote{Rationing at posted prices yields a fixed-price equilibrium in the sense of \citet{dreze1975} and \citet{benassy1975}, in which prices are posted, agents trade under quantity constraints, and the constraints are consistent. The allocation can equally be read as market clearing at pseudo-prices, namely the prices that the existing trading pattern implies and at which an unconstrained buyer would choose its rationed trades. These pseudo-prices are the virtual prices of \citet{rothbarth1941} and \citet{nearyroberts1980}. Only relative pseudo-prices within each connected component of the network of trading links are determined, and no buyer pays a pseudo-price, since every trade is made at its posted price.}

Not all of the displaced trade re-forms. The unrecovered share $\tau\in[0,1]$ is the fraction of the trade displaced on a severed link that finds no counterpart. Substitution runs sector by sector, since a buyer replaces a lost supplier with another supplier of the same input. Buyer $b$'s surviving sector-$r$ budget, as a share of its spending $w_b$, and seller $i$'s surviving capacity are
\begin{align}
t^{(\omega)}_{rb}
&:=\alpha_{rb}-\tau(1-\delta)\!\!\sum_{i\in\mathcal N_r:\,i\to b\text{ severed}}\!\!a_{ib}
\label{eq:m-coltarget}\\
g^{(\omega)}_i
&:=\underbrace{s_i-f^{o}_i}_{\text{benchmark sales to producers and households}}-\;\tau(1-\delta)\!\!\sum_{b:\,i\to b\text{ severed}}\!\!Z_{ib}.
\label{eq:m-rowtarget}
\end{align}
The subtracted terms run over severed links only, so a buyer or a seller that $\omega$ leaves alone keeps its benchmark budget or capacity.\footnote{The capacity counts only the sales that the trade table records buyer by buyer, so the residual final demand $f^{o}_i$, which no sanction reaches, enters neither a budget nor a capacity, and no supplier's capacity exceeds its pre-sanction sales.} A sanction barring Russian coal, for instance, leaves the coal budget of a buyer that never bought Russian coal exactly as it was.

Within each affected sector the budgets and capacities meet in an exchange at fixed prices, in which a supplier whose customers would overdraw its capacity rations them and a supplier that its customers would leave with spare capacity is rationed by them. The post-sanction trading pattern is the allocation of that exchange, which gives the input coefficients of the trade table after the sanction. We call it the effective demand.\footnote{We use the term in the sense of \citet{benassy1975}, for whom the effective demand of an agent is the demand that the agent expresses once the quantity constraints it faces are taken into account, as distinct from the notional demand of its unconstrained plans.} Writing $Z_{\omega,ib}$ for the flow the exchange places on the link $i\to b$, the effective demand coefficient of the link is that flow divided by the buyer's benchmark spending,
\begin{equation}
a^{(\omega)}_{ib}:=\frac{Z_{\omega,ib}}{w_b}.
\label{eq:m-balanced-coeff}
\end{equation}
The coefficients form the producer block $\mathbf A_\omega=[a^{(\omega)}_{ij}]$ and the household block $\mathbb A_\omega=[a^{(\omega)}_{ic}]$, the effective demand matrices, with their dependence on $\tau$ and $\delta$ suppressed in the notation. A sector block with no severed link stays at benchmark, $a^{(\omega)}_{ib}=a_{ib}$.

The exchange settles to a two-sided proportional rationing of the disrupted flows,
\begin{equation}
a^{(\omega)}_{ib}=u_i\,\check a_{ib}\,v_{rb}.
\label{eq:m-reroute}
\end{equation}
Here the seller scalings $u_i$ and the buyer scalings $v_{rb}$ are positive, on the rows and columns with a positive entry in the benchmark table. The seller scaling brings each seller's sales to its capacity. The buyer scaling brings each buyer's sector-$r$ purchases to its budget, one scaling per sector and not one in common, since inputs are complements and a buyer short of coal bids for coal and not for steel. Note that the reciprocal of seller $i$'s scaling is the pseudo-price of its good, $\tilde p_i:=1/u_i$.\footnote{The exchange needs no elasticity of substitution between suppliers, since any within-sector elasticity returns the same allocation and changes only the pseudo-prices, which become a power of the seller scalings rather than their reciprocals (Lemma~\ref{lem:m-curvature}). That invariance is an instance of the central point of the disequilibrium literature, namely that binding quantity constraints, not preferences, determine realized trades \citep{barrogrossman1971,malinvaud1977}.} A settled allocation is a fixed point of the rationing, an allocation that no seller and no buyer rescales further. We now record that the settled allocation exists and is unique.

\begin{lemma}[Existence and uniqueness of the rationing fixed point]
\label{lem:m-fixedpoint}
On each sector block, exactly one allocation of the form \eqref{eq:m-reroute} with positive scalings meets every capacity \eqref{eq:m-rowtarget} and every budget \eqref{eq:m-coltarget}.
\end{lemma}

\noindent Proof in Appendix~\ref{app:m-wellposed}.

Since the buyer index $b$ runs over producers and country-households alike, the lemma defines both effective demand matrices, $\mathbf A_\omega$ and $\mathbb A_\omega$, at once.

\subsection{From effective demand to production}
\label{subsec:model_ces}

We have so far established that fixed-price rationing settles on a unique pattern of effective demand, but we have said nothing about the scale of activity at which that demand is expressed. The way a sanction affects output depends on that scale as well as on the effective demand, and the effective demand is consistent with many levels of economic activity in each country-sector. To pin down the level of economic activity we need to characterize production.

We do not make any strong assumption about the production function of a country-sector. We make a weaker assumption instead, on the curvature of the loss after a sanction. We assume that the benchmark input bundle was the efficient one at the fixed pre-sanction equilibrium prices, so it remains the reference point after the sanction. To characterize the loss in production that remains after trade is redirected, we begin by defining the fraction that each country-sector seller meets of the effective demand it faces. We assume that each supplier delivers the same fraction of the effective demand of each of its buyers, producers and households alike, rationing them pro rata. We write $\mathbf h\in[0,1]^N$ for the supplier-side fulfillment vector, whose component $h_i$ is that common fraction at supplier $i$.

Within each input sector, we treat deliveries from different sellers as fungible. What matters for buyer $j$'s production is the total it receives from sector $r$, relative to its benchmark intensity $\alpha_{rj}$, over the sectors $\mathcal R_j$ it uses.\footnote{A household places orders and takes deliveries but produces nothing, so the retention ratios and output ratios that follow are defined for producers alone, and a severed household purchase creates no direct producer input shortfall.} We define the ratio of buyer $j$'s delivered sector-$r$ input intensity to its benchmark sector-$r$ input intensity as follows,
\begin{equation}
\theta^{(\omega)}_{rj}(\mathbf h)
\;:=\;
\frac{\sum_{i\in\mathcal N_r}h_i\,a^{(\omega)}_{ij}}{\alpha_{rj}}.
\label{eq:m-theta}
\end{equation}
We call $\theta^{(\omega)}_{rj}(\mathbf h)$ the realized sectoral input-retention ratio. A buyer's ratio lies in $[0,1]$ (Lemma~\ref{lem:m-kernel} in Appendix~\ref{app:m-coreproofs}), falling short of one when its effective demand already falls short of its benchmark intensity, and again when its suppliers fail to deliver. Setting $\mathbf h=\mathbf 1$ isolates the first shortfall, $\theta^{(\omega)}_{rj}(\mathbf 1)=\sum_{i\in\mathcal N_r}a^{(\omega)}_{ij}/\alpha_{rj}$, the buyer's surviving budget share divided by its benchmark share. Fulfillment adds the second wherever a seller cannot produce what the effective demand facing it calls for. A German--Russian sanction, for instance, would leave the German steel sector short of coal on two distinct counts. The first is the barring of coal flows from Russia, which trims the coal entries in its column of effective demand and leaves the first shortfall $\theta^{(\omega)}_{rj}(\mathbf 1)$ below one. The second is that the Australian or Indonesian coal it turns to is itself rationed after the sanction and arrives at $h_i<1$.

The retention ratios of producer $j$ combine into one index of the production loss that arises when its intermediate inputs depart from the proportions that were optimal before the sanction,
\begin{equation}
\kappa^{(\omega)}_j(\mathbf h)
\;:=\;
\biggl(\sum_{r\in\mathcal R_j}\alpha_{rj}
\bigl(\theta^{(\omega)}_{rj}(\mathbf h)\bigr)^{\rho}\biggr)^{1/\rho}.
\label{eq:m-kappa}
\end{equation}
Here $\rho\in[\underline\rho,\overline\rho]\subset(-\infty,0)$ is the exponent of the CES function with which the intermediate inputs are combined, and we call $\kappa^{(\omega)}_j(\mathbf h)$ the intermediate input efficiency factor. A larger $|\rho|$ makes an uneven sectoral bundle more costly.\footnote{The factor lies in $[0,1]$, since a weighted power mean of the retention ratios lies between their smallest and largest values. If $\theta^{(\omega)}_{rj}(\mathbf h)=0$ for any $r\in\mathcal R_j$, we set $\kappa^{(\omega)}_j(\mathbf h)=0$ and the output ratio is zero as well, the continuous values at that boundary, so a producer stripped of its intermediate composite produces nothing. The boundary $\kappa^{(\omega)}_j=0$ cannot occur at $\mathbf h>\mathbf 0$ because positive attenuation leaves every used sector with positive effective demand.}

We take the curvature of the loss to be that of a nested CES \citep{sato1967}, whose outer nest combines the intermediate composite and the non-intermediate block with exponent $\varrho\in[\underline\varrho,\overline\varrho]\subset(-\infty,0)$.\footnote{Online Appendix~\ref{app:m-local-efficiency} distinguishes this fulfillment response from a production-feasibility condition in realized input levels. Lemma~\ref{lem:m-local-form} in Online Appendix~\ref{app:m-form-robustness} bounds the local effect of changing the response, and for a smooth alternative with the same benchmark value and first derivatives the difference between the two responses is second order in the displacement of the bundle.} More specifically, producer $j$'s output ratio at the deliveries $\mathbf h$ is
\begin{equation}
\lambda^{(\omega)}_j(\mathbf h)
\;:=\;
\Bigl[\beta_j+(1-\beta_j)\,\bigl(\kappa^{(\omega)}_j(\mathbf h)\bigr)^{\varrho}\Bigr]^{1/\varrho}.
\label{eq:m-lambda-rule}
\end{equation}
The weight $\beta_j$ is the leakage parameter of Section~\ref{subsec:model_benchmark}, the benchmark share of spending on the non-intermediate block. We assume that this spending purchases each country-sector a specific input $\bar v_j=\beta_js_j$ that does not move across country-sectors.\footnote{The block collects the non-tradable factors of production, namely labor, land, and local capital. Their immobility across countries is permanent rather than a matter of horizon, and their immobility across sectors is the short run itself. The specific-factors tradition of trade theory gives this short-run specificity to capital and land \citep{jones1971,neary1978}, and we extend it to labor.} Read $\lambda^{(\omega)}_j(\mathbf h)$ as the fraction of each of its orders that country-sector $j$ can fill with the bundle the deliveries $\mathbf h$ bring it. Since the sanction does not alter $\beta_j$ and $\kappa^{(\omega)}_j$ lies in $[0,1]$, the fraction $\lambda^{(\omega)}_j$ lies in $[0,1]$ as well.\footnote{At the benchmark the elasticity of the output ratio with respect to the efficiency factor equals the intermediate cost share $1-\beta_j$ \citep{hulten1978}. Away from the benchmark the elasticity climbs toward one, since the intermediate composite becomes more essential as it degrades (Lemma~\ref{lem:m-outer} in Appendix~\ref{app:m-coreproofs}).} Across producers, these output ratios form the vector $\boldsymbol\lambda^{(\omega)}(\mathbf h)$.

The formal statements of the paper are made under the conditions set so far, a sanction in $\Omega$, the positive attenuation \eqref{eq:m-leak}, an unrecovered share in $[0,1]$, the two curvature parameters in their intervals, leakages in $(0,1)$, and a connected network of benchmark trading links, and they do not repeat them. A statement that needs more, a positive unrecovered share, an interior equilibrium, or a pair trading in both directions, says so.

\subsection{The sanction equilibrium and the loss vector}
\label{subsec:model_counterfactual}
We have now established that the redirection of trade has a fixed point, and we have characterized how the redirected trade translates into levels of production through the curvature of the loss. What remains is to establish that the levels of production the redirection generates have a fixed point of their own. Write $\mathbf f_\omega:=\mathbb A_\omega\mathbf F+\mathbf f^{o}$ for the effective final demand.

\begin{definition}[Sanction equilibrium]
\label{def:m-sanction-equilibrium}
A pair $(\mathbf x^*,\mathbf h^*)\in\mathbb R_+^N\times[0,1]^N$ is a sanction equilibrium if
\begin{equation}
\mathbf x^*
=\boldsymbol\lambda^{(\omega)}(\mathbf h^*)\circ\mathbf f_\omega
+\diag(\mathbf h^*)\,\mathbf A_\omega(\mathbf I-\boldsymbol\beta)\,\mathbf x^*,
\label{eq:m-equilibrium}
\end{equation}
and the fulfillment rates satisfy $\mathbf h^*=\boldsymbol\lambda^{(\omega)}(\mathbf h^*)$.
\end{definition}

The condition $\mathbf h^*=\boldsymbol\lambda^{(\omega)}(\mathbf h^*)$ means that each supplier delivers the fraction of the effective demand it faces that its degraded input bundle lets it produce. At activity $\mathbf x^*$, suppliers face the effective intermediate demand $\mathbf A_\omega(\mathbf I-\boldsymbol\beta)\mathbf x^*$ and the effective final demand $\mathbf f_\omega$.

Starting from full fulfillment, $\mathbf h^{(0)}=\mathbf 1$, we iterate $\mathbf x^{(n)}=\mathbf s(\mathbf h^{(n)})$, the unique nonnegative solution of the activity equation at $\mathbf h^{(n)}$, and $\mathbf h^{(n+1)}=\boldsymbol\lambda^{(\omega)}(\mathbf h^{(n)})$, and we call this sequence the benchmark path.\footnote{Complete collapse, $(\mathbf 0,\mathbf 0)$, is always a sanction equilibrium, since if no supplier delivers, no buyer receives the inputs needed to restore production, so the equilibrium equations do not select an outcome by themselves.} We now establish that the path converges and that its limit, the benchmark-selected equilibrium $(\mathbf s_\omega,\mathbf h_\omega)$, is a sanction equilibrium in the sense of Definition~\ref{def:m-sanction-equilibrium}.

\begin{proposition}[Equilibrium, selection, and interiority]
\label{prop:m-monotone}
A sanction equilibrium exists, and every sanction equilibrium satisfies $\mathbf 0\le\mathbf x^*\le\mathbf s$, so that no country-sector produces more than at the benchmark.
\begin{enumerate}
\item[\textnormal{(i)}] The benchmark path is entrywise nonincreasing and converges, $\mathbf h^{(n)}\downarrow\mathbf h_\omega$ and $\mathbf x^{(n)}\to\mathbf s_\omega:=\mathbf s(\mathbf h_\omega)$, and its limit $(\mathbf s_\omega,\mathbf h_\omega)$ is a sanction equilibrium.
\item[\textnormal{(ii)}] The limit is the equilibrium of least loss. Its fulfillment vector $\mathbf h_\omega$ is the greatest fixed point of the fulfillment rule, and every sanction equilibrium $(\mathbf x^*,\mathbf h^*)$ satisfies $\mathbf h^*\le\mathbf h_\omega$ and $\mathbf x^*\le\mathbf s_\omega$.
\item[\textnormal{(iii)}] If the sanction satisfies the interiority condition
\begin{equation}
\kappa^{(\omega)}_i(\mathbf 1)
\;>\;
(1-\beta_i)^{-1/\varrho}
\qquad\text{for every active }i,
\label{eq:m-cushion-cond}
\end{equation}
then $\mathbf h_\omega\ge c_\omega\mathbf 1$ for a positive constant $c_\omega$, and the selected equilibrium lies bounded away from collapse.
\end{enumerate}
\end{proposition}

\noindent Proof in Appendix~\ref{app:m-proof-monotone}.

When a producer's input bundle degrades, its deliveries fall, and the bundles of the producers it supplies degrade in turn. Condition~\eqref{eq:m-cushion-cond} halts this spiral short of zero production and guarantees an interior equilibrium.\footnote{The condition is the nonlinear, post-sanction counterpart of the Hawkins--Simon productivity condition for a viable input--output economy \citep{hawkinssimon1949}, and the proof of Proposition~\ref{prop:m-monotone} obtains the interior equilibrium as a special case of the subsolution argument of \citet{krasnoselskii1964}. The benchmark economy satisfies the linear Hawkins--Simon form through leakage (Lemma~\ref{lem:m-benchmark}).} The condition sets against each other the two forces that drive and stem the spiral. The stemming force is the cushion of the never-rationed non-intermediate block, whose factor $(1-\beta_i)^{1/\varrho}$ exceeds one and falls to one in the Leontief limit, where the condition fails wherever the intermediate-input bundle has degraded at all (Online Appendix~\ref{app:m-correction}). The driving force is the degradation of the bundle, which the intermediate input efficiency factor $\kappa^{(\omega)}_i(\mathbf 1)$ records as the efficiency that producer $i$'s intermediate input bundle retains when every supplier delivers in full the effective demand it faces. Since the rationed exchange replaces the severed purchases only in part, this efficiency falls short of one for a producer that has lost purchases. The condition is sufficient and not necessary.\footnote{On the 2022 trade table every bilateral severance we compute settles at an interior equilibrium, including the few severances at which the condition fails (Online Appendix~\ref{app:m-proofs}).}

Comparing the equilibrium activity $\mathbf s_\omega$ with the benchmark activity $\mathbf s$ defines the loss vector
\begin{equation}
\boldsymbol\Delta\mathbf s_\omega:=\mathbf s-\mathbf s_\omega.
\label{eq:m-loss}
\end{equation}
The loss vector is nonnegative by Proposition~\ref{prop:m-monotone}, and it has an exact anatomy that shows which part of a sanction does the damage. Write $\boldsymbol\xi_\omega$ for the vector of lost sales, the sales the capacity \eqref{eq:m-rowtarget} strips from each seller,
\begin{equation}
\xi_{\omega,i}:=\tau(1-\delta)\!\!\sum_{b:\,i\to b\text{ severed}}\!\!Z_{ib}
\;=\;s_i-f^{o}_i-g^{(\omega)}_i.
\label{eq:m-lost-sales}
\end{equation}

We now decompose the loss in economic activity into the output a producer forgoes because its input bundle has degraded and the severed sales $\boldsymbol\xi_\omega$ that the exchange does not recover.

\begin{proposition}[Anatomy of the loss]
\label{prop:m-anatomy}
At every sanction equilibrium $(\mathbf x^*,\mathbf h^*)$,
\begin{equation}
\mathbf s-\mathbf x^*
\;=\;
\bigl[\mathbf I-\diag(\mathbf h^*)\mathbf A_\omega(\mathbf I-\boldsymbol\beta)\bigr]^{-1}
\Bigl[(\mathbf 1-\mathbf h^*)\circ\mathbf s+\mathbf h^*\circ\boldsymbol\xi_\omega\Bigr].
\label{eq:m-anatomy}
\end{equation}
Both bracketed vectors are nonnegative, and so is the inverse.
\end{proposition}

\noindent Proof in Appendix~\ref{app:m-proof-anatomy}.

In \eqref{eq:m-anatomy}, the productivity channel is $(\mathbf 1-\mathbf h^*)\circ\mathbf s$, the unfilled fraction of benchmark activity, and the lost-market channel is $\mathbf h^*\circ\boldsymbol\xi_\omega$, the lost sales delivered at the seller's own fulfillment rate.\footnote{The unrecovered share of a severed household purchase enters the lost sales $\boldsymbol\xi_\omega$ of its seller and propagates through the production network, potentially back to suppliers in the buying country. The effective final demand $\mathbf f_\omega$ does not appear in \eqref{eq:m-anatomy}, since it cancels at a sanction equilibrium (Appendix~\ref{app:m-proof-anatomy}). At $\tau=0$ no activity is lost (Lemma~\ref{lem:m-frictionless} in Appendix~\ref{app:m-coreproofs}), and Online Appendix~\ref{oa:loss-anatomy} splits each country's vulnerability between the productivity channel and the lost-market channel.}

\section{International Trade Vulnerabilities}
\label{sec:statistics}

We now compute the vulnerabilities of countries to sanctions by comparing the benchmark and post-sanction fixed-price general equilibrium levels of economic activity.

\subsection{Vulnerability and bilateral power}
\label{subsec:vulnerability_power}

Let
\(\mathbbm{1}_c\in\{0,1\}^N\) be the indicator vector of country \(c\),
with \((\mathbbm{1}_c)_i=1\) if \(\mathcal C(i)=c\) and \(0\) otherwise. Here \(\mathbbm{1}_c\) acts as a row, so
\(\mathbbm{1}_c\mathbf x\) sums any country-sector vector
\(\mathbf x\) over the sectors of \(c\), and
\(Q(c)=\mathbbm{1}_c\mathbf s\).

Under sanction \(\omega\), the
post-sanction size of country \(c\) is
\begin{equation}
Q_\omega(c)
:=
\mathbbm{1}_c\mathbf s_\omega.
\label{eq:counterfactual_country_size}
\end{equation}
The loss of country $c$ sums the loss vector \eqref{eq:m-loss} over its sectors,
\begin{equation}
Q(c)-Q_\omega(c)
=
\mathbbm{1}_c\boldsymbol\Delta\mathbf s_\omega.
\label{eq:delta_country_size}
\end{equation}

The unrecovered share \(\tau\), the two CES exponents \(\rho\) and
\(\varrho\), the barred-link attenuation \(\delta\), and the cost shares enter only
through the post-sanction equilibrium \(\mathbf s_\omega\). We hold them at a fixed calibration, so the losses vary only with the trade severance.

\begin{definition}[Country vulnerability]
\label{def:country_vulnerability}
The vulnerability of affected country \(c\) under sanction \(\omega\) is its
own-size-normalized loss,
\begin{equation}
\gamma^{(\omega)}_c
:=
\frac{Q(c)-Q_\omega(c)}{Q(c)}
=
\frac{\mathbbm{1}_c\boldsymbol\Delta\mathbf s_\omega}
{\mathbbm{1}_c\mathbf s}.
\label{eq:gamma_general}
\end{equation}
\end{definition}

We compute this vulnerability for the bilateral sanction \(c'\notnotleftrightarrow c\) of Section~\ref{subsec:model_benchmark} between the affected country \(c\) and a partner \(c'\), which bars every seller sector in either country from supplying any producer or the household of the other. The sanction is symmetric, \(c'\notnotleftrightarrow c=c\notnotleftrightarrow c'\), and substituting it into \eqref{eq:gamma_general} gives the two vulnerabilities \(\gamma^{(c'\notnotleftrightarrow c)}_c\) and \(\gamma^{(c'\notnotleftrightarrow c)}_{c'}\). Both come from the same post-sanction fixed point problem, and each is normalized by its own country's size.\footnote{The denominator is gross economic activity and not value added. The value added of a country is roughly its activity weighted by the leakage $\beta_j$ of each country-sector, and the leakage is smaller in economies carrying heavy processing trade, so a value-added denominator would raise the measured vulnerability of such an economy.} Normalizing this way makes the vulnerability comparable across sanctions and across countries of different size.

The bilateral experiment measures how the cost of breaking the relationship divides between the two countries. One partner may rely on the relationship for a larger share of its economic activity, may depend on it for inputs that are harder to replace or for customers that are harder to find, or may occupy a more exposed position in the production network. The trade severance is common to the two countries. But the dependence it reveals need not be. That asymmetry is what \citeauthor{hirschman1945} took to be the source of power, since the country that would lose little can threaten the break more readily. We compare the two countries' own-size-normalized losses under the same bilateral
trade severance and collect
the signed comparisons in a matrix. The
resulting matrix maps how asymmetric dependence, and with it potential
political leverage, is distributed across the world economy.

We measure the power asymmetry of a pair, or simply its asymmetry, by the bounded signed
ratio
\begin{equation}
\mathcal H_{cc'}
\;:=\;
\frac{\gamma^{(c'\notnotleftrightarrow c)}_c-\gamma^{(c'\notnotleftrightarrow c)}_{c'}}{\gamma^{(c'\notnotleftrightarrow c)}_c+\gamma^{(c'\notnotleftrightarrow c)}_{c'}},
\label{eq:bounded-Hirschman}
\end{equation}
whenever the combined loss $\gamma^{(c'\notnotleftrightarrow c)}_c+\gamma^{(c'\notnotleftrightarrow c)}_{c'}$ is positive and $c\ne c'$. We set $\mathcal H_{cc'}=0$ otherwise, both on the diagonal $c=c'$ and at a zero-loss pair. By construction
$\mathcal H_{cc'}\in[-1,1]$ and $\mathcal H_{c'c}=-\mathcal H_{cc'}$. A positive value of $\mathcal H_{cc'}$ means that country $c$ is more vulnerable than country $c'$ in their bilateral relationship, so the relationship tilts against country $c$ and in favor of country $c'$. A negative value reverses the two roles, so that country $c'$ is the more vulnerable of the two and the relationship tilts in favor of country $c$. We call the more vulnerable of the two the weaker side of the relationship, and the other the favorable side, the more powerful of the two.
When the combined loss is positive, $(1+\mathcal H_{cc'})/2$
is country $c$'s share of the pair's combined proportional loss. The bounded signed asymmetries collect into what we call the Hirschman matrix
\begin{equation}
\boldsymbol{\mathcal H}
\;:=\;
\bigl[\mathcal H_{cc'}\bigr]_{c,c'=1}^{C}.
\label{eq:hirschman-matrix}
\end{equation}
Note that the Hirschman matrix is skew-symmetric, $\boldsymbol{\mathcal H}^{\top}=-\boldsymbol{\mathcal H}$, because exchanging the two countries in \eqref{eq:bounded-Hirschman} changes the sign of the numerator and leaves the denominator unchanged. This diagonal convention is the only one consistent with that reversal, since $\mathcal H_{cc}=-\mathcal H_{cc}$ forces $\mathcal H_{cc}=0$. The strict upper triangle determines the whole matrix. Averaging over its $C(C-1)/2$ entries, one for each unordered pair, gives us a scalar measure of asymmetry. 

\begin{definition}[Hirschman asymmetry index]
\label{def:hirschman-matrix}
The Hirschman asymmetry index is
\begin{equation}
\Psi(\boldsymbol{\mathcal H})
\;:=\;
\frac{2}{C(C-1)}
\sum_{1\le c<c'\le C}|\mathcal H_{cc'}|
\;\in\;[0,1].
\label{eq:H_asymmetry}
\end{equation}
\end{definition}

The index $\Psi$ is the average absolute bilateral asymmetry, and it measures how unevenly power is distributed within the world's trading relationships. The index equals zero exactly when every pair is balanced ($\mathcal H_{cc'}=0$ for all
$c<c'$, equivalently $\gamma^{(c'\notnotleftrightarrow c)}_c=\gamma^{(c'\notnotleftrightarrow c)}_{c'}$ on every pair), and it equals one exactly when
every pair is maximally asymmetric.

The index is built from the loss ratio of each pair, and the shape of that ratio governs how far any disturbance can move the asymmetries it averages. For positive vulnerabilities, we write the asymmetry as a function of their ratio. For a positive ratio $r$ the normalized difference is
\begin{equation}
\psi(r):=\frac{r-1}{r+1}\in(-1,1),
\qquad\text{so that}\qquad
\frac{x-y}{x+y}=\psi\!\left(\frac{x}{y}\right)\quad\text{for }x,y>0.
\label{eq:m-psi}
\end{equation}
The normalized difference increases strictly in $r$, vanishes at $r=1$, and changes sign when the ratio is inverted.\footnote{The normalized difference is half the growth-rate measure of \citet{davishaltiwangerschuh1996}, which divides the difference of two quantities by their mean rather than by their sum.} For a pair on which both trading partners lose, the loss ratio is
\begin{equation}
r_{cc'}
\;:=\;
\frac{\gamma^{(c'\notnotleftrightarrow c)}_c}{\gamma^{(c'\notnotleftrightarrow c)}_{c'}}.
\label{eq:m-logratio}
\end{equation}

\begin{lemma}[The asymmetry as a damped log loss ratio]
\label{lem:m-rigidity}
Fix a pair $\{c,c'\}$ whose two vulnerabilities are positive.
\begin{enumerate}
\item[\textnormal{(i)}] The asymmetry is the normalized difference of the loss ratio, $\mathcal H_{cc'}=\psi\bigl(r_{cc'}\bigr)$.
\item[\textnormal{(ii)}] Along any path on which both vulnerabilities stay positive, $d\mathcal H_{cc'}=\tfrac12\bigl(1-\mathcal H_{cc'}^{2}\bigr)\,d\log r_{cc'}$.
\item[\textnormal{(iii)}] Any perturbation carrying the pair to positive vulnerabilities satisfies
\begin{equation}
\bigl|\Delta\mathcal H_{cc'}\bigr|
\;\le\;
\tfrac12\bigl|\Delta\log r_{cc'}\bigr|
\;\le\;
\tfrac12\Bigl(\bigl|\Delta\log\gamma^{(c'\notnotleftrightarrow c)}_c\bigr|
+\bigl|\Delta\log\gamma^{(c'\notnotleftrightarrow c)}_{c'}\bigr|\Bigr),
\label{eq:m-rigidity}
\end{equation}
and a perturbation that scales both vulnerabilities by one common factor leaves $\mathcal H_{cc'}$ exactly where it was.
\end{enumerate}
\end{lemma}

\noindent Proof in Appendix~\ref{app:m-proof-rigidity}.

By Lemma~\ref{lem:m-rigidity}, the asymmetry cancels whatever the losses of the two trading partners share. A disturbance that raises both losses in the same proportion moves no asymmetry at all. The asymmetry responds only to the part of a disturbance that falls unevenly on the two trading partners. The lemma also shows that the local rate $\tfrac12(1-\mathcal H_{cc'}^{2})$ falls to zero as the asymmetry of a pair approaches plus or minus one, so the asymmetries that are already extreme are the hardest to move. A relationship at $\mathcal H_{cc'}=0.9$ moves by about a fifth of what an evenly balanced one would move under the same relative disturbance.

\subsection{Estimates of country vulnerabilities}
\label{subsec:vulnerability_estimates}


The vulnerabilities we report come from one benchmark network, the 2022 OECD
inter-country input--output table, with $C=80$ economies and $S=50$ industries,
giving $4{,}000$ country-sectors once we drop the rest-of-world
aggregate.\footnote{The matrix calculations run on the active set left after omitting
the $51$ zero-output country-sectors of Section~\ref{subsec:model_benchmark}, and the
benchmark identity \eqref{eq:m-benchmark} then holds by construction. Online
Appendix~\ref{app:m-empirics} records the rest of the processing, which consists of the
rest-of-world closure, the inventory treatment, and the capping of two buyer columns
at $w_j=0.98s_j$. Seventy of the active country-sectors, namely the industry of households as employers in the seventy economies where that industry is active, buy no intermediate inputs and sell only to the household of their own country. Since no trade severance reaches these seventy country-sectors, their output stays at its benchmark value in every equilibrium and counts in the size of their country. The model of Section~\ref{sec:model}, in which every producer buys intermediate inputs, describes the other $3{,}879$ of the $3{,}949$ active country-sectors.} The estimates form a single object, an $80\times80$ matrix of
own-size-normalized losses $\gamma^{(c'\notnotleftrightarrow c)}_c$.

For every unordered pair of countries we solve the benchmark-selected
equilibrium under the bilateral trade severance $c'\notnotleftrightarrow c$, which gives $3{,}160$ counterfactual equilibria. All $3{,}160$
converge to an interior point (Online Appendix~\ref{app:m-proofs} reports the convergence and interiority
diagnostics). The baseline calibration sets the unrecovered
share at $\tau=0.30$, both CES exponents at $\rho=\varrho=-1$, and the
barred-link attenuation at $\delta=0.10$. This CES baseline lies within the short-run complementarity range that the
disruption literature estimates for intermediate inputs. Table~\ref{tab:calibration} in Online
Appendix~\ref{app:m-tau} collects the baseline values and the evidence behind
each.\footnote{The vulnerabilities move with the parameters, climbing near-linearly in the unrecovered share $\tau$, scaling with the displaced share $1-\delta$, and moving only mildly in the inner CES exponent (Figure~\ref{fig:s4_sensitivity} in Online Appendix~\ref{app:m-empirics}), while the power asymmetries hardly move, as Lemma~\ref{lem:m-rigidity} leads one to expect. Over the thirteen-point parameter grid of Online Appendix~\ref{sec:robustness}, at least $99\%$ of the $3{,}160$ pairs keep their baseline asymmetry sign, the trade-weighted power ranking of the eighty economies rank-correlates with the baseline ranking at $0.999$ or above, and the asymmetry index $\Psi$ stays within $[0.581,0.584]$. The ordering of power also changes little when prices adjust. Online Appendix~\ref{app:flex} solves the same trade severances in a flexible-price economy calibrated to the same table, in which price adjustment relieves the more powerful side of a relationship by more than it relieves the less powerful side, so that the tilt in power between the two sides widens more often than it narrows.}

Table~\ref{tab:s4_top_bilateral} lists the twenty-five bilateral trade severances that
impose the largest joint loss on the two countries involved.
The largest is the United States--China severance, which would erase about $0.5\%$ of world gross economic
activity. China loses $1.4\%$ of its economic activity against $0.6\%$ for
the United States, an asymmetry $\mathcal H_{cc'}$ of $0.42$ on a scale bounded
by one, and Table~\ref{tab:s4_headline} attributes nearly two fifths of the
Chinese loss to its severed final sales. Canada--United States ranks second at
about $0.4\%$, with the proportional burden falling almost entirely on Canada.
Canada loses $9.3\%$ of its economic activity against $1.0\%$ for the United
States, an asymmetry of $0.80$.
Joint loss and asymmetry are distinct quantities that happen to
move together here. The largest trade severances arise where a very large economy meets
a very dependent one, so all but four of these twenty-five pairs have a loss ratio above three. Korea loses $5.6\%$ of its economic activity to a trade severance
with China against China's $0.9\%$, and Mexico loses $10.9\%$ against the United States'
$0.6\%$. The four exceptions all join two of the largest economies.

\begin{table}[!htbp]
\centering
\footnotesize
\begin{adjustbox}{max width=\textwidth}
\begin{tabular}{lrrrr}
\toprule
Bilateral severance $c\,\notnotleftrightarrow\,c'$ & $\gamma^{(c\notnotleftrightarrow c')}_c$ (\%) & $\gamma^{(c\notnotleftrightarrow c')}_{c'}$ (\%) & $\mathcal H_{cc'}$ & Joint loss (\% of world) \\
\midrule
China--United States & 1.37 & 0.56 & 0.42 & 0.45 \\
Canada--United States & 9.30 & 1.01 & 0.80 & 0.42 \\
China--Korea & 0.86 & 5.55 & -0.73 & 0.33 \\
China--Japan & 0.89 & 2.57 & -0.49 & 0.32 \\
Mexico--United States & 10.88 & 0.62 & 0.89 & 0.29 \\
Australia--China & 3.28 & 0.75 & 0.63 & 0.23 \\
China--Chinese Taipei & 0.63 & 7.43 & -0.84 & 0.21 \\
China--Vietnam & 0.44 & 11.49 & -0.93 & 0.20 \\
China--Germany & 0.56 & 1.51 & -0.46 & 0.19 \\
Brazil--China & 2.76 & 0.49 & 0.70 & 0.17 \\
China--Russia & 0.49 & 1.85 & -0.58 & 0.15 \\
Germany--United States & 1.86 & 0.26 & 0.75 & 0.14 \\
Japan--United States & 1.91 & 0.24 & 0.77 & 0.14 \\
China--Indonesia & 0.41 & 2.82 & -0.74 & 0.13 \\
United Kingdom--United States & 2.41 & 0.26 & 0.81 & 0.13 \\
China--Saudi Arabia & 0.44 & 2.27 & -0.67 & 0.12 \\
China--Singapore & 0.36 & 5.83 & -0.88 & 0.12 \\
Korea--United States & 3.26 & 0.20 & 0.89 & 0.12 \\
Germany--France & 1.36 & 2.04 & -0.20 & 0.11 \\
United Arab Emirates--China & 4.29 & 0.38 & 0.84 & 0.11 \\
India--United States & 1.76 & 0.21 & 0.79 & 0.11 \\
Ireland--United States & 7.96 & 0.24 & 0.94 & 0.10 \\
China--Malaysia & 0.27 & 5.75 & -0.91 & 0.09 \\
China--India & 0.23 & 1.18 & -0.67 & 0.09 \\
China--Thailand & 0.26 & 5.13 & -0.90 & 0.09 \\
\bottomrule
\end{tabular}
\end{adjustbox}
\caption{The twenty-five largest bilateral trade severances at the baseline, ranked by the joint loss, the sum of the two trading partners' absolute losses in percent of world gross economic activity. The two trading partners' losses are in percent of own benchmark activity, and $\mathcal H_{cc'}$ is the asymmetry \eqref{eq:bounded-Hirschman}, positive when country $c$ is the weaker side. Baseline calibration $\tau=0.30$, $\rho=-1$, $\varrho=-1$, $\delta=0.10$.}
\label{tab:s4_top_bilateral}
\end{table}

Table~\ref{tab:s4_headline} reports the headline pairs together with the anatomy of Proposition~\ref{prop:m-anatomy}. The last two columns of the table give the share of each trading partner's loss that runs through the lost-market channel, and within it the part attributable to severed final sales at the solved equilibrium. In the first row three fifths of the Chinese loss runs through the lost-market channel, and severed final sales alone account for $38\%$ of the loss, so China loses through its customers, above all the American households that buy its consumer goods. The American loss runs mostly through the productivity channel, the input bundles its producers can no longer complete, and only a tenth of it through severed final sales. The same division recurs wherever a country that exports final goods meets a large buyer of them. Mexico against the United States and China against India lose about a third of their loss through severed final sales. India loses three quarters of its loss through its input bundles, and the United States loses half of its loss against Mexico through its input bundles. Chinese Taipei against China is the intermediate-goods case. Two thirds of its loss is lost market, and severed final sales account for only $16$ points of it, since what it sells to China is inputs.
\begin{table}[!htbp]
\centering
\footnotesize
\begin{adjustbox}{max width=\textwidth}
\begin{tabular}{lrrrrr}
\toprule
Pair $c$ / $c'$ & $\gamma^{(c\notnotleftrightarrow c')}_c$ & $\gamma^{(c\notnotleftrightarrow c')}_{c'}$ & $\mathcal H_{cc'}$ & Lost-market share, $c$ (final) & Lost-market share, $c'$ (final) \\
\midrule
United States / China & 0.56 & 1.37 & $-0.42$ & 38 (10) & 60 (38) \\
China / Chinese Taipei & 0.63 & 7.43 & $-0.84$ & 18 (6) & 68 (16) \\
United States / Russia & 0.05 & 0.62 & $-0.84$ & 35 (15) & 53 (15) \\
Germany / Russia & 0.57 & 0.95 & $-0.25$ & 30 (14) & 56 (15) \\
Japan / China & 2.57 & 0.89 & $+0.49$ & 47 (12) & 40 (19) \\
India / China & 1.18 & 0.23 & $+0.67$ & 24 (12) & 69 (30) \\
United States / Mexico & 0.62 & 10.88 & $-0.89$ & 50 (21) & 57 (35) \\
Germany / China & 1.51 & 0.56 & $+0.46$ & 50 (17) & 46 (24) \\
\bottomrule
\end{tabular}
\end{adjustbox}
\caption{Losses in percent of own benchmark activity under the bilateral trade severance, the asymmetry $\mathcal H_{cc'}$ of \eqref{eq:bounded-Hirschman}, positive when country $c$ is the weaker side, and the share of each trading partner's loss that runs through the lost-market channel of Proposition~\ref{prop:m-anatomy}, with the part due to severed final sales in parentheses. Baseline calibration $\tau=0.30$, $\rho=-1$, $\varrho=-1$, $\delta=0.10$.}
\label{tab:s4_headline}
\end{table}

Table~\ref{tab:s4_most_exposed} turns from relationships to countries, asking
for each economy the worst that a single partner could do to it. The results
are stark for small, specialized economies. Belarus heads the list, facing a
$14.2\%$ loss of gross economic activity from a trade severance with Russia, and Vietnam comes
next at $11.5\%$ against China. Mexico and Canada, at $10.9\%$ and $9.3\%$
against the United States, come third and fourth. Severed final sales account for
$35\%$ of the Mexican loss and $17\%$ of the Canadian loss, since much of what
each sells across the border goes to American households rather than to
American factories. The Democratic Republic of the
Congo comes fifth. A trade severance with China, which buys its copper and cobalt and
supplies much of what its mining sector uses, would cost it close to $9\%$ of gross economic activity. China is the worst-case partner for eleven of the
twenty-five, and Germany and the United States account for most of the
remainder. The worst-case losses of the twenty-five most exposed economies run
from $4.7\%$ to $14.2\%$ of economic activity, against $1.0\%$ for the
United States and $1.4\%$ for China against their own worst-case partners. 

\begin{table}[!htbp]
\centering
\small
\begin{adjustbox}{max width=\textwidth}
\begin{tabular}{lrrr}
\toprule
Country & $\max_{c'} \gamma^{(c\notnotleftrightarrow c')}_c$ (\%) & Worst-case partner & Mean $\gamma^{(c\notnotleftrightarrow c')}_c$ (\%) \\
\midrule
Belarus & 14.2 & Russia & 0.3 \\
Vietnam & 11.5 & China & 0.5 \\
Mexico & 10.9 & United States & 0.2 \\
Canada & 9.3 & United States & 0.2 \\
Democratic Republic of the Congo & 8.9 & China & 0.2 \\
Luxembourg & 8.6 & Germany & 0.5 \\
Ireland & 8.0 & United States & 0.4 \\
Laos & 7.6 & Thailand & 0.3 \\
Cambodia & 7.6 & China & 0.3 \\
Chinese Taipei & 7.4 & China & 0.3 \\
Austria & 6.5 & Germany & 0.3 \\
Hungary & 6.2 & Germany & 0.4 \\
Czechia & 6.1 & Germany & 0.3 \\
Malta & 6.0 & United Kingdom & 0.4 \\
Singapore & 5.8 & China & 0.5 \\
Malaysia & 5.7 & China & 0.3 \\
Korea & 5.6 & China & 0.3 \\
Costa Rica & 5.3 & United States & 0.2 \\
Thailand & 5.1 & China & 0.3 \\
Portugal & 5.1 & Spain & 0.3 \\
Angola & 5.0 & China & 0.2 \\
Poland & 4.9 & Germany & 0.3 \\
Slovakia & 4.8 & Germany & 0.4 \\
Chile & 4.8 & China & 0.2 \\
Hong Kong, China & 4.7 & China & 0.2 \\
\bottomrule
\end{tabular}
\end{adjustbox}
\caption{The twenty-five economies with the largest worst-case loss. The mean column averages over partners with positive bilateral trade only. Baseline calibration $\tau=0.30$, $\rho=-1$, $\varrho=-1$, $\delta=0.10$.}
\label{tab:s4_most_exposed}
\end{table}

\begin{figure}[!htbp]
\centering
\includegraphics[width=0.62\linewidth]{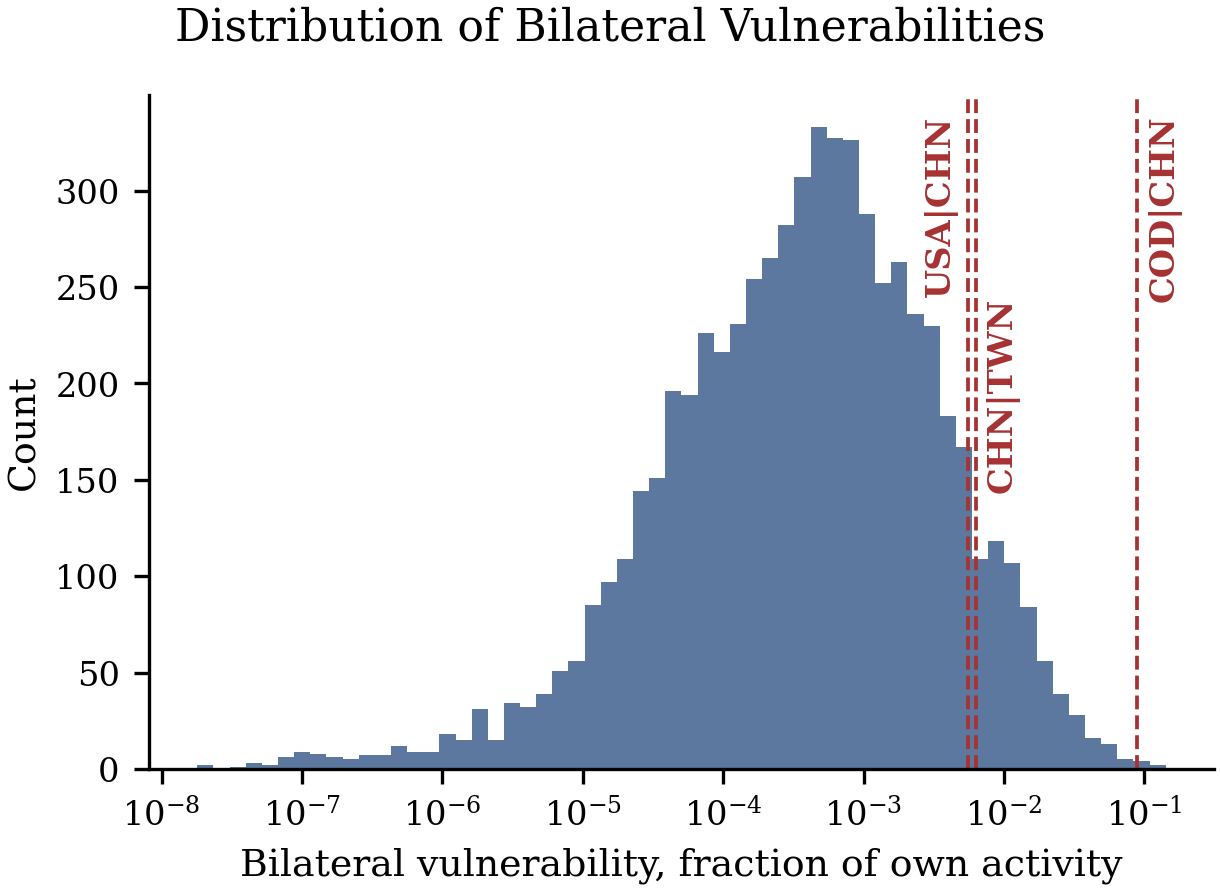}
\caption{The distribution of the $6{,}320$ bilateral vulnerabilities $\gamma^{(c\notnotleftrightarrow c')}_c$, one for each of the two trading partners of each of the $3{,}160$ trade severances, on a logarithmic scale. Dashed lines mark selected pairs. Baseline calibration $\tau=0.30$, $\rho=-1$, $\varrho=-1$, $\delta=0.10$.}
\label{fig:s4_distribution}
\end{figure}

Figure~\ref{fig:s4_distribution} plots on a logarithmic scale the $6{,}320$ bilateral vulnerabilities the $3{,}160$ trade severances produce, one for each of the two trading partners of each severance. The distribution spans nearly seven orders of magnitude, from little more than a millionth of a percent to the $14.2\%$ Belarus faces against Russia. A single world network, cut one
relationship at a time, produces losses running from the negligible to the
catastrophic.

\begin{figure}[!htbp]
\centering
\includegraphics[width=0.9\linewidth]{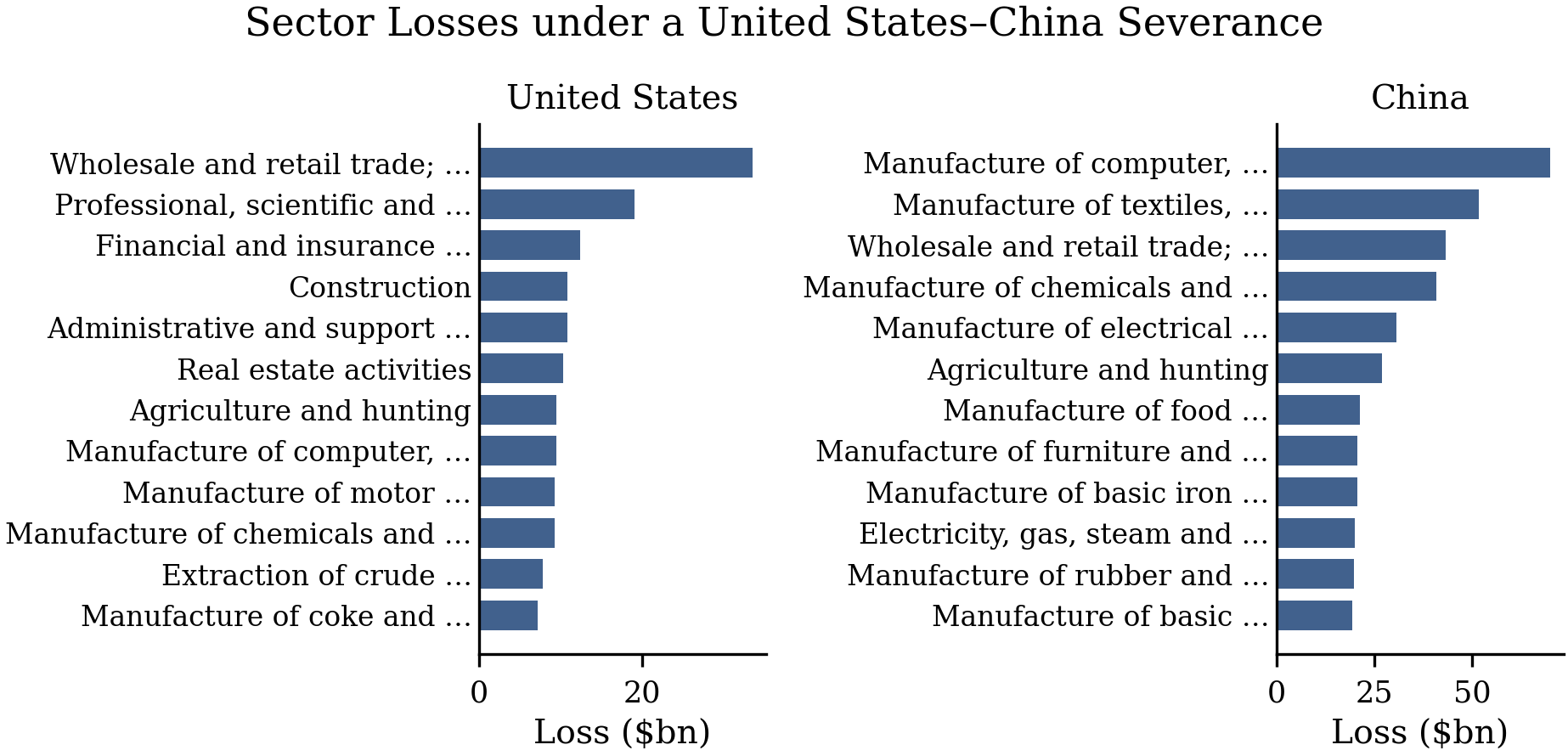}
\caption{Each country's losses by sector under the United States--China trade severance, in billions of 2022 US dollars.}
\label{fig:s4_usa_chn}
\end{figure}

Within an economy, the activity losses from a trade severance concentrate in particular sectors. Figure~\ref{fig:s4_usa_chn} decomposes the United States--China trade severance by sector for each of the two countries. China's largest losses fall on its export sectors, computers and electronics first, textiles and apparel second, with wholesale and retail trade and chemicals close behind. Computers and electronics are an input to producers everywhere and textiles and apparel are sold to
households, and both lose their largest foreign market without being able to replace all of it. The American losses have a different sectoral composition. They fall on wholesale and retail trade, professional services, and finance, the sectors that move, broker, and finance the flow across the Pacific rather than the factories at either end of it.

\subsection{The Hirschman matrix of bilateral power}
\label{subsec:hirschman_matrix}

A bilateral trade severance costs the two countries different fractions of their own
activity, and the Hirschman matrix $\boldsymbol{\mathcal H}$ collects those
differences over the pairs of
Section~\ref{subsec:vulnerability_estimates}. The United States is the more powerful side of every relationship it enters, and China in all but one. The middle powers hold the favorable side against most partners and often the weaker side against their largest partners, and the advantage declines steadily toward the smaller open economies. We compute every object at the baseline parameters over the comparison set, the $3{,}160$ country pairs carrying positive bilateral trade and positive bilateral loss.

Averaged over that set, the absolute asymmetry index $\Psi(\boldsymbol{\mathcal H})$ of Definition~\ref{def:hirschman-matrix} is $0.58$, well above the midpoint of the index's unit interval. An asymmetry of $0.58$ is a loss ratio of nearly four, the ratio $r$ at which $\psi(r)=0.58$, and the median pair has a loss ratio of $4.4$ (Online Appendix~\ref{oa:asymmetry-distribution}). Shares read more
easily than ratios, and $(1+|\mathcal H_{cc'}|)/2$ is the fraction of a pair's
combined loss borne by its weaker member. On that scale the median relationship in the world economy puts $81\%$ of the damage on one of its two members. The share is $93\%$ at the upper quartile and $98\%$ at the top decile, and even at the first quartile it is $67\%$. Balanced dependence is rare, and only one pair in seven keeps the weaker member's share at or below $60\%$, a loss ratio of three to two. The same asymmetry makes the distribution of power hard to disturb, since by Lemma~\ref{lem:m-rigidity}\textnormal{(ii)} an asymmetric pair moves less than a balanced one under the same relative disturbance.\footnote{Across the $3{,}160$ pairs the local rate $\tfrac12(1-\mathcal H_{cc'}^{2})$ of Lemma~\ref{lem:m-rigidity}\textnormal{(ii)} has a median of $0.30$ and quartiles of $0.14$ and $0.44$, against one half for a balanced pair, and the rate lies below a tenth on nearly a fifth of the pairs.}

\begin{figure}[!htbp]
\centering
\scalebox{0.8}{\includegraphics[width=0.82\linewidth]{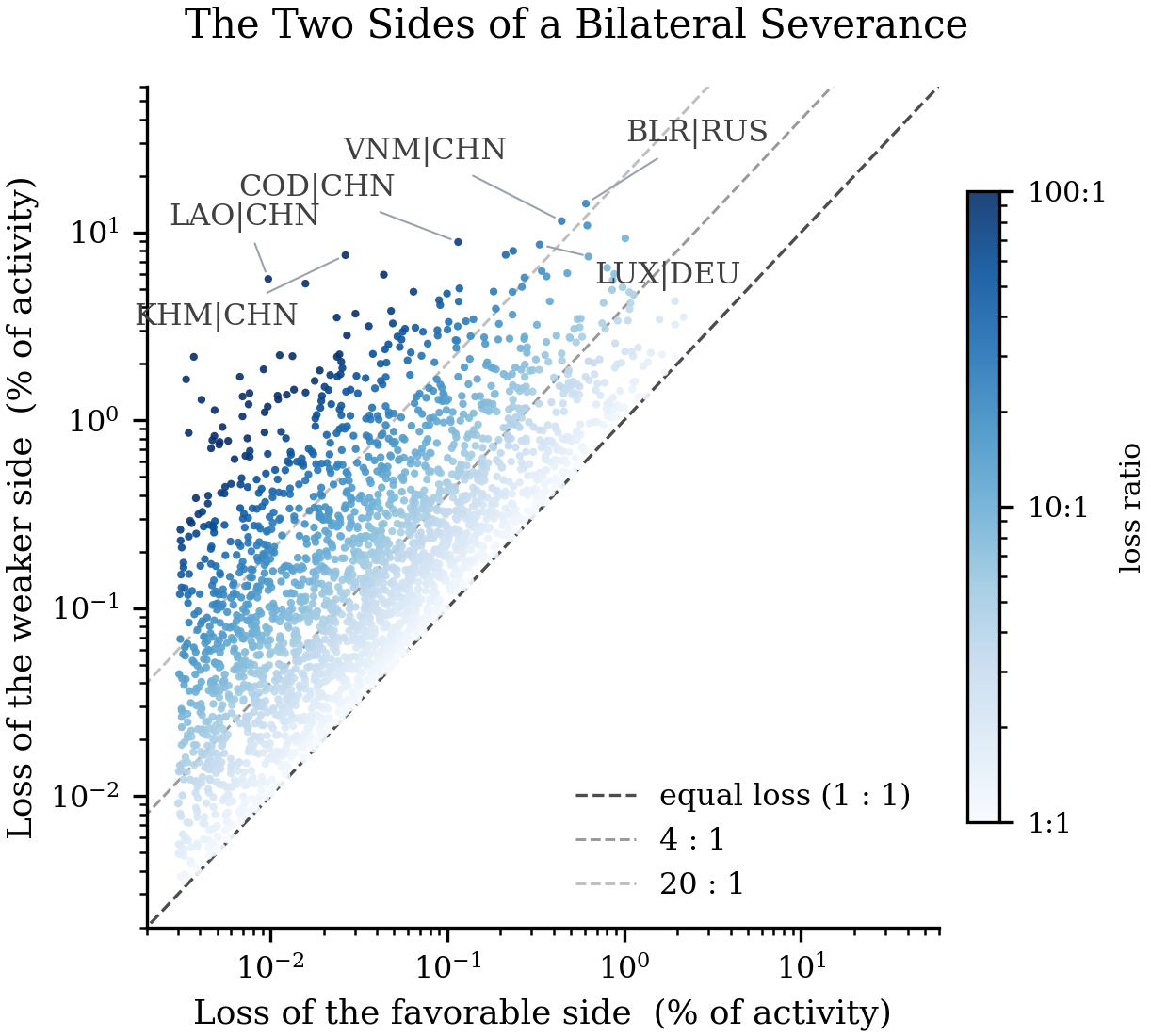}}
\caption{Each trading partner's loss as a percentage of economic activity, on log scales, the smaller loss on the horizontal axis. The guides mark loss ratios of four and twenty, and color is the loss ratio. Plotted are the $2{,}507$ pairs on which both trading partners lose more than $0.003\%$ of activity, whose median loss ratio is $3.7$ against $4.4$ over all $3{,}160$ pairs.}
\label{fig:s6_not_balanced}
\end{figure}

Figure~\ref{fig:s6_not_balanced} plots the two trading partners of each trade severance against
each other, the larger of the two proportional losses against the smaller, over the pairs on which both trading partners lose more than a rounding error. The points lie far above the diagonal of equal loss, and the most one-sided
trade severances have loss ratios above twenty. The pairs furthest from the diagonal are small economies
bound to a single large partner. A trade severance with China would cost Vietnam, Cambodia, Laos, or the Democratic Republic of the Congo several percent of their economic activity and would cost China almost nothing. 

Table~\ref{tab:s6_power_ranking} summarizes each country's row of the matrix by
its mean bilateral advantage, both unweighted and weighted by bilateral trade. The
United States heads both columns, more powerful in all
seventy-nine relationships at a trade-weighted mean advantage of $0.82$, and
China follows at $0.59$, more powerful in seventy-eight. Germany is some distance behind at
$0.34$, and the next five economies lie in a band from $0.06$ to $0.17$. 

The unweighted and weighted columns diverge for the middle powers. Japan's unweighted advantage of $0.69$ falls to $0.06$ once we
weight by trade. Japan is more powerful than the partners it trades little with, and
its two largest relationships, with China and the United States, place it on the
weaker side.

The bottom of the table holds the most asymmetrically attached economies, which are not always the smallest. They include Vietnam,
Mexico, Chinese Taipei, and Canada, all large traders whose trade runs
overwhelmingly through a single dominant partner.

\begin{table}[!htbp]
\centering
\small
\begin{adjustbox}{max width=\textwidth}
\begin{tabular}{lrrrr}
\toprule
Country & Mean advantage $\overline{-\mathcal H_{c\cdot}}$ & Trade-weighted & Favorable pairs & Mean $\gamma^{(c\notnotleftrightarrow c')}_c$ (\%) \\
\midrule
United States & 0.94 & 0.82 & 79/79 & 0.07 \\
China & 0.87 & 0.59 & 78/79 & 0.16 \\
Germany & 0.72 & 0.34 & 77/79 & 0.25 \\
France & 0.63 & 0.17 & 73/79 & 0.21 \\
United Kingdom & 0.67 & 0.13 & 74/79 & 0.19 \\
Russia & 0.61 & 0.13 & 72/79 & 0.14 \\
India & 0.66 & 0.11 & 73/79 & 0.16 \\
Japan & 0.69 & 0.06 & 76/79 & 0.17 \\
Italy & 0.54 & 0.03 & 68/79 & 0.21 \\
Spain & 0.46 & -0.08 & 64/79 & 0.21 \\
Australia & 0.55 & -0.11 & 70/79 & 0.14 \\
Saudi Arabia & 0.47 & -0.16 & 67/79 & 0.16 \\
Netherlands & 0.38 & -0.19 & 61/79 & 0.29 \\
Türkiye & 0.36 & -0.22 & 61/79 & 0.18 \\
Korea & 0.51 & -0.23 & 66/79 & 0.29 \\
Indonesia & 0.43 & -0.23 & 63/79 & 0.15 \\
Brazil & 0.48 & -0.25 & 66/79 & 0.15 \\
Poland & 0.25 & -0.27 & 54/79 & 0.31 \\
Belgium & 0.24 & -0.34 & 54/79 & 0.35 \\
Switzerland & 0.26 & -0.40 & 54/79 & 0.30 \\
Singapore & 0.27 & -0.41 & 57/79 & 0.46 \\
Canada & 0.60 & -0.45 & 72/79 & 0.19 \\
Chinese Taipei & 0.30 & -0.50 & 60/79 & 0.33 \\
Mexico & 0.54 & -0.57 & 69/79 & 0.23 \\
Vietnam & 0.06 & -0.67 & 41/79 & 0.49 \\
\bottomrule
\end{tabular}
\end{adjustbox}
\caption{The twenty-five largest economies by benchmark gross activity, ordered by the trade-weighted advantage. Advantage is the negated asymmetry $-\mathcal H_{c\cdot}$, so a positive
entry is the favorable side. The last column is the country's mean own
vulnerability across its partners. Baseline calibration $\tau=0.30$, $\rho=-1$, $\varrho=-1$, $\delta=0.10$.}
\label{tab:s6_power_ranking}
\end{table}

\section{Power and Economic Size}
\label{sec:power_size}

We have so far established that the distribution of power across the trading nations of the world is highly unequal, and that the two largest economies are more powerful than everyone else. It is natural to read these two facts as saying that power is size. We now ask how much of that reading is supported by the data when sieved through our model. Recall that power compares what each trading partner loses between two fixed points, i.e., the pre- and post-sanction equilibria, as a proportion of their respective pre-sanction levels of economic activity. Since a lost unit of economic activity is a smaller fraction for a larger economy, the normalization to the pre-sanction level of economic activity by itself causes power to increase with size. We call this the `base effect'. What is more intriguing is how size influences power by shaping the absolute difference between the pre- and post-sanction equilibrium levels of economic activity. This difference depends on who buys from whom through the production linkages behind each trading partner's trade. We therefore call the ratio of the two trading partners' dollar losses the `linkage effect'. It is the part of power that can, in principle, run orthogonal to size. This is because nothing in the model lets the sizes of economies pin down the trade network, i.e., many different patterns of trade flows are capable of generating the same sizes (Lemma~\ref{lem:m-fiber} in Appendix~\ref{app:m-proof-fiber}).\footnote{The 2022 trade table has about $9.8$ million positive intermediate trade flows and only about $7{,}900$ totals for them to respect, one for each active country-sector as a seller and one as a buyer. The flows can therefore be rearranged in nearly as many independent ways as there are flows, and every such rearrangement of who buys from whom keeps every size where it is.} And these varied patterns of trade need not be neutral to the effect of a sanction on an economy, let alone to the relative effects of a sanction on two economies. A sanction bars particular trade flows, and two networks that generate the same sizes can place very different flows on the barred links.

In what follows, we first develop the analytical machinery that relates size to power within our model. In Section~\ref{subsec:power_size}, we decompose each trading partner's vulnerability into the sum of import and export elasticities with respect to a partner and a second-order interaction of the two flows. What matters for these elasticities is not merely how much a country trades with its partner but how much of its home production depends on that trade, through the backward linkages behind its exports and the forward linkages from its imports. We then use this machinery, in Section~\ref{subsec:power_size_base}, to measure the base effect and the linkage effect on the 2022 trade table. In Section~\ref{subsec:comparable_size}, we ask how far size remains the measure of power among economies of comparable size. Throughout this section and the next the maintained assumptions of Section~\ref{sec:model} are in force, the unrecovered share $\tau$ is positive, the selected equilibrium of every severance named is interior, and every pair named trades in both directions.

\subsection{A decomposition of trade vulnerability}
\label{subsec:power_size}

Recall that a trade severance between country $c$ and a partner $c'$ bars two flows, namely what country $c$ sells to the partner and what it buys from the partner. Write $X_{cc'}$ for the severed sales of country $c$ and $M_{cc'}$ for its severed purchases. The lost sales of country $c$ sum to $\tau(1-\delta)X_{cc'}$, the unrecovered part of its severed sales. And the lost sales of the partner's export sectors, which are the lost purchases of country $c$, sum to $\tau(1-\delta)M_{cc'}$ (Lemma~\ref{lem:m-injuries} in Appendix~\ref{app:m-content}). Write $L_c:=Q(c)-Q_\omega(c)$ for the dollar loss of country $c$, $L^{\downarrow}_c$ for its export loss, namely what it loses through its lost sales, and $L^{\uparrow}_c$ for its import loss, namely what it loses through its lost purchases. It turns out that $L^{\downarrow}_c$ and $L^{\uparrow}_c$ can each be neatly written as the sum of the two channels of Proposition~\ref{prop:m-anatomy}, namely a lost-market channel, which collects the lost sales of export sectors, and a productivity channel, which collects the production shortfalls of producers left short of inputs,
\begin{equation}
\begin{aligned}
L^{\downarrow}_c&=\underbrace{\sum_{i\in c} m^{*}_{c,i}\,h_{\omega,i}\,\xi_{\omega,i}}_{\text{lost-market channel}}+\underbrace{\sum_{i} m^{*}_{c,i}\,\bigl(1-h^{(c')}_{\omega,i}\bigr)\,s_i}_{\text{productivity channel}},\\
L^{\uparrow}_c&=\underbrace{\sum_{i\in c'} m^{*}_{c,i}\,h_{\omega,i}\,\xi_{\omega,i}}_{\text{lost-market channel}}+\underbrace{\sum_{i} m^{*}_{c,i}\,\bigl(1-h^{(c)}_{\omega,i}\bigr)\,s_i}_{\text{productivity channel}},
\end{aligned}
\label{eq:m-two-parts}
\end{equation}
where $\xi_{\omega,i}$ is the lost sales of country-sector $i$ of \eqref{eq:m-lost-sales}, counted at its fulfillment rate $h_{\omega,i}$ (Appendix~\ref{app:m-content}), and $m^{*}_{c,i}$ is the content of country $c$ in a dollar of country-sector $i$'s output.\footnote{The content of country $c$ in a dollar of a producer's output is the activity that this dollar generates in country $c$, which is also what country $c$ loses when the dollar is lost. To make the dollar, the producer orders inputs, the suppliers of these inputs order inputs in turn, and so on up the chain. Every producer places its orders in the proportions of the effective demand ($\mathbf A_\omega$), and every supplier fills them at its equilibrium fulfillment rate ($\mathbf h_\omega$). The content adds up what the producers of country $c$ supply along this chain of orders, together with the dollar itself when the producer belongs to country $c$. A dollar of Mexican car output, for instance, generates in Mexico the dollar itself together with the Mexican parts behind it, while in the United States it generates only the American parts behind it. Since the content counts the car, its parts, and the inputs behind the parts each at full value, it is not a share of the dollar of output. It can exceed one, and it is at least one for the producers of country $c$ itself (Lemma~\ref{lem:m-content} in Appendix~\ref{app:m-content}).} 

In the productivity channel, $\mathbf h^{(c)}_\omega$ are the fulfillment rates that would arise if only the buyer sectors of country $c$ had lost their inputs, and $\mathbf h^{(c')}_\omega$ are the fulfillment rates that would arise if only the buyer sectors of the partner had lost theirs. We call them the one-sided fulfillment rates (Lemma~\ref{lem:m-onesided} in Appendix~\ref{app:m-content}), and we obtain each by solving the fulfillment fixed point once more at the realized effective demand ($\mathbf A_\omega$). Since the lost sales of country $c$ starve the buyer sectors of country $c'$, the production shortfall that these lost sales cause on their own at country-sector $i$ is $(1-h^{(c')}_{\omega,i})s_i$. And since the lost purchases of country $c$ starve its own buyer sectors, the production shortfall that they cause on their own is $(1-h^{(c)}_{\omega,i})s_i$. The productivity channels run over producer sectors of all countries because a production shortfall percolates through the world trade network.

It is worth noting that when the lost sales of country $c$ starve the buyer sectors of country $c'$, the production shortfalls travel through the world trade network and can come back to hurt producers of country $c$ through a decrease in the availability of their inputs. A loss of the `import' kind, namely a production shortfall for want of inputs, can therefore be triggered by the cessation of exports. Such a production shortfall is counted inside the export loss $L^{\downarrow}_c$, by the terms $m^{*}_{c,i}(1-h^{(c')}_{\omega,i})s_i$ of its productivity channel in which country-sector $i$ belongs to country $c$. Since the one-sided fulfillment rates $\mathbf h^{(c')}_\omega$ solve the fulfillment fixed point of the whole world trade network, they record a production shortfall at every producer that the shortfalls of the buyer sectors of country $c'$ reach, in country $c$ as in every other country. The same holds for the lost purchases of country $c$ and its import loss $L^{\uparrow}_c$. When the export sectors of country $c'$ lose their sales to country $c$, they order fewer inputs, and the producers of country $c$ lose whatever part of these orders they filled, directly or through other suppliers. The cessation of imports can therefore trigger a loss of the `export' kind, namely a fall in orders, and the lost-market channel of the import loss counts this fall through the content $m^{*}_{c,i}$ of country $c$ in the output of the export sectors of country $c'$.

Write $\ell^{\downarrow}_c:=L^{\downarrow}_c/[\tau(1-\delta)X_{cc'}]$ for the export multiplier of country $c$, the activity it loses per dollar of lost sales, and $\ell^{\uparrow}_c:=L^{\uparrow}_c/[\tau(1-\delta)M_{cc'}]$ for its import multiplier, the activity it loses per dollar of lost purchases. Suppose, for example, that country $c$ sells cars to its partner and buys engines from it. A lost dollar of car sales idles the car plants of country $c$ and, upstream, the domestic suppliers of their steel, glass, and seats, and the export multiplier of country $c$ counts this lost activity per lost dollar. A lost dollar of engine purchases idles the same car plants for want of an input they cannot fully replace in the short run, and with them the same domestic suppliers and the domestic customers the plants no longer serve in full, so the import multiplier of country $c$ exceeds its export multiplier. Had country $c$ bought finished cars from its partner instead, a lost dollar of purchases would idle no production at home, and its import multiplier would be small. More generally, a producer's lost intermediate purchase costs its country at least as much per unrecovered dollar as its lost sale costs it through the lost-market channel.\footnote{Per unrecovered dollar, a producer's lost sale costs its country, through the lost-market channel, at most the content of the country in that producer's output, while its lost intermediate purchase degrades its input bundle and costs at least that content (Lemma~\ref{lem:m-ordering} in Appendix~\ref{app:m-ordering}). Both costs differ across countries, and Online Appendix~\ref{oa:diversification} compares two countries alike in all but their trade with a partner. The comparison shows that diversifying purchases lowers both what a country buys from the partner and the cost of an uneven shortfall across inputs, while diversifying sales lowers only the orders at risk.}

The dollar loss $L_c$ of country $c$ is in general not the sum of its export loss $L^{\downarrow}_c$ and its import loss $L^{\uparrow}_c$. The complementarity between inputs in production means that a producer facing two input shortages at once does not lose the sum of what each shortage would cost it on its own. The reduction in a producer's output due to the shortage of one input depends on how much of its other inputs is available. And insofar as the cessation of exports can interact with the cessation of imports in generating shortages of multiple inputs at the same producer, their interaction matters for the dollar loss of country $c$. We accordingly write the dollar loss as $L_c=L^{\downarrow}_c+L^{\uparrow}_c+L^{\times}_c$, and call the remainder $L^{\times}_c$ the interaction of the two flows.\footnote{A producer short of one input on both counts loses more than the sum of what it would lose on each count alone, for the loss from a lost input grows faster than the input lost. A producer short of two different inputs loses less than the sum when $1-\rho\ge\beta_j(1-\varrho)$, as when the two exponents are equal, for each shortage then leaves less output for the other to idle. When the inequality fails, as it does when the exponent $\varrho$ of the outer nest lies far enough below the exponent $\rho$ of the intermediate inputs, sufficiently small shortages of two different inputs cost the producer more than the sum as well. The interaction therefore has no general sign.} The interaction is of second order in the input shortages that the two flows plant at the buyer sectors of the two countries, while the two productivity channels of \eqref{eq:m-two-parts} are of first order (Lemma~\ref{lem:m-interaction} in Appendix~\ref{app:m-content}). Since the interaction belongs to the two flows jointly, we measure it per dollar of the pair's lost trade, and write $\ell^{\times}_c:=L^{\times}_c/[\tau(1-\delta)(X_{cc'}+M_{cc'})]$ for the interaction rate of country $c$. Weighting each multiplier by the share of country $c$'s activity that its severed flow accounts for, and the interaction rate by the share that the pair's whole severed trade accounts for, gives Proposition~\ref{prop:m-exposure}.

\begin{proposition}[Vulnerability as the sum of two elasticities and an interaction]
\label{prop:m-exposure}
Suppose that the unrecovered share $\tau$ is positive and that countries $c$ and $c'$ trade in both directions. The vulnerability of country $c$ under the trade severance $c'\notnotleftrightarrow c$ is
\begin{equation}
\gamma^{(c'\notnotleftrightarrow c)}_c
=\tau(1-\delta)\Bigl[\ell^{\downarrow}_c\,\frac{X_{cc'}}{Q(c)}+\ell^{\uparrow}_c\,\frac{M_{cc'}}{Q(c)}+\ell^{\times}_c\,\frac{X_{cc'}+M_{cc'}}{Q(c)}\Bigr].
\label{eq:m-exact-exposure}
\end{equation}
The vulnerability of country $c'$ is given by the same equation with the multipliers, the interaction rate, and the size of country $c'$, its severed sales being $M_{cc'}$ and its severed purchases $X_{cc'}$. If the export multiplier and the import multiplier of country $c'$, each with its interaction rate added, are positive, the ratio of the dollar loss $L_c$ of country $c$ to the dollar loss $L_{c'}$ of country $c'$ satisfies
\begin{equation}
\min\Bigl\{\frac{\ell^{\downarrow}_c+\ell^{\times}_c}{\ell^{\uparrow}_{c'}+\ell^{\times}_{c'}},\frac{\ell^{\uparrow}_c+\ell^{\times}_c}{\ell^{\downarrow}_{c'}+\ell^{\times}_{c'}}\Bigr\}
\le\frac{L_c}{L_{c'}}
\le\max\Bigl\{\frac{\ell^{\downarrow}_c+\ell^{\times}_c}{\ell^{\uparrow}_{c'}+\ell^{\times}_{c'}},\frac{\ell^{\uparrow}_c+\ell^{\times}_c}{\ell^{\downarrow}_{c'}+\ell^{\times}_{c'}}\Bigr\}.
\label{eq:m-size-bound}
\end{equation}
\end{proposition}

\noindent Proof in Appendix~\ref{app:m-proof-exposure}.

Proposition~\ref{prop:m-exposure} tells us that the vulnerability of a country to a trading partner is the share of its activity that it trades with that partner, weighted flow by flow by what a lost dollar of that trade costs it. More specifically, the bracket of \eqref{eq:m-exact-exposure} sums the export elasticity and the import elasticity of country $c$ and the interaction term. The export elasticity is the elasticity of country $c$'s activity with respect to its exports to the partner, namely the proportional loss of activity through that flow per proportional loss of the flow, and the import elasticity is the same elasticity with respect to its imports from the partner. The trade severance sets the proportional loss of both flows at $\tau(1-\delta)$. Each elasticity is the product of a multiplier and a trade share, with the export share $X_{cc'}/Q(c)$ being the fraction of country $c$'s activity that it sells to the partner, and the import share $M_{cc'}/Q(c)$ being what it buys from the partner as a fraction of the same activity.

Proposition~\ref{prop:m-exposure} also tells us that power can depart from size only as far as a lost dollar of trade costs one trading partner more than it costs the other. To see how, recall that the vulnerability of each trading partner is its dollar loss divided by its own activity, so that the loss ratio $r_{cc'}$ of \eqref{eq:m-logratio} is the product of the base effect, which is the pair's size ratio, and the linkage effect, which is the ratio of the two dollar losses. Country $c$ loses the flow $X_{cc'}$ as sales and country $c'$ loses it as purchases, and the two countries lose the flow $M_{cc'}$ with the roles exchanged. By Proposition~\ref{prop:m-exposure} we can therefore write the loss ratio as
\begin{equation}
r_{cc'}
=\underbrace{\frac{Q(c')}{Q(c)}}_{\text{base effect}}\cdot\underbrace{\frac{L_c}{L_{c'}}}_{\text{linkage effect}}
=\frac{Q(c')}{Q(c)}\cdot\frac{(\ell^{\downarrow}_c+\ell^{\times}_c)\,X_{cc'}+(\ell^{\uparrow}_c+\ell^{\times}_c)\,M_{cc'}}{(\ell^{\uparrow}_{c'}+\ell^{\times}_{c'})\,X_{cc'}+(\ell^{\downarrow}_{c'}+\ell^{\times}_{c'})\,M_{cc'}}.
\label{eq:m-size-split}
\end{equation}
The numerator and the denominator of the linkage effect contain the same two flows, and differ only in the multipliers that weigh them. At given multipliers, the sizes of the two economies do not appear in the linkage effect, and the volume of the pair's trade cancels from it, since doubling both flows doubles both dollar losses. And the multipliers themselves depend on the levels of activity only through shares, so that scaling every level of the trade table by a common factor leaves them unchanged (Lemmas~\ref{lem:m-scale} and~\ref{lem:m-rates-shares} in Appendix~\ref{app:m-proof-exposure}).\footnote{The sourcing shares of a producer on its suppliers at home add up to its home share, namely the fraction of its intermediate inputs that it buys at home. At given fulfillment rates, moving a purchase from a supplier abroad to a supplier at home raises the content of the country in the producer that moves the purchase, provided that a dollar of orders at home returns more of the country's activity than a dollar of orders abroad. Neither the export multiplier nor the import multiplier of the country falls with such a move (Lemma~\ref{lem:m-homechain}(i) in Appendix~\ref{app:m-proof-homechain}). When the producers of country $c$ share the leakage $\beta_c$ and the home share $\varphi_c$, the content of country $c$ in each of them is at least $1/(1-\varphi_c(1-\beta_c))$ at the benchmark orders and full delivery (Lemma~\ref{lem:m-homechain}(ii)).} If a lost dollar of trade cost the two trading partners the same activity on each flow, the linkage effect would be one and the loss ratio would be the size ratio. Power would then be a matter of size alone.

The bound \eqref{eq:m-size-bound} says how far the multipliers can take the linkage effect from one. Put simply, if country $c$ exported to country $c'$ but did not import from it, then every lost dollar of their trade would be a lost sale for country $c$ and a lost purchase for country $c'$, and the linkage effect would be the export multiplier of country $c$ divided by the import multiplier of country $c'$. If country $c$ instead imported from country $c'$ but did not export to it, the linkage effect would be the import multiplier of country $c$ divided by the export multiplier of country $c'$. The two cases give the two ratios of \eqref{eq:m-size-bound}, each multiplier carrying its interaction rate. The linkage effect of a pair that trades in both directions lies between these two ratios, since it is an average of them, each weighted by the dollar loss of country $c'$ on the flow of that ratio, which the proposition takes to be positive.\footnote{The dollar loss of a country on a flow is never negative, since each multiplier with its interaction rate added is a nonnegative combination of the contents of the country (Lemma~\ref{lem:m-onesided}(ii) in Appendix~\ref{app:m-content}). This dollar loss can be zero, as when all that country $c'$ buys on a flow is final goods, whose loss idles no production at home, and the content of country $c'$ in the output of the sellers of that flow is zero. The multiplier ratio on that flow then has a zero denominator, and only the ratio on the other flow bounds the linkage effect, from below.} The division of the pair's trade between the two flows can move the linkage effect from one multiplier ratio toward the other, and no further. An economy twice the size of its trading partner is therefore the more powerful of the two unless a lost dollar of its sales or of its purchases costs it at least twice what that very dollar costs the partner.

\subsection{Measuring the base effect and the linkage effect}
\label{subsec:power_size_base}

We now measure the base effect and the linkage effect on the solved equilibria of the $3{,}160$ trade severances.\footnote{On these equilibria the interaction of Proposition~\ref{prop:m-exposure} is negligible next to what a trading partner loses through the production shortfalls of the equilibrium. The interaction has a median of about one millionth of that loss over the trading partners of the pairs, and it is at most $2.2\%$ of that loss on the pairs that trade at least a million dollars (Online Appendix~\ref{oa:exact-decomposition}).} The sizes of the eighty economies span nearly five orders of magnitude, from $0.0005\%$ of world gross economic activity for S\~{a}o Tom\'{e} and Pr\'{\i}ncipe to about $24\%$ for the United States. The size ratio is about $5$ on the median pair and exceeds $50$ on one pair in ten. By contrast, the two trading partners' dollar losses, whose ratio is the linkage effect, differ by a factor of about $1.3$ on the median pair and by more than $2.4$ on only one pair in ten. On $2{,}792$ of the $3{,}160$ trading pairs the two trading partners differ more in size than in the dollars they lose. Naturally then, on every one of these pairs the larger economy is the more powerful side, whichever way the linkage effect runs.

On the other $368$ trading pairs the two trading partners differ at least as much in the dollars they lose as in size, and on $173$ of them the linkage effect favors the larger economy, thereby adding to the size advantage. The power relation between the United States and China is a case in point. The United States is only slightly larger than China. What makes the United States more powerful is not its size but the fact that a lost dollar of trade between them costs China about two and a half times the activity that it costs the United States. The severance therefore costs China about $1.4\%$ of its activity and the United States about $0.6\%$ of its.

\begin{figure}[!htbp]
\centering
\includegraphics[width=0.72\linewidth]{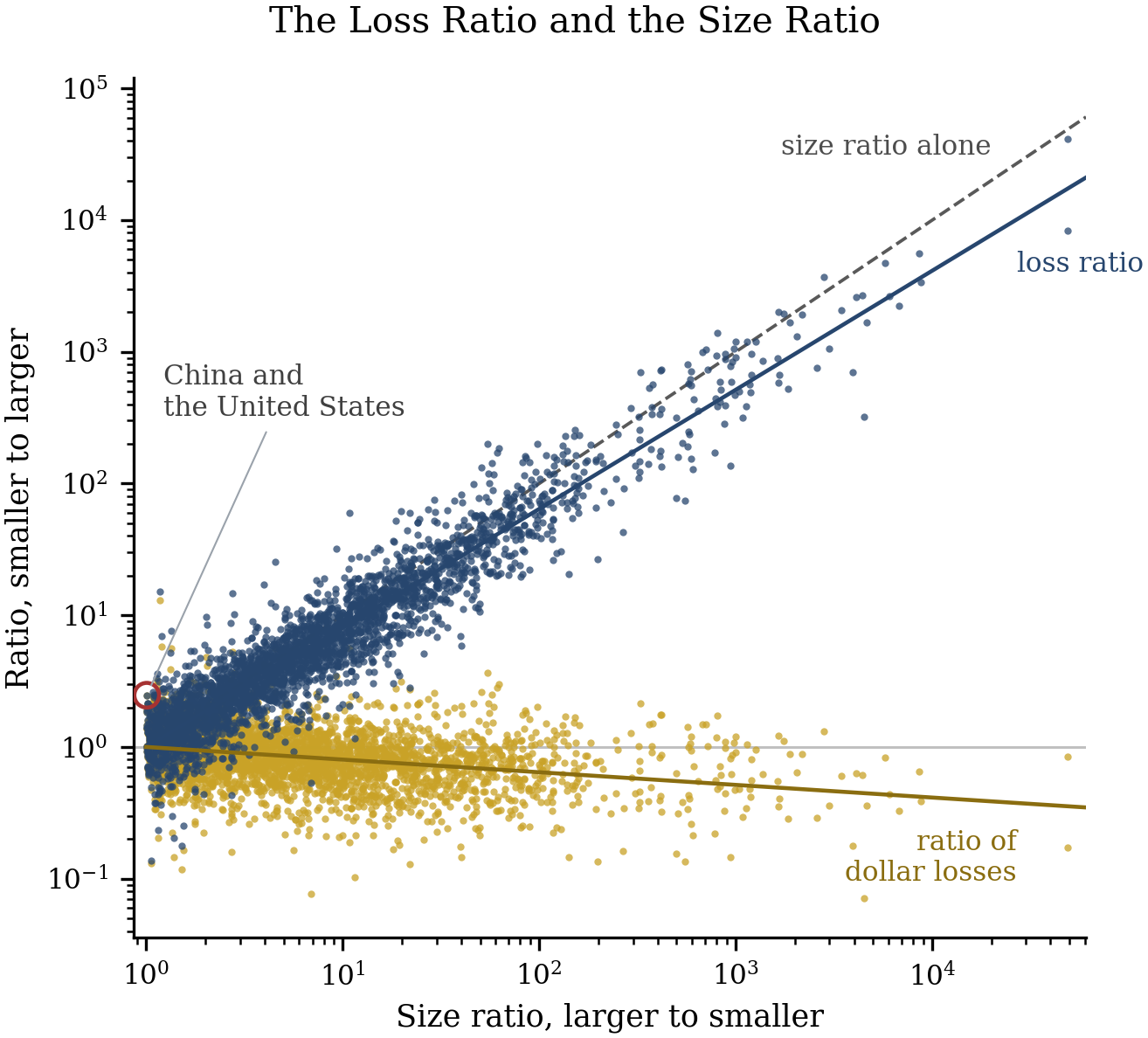}
\caption{Each of the $3{,}160$ pairs appears as two points, with the smaller economy as country $c$. The dark point is the loss ratio $r_{cc'}$ and the light point is the linkage effect $L_c/L_{c'}$, the ratio of the two dollar losses, both against the base effect $Q(c')/Q(c)$, the size ratio of the pair, and both on log scales. The dashed diagonal, marked `size ratio alone', is the line on which the loss ratio equals the size ratio and the linkage effect is one. The two solid lines are least-squares fits in logs through the origin, with slopes $0.90$ for the loss ratio and $-0.10$ for the linkage effect. The thin horizontal line marks equal dollar losses for the light points and equal proportional losses for the dark points, so a dark point below it is a pair whose smaller economy is the more powerful side. The circled pair at the left edge is the United States and China.}
\label{fig:size_decomposition}
\end{figure}

On the other $195$ of the $368$ pairs the linkage effect runs against the larger economy \emph{and} overturns the base effect, thereby making the smaller economy more powerful.\footnote{The bound \eqref{eq:m-size-bound} says which of these verdicts depend on how a pair divides its trade between the two flows. Orient a pair with the smaller economy first, so that by \eqref{eq:m-size-split} the larger economy is the more powerful side when the linkage effect exceeds the reciprocal of the size ratio. Since the linkage effect lies between the two multiplier ratios, the verdict does not depend on the division of trade, at these multipliers, when both ratios lie on the same side of that reciprocal. Both exceed it on $2{,}814$ of the $3{,}160$ pairs, where the larger economy is the more powerful side, and both fall short of it on $65$ pairs, where the smaller economy is the more powerful side. On the other $281$ pairs the reciprocal of the size ratio lies between the two multiplier ratios and the division of trade decides, for the larger economy on $151$ pairs and for the smaller economy on $130$.} Australia and Korea are one of these pairs. Korea's economy is a third larger than Australia's, but what Korea buys from Australia is mostly oil and gas, coal, and ores, all of them inputs whose loss idles Korean production. Every lost dollar of these purchases costs Korea about $4.6$ dollars of activity, where the same lost dollar of sales costs Australia about $1.7$. The severance costs Korea about $1.2\%$ of its activity and Australia about $0.8\%$ of its.

Figure~\ref{fig:size_decomposition} plots the base effect and the linkage effect of each of the $3{,}160$ pairs, together with its loss ratio. Taking country $c$ to be the smaller economy of the pair, the x-axis of the figure gives the base effect $Q(c')/Q(c)$, and the y-axis gives the linkage effect $L_c/L_{c'}$ as a light point and the loss ratio $r_{cc'}$ as a dark point. Note that the ratio on the x-axis has the larger economy in the numerator, while the two ratios on the y-axis have the smaller economy in the numerator, which is how \eqref{eq:m-size-split} writes the three ratios. Both axes are in logs. The light points keep within a band around the horizontal line, meaning that the ratio of the two dollar losses changes little with the disparity in size between the two economies. The line fitted through the light points slopes slightly downward, since the larger economy loses more dollars than its partner on two pairs in three, and the more so the further apart the two economies are in size. The linkage effect thereby offsets about a tenth of the base effect.\footnote{The loss ratio accordingly rises with the size ratio at a slope of $0.90$ in logs rather than one. The larger economy tends to lose more dollars because it has the higher export multiplier on $71\%$ of the pairs and the higher import multiplier on $65\%$. The export sectors of the larger economy tend to source more of their inputs at home, and more of what the larger economy buys from its partner is inputs. These tendencies are features of the 2022 trade table, and the model does not require them (Online Appendix~\ref{oa:linkage-size}).} The dark points keep close to the dashed diagonal, on which they would lie if size were all that mattered, meaning that the ratio of the two losses as fractions of activity grows almost as fast as the disparity in size. The dark points fall below the horizontal line on only $195$ pairs, which is to say that on those pairs the smaller economy loses the lesser fraction of its activity and is therefore the more powerful side. Overall, the larger economy is more powerful in $94\%$ of the pairs.\footnote{Rearranging the flows of trade while retaining every size, which Lemma~\ref{lem:m-fiber} makes possible in abundance, moves the losses far more than the power relations. A flattening of the intermediate flows that raises the median vulnerability by $57\%$ moves each country's share of the pair's combined proportional loss by a median of only $1.4$ percentage points, which is $0.028$ in the units of the asymmetry, and leaves the asymmetry index $\Psi$ at $0.58$. Power reverses in none of the eight headline pairs even when their own trade is frozen and two thirds of the world's flows move (Online Appendix~\ref{oa:fiber}).}

\subsection{Power among economies of comparable size}
\label{subsec:comparable_size}

Though size is by and large the measure of power for the world economy as a whole, it is much less so among economies of comparable size. For instance, there are $746$ pairs of economies whose sizes are within a factor of two of one another, and the smaller economy is the more powerful side on $23\%$ of them. When we look at the $1{,}131$ pairs within a factor of three, the number drops to $17\%$, which is still substantial and far from the $6\%$ across all the $3{,}160$ pairs. The reason is that the base effect between two such economies is small, and a linkage effect that runs in the opposite direction is then often strong enough to overwhelm it (Online Appendix~\ref{oa:linkage-size}). Australia and Korea, Ireland and Belgium, and Saudi Arabia and Indonesia are cases in point.

\begin{figure}[!htbp]
\centering
\includegraphics[width=\linewidth]{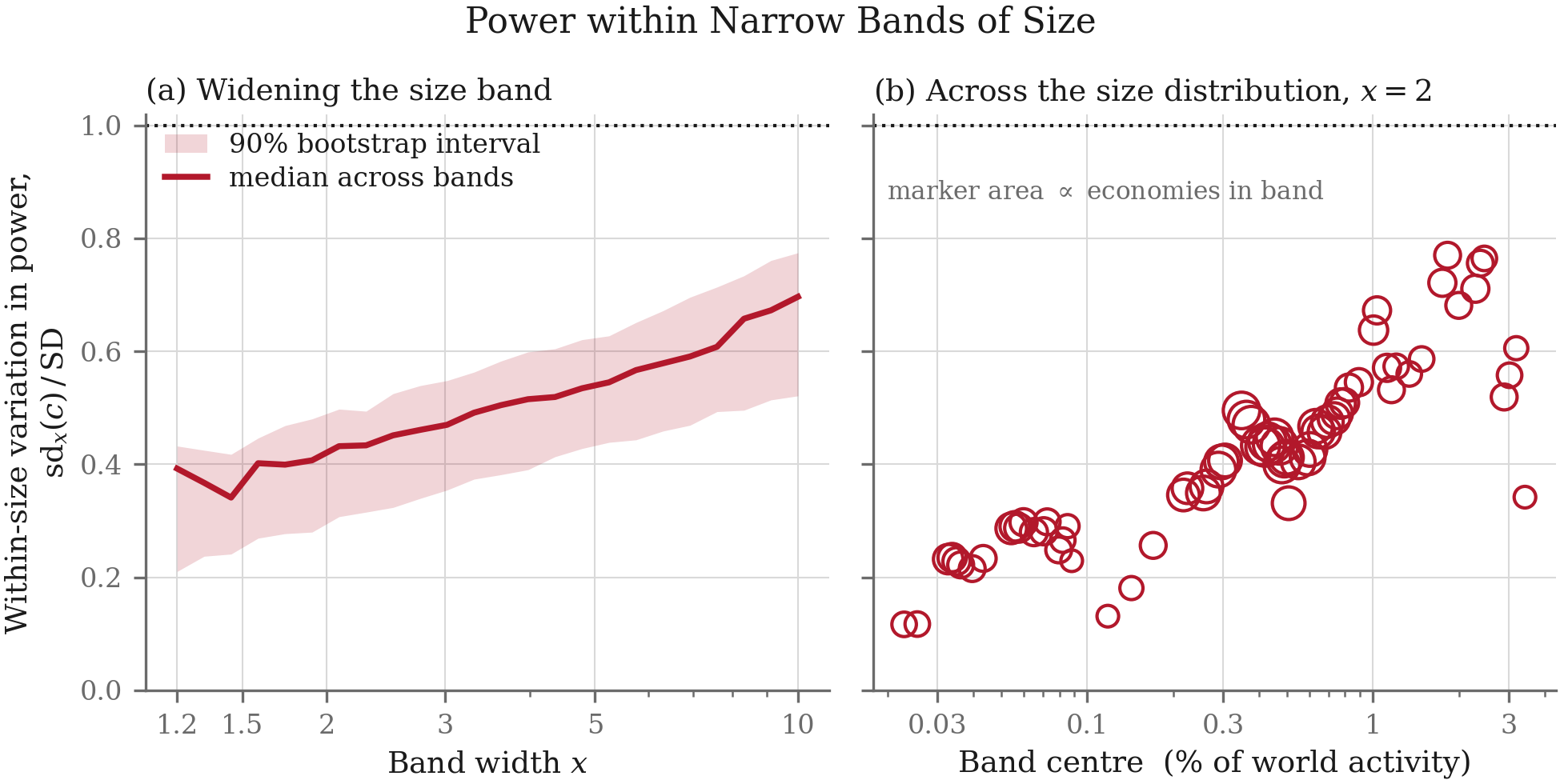}
\caption{The within-size variation in power. A size band of width $x$ starts at one economy and holds every economy up to $x$ times larger. The y-axis of both panels gives the within-size variation in power $\mathrm{sd}_x(c)/\mathrm{SD}$ of a band, namely the standard deviation of the trade-weighted mean advantage within the band divided by its standard deviation across all seventy-seven economies, as in \eqref{eq:m-within-size}, and the dotted line marks a ratio of one. Panel~(a) plots the median of this ratio over the bands against the width $x$, with the $90\%$ bootstrap interval of the median shaded. Panel~(b) is the cross-section of Panel~(a) at a width of two, and plots the ratio of each band of that width against the size at the centre of the band, with the area of a circle proportional to the number of economies in the band. S\~{a}o Tom\'{e} and Pr\'{\i}ncipe, China, and the United States are excluded throughout. Baseline calibration $\tau=0.30$, $\rho=-1$, $\varrho=-1$, $\delta=0.10$.}
\label{fig:size_window}
\end{figure}

Figure~\ref{fig:size_window} illustrates how the variation in power across economies of comparable size depends on how close in size they are and on how large they are. Recall that the asymmetry $\mathcal H_{cc'}$ of \eqref{eq:bounded-Hirschman} lies between $-1$ and $1$ and is negative when country $c$ is the more powerful side of its relationship with country $c'$. Recall also that Section~\ref{subsec:hirschman_matrix} summarizes the power of country $c$ over its trading partners taken together by its trade-weighted mean advantage, namely $\mathcal V_c:=-\sum_{c'\ne c}\mu_{cc'}\mathcal H_{cc'}$, where $\mu_{cc'}$ is the share of the trade of country $c$, sales and purchases together, that is with country $c'$. The advantage $\mathcal V_c$ is thus a number between $-1$ and $1$ that says how much power country $c$ holds, and not merely on which side of each pair it stands.

A size band of width $x$ starts at one economy and holds every economy up to $x$ times larger, so that the width says how far apart in size two economies may be and still count as comparable. At a width of two, for instance, the band that starts at Brazil holds Brazil, Canada, Russia, Italy, Korea, France, the United Kingdom, and India. For the size band of width $x$ that starts at country $c$, we write $\mathrm{sd}_x(c)$ for the standard deviation of the advantage over the economies of the band, and $\mathrm{SD}$ for its standard deviation over all the economies, namely
\begin{equation}
\begin{aligned}
\mathrm{sd}_x(c)&:=\Bigl[\frac{1}{n_x(c)-1}\sum_{c'\,:\,Q(c)\le Q(c')\le x\,Q(c)}\bigl(\mathcal V_{c'}-\overline{\mathcal V}_x(c)\bigr)^{2}\Bigr]^{1/2},\\
\mathrm{SD}&:=\Bigl[\frac{1}{n-1}\sum_{c'}\bigl(\mathcal V_{c'}-\overline{\mathcal V}\bigr)^{2}\Bigr]^{1/2}.
\end{aligned}
\label{eq:m-within-size}
\end{equation}
Here $n_x(c)$ is the number of economies in the band and $\overline{\mathcal V}_x(c)$ is their mean advantage, while $n$ and $\overline{\mathcal V}$ are the number and the mean advantage of all the economies.\footnote{We set aside S\~{a}o Tom\'{e} and Pr\'{\i}ncipe, which no band can pair with more than a handful of others, and China and the United States, which hold the favorable side against nearly everyone, so that the comparison runs over seventy-seven economies. We also drop any band that holds fewer than five economies, which leaves seventy-two bands at a width of two, with ten economies in the median band (Online Appendix~\ref{oa:power_size_local}).} We call the ratio $\mathrm{sd}_x(c)/\mathrm{SD}$ the within-size variation in power of the band. For the band that starts at Brazil, the advantages of the eight economies run from $-0.45$ for Canada to $0.17$ for France, and their standard deviation is $0.23$ against $0.31$ across all the economies, so that the within-size variation in power of this band is about three quarters. A ratio close to zero means that the economies of a band hardly differ in power, and a ratio of one that they differ in power as much as all the economies do.

Figure~\ref{fig:size_window}a plots the within-size variation in power against the width of the band. Its x-axis gives the width $x$, and its y-axis gives the median of the ratio $\mathrm{sd}_x(c)/\mathrm{SD}$ over the bands of that width, one band starting at each country $c$ in turn. The shaded region is the $90\%$ bootstrap interval of this median, which we obtain by redrawing the seventy-seven economies with replacement a thousand times and recomputing the median each time, so that the region holds nine in ten of the recomputed medians. The shaded region thereby shows how far the plotted curve depends on which economies happen to be in the sample. Were size all that mattered for power, the economies of a narrow band would hardly differ in power, i.e., the curve and its shaded region would lie close to zero at the left of the figure and would climb toward one only as the band widens. The within-size variation in power is instead about two fifths among economies within a factor of two of one another in size, where the shaded region runs from about three tenths to one half. It rises only to about half among economies within a factor of three, and to about seven tenths among economies within a factor of ten. Note therefore that even among a pool of economies whose sizes are within an order of magnitude of one another, size does not explain all the variation in power. And even at the narrowest band that Figure~\ref{fig:size_window}a considers, a width of $1.2$ at the left end of its x-axis, the within-size variation in power is about two fifths, far from zero: bringing the economies of a band closer in size does not make them more alike in power.

Figure~\ref{fig:size_window}b is a cross-section of Figure~\ref{fig:size_window}a at the width $x=2$. Where Figure~\ref{fig:size_window}a shows only the median of the ratio $\mathrm{sd}_2(c)/\mathrm{SD}$ over the seventy-two bands of that width, which is about two fifths, Figure~\ref{fig:size_window}b shows the seventy-two ratios themselves, so that we can see where in the size distribution economies of comparable size differ most in power. Its x-axis gives the size at the centre of a band as a percentage of world gross economic activity, and its y-axis gives the within-size variation in power of that band. More specifically, write $\overline Q_2(c)$ for the size at the centre of the band that starts at country $c$, namely the geometric mean of the sizes of the economies of the band. This band appears in the figure as the circle at the point
\[
\Bigl(\overline Q_2(c),\ \frac{\mathrm{sd}_2(c)}{\mathrm{SD}}\Bigr).
\]
There is one such circle for each of the seventy-two countries $c$ at which a band starts, with the bands of the small economies to the left and the bands of the large economies to the right. The area of a circle is proportional to the number $n_2(c)$ of economies in the band, which runs from five to twenty. The band that starts at Brazil, for instance, is the circle at $2.4\%$ on the x-axis and about three quarters on the y-axis. Were size all that mattered for power, every circle would lie close to zero, and were economies of comparable size to differ in power as much at one size as at another, the circles would lie along a horizontal line.

The circles instead rise from left to right, from about a tenth in the bands of the smallest economies to about three quarters in the bands centred near $2\%$. Only the last four bands on the right, which reach up to Germany and Japan, lie lower again. The reason for this pattern is that the economies at either end of the size distribution are alike in power. The smallest economies hold the weaker side against nearly every partner, and the largest of the seventy-seven economies hold the favorable side against most of theirs. An economy near $2\%$ of world gross economic activity, by contrast, trades with larger and with smaller economies, and its advantage depends on which of them it trades with. Canada, whose trade runs overwhelmingly through one much larger partner, is at the bottom of the band that starts at Brazil, and France is at its top. The differences in size that remain within a band account for almost none of this variation. What explains most of the within-size variation in power instead is how large the trade of an economy with each of its partners is next to its own activity, which is to say the export share and the import share of Proposition~\ref{prop:m-exposure} (Online Appendix~\ref{oa:power_size_local}).

\FloatBarrier

\section{Power and the Balance of Trade}
\label{sec:power_balance}

Whether a country that runs a trade surplus with a partner thereby holds power over it is a question with a long, somewhat embarrassing history.\footnote{The surplus-as-strength reading is as old as mercantilism \citep{mun1664}, and its classic rebuttal nearly as old \citep{hume1752}. The modern policy revival reads a trade deficit as a drag on the nation's growth and traces it to the deficits with a handful of named partners \citep{navarroross2016}. The reply has been that a trade balance is no measure of a nation's strength, since a surplus can accompany weakness and a deficit can accompany strength \citep{krugman1996}. No side of these exchanges had a measure of power itself, namely the losses that a trade severance would inflict on each trading partner, and the Hirschman matrix $\boldsymbol{\mathcal H}$ supplies one.} On the 2022 trade table, the trade balance confers almost no power. The country with the trade surplus is the more powerful side on $54\%$ of the $3{,}160$ pairs, a mere coin flip. Erasing a pair's trade imbalance moves the power asymmetry between its two trading partners, which runs from minus one to one, by $0.01$ at the median across the pairs. This movement is to be set against a median absolute asymmetry of $0.63$. Why this is so has to do with the narrow mechanism by which a trade surplus can generate power. Section~\ref{subsec:power_surplus} explicates this mechanism, and Section~\ref{subsec:balance_table} measures how far the trade imbalance moves power on the solved equilibria of the $3{,}160$ trade severances.

\subsection{How the trade balance influences economic power}
\label{subsec:power_surplus}

We define a pair's trade balance on the flows the trade severance bars, namely the purchases $M_{cc'}$ and the sales $X_{cc'}$ of country $c$ (Section~\ref{subsec:power_size}), positive when country $c$ runs a deficit with country $c'$, as
\begin{equation}
\mathcal B_{cc'}:=\frac{M_{cc'}-X_{cc'}}{M_{cc'}+X_{cc'}}.
\label{eq:direct_trade_asymmetry}
\end{equation}
The trade balance is zero when the two flows are equal. We balance a pair's trade by lowering the sales of the country with the trade surplus and raising its purchases by the same amount, so that the two flows meet halfway and each country sells the other $(X_{cc'}+M_{cc'})/2$ dollars. Naturally, the gross volume of the pair's trade is unchanged.\footnote{The pair's trade could also be balanced at any other level between the two flows, with each country selling the other $\zeta M_{cc'}+(1-\zeta)X_{cc'}$ dollars for a weight $\zeta\in(0,1)$, the two flows meeting halfway at $\zeta=1/2$. A larger volume of balanced trade means more lost dollars and a larger loss for each country, but in the same proportion for both. Since we work with the ratio of the two countries' losses at balanced trade, the weight $\zeta$ cancels and nothing changes (Appendix~\ref{app:m-proof-decomposition}).}

Suppose for a moment that the pair's trade were balanced in this way with the sizes of the two economies unchanged, and that the export multipliers, the import multipliers, and the interaction rates of Proposition~\ref{prop:m-exposure} remained at the values the solved equilibrium returns. Each trading partner would then lose as many dollars of sales as of purchases. The trade imbalance tilts the lost dollars of the country with the trade deficit toward purchases and those of the country with the trade surplus toward sales. More specifically, for every two dollars of lost trade, country $c$ loses $1+\mathcal B_{cc'}$ dollars of purchases and $1-\mathcal B_{cc'}$ dollars of sales, and country $c'$ loses $1-\mathcal B_{cc'}$ dollars of purchases and $1+\mathcal B_{cc'}$ dollars of sales.

By Proposition~\ref{prop:m-exposure} a lost dollar of sales costs country $c$ its export multiplier plus its interaction rate in dollars of activity ($\ell^{\downarrow}_c+\ell^{\times}_c$), and a lost dollar of purchases costs it its import multiplier plus the same rate ($\ell^{\uparrow}_c+\ell^{\times}_c$). Two dollars of lost trade at balanced trade, one of sales and one of purchases, would therefore cost country $c$ a total of $\Sigma_c:=\ell^{\downarrow}_c+\ell^{\uparrow}_c+2\ell^{\times}_c$ dollars of activity. And country $c'$ would lose $\Sigma_{c'}$ dollars of activity for the same two dollars. The two sums $\Sigma_c$ and $\Sigma_{c'}$ are set by the composition of each trading partner's trade with the other, whether it be inputs or final goods, and by how much of its own production lies upstream of its sales and downstream of its purchases.

Put simply, the tilt in the trade balance matters only as far as a lost purchase costs a country differently from a lost sale. We measure that difference by the excess of the import multiplier of country $k$ over its export multiplier relative to $\Sigma_k$,
\begin{equation}
\mathcal G_k:=\frac{\ell^{\uparrow}_k-\ell^{\downarrow}_k}{\Sigma_k}\in[-1,1]
\qquad(k\in\{c,c'\}).
\label{eq:m-gap}
\end{equation}
We call $\mathcal G_k$ the `import premium' of country $k$. The import premium is positive when country $k$ loses more per dollar of lost purchases than per dollar of lost sales. With the tilt, the two dollars of lost trade cost country $c$ a total of $\Sigma_c+\mathcal B_{cc'}(\ell^{\uparrow}_c-\ell^{\downarrow}_c)$ dollars of activity, which is $\Sigma_c(1+\mathcal B_{cc'}\mathcal G_c)$ by \eqref{eq:m-gap}, and they cost country $c'$ a total of $\Sigma_{c'}(1-\mathcal B_{cc'}\mathcal G_{c'})$ dollars. Since the pair's lost trade is $\tau(1-\delta)(X_{cc'}+M_{cc'})$ dollars, and each cost is for two of these dollars, we can write the dollar losses of the two countries (Appendix~\ref{app:m-proof-decomposition}) as
\begin{equation}
\begin{aligned}
L_c&=\tfrac12\,\tau(1-\delta)\,(X_{cc'}+M_{cc'})\,\Sigma_c\bigl(1+\mathcal B_{cc'}\mathcal G_c\bigr),\\
L_{c'}&=\tfrac12\,\tau(1-\delta)\,(X_{cc'}+M_{cc'})\,\Sigma_{c'}\bigl(1-\mathcal B_{cc'}\mathcal G_{c'}\bigr).
\end{aligned}
\label{eq:m-balanced-loss}
\end{equation}

Dividing the first line of \eqref{eq:m-balanced-loss} by the second, we can write the linkage effect $L_c/L_{c'}$ of Section~\ref{subsec:power_size} as
\[
\frac{L_c}{L_{c'}}
=\frac{\Sigma_c}{\Sigma_{c'}}\cdot\underbrace{\frac{1+\mathcal B_{cc'}\,\mathcal G_c}{1-\mathcal B_{cc'}\,\mathcal G_{c'}}}_{\text{trade-imbalance effect}}.
\]
When the trade balance $\mathcal B_{cc'}$ is zero the second ratio is one, so the first ratio is the linkage effect at balanced trade. We can therefore write the loss ratio as the product of the base effect of \eqref{eq:m-size-split}, the linkage effect at balanced trade, and the trade-imbalance effect.

\begin{proposition}[The trade balance and the loss ratio]
\label{prop:m-decomposition}
Suppose that the unrecovered share $\tau$ is positive and that countries $c$ and $c'$ trade in both directions. The loss ratio is the product of three ratios,
\begin{equation}
r_{cc'}
=\underbrace{\frac{Q(c')}{Q(c)}\cdot\frac{\Sigma_c}{\Sigma_{c'}}}_{r^{\mathrm b}_{cc'}}\cdot\underbrace{\frac{1+\mathcal B_{cc'}\,\mathcal G_c}{1-\mathcal B_{cc'}\,\mathcal G_{c'}}}_{\textnormal{trade-imbalance effect}}.
\label{eq:m-exact-decomposition}
\end{equation}
The product $r^{\mathrm b}_{cc'}$ of the first two ratios is the loss ratio of the pair at balanced trade.
\end{proposition}

\noindent Proof in Appendix~\ref{app:m-proof-decomposition}.

At given multipliers the trade balance $\mathcal B_{cc'}$ enters \eqref{eq:m-exact-decomposition} only through the trade-imbalance effect, and there it is always multiplied by an import premium ($\mathcal G_c$ or $\mathcal G_{c'}$).\footnote{If the rationed exchange left unrecovered a larger share of one flow than of the other, the lost dollars of both countries would tilt toward that flow, as a trade imbalance tilts them. The loss ratio would then be \eqref{eq:m-exact-decomposition} with the normalized difference of the lost purchases and the lost sales of country $c$ in place of the trade balance (Lemma~\ref{lem:oa-unequal-shares} in Online Appendix~\ref{oa:unequal-shares}). A flow that is harder to redirect therefore weighs on its buyers and on its sellers alike.} Since a premium can be of either sign, a trade surplus does not by itself add to a country's power. A positive import premium of the country with the trade deficit raises what a dollar of lost trade costs it, and a positive premium of its partner lowers what such a dollar costs the partner, each by a fraction equal to the trade balance times the premium. The trade imbalance therefore works against the country with the trade deficit exactly when the two import premia sum to a positive number ($\mathcal G_c+\mathcal G_{c'}>0$).

\subsection{Measuring the effect of the trade balance on power}
\label{subsec:balance_table}

We now measure the relation between power and the trade balance on the solved equilibria of the $3{,}160$ trade severances, at the export multipliers, the import multipliers, and the interaction rates those equilibria return. More specifically, we measure the effect of the trade imbalance on a pair by the normalized difference \eqref{eq:m-psi} of its loss ratio $r_{cc'}$ and its loss ratio $r^{\mathrm b}_{cc'}$ at balanced trade. Since by Proposition~\ref{prop:m-decomposition} the ratio of the two loss ratios is the trade-imbalance effect, we can write the normalized difference as
\begin{equation}
\frac{r_{cc'}-r^{\mathrm b}_{cc'}}{r_{cc'}+r^{\mathrm b}_{cc'}}
=\frac{\mathcal B_{cc'}\,(\mathcal G_c+\mathcal G_{c'})}{2+\mathcal B_{cc'}\,(\mathcal G_c-\mathcal G_{c'})}.
\label{eq:m-balance-transform}
\end{equation}
This normalized difference is the asymmetry that the trade imbalance alone would give a pair whose two trading partners lose equal shares of their activity at balanced trade.

We find that the trade balance moves power very little.\footnote{Regressions of power on the trade balance tell the same story. Online Appendix~\ref{oa:balance-robustness} runs a grid of them, crossing five measures of the trade imbalance with three weightings of the pairs and three network-position controls, and the trade balance adds a median of one thousandth to the explained variance of power. The trade balance moves power as little when we erase each pair's trade imbalance from the trade table, one pair at a time, and solve the trade severance again, which lets the multipliers and the interaction rates respond. The power asymmetries then move by a median of $0.010$ (Online Appendix~\ref{oa:balanced-trade}).} The effect of the trade imbalance has a median absolute value of only $0.025$. In fact, the trade imbalance moves the power asymmetry of a pair away from its value at balanced trade by only about $0.01$ at the median, whereas the power asymmetry itself has a median absolute value of $0.63$.\footnote{Since by Lemma~\ref{lem:m-rigidity}\textnormal{(ii)} an asymmetric pair moves less under the same relative disturbance than a pair whose two trading partners lose equal shares of their activity, the median of $0.025$ exceeds the $0.011$ by which the trade imbalance moves the asymmetries themselves at the median.}

\begin{figure}[!htbp]
\centering
\includegraphics[width=0.62\linewidth]{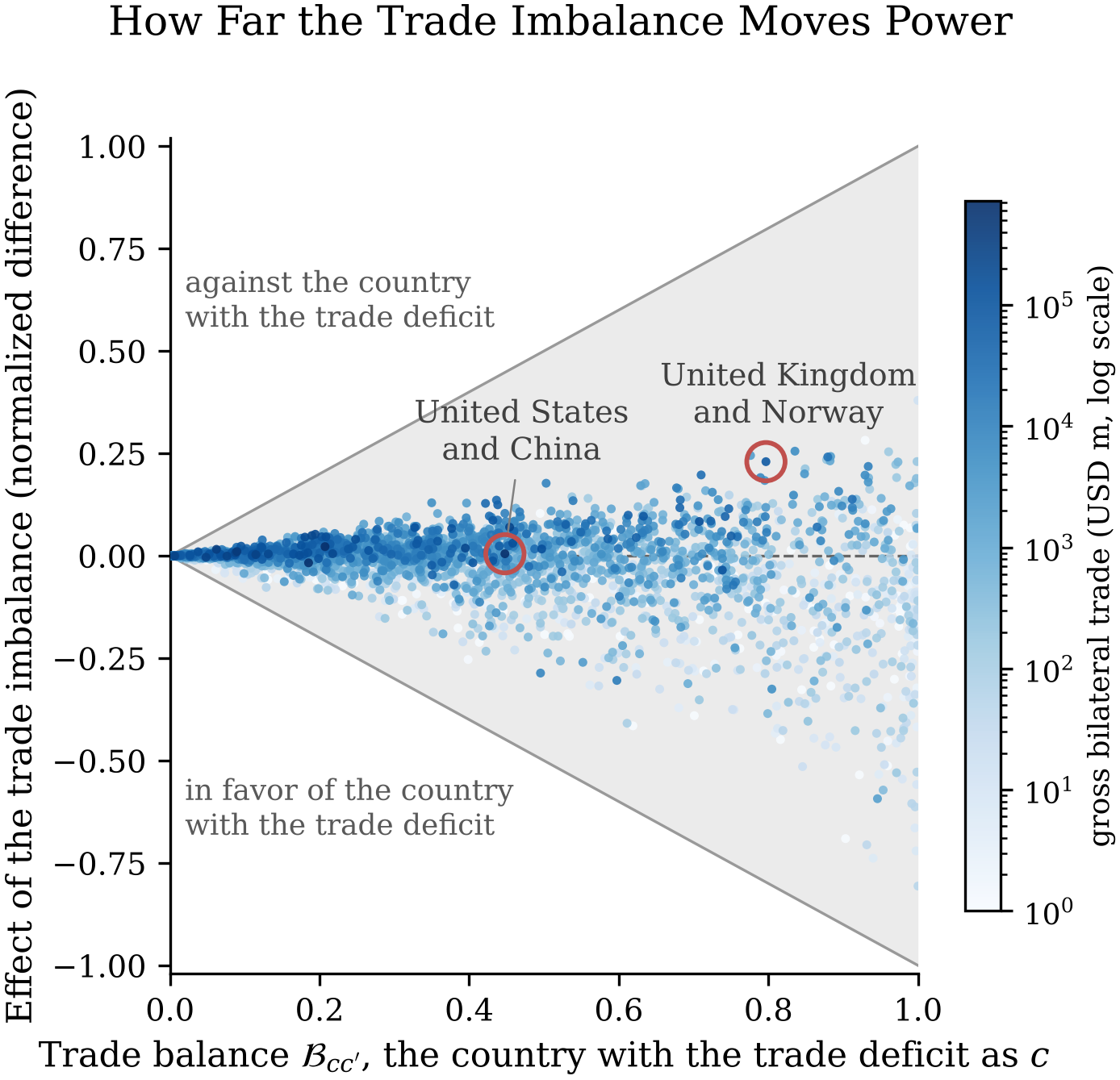}
\caption{The effect of the trade imbalance on each pair, measured by the normalized difference \eqref{eq:m-balance-transform}, against its trade balance $\mathcal B_{cc'}$ \eqref{eq:direct_trade_asymmetry}, with each pair oriented so that the country with the trade deficit is country $c$. A positive value works against the country with the trade deficit, and the dashed horizontal line marks an effect of zero. The shaded triangle is the largest effect that any import premia could give, which is the trade balance itself, reached only when the two premia are equal and of absolute value one. Each point is one of the $3{,}160$ pairs, color is gross bilateral trade on a log scale, and the circled pairs are the United States and China and the United Kingdom and Norway (Online Appendix~\ref{oa:largest-deficits}).}
\label{fig:s6_balance_effect}
\end{figure}

Figure~\ref{fig:s6_balance_effect} plots the effect of the trade imbalance on each of the $3{,}160$ pairs against its trade balance. Taking country $c$ to be the country with the trade deficit, the x-axis of the figure gives the trade balance $\mathcal B_{cc'}$, which then runs from zero, where the two flows of the pair are equal, to one, where its trade runs entirely one way. The y-axis gives the effect of the trade imbalance, which is positive when the trade imbalance works against the country with the trade deficit and negative when it works in its favor. The edges of the shaded triangle are where the effect of the trade imbalance is as large as the trade balance itself, which by \eqref{eq:m-balance-transform} it is when the two import premia are equal and of absolute value one. Were a lost dollar of purchases all that cost a trading partner activity, both import premia would be one and the points would lie along the upper edge of the triangle, and were a lost dollar of sales all that cost it activity, the points would lie along the lower edge. Which is to say, the upper edge is where a trade surplus would generate power to the full extent of the trade balance, and the lower edge is where a trade deficit would.

The points instead keep within a narrow band around zero, far short of the edges of the triangle. The trade imbalance also works about as often in favor of the country with the trade deficit as against it, namely on $1{,}653$ pairs against $1{,}507$.\footnote{The larger effects belong mostly to small pairs whose trade runs almost entirely one way. On $192$ pairs the effect of the trade imbalance exceeds $0.2$ in absolute value, about a third of the median asymmetry, and $176$ of these pairs trade less than a billion dollars a year. Since the import premia fall as low as minus one but nowhere rise above $0.6$, the points of Figure~\ref{fig:s6_balance_effect} fan out further below zero than above it, and on more than nine in ten of the $192$ pairs the trade imbalance works in favor of the country with the trade deficit. On these pairs the country with the trade deficit buys mostly final goods from its partner, with inputs making up only a quarter of its purchases at the median, and has a median import premium of $-0.5$. The largest such pair is that of the United States with Bangladesh, where the United States buys four times what it sells and nine-tenths of its purchases are final goods, so that the American import premium is as low as $-0.75$ and the effect of the trade imbalance is $0.30$ in favor of the United States.}

And the decomposition of Section~\ref{subsec:power_surplus} tells us exactly why the trade balance does so little in shaping the power relations between nations. By \eqref{eq:m-balance-transform} the trade balance moves power only through the import premia of the two trading partners. In absolute value the effect of the trade imbalance never exceeds the trade balance times the larger of the two import premia,\footnote{The bound is derived in Appendix~\ref{app:m-proof-decomposition}, which also shows that for a pair with a nonzero trade imbalance it is reached only when the two premia are equal.}
\begin{equation}
\frac{|r_{cc'}-r^{\mathrm b}_{cc'}|}{r_{cc'}+r^{\mathrm b}_{cc'}}
\le|\mathcal B_{cc'}|\max\bigl\{|\mathcal G_c|,|\mathcal G_{c'}|\bigr\}.
\label{eq:m-balance-bound}
\end{equation}
In the data, the trade balance is sizeable, with a median absolute value of $0.35$ across the pairs, which is the value at which one trading partner sells the other about twice what it buys in return. But the import premia are small, with a median absolute value of $0.12$, and the bound \eqref{eq:m-balance-bound} on the effect of the trade imbalance has a median of only $0.06$, about a tenth of the median asymmetry.

The import premia are small because of what the two trading partners buy from each other. Recall that the import premium of a trading partner measures how much more a lost dollar of purchases costs it than a lost dollar of sales. A lost dollar of purchased inputs costs a trading partner far more than a lost dollar of sales, at a median of $2.7$ against $1.6$ dollars of activity, whereas a lost dollar of purchased final goods idles no production at home. A trading partner's premium therefore rises with the share of inputs in what it buys from the other, and turns from negative to positive where inputs make up about $60\%$ of its purchases. The median trading partner buys $61\%$ inputs, and so loses about as much on a lost dollar of purchases as on a lost dollar of sales. Simply put, the flows of goods in the world economy are not organized in such a way that the cessation of imports, through its effects on downstream production, hurts a country more than the cessation of exports, through its effects on upstream production, or the reverse. And therefore neither a trade deficit nor a trade surplus confers power in and of itself.

The largest bilateral deficit in dollars, that of the United States with China, is a case in point. The effect of the trade imbalance on this pair is only $0.005$, for the American import premium of $-0.24$ and the Chinese premium of $0.26$ are nearly equal in size and opposite in sign.\footnote{In Online Appendix~\ref{oa:largest-deficits}, we examine the twenty largest bilateral deficits in dollars of the 2022 trade table, the American deficit with China among them.} Among the twenty largest bilateral deficits in dollars the effect of the trade imbalance is largest on that of the United Kingdom with Norway, at $0.23$, a little over a third of the median asymmetry. Britain buys from Norway mostly oil and gas, both of them inputs, so the British import premium is as high as $0.54$, and since the Norwegian premium of $0.13$ is positive as well, the trade imbalance works against Britain. On the other nineteen of these deficits the effect of the trade imbalance stays below $0.1$, which is less than a sixth of the median asymmetry.
\FloatBarrier

\section{An Empirical Assessment of Our Measure of Power}
\label{sec:validation}

Power in human affairs is notoriously difficult to measure, not because it is a rare or subtle phenomenon but precisely because it is pervasive and often quietly so. A blunt, perhaps crude, expression of economic power is the initiation of a sanction, and we use it to examine the empirical significance of our measure of power in international trade. The state that initiates a sanction is presumably the more powerful of the two, for the weaker side has little reason to open a contest it expects to lose. The test is coarse, but it can be carried out on the historical record.

Every act of sanction has a date, an initiator, and a target. We run the model on the world trade table of that date and compute the power of the initiator and of the target as they stood when the decision was taken.  Note that the initiator of sanctions is seldom a single state, so the unit of the test is the decision, and the actor is the group that made it. We sever every link between the acting group and the target group, in both directions, in every sector, and for final and intermediate use alike, and compare the two groups' proportional activity losses as Definition~\ref{def:country_vulnerability} compares two countries. When each group is a single country, the group trade severance is the bilateral severance of Section~\ref{sec:statistics}, so the test and the Hirschman matrix measure power on one scale.

We run the test on two records of sanctions. Our own record holds $45$ decisions to restrict goods trade, $36$ initiations and $9$ retaliations, taken between 1998 and 2022 with every sender and every target among the $80$ economies of the world trade table. We built it from official program lists, government releases, and documented case studies. The other record is the Global Sanctions Data Base (GSDB) of \citet{felbermayr2020gsdb,gsdbR4}, which holds $1{,}547$ sanction cases imposed between 1949 and 2023 and, grouped by target and starting year, gives $185$ sanctions episodes inside the benchmark span. Our record takes the decision as its unit and the group of governments that took it as the actor, distinguishes retaliations from initiations, and admits the informal instruments that formal program lists do not carry, sanitary bans and customs slowdowns among them. Our record is also more recent, with a median decision in 2018 against 2014 for the GSDB episodes. Online Appendix~\ref{oa:datasets} gives the inclusion rules, the sources, and the overlap between the two records.

Table~\ref{tab:val-power-groups} reports the results. The sender was the more powerful side in 32 of the 36 initiations and in 3 of the 9 retaliations. Had each initiator picked its side by the toss of a coin, the chance of landing on the more powerful side in 32 or more of the 36 initiations would be about one in a million.\footnote{The result does not rest on the informal measures. The initiator is the more powerful side in 21 of the 22 formal initiations, those announced in a legal instrument, and in 11 of the 14 informal ones (Online Appendix~\ref{oa:datasets}).} The power asymmetry between initiator and target is also large, the median initiation putting $94\%$ of the combined loss on the target. 

Four initiations come from the weaker side. In two of them the asymmetry is so close to zero that the pair is all but balanced, Saudi Arabia against Canada in 2018 at $\mathcal H = -0.06$ and China against Japan in 2010 at $-0.08$, so neither initiator was meaningfully the weaker side. In the other two the initiator was the weaker side by a wide margin, India against China in 2020 at $-0.65$ and Pakistan against India in 2019 at $-0.81$. Every one of the four exceptions followed a territorial or other political shock.

The coin calculation is descriptive and is not a model of who initiates, since sender identities are far from random and the distribution of power is itself skewed (Section~\ref{subsec:hirschman_matrix}). The United States and China, which account for 17 of the 36 initiations, are more powerful than nearly every other economy, so the sign test has little room to fail when they act. But the other senders, for whom it has room to fail, also acted from the more powerful side of the relationship they chose in all but three of their initiations.\footnote{On the 2022 trade table the United States is more powerful than all 79 other economies and China than 78. The sign test can still fail on the trade table of an earlier year, and it does once, for China against Japan in 2010. Online Appendix~\ref{oa:datasets} compares, for the United States and for China, the advantage each holds over the targets it chose with the advantage it holds over its other partners, and neither concentrates its choices where its advantage is greatest.}

Six of the nine retaliations are aimed back at the coercing side, by Russia in 2014 and again with its 2022 counter-bans, China in 2018, Korea in 2019, Ukraine in 2016, and Belarus in 2022, and all six respond from the weaker side. Retaliation from the weaker side is to be expected, since a state retaliates when a stronger one has already moved against it. The three retaliations from strength are Russia's 2015 fish embargo on Latvia and Estonia, its 2016 food embargo on Ukraine, and its 2022 gas cutoff of Poland and Bulgaria. In each case Russia targeted one or two members of the coalition that had sanctioned it, and it was the more powerful side only because it picked those smaller adversaries and not the coalition itself. 

The result holds up on the Global Sanctions Data Base, whose episodes group cases by target and starting year and do not record the acting unit that our own record takes. The sender was the more powerful side in $170$ of the $185$ episodes. Splitting by sender type leaves the result unchanged, at $22$ of the $24$ coalition episodes and $148$ of the $161$ single-sender episodes, so the result does not
turn on how cases were grouped. Nor does it turn on the American cases. Excluding every episode with the United States among the senders leaves $123$ of $138$, and excluding China as well leaves $94$ of $109$. The GSDB panel of Table~\ref{tab:val-power-groups} collects these counts next to our own. 

\begin{figure}[!htbp]
\centering
\includegraphics[width=0.72\linewidth]{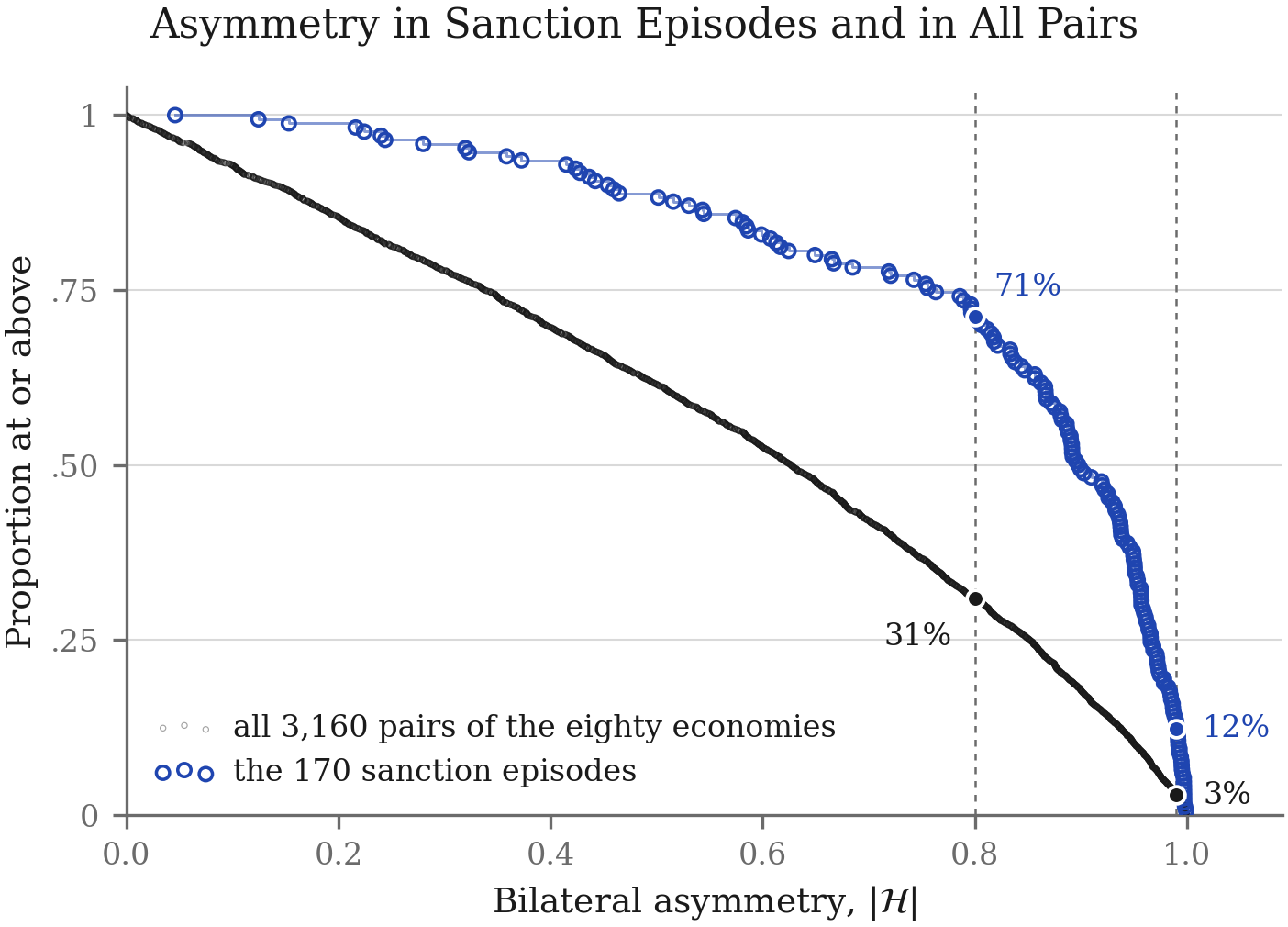}
\caption{Small black circles are all $3{,}160$ pairs at the favorable side's
$|\mathcal H|$, and larger blue circles are the $170$ GSDB episodes \citep{gsdbR4}
whose sender held the favorable side. Dotted rules mark $0.8$ and $0.99$. Baseline calibration $\tau=0.30$, $\rho=-1$, $\varrho=-1$, $\delta=0.10$.}
\label{fig:val-gsdb-ccdf}
\end{figure}

The count shows which side of a relationship coerces, and it leaves open how asymmetric the coerced relationships are. Figure~\ref{fig:val-gsdb-ccdf}
compares two distributions, the $170$ episodes whose sender held the
favorable side against all $3{,}160$ pairs of the eighty economies. Each curve gives, for every level of
asymmetry, the share of its own sample at or above that level. The episode curve lies above the curve for all pairs everywhere, and the gap between the two curves widens as the asymmetry rises. Relationships above $|\mathcal H|=0.8$ account for $71\%$ of the episodes against $31\%$ of all pairs, and above $0.99$ the shares are $12\%$ and $3\%$. Coercion takes place in the extreme tail of a distribution that is already asymmetric. The comparison sets the episodes against all pairs of the eighty economies, so it measures where coercion falls in the world distribution of power.

\begin{table}[!htbp]
\centering
\small
\begin{tabular}{lrrr}
\toprule
& Decisions & Sender favorable & Median $\mathcal H$ \\
\midrule
All decisions & 45 & 35 & $+0.84$ \\
\quad Initiations & 36 & 32 & $+0.87$ \\
\quad Retaliations & 9 & 3 & $-0.50$ \\
\addlinespace[0.35em]
Formal measures & 30 & 23 & $+0.86$ \\
Informal measures & 15 & 12 & $+0.80$ \\
Standard instruments only & 38 & 29 & $+0.82$ \\
\addlinespace[0.35em]
\multicolumn{4}{l}{\emph{The same test on the Global Sanctions Data Base}} \\
GSDB episodes & 185 & 170 & $+0.89$ \\
\quad Coalition senders & 24 & 22 & $+0.99$ \\
\quad Single senders & 161 & 148 & $+0.87$ \\
\quad Senders other than the United States & 138 & 123 & $+0.85$ \\
\midrule
\multicolumn{4}{l}{\emph{The four initiations from the weaker side}} \\
Saudi Arabia against Canada, 2018 & & & $-0.06$ \\
China against Japan, 2010 & & & $-0.08$ \\
India against China, 2020 & & & $-0.65$ \\
Pakistan against India, 2019 & & & $-0.81$ \\
\bottomrule
\end{tabular}
\caption{$\mathcal H$ is positive when the target is the weaker side. The
standard-instrument row removes the tariff and preference-withdrawal decisions
whose classification is debated. Online Appendix~\ref{oa:datasets} sets out the inclusion rules, the provenance of each decision, and where the full decision list is deposited. Baseline calibration $\tau=0.30$, $\rho=-1$, $\varrho=-1$, $\delta=0.10$.}
\label{tab:val-power-groups}
\end{table}

\FloatBarrier
\section{Concluding Remarks}
\label{sec:conclusion}

Every trading relationship distributes, alongside its mutual benefits, an unequal capacity to harm, and the side that can better do without the relationship holds power over the side that cannot.  The practice of strategically using trade for leverage is far older than the science of economics.  The \emph{Guanzi}, for instance, records Qi buying up the weapons of the rival state of Hengshan until its people abandoned farming for the forge, then cornering the grain they no longer grew so as to exert complete control over their material lives. Germany in the 1930s bound the agrarian economies of southeastern Europe to its market on the same principle, cultivating their dependence on it as the sole market for produce and the sole supplier of industrial goods, while Germany itself had diversified among several such agrarian nations \citep{ellis1941,basch1943}. The idea of asymmetric dependence and what it could mean for the wealth of nations is centuries old, the most lucid rendition of which is perhaps Hirschman's \citeyearpar{hirschman1945} book \emph{National Power and the Structure of Foreign Trade}. In this paper, we formalized the idea and measured the distribution of power in the world economy using a general equilibrium model of the world trade network. Our model combines two ingredients, each vital to the question of power. The first is the ability of buyers to find new sellers of inputs and of sellers to find new buyers of their goods in the aftermath of a sanction. Trade dependence, particularly its most asymmetrical form, emerges from flows that cannot be substituted. Letting the network shape how the severed trade is replaced is of first importance in modeling sanctions. The second ingredient is the complementarity of inputs in production. The loss of an input in one part of the economy damages production downstream. Sanctions can damage nations that were no part of the original target. Complementarity among inputs interacts with the two-sided adjustment of the trade network, and together they determine the sanction equilibrium and the distribution of its losses. The costs we measure are declines in production, not the transport costs of a detour, and they arise primarily because diverting flows from one nation to another in response to a sanction does not keep input bundles from being distorted.\footnote{The classical argument runs the other way, holding that the welfare cost of a small price distortion is second order, a quadratic form in the size of the shock \citep{harberger1964}. Some economists have argued that the loss scales with the affected expenditure share, which is to say that a country that trades little with the sanctioned partner loses little \citep{arkolakis2012,costinotrodriguez2014}. In this view trade flows, and therefore production, are by and large diverted from one nation to another rather than lost. Neither argument holds in a world where input complementarity propagates through the production network \citep{baqaeefarhi2019}.}

The distribution of power among the trading nations of the world is highly skewed, and a handful of economies are more powerful than nearly everyone. We map the bilateral asymmetries of all nations into an index that runs from balance at zero to complete asymmetry at one. The world stands close to six tenths. In the median trade relationship, a trade severance costs the weaker member more than four times as much as it costs the stronger one. The worst vulnerabilities are not small. A trade severance with Russia would cost Belarus over a tenth of its economic activity, and one with China would cost Vietnam the same, against six tenths of one percent or less on the other side. China and the United States, at this point in history, sit in the more balanced half of the world's trading relationships, with the United States on the more powerful side, since a trade severance would cost it less than half of what it would cost China. Here, as elsewhere in the world economy, the trade balance does not carry much information about power. 

Much of the skew in the distribution of power follows economic size. The same severed dollars are a smaller fraction of a larger economy's activity, and this base effect accounts for nearly all of the relation between size and power. A lost dollar of trade costs the larger economy slightly more, on average, than it costs the smaller one. This is because more of the supply chain behind the lost dollar is at home, and more of what the larger economy buys from its partner is inputs. But size is a much weaker measure of power among economies of comparable size, and the smaller economy is the more powerful side on nearly one in four of the pairs of economies within a factor of two of one another in size. The measure also accords with the historical record of economic coercion. In thirty-two of the thirty-six sanctions initiated between 1998 and 2022 that we collected, and in nine of every ten episodes of the Global Sanctions Data Base, the initiator was the more powerful side of the relationship.

\newpage
\bibliography{ref}

@article{baqaeefarhi2019,
  author  = {Baqaee, David Rezza and Farhi, Emmanuel},
  title   = {The Macroeconomic Impact of Microeconomic Shocks: Beyond Hulten's Theorem},
  journal = {Econometrica},
  year    = {2019},
  volume  = {87},
  number  = {4},
  pages   = {1155--1203}
}

@article{farrell2019,
  author  = {Farrell, Henry and Newman, Abraham L.},
  title   = {Weaponized Interdependence: How Global Economic Networks Shape State Coercion},
  journal = {International Security},
  year    = {2019},
  volume  = {44},
  number  = {1},
  pages   = {42--79}
}

@book{hirschman1945,
  author    = {Hirschman, Albert O.},
  title     = {National Power and the Structure of Foreign Trade},
  publisher = {University of California Press},
  year      = {1945}
}

@book{baldwin1985,
  author    = {Baldwin, David A.},
  title     = {Economic Statecraft},
  publisher = {Princeton University Press},
  year      = {1985}
}

@book{keohanenye1977,
  author    = {Keohane, Robert O. and Nye, Joseph S.},
  title     = {Power and Interdependence: World Politics in Transition},
  year      = {1977},
  publisher = {Little, Brown},
  address   = {Boston}
}

@article{sinkhorn1967concerning,
  author  = {Sinkhorn, Richard and Knopp, Paul},
  title   = {Concerning Nonnegative Matrices and Doubly Stochastic Matrices},
  journal = {Pacific Journal of Mathematics},
  year    = {1967},
  volume  = {21},
  number  = {2},
  pages   = {343--348}
}

@book{bacharach1970,
  author    = {Bacharach, Michael},
  title     = {Biproportional Matrices and Input--Output Change},
  publisher = {Cambridge University Press},
  address   = {Cambridge},
  year      = {1970}
}

@book{stonebrown1962,
  author    = {Stone, Richard and Brown, Alan},
  title     = {A Computable Model of Economic Growth},
  publisher = {Chapman and Hall},
  address   = {London},
  year      = {1962},
  note      = {A Programme for Growth, vol. 1}
}

@article{hulten1978,
  author  = {Hulten, Charles R.},
  title   = {Growth Accounting with Intermediate Inputs},
  journal = {Review of Economic Studies},
  year    = {1978},
  volume  = {45},
  number  = {3},
  pages   = {511--518}
}

@article{sato1967,
  author  = {Sato, Ryuzo},
  title   = {A Two-Level Constant-Elasticity-of-Substitution Production Function},
  journal = {Review of Economic Studies},
  year    = {1967},
  volume  = {34},
  number  = {2},
  pages   = {201--218},
  doi     = {10.2307/2296809}
}

@article{clayton2026geoeconomics,
  author  = {Clayton, Christopher and Maggiori, Matteo and Schreger, Jesse},
  title   = {A Framework for Geoeconomics},
  journal = {Econometrica},
  year    = {2026},
  volume  = {94},
  number  = {1},
  pages   = {105--136}
}

@article{ghironi2025sanctions,
  author  = {Ghironi, Fabio and Kim, Daisoon and Ozhan, Galip Kemal},
  title   = {International Trade and Macroeconomic Dynamics with Sanctions},
  journal = {Journal of Monetary Economics},
  year    = {2025}
}

@article{felbermayr2020gsdb,
  author  = {Felbermayr, Gabriel and Kirilakha, Aleksandra and Syropoulos, Constantinos and Yalcin, Erdal and Yotov, Yoto V.},
  title   = {The Global Sanctions Data Base},
  journal = {European Economic Review},
  year    = {2020},
  volume  = {129},
  pages   = {103561}
}

@book{krasnoselskii1964,
  author    = {Krasnosel'skii, Mark A.},
  title     = {Positive Solutions of Operator Equations},
  year      = {1964},
  publisher = {P. Noordhoff},
  address   = {Groningen}
}

@article{hawkinssimon1949,
  author    = {Hawkins, David and Simon, Herbert A.},
  title     = {Note: Some Conditions of Macroeconomic Stability},
  journal   = {Econometrica},
  year      = {1949},
  volume    = {17},
  number    = {3/4},
  pages     = {245--248}
}

@article{bachmann2022whatif,
  author  = {Bachmann, R{\"u}diger and Baqaee, David and Bayer, Christian and Kuhn, Moritz and L{\"o}schel, Andreas and Moll, Benjamin and Peichl, Andreas and Pittel, Karen and Schularick, Moritz},
  title   = {What If? The Macroeconomic and Distributional Effects for Germany of a Stop of Energy Imports from Russia},
  journal = {Economica},
  year    = {2024},
  volume  = {91},
  number  = {364},
  pages   = {1157--1200}
}

@article{crozethinz2020,
  author  = {Crozet, Matthieu and Hinz, Julian},
  title   = {Friendly Fire: The Trade Impact of the Russia Sanctions and Counter-Sanctions},
  journal = {Economic Policy},
  year    = {2020},
  volume  = {35},
  number  = {101},
  pages   = {97--146}
}

@article{joshi2024networks,
  author  = {Joshi, Sumit and Mahmud, Ahmed Saber and Nandy, Abhinaba and Sarangi, Sudipta},
  title   = {Sanctions in Directed Trade Networks},
  journal = {Review of International Economics},
  year    = {2024},
  volume  = {32},
  number  = {1},
  pages   = {72--108}
}

@article{imbspauwels2024,
  author  = {Imbs, Jean and Pauwels, Laurent},
  title   = {An Empirical Approximation of the Effects of Trade Sanctions with an Application to Russia},
  journal = {Economic Policy},
  year    = {2024},
  volume  = {39},
  number  = {117},
  pages   = {159--200}
}

@techreport{prebisch1950,
  author      = {Prebisch, Ra\'ul},
  title       = {The Economic Development of Latin America and Its Principal Problems},
  institution = {United Nations Economic Commission for Latin America},
  year        = {1950}
}

@article{singer1950,
  author  = {Singer, Hans W.},
  title   = {The Distribution of Gains between Investing and Borrowing Countries},
  journal = {American Economic Review},
  year    = {1950},
  volume  = {40},
  number  = {2},
  pages   = {473--485}
}

@article{gsdbR4,
  author  = {Yalcin, Erdal and Kirilakha, Aleksandra and Syropoulos, Constantinos and Yotov, Yoto V.},
  title   = {The Global Sanctions Data Base---Release 4: The Heterogeneous Effects of the Sanctions on {R}ussia},
  journal = {The World Economy},
  year    = {2025},
  volume  = {48},
  number  = {5},
  pages   = {1121--1148}
}

@book{ellis1941,
  author    = {Ellis, Howard S.},
  title     = {Exchange Control in Central Europe},
  publisher = {Harvard University Press},
  address   = {Cambridge, MA},
  year      = {1941}
}

@book{basch1943,
  author    = {Basch, Anton\'{i}n},
  title     = {The Danube Basin and the German Economic Sphere},
  publisher = {Columbia University Press},
  address   = {New York},
  year      = {1943}
}

@article{harberger1964,
  author  = {Harberger, Arnold C.},
  title   = {The Measurement of Waste},
  journal = {American Economic Review},
  volume  = {54},
  number  = {3},
  pages   = {58--76},
  note    = {Papers and Proceedings},
  year    = {1964}
}

@article{arkolakis2012,
  author  = {Arkolakis, Costas and Costinot, Arnaud and Rodr{\'i}guez-Clare, Andr{\'e}s},
  title   = {New Trade Models, Same Old Gains?},
  journal = {American Economic Review},
  volume  = {102},
  number  = {1},
  pages   = {94--130},
  year    = {2012}
}

@article{dreze1975,
  author  = {Dr\`{e}ze, Jacques H.},
  title   = {Existence of an Exchange Equilibrium under Price Rigidities},
  journal = {International Economic Review},
  year    = {1975},
  volume  = {16},
  number  = {2},
  pages   = {301--320}
}

@article{benassy1975,
  author  = {B\'{e}nassy, Jean-Pascal},
  title   = {Neo-{K}eynesian Disequilibrium Theory in a Monetary Economy},
  journal = {The Review of Economic Studies},
  year    = {1975},
  volume  = {42},
  number  = {4},
  pages   = {503--523}
}

@article{barrogrossman1971,
  author  = {Barro, Robert J. and Grossman, Herschel I.},
  title   = {A General Disequilibrium Model of Income and Employment},
  journal = {The American Economic Review},
  year    = {1971},
  volume  = {61},
  number  = {1},
  pages   = {82--93}
}

@book{malinvaud1977,
  author    = {Malinvaud, Edmond},
  title     = {The Theory of Unemployment Reconsidered},
  publisher = {Basil Blackwell},
  address   = {Oxford},
  year      = {1977}
}

@article{nearyroberts1980,
  author  = {Neary, J. Peter and Roberts, Kevin W. S.},
  title   = {The Theory of Household Behaviour under Rationing},
  journal = {European Economic Review},
  year    = {1980},
  volume  = {13},
  number  = {1},
  pages   = {25--42}
}

@techreport{liuyang2026power,
  author      = {Liu, Ernest and Yang, David Y.},
  title       = {International Power},
  institution = {National Bureau of Economic Research},
  type        = {Working Paper},
  number      = {34006},
  year        = {2025}
}

@article{hausmannschetter2024,
  author  = {Hausmann, Ricardo and Schetter, Ulrich and Yildirim, Muhammed A.},
  title   = {On the Design of Effective Sanctions: The Case of Bans on Exports to Russia},
  journal = {Economic Policy},
  year    = {2024},
  volume  = {39},
  number  = {117},
  pages   = {109--153}
}

@incollection{costinotrodriguez2014,
  author    = {Costinot, Arnaud and Rodr{\'i}guez-Clare, Andr{\'e}s},
  title     = {Trade Theory with Numbers: Quantifying the Consequences of Globalization},
  booktitle = {Handbook of International Economics},
  volume    = {4},
  editor    = {Gopinath, Gita and Helpman, Elhanan and Rogoff, Kenneth S.},
  publisher = {Elsevier},
  year      = {2014},
  pages     = {197--261}
}

@incollection{jones1971,
  author    = {Jones, Ronald W.},
  title     = {A Three-Factor Model in Theory, Trade, and History},
  booktitle = {Trade, Balance of Payments and Growth: Papers in International Economics in Honor of Charles P. Kindleberger},
  editor    = {Bhagwati, Jagdish N. and Jones, Ronald W. and Mundell, Robert A. and Vanek, Jaroslav},
  publisher = {North-Holland},
  address   = {Amsterdam},
  year      = {1971},
  pages     = {3--21}
}

@article{neary1978,
  author  = {Neary, J. Peter},
  title   = {Short-Run Capital Specificity and the Pure Theory of International Trade},
  journal = {Economic Journal},
  year    = {1978},
  volume  = {88},
  number  = {351},
  pages   = {488--510}
}

@book{mun1664,
  author    = {Mun, Thomas},
  title     = {England's Treasure by Forraign Trade},
  publisher = {Thomas Clark},
  address   = {London},
  year      = {1664}
}

@incollection{hume1752,
  author    = {Hume, David},
  title     = {Of the Balance of Trade},
  booktitle = {Political Discourses},
  publisher = {A. Kincaid and A. Donaldson},
  address   = {Edinburgh},
  year      = {1752}
}

@book{krugman1996,
  author    = {Krugman, Paul},
  title     = {Pop Internationalism},
  publisher = {MIT Press},
  address   = {Cambridge, MA},
  year      = {1996}
}

@misc{navarroross2016,
  author       = {Navarro, Peter and Ross, Wilbur},
  title        = {Scoring the {Trump} Economic Plan: Trade, Regulatory, and Energy Policy Impacts},
  howpublished = {Campaign white paper},
  year         = {2016}
}

@article{rothbarth1941,
  author  = {Rothbarth, Erwin},
  title   = {The Measurement of Changes in Real Income under Conditions of Rationing},
  journal = {Review of Economic Studies},
  year    = {1941},
  volume  = {8},
  number  = {2},
  pages   = {100--107}
}

@book{davishaltiwangerschuh1996,
  author    = {Davis, Steven J. and Haltiwanger, John C. and Schuh, Scott},
  title     = {Job Creation and Destruction},
  publisher = {MIT Press},
  address   = {Cambridge, MA},
  year      = {1996}
}
\bibliographystyle{aea}

\newpage
\FloatBarrier
\appendix
\numberwithin{equation}{section}
\numberwithin{theorem}{section}
\numberwithin{proposition}{section}
\numberwithin{lemma}{section}
\numberwithin{corollary}{section}
\numberwithin{definition}{section}

\section{Proofs of the Main Results}
\label{app:m-coreproofs}

This appendix proves the formal claims of the paper, and it states and proves the auxiliary results those claims rest on, in the order the argument uses them. All statements are restricted to the active country-sector set of Section~\ref{subsec:model_benchmark} and are made under the standing conditions of Section~\ref{sec:model}. These conditions are a sanction in $\Omega$, the positive attenuation \eqref{eq:m-leak}, an unrecovered share in $[0,1]$, the two curvature parameters in their intervals, leakages in $(0,1)$, and a connected benchmark support. A statement that needs more says so. The Online Appendix carries the local representation of the technology, the conditional-uniqueness and convergence-rate results, the flexible-price economy, and the calibration. Throughout, $G_j(\kappa):=\bigl[\beta_j+(1-\beta_j)\,\kappa^{\varrho}\bigr]^{1/\varrho}$ denotes the outer layer of the output ratio \eqref{eq:m-lambda-rule}, with $G_j(0):=0$ its continuous value at the boundary, so that $\lambda^{(\omega)}_j(\mathbf h)=G_j(\kappa^{(\omega)}_j(\mathbf h))$.

\subsection{Coordinate and aggregation conventions}

We state formally here the coordinate conventions of Section~\ref{subsec:model_benchmark} and the sector aggregation operator, both of which the proofs of this appendix use throughout. The sector aggregation operator is built from the partition of country-sectors into sectors alone, so it does not depend on the sanction.

\begin{definition}[Country-sectors and their coordinates]
\label{def:m-country-sector}
A country-sector $i\in\{1,\dots,N\}$ is one active
pair in the country-by-sector network. Two projections return its coordinates,
\[
\mathcal C(i)\in\{1,\dots,C\}\ \ \text{(the country of $i$)},
\qquad
\mathcal S(i)\in\{1,\dots,S\}\ \ \text{(the sector of $i$)}.
\]
The pair map $i\mapsto\bigl(\mathcal C(i),\mathcal S(i)\bigr)$ is injective
into $\{1,\dots,C\}\times\{1,\dots,S\}$. Before we drop inactive country-sectors the map
is a bijection and the full network has $CS$ pairs. On the retained set, the country-sectors of a country and those of a sector are the level sets
\[
\{\,i:\mathcal C(i)=c\,\}
\qquad\text{and}\qquad
\mathcal N_r:=\{\,i:\mathcal S(i)=r\,\}.
\]
\end{definition}

\begin{definition}[Sector aggregation operator]
\label{def:m-aggregate}
The sector aggregation operator $\mathcal F$ adds the entries of a country-sector vector $\mathbf x$ over each sector,
\begin{equation}
\mathcal F(\mathbf x)_r:=\sum_{i\in\mathcal N_r}x_i.
\label{eq:m-aggregate}
\end{equation}
The operator collapses the $N$ country-sectors onto the $S$ sectors. On a matrix it acts column by column, so that $\mathcal F(\mathbf B)_{rj}=\sum_{i\in\mathcal N_r}b_{ij}$, and the benchmark sector share of
Section~\ref{subsec:model_benchmark} is $\alpha_{rj}=\mathcal F(\mathbf A)_{rj}$.
\end{definition}

\subsection{Proof of Lemma~\ref{lem:m-rigidity}}
\label{app:m-proof-rigidity}

\begin{proof}
Write $x:=\gamma^{(c'\notnotleftrightarrow c)}_c>0$ and $y:=\gamma^{(c'\notnotleftrightarrow c)}_{c'}>0$, so that the loss ratio \eqref{eq:m-logratio} is $r_{cc'}=x/y$ and the asymmetry \eqref{eq:bounded-Hirschman} is $\mathcal H_{cc'}=(x-y)/(x+y)$.

For part \textnormal{(i)}, divide the numerator and the denominator of the asymmetry by $y$, which gives $\mathcal H_{cc'}=(x/y-1)/(x/y+1)=\psi(r_{cc'})$ by \eqref{eq:m-psi}.

For part \textnormal{(ii)}, abbreviate $r_{cc'}$ to $r$, so that $\mathcal H_{cc'}=\psi(r)$ and $dr=r\,d\log r$ along the path. Since $\psi'(r)=2/(r+1)^{2}$ and $1-\psi(r)^{2}=4r/(r+1)^{2}$, the chain rule gives
\[
d\mathcal H_{cc'}
=r\,\psi'(r)\,d\log r
=\frac{2r}{(r+1)^{2}}\,d\log r
=\tfrac12\bigl(1-\mathcal H_{cc'}^{2}\bigr)\,d\log r.
\]

For part \textnormal{(iii)}, the asymmetry $\psi(r)$ has the derivative $\tfrac12(1-\psi(r)^{2})$ in $\log r$ by part \textnormal{(ii)}, which lies in $(0,\tfrac12]$ on the whole line, so the mean value theorem gives $|\Delta\mathcal H_{cc'}|\le\tfrac12|\Delta\log r_{cc'}|$. The log loss ratio is a difference, $\Delta\log r_{cc'}=\Delta\log x-\Delta\log y$, and the triangle inequality gives the second bound of \eqref{eq:m-rigidity}. A perturbation carrying $(x,y)$ to $(\varsigma x,\varsigma y)$ for a common $\varsigma>0$ leaves the loss ratio $r_{cc'}=x/y$ unchanged, so it leaves $\mathcal H_{cc'}$ unchanged. This proves the lemma.
\end{proof}

\subsection{The block-wise program for balancing disrupted input flows}
Throughout this appendix a column index $j$ runs over the buyers of Section~\ref{subsec:model_benchmark}, the producers and the households, with spending $w_j$ equal to $(1-\beta_j)s_j$ for a producer and to $F^{c}$ for the household of country $c$, and with sourcing shares $a_{ij}$ read off $\mathbf A$ or $\mathbb A$ accordingly. The producer block of the balanced matrix is $\mathbf A_\omega$ and the household block $\mathbb A_\omega$, and only the producer block enters the propagation of activity.

Fix an input sector $r$. The block-wise program balances the disrupted flow
block $\check{\mathbf Z}^{(r)}=[\check a_{ij}w_j]_{\mathcal S(i)=r}$. For buyer
$j$, column $j$ of the disrupted coefficient block
$\check{\mathbf A}^{(r)}=[\check a_{ij}]$ contains the disrupted sourcing
coefficients. Multiplying that column by the buyer's benchmark spending
$w_j$ gives column $j$ of
$\check{\mathbf Z}^{(r)}$. The rows of the disrupted flow block correspond to
the producers $\mathcal N_r:=\{i:\mathcal S(i)=r\}$ of input $r$, and its
columns correspond to buyers $j$. The flow and coefficient blocks have the same
support. The balanced block must meet the seller row targets
$g^{(\omega)}_i$ of \eqref{eq:m-rowtarget} and the buyer flow column
targets $t^{(\omega)}_{rj}w_j$ built from \eqref{eq:m-coltarget}.

Before scaling, we remove zero-margin rows and columns and define
\[
\mathcal I_r^{\omega}:=\{i\in\mathcal N_r:g^{(\omega)}_i>0\},
\qquad
\mathcal J_r^{\omega}:=\{j:t^{(\omega)}_{rj}w_j>0\}.
\]
We apply both the flow-balancing program and the Sinkhorn iteration only to
$\mathcal I_r^{\omega}\times\mathcal J_r^{\omega}$ and restore omitted rows
and columns as zeros afterward. If either active set is empty, margin
consistency implies that both active sets are empty, so there is no positive
flow block to balance. Removing zero-margin rows and columns does not alter any
positive flow, and it avoids undefined $0/0$ updates. The remaining targets
are strictly positive, at or below the corresponding benchmark margins,
seller sales for rows and buyer-sector input purchases for columns, and
consistent in value within the block:
\begin{equation}
\sum_{i\in\mathcal I_r^{\omega}}g^{(\omega)}_i
\;=\;
\sum_{j\in\mathcal J_r^{\omega}}t^{(\omega)}_{rj}\,w_j.
\label{eq:m-margincons}
\end{equation}
Both sums equal the block's total benchmark flow minus $\tau(1-\delta)$ times
its benchmark flow on severed links. Dividing any feasible flow column by $w_j$
gives the prescribed buyer-sector coefficient total
$t^{(\omega)}_{rj}$.

These margins define a transportation polytope on the disrupted support. The
balanced flow block is the information projection (I-projection) of the
disrupted flow pattern onto this polytope:
\begin{equation}
\mathbf Z^{(r)}_\omega
\;=\;
\arg\min_{X\ge 0,\ \operatorname{supp}X\subseteq
\operatorname{supp}\check{\mathbf A}^{(r)}_{\mathcal I_r^{\omega}\times\mathcal J_r^{\omega}}}
\ \sum_{\substack{i\in\mathcal I_r^{\omega},\,j\in\mathcal J_r^{\omega}:\\
\check a_{ij}>0}}
X_{ij}\,\log\frac{X_{ij}}{\check a_{ij}w_j}
\;\;\text{s.t.}\;\;
X\mathbf 1=\mathbf g^{(\omega)}_{\mathcal I_r^{\omega}},\ \
X^{\!\top}\mathbf 1=(\mathbf t^{(\omega)}\!\circ\mathbf w)_{\mathcal J_r^{\omega}}.
\label{eq:m-balprog}
\end{equation}
We use the convention $0\log(0/q)=0$ for $q>0$. The omitted linear terms
$-X_{ij}+\check a_{ij}w_j$ are constant after summation on the prescribed
margin set, so including them gives the same minimizer.
The coefficient block is recovered as
$a^{(\omega)}_{ij}=(\mathbf Z^{(r)}_\omega)_{ij}/w_j$ on the active block and
set to zero on omitted rows and columns.

The optimizer of \eqref{eq:m-balprog} has the biproportional form
$X_{ij}=u_i\,\check a_{ij}w_j\,v_j$ with positive active-block scalings
$(u_i,v_j)$. Dividing by $w_j$ gives the coefficient
$a^{(\omega)}_{ij}=X_{ij}/w_j=u_i\check a_{ij}v_j$ in
\eqref{eq:m-reroute}. Within each block problem, we hold the sector index $r$
fixed and write $v_j:=v_{rj}$. The program in \eqref{eq:m-balprog} contains
no transport cost. In this appendix, the disrupted benchmark flow pattern
$\check{\mathbf Z}^{(r)}$ serves as the base measure. The balanced flow block
is therefore the relative-entropy-minimizing adjustment consistent with the
prescribed margins.
Equivalently, the solution is the unique biproportional adjustment of the disrupted block
that satisfies both margins. Imposing only one margin recovers the corresponding closed-form column or row rescaling, the demand-oriented and the supply-oriented one-sided rules of Online Appendix~\ref{subsec:recomposition}.

In flow units, write
$\check Z_{ij}:=\check a_{ij}w_j$ for the disrupted flow kernel and
$Z_{\omega,ij}:=a^{(\omega)}_{ij}w_j$ for the balanced standing-order flow. All Sinkhorn
iterations act on these flow quantities, with row targets
$g^{(\omega)}_i$ and column targets $t^{(\omega)}_{rj}w_j$.

\subsection{Existence, uniqueness, and regularity of the balanced flow blocks}

\begin{proposition}[Existence, uniqueness, and regularity of the balanced flow blocks]
\label{prop:m-balancing}
For each sector $r$, after deleting the
zero-margin rows and columns, the flow-balancing program
\eqref{eq:m-balprog} has a unique
optimizer $\mathbf Z^{(r)}_\omega$. Extended by zeros to the full block, the
associated coefficient block $\mathbf A^{(r)}_\omega$ has the biproportional
form $a^{(\omega)}_{ij}=u_i\check a_{ij}v_{rj}$ on the active support, with
strictly positive active-block scalings. These scalings are unique up to the
reciprocal transformation
$(u,v)\mapsto(\chi u,\chi^{-1}v)$ on each connected component of the
support-reduced bipartite support graph. The optimizer is nonnegative,
supported on $\operatorname{supp}\check{\mathbf A}^{(r)}$, and meets the
prescribed seller row and buyer column targets. After one gauge is fixed in
each support component, the active-block scalings, and hence the optimizer,
depend $C^\infty$-smoothly on target margins admitting a feasible flow strictly positive on the fixed support.
In particular, $\mathbf A_\omega$ is smooth in $\tau$ on $[0,1]$ and is
independent of both CES exponents $(\rho,\varrho)$. In flow units,
positive attenuation gives
\[
g^{(\omega)}_i\ge\delta\sum_j a_{ij}w_j,
\qquad
 t^{(\omega)}_{rj}w_j\ge\delta\,\alpha_{rj}w_j
 \quad (r\in\mathcal R_j).
\]
At $\delta=0$, after zero-margin rows and columns are removed, a nonempty
restricted transportation polytope gives a unique optimizer. A representation
by finite strictly positive scalings additionally requires a feasible point
strictly positive on every retained support edge. Smoothness along a $\tau$-path additionally requires a
fixed retained support and persistent strict feasibility.
\end{proposition}

\begin{proof}
Fix a sector $r$ and consider the flow transportation polytope on its
support-reduced block,
\[
\mathcal U:=\bigl\{X\ge0:\operatorname{supp}X\subseteq
\operatorname{supp}\check{\mathbf A}^{(r)}_{\mathcal I_r^{\omega}\times\mathcal J_r^{\omega}},\ 
X\mathbf 1=(g^{(\omega)}_i)_{i\in\mathcal I_r^{\omega}},\ 
X^{\!\top}\mathbf 1=(t^{(\omega)}_{rj}w_j)_{j\in\mathcal J_r^{\omega}}\bigr\}.
\]
The polytope $\mathcal U$ is compact and convex. The following flow matrix
proves that it is nonempty:
\[
X^{*}_{ij}
\;:=\;
\bigl[1-\tau(1-\delta)\bigr]\,a_{ij}w_j\quad\text{on severed links},
\qquad
X^{*}_{ij}\;:=\;a_{ij}w_j\quad\text{otherwise}.
\]
On the support-reduced block, this matrix is strictly positive on
$\operatorname{supp}(\mathbf A^{(r)})
=\operatorname{supp}(\check{\mathbf A}^{(r)})$, since $1-\tau(1-\delta)\ge\delta>0$, and positive attenuation needs no separate support condition. The matrix meets the positive flow row
and column targets by inspection. Margin consistency
\eqref{eq:m-margincons} alone would not suffice on a restricted support. The
explicit matrix $X^*$ verifies the stronger prescribed-margin support
condition needed here.

The relative entropy
$\sum_{ij}X_{ij}\log\!\bigl(X_{ij}/(\check a_{ij}w_j)\bigr)$ is strictly
convex on $\mathcal U$, so the optimizer exists and is unique. The optimizer
is strictly positive on the retained support. Suppose a feasible optimizer had
$X_{ij}=0$ on a retained edge. Moving along the feasible line segment from
that optimizer toward $X^*$ would have directional derivative $-\infty$ at
that entry. Stationarity at the interior optimizer gives
$X_{ij}=u_i\check a_{ij}w_jv_{rj}$ with $u_i,v_{rj}>0$ on the active block.
Division by $w_j>0$ then gives
$a^{(\omega)}_{ij}=u_i\check a_{ij}v_{rj}$ there. Zero-margin rows and columns
are restored as zeros.

Let the bipartite support graph have connected components
$\mathcal G_1,\ldots,\mathcal G_K$. On each component, adding a constant to
the row log-scalings and subtracting it from the column log-scalings leaves
$X$ unchanged. No other gauge freedom remains. Hence the scalings are unique
up to one reciprocal scalar per component, while $X$ itself is unique.

To establish regularity of the optimizer in the target margins, write the
stationarity equations in $(\log\mathbf u,\log\mathbf v)$. Their Jacobian is congruent to
the weighted Laplacian of the bipartite support graph with edge weights
$X_{ij}$. Its kernel consists
exactly of the $K$ componentwise gauge directions. After fixing one gauge in
each component and restricting target perturbations to the compatible-margin
manifold, the Jacobian is nonsingular. The implicit function theorem therefore
gives $C^\infty$ dependence on strictly feasible targets on the fixed support. The
model's targets are affine in $\tau$, and the explicit $X^*(\tau)$ preserves
componentwise compatibility for every $\tau\in[0,1]$. The optimizer is
therefore smooth along the maintained parameter path. At $\delta=0$, compactness and strict convexity still give a unique optimizer whenever the restricted feasible set is nonempty. The same positivity and stationarity arguments give finite positive scalings when a feasible point is positive on every retained edge, and such a point is necessary because finite positive scalings preserve the kernel's support. This proves the proposition.
\end{proof}

\subsection{The benchmark equilibrium and its propagation bound}

\begin{lemma}[Benchmark activity vector]
\label{lem:m-benchmark}
The benchmark propagation matrix $\mathbf A(\mathbf I-\boldsymbol\beta)$ satisfies
\[
\|\mathbf A(\mathbf I-\boldsymbol\beta)\|_1=\max_j(1-\beta_j)<1.
\]
The matrix $\mathbf I-\mathbf A(\mathbf I-\boldsymbol\beta)$ is therefore invertible. The
benchmark equation
$\mathbf s=\mathbf f+\mathbf A(\mathbf I-\boldsymbol\beta)\,\mathbf s$ then has
the unique nonnegative solution
\[
\mathbf s=\bigl(\mathbf I-\mathbf A(\mathbf I-\boldsymbol\beta)\bigr)^{-1}\mathbf f.
\]
Moreover $\mathbf s\ge\mathbf f$.
\end{lemma}
\begin{proof}

Each column of \(\mathbf A(\mathbf I-\boldsymbol\beta)\) sums to \(1-\beta_j\),
so
\[
\|\mathbf A(\mathbf I-\boldsymbol\beta)\|_1=\max_j(1-\beta_j)<1.
\]
Since any induced norm bounds the spectral radius $\spr(\cdot)$, it follows that
\(\spr\bigl(\mathbf A(\mathbf I-\boldsymbol\beta)\bigr)<1\), and the Neumann series
\(\sum_{t\ge0}\bigl(\mathbf A(\mathbf I-\boldsymbol\beta)\bigr)^t\) converges to
\(\bigl(\mathbf I-\mathbf A(\mathbf I-\boldsymbol\beta)\bigr)^{-1}\). The
benchmark equation
\(\mathbf s=\mathbf f+\mathbf A(\mathbf I-\boldsymbol\beta)\mathbf s\) therefore
has the unique solution
\[
\mathbf s
=\bigl(\mathbf I-\mathbf A(\mathbf I-\boldsymbol\beta)\bigr)^{-1}\mathbf f
=\sum_{t\ge0}\bigl(\mathbf A(\mathbf I-\boldsymbol\beta)\bigr)^t\mathbf f.
\]
Since \(\mathbf A(\mathbf I-\boldsymbol\beta)\ge\mathbf 0\) and
\(\mathbf f\ge\mathbf 0\), every term of the series is nonnegative, so
\(\mathbf s\ge\mathbf 0\), and \(\mathbf A(\mathbf I-\boldsymbol\beta)\mathbf s\ge\mathbf 0\)
gives \(\mathbf s\ge\mathbf f\). This proves the lemma.
\end{proof}

\subsection{Structure of the sanctioned propagation matrix}

\begin{lemma}[Propagation under the sanction]
\label{lem:m-kernel}
The producer block
$\mathbf A_\omega$ is nonnegative and supported on
$\operatorname{supp}(\check{\mathbf A})$, and the household block $\mathbb A_\omega$ is nonnegative and supported on $\operatorname{supp}(\check{\mathbb A})$. Writing $\mathcal F$ for the sector
aggregation operator that adds a country-sector vector over each sector,
Definition~\ref{def:m-aggregate}, the sector-$r$ block of an affected
buyer $j$'s column sums to
\begin{equation}
\sum_{i\in\mathcal N_r}a^{(\omega)}_{ij}
=\mathcal F(\mathbf A_\omega)_{rj}
=\mathcal F(\mathbf A)_{rj}-\tau(1-\delta)\!\!\sum_{i\in\mathcal N_r:\,i\to j\text{ severed}}\!\!a_{ij},
\label{eq:m-blocksum}
\end{equation}
if sector $r$ retains a supplier for buyer $j$ and to zero otherwise, so every column
of $\mathbf A_\omega$ sums to at most one. Consequently
$\mathbf A_\omega(\mathbf I-\boldsymbol\beta)$ is nonnegative and
column substochastic, with
\[
\|\mathbf A_\omega(\mathbf I-\boldsymbol\beta)\|_1
\le\|\mathbf A(\mathbf I-\boldsymbol\beta)\|_1<1. 
\]
Moreover the supplier side sales obey
$\sum_b a^{(\omega)}_{ib}\,w_b=g^{(\omega)}_i\le s_i-f^{o}_i$, the sum running over producers and households, so the rationed exchange places no seller beyond its benchmark sales to producers and households. In the notation of the effective final demand $\mathbf f_\omega$, the row identity reads $[\mathbf A_\omega(\mathbf I-\boldsymbol\beta)\mathbf s]_i+f_{\omega,i}=s_i-\xi_{\omega,i}$.
\end{lemma}
\begin{proof}

A sector block with no severed link
stays at benchmark, so its entries are nonnegative, its buyer columns sum to
their benchmark sector mass, and its seller flow rows sum to
\(\sum_b a_{ib}w_b=s_i-f^{o}_i\) by \eqref{eq:m-flows}, leaving nothing further to check there. The work
is therefore all in the affected sector-\(r\) blocks. On such a block the fixed-price equilibrium with rationing of Definition~\ref{def:oa-rationed-exchange} exists by Proposition~\ref{prop:m-balancing}. Its allocation carries strictly positive scalings \((u_i)_{i\in\mathcal I_r^{\omega}}\) and \((v_{rj})_{j\in\mathcal J_r^{\omega}}\) on the active rows and columns, and it meets the row targets \eqref{eq:m-rowtarget} and column targets \eqref{eq:m-coltarget} exactly. The four claims of the lemma can then be taken
in turn.

Nonnegativity and support of the two blocks come first. On the block,
\(a^{(\omega)}_{ij}=u_i\,\check a_{ij}\,v_{rj}\) with \(u_i,v_{rj}>0\) and
\(\check a_{ij}\ge0\), so \(a^{(\omega)}_{ij}\ge0\), and
\(a^{(\omega)}_{ij}>0\) exactly where \(\check a_{ij}>0\), for producer and household columns alike. Hence
\(\mathbf A_\omega\ge\mathbf 0\) and
\(\operatorname{supp}(\mathbf A_\omega)=\operatorname{supp}(\check{\mathbf A})\), and the household block satisfies \(\mathbb A_\omega\ge\mathbf 0\) and \(\operatorname{supp}(\mathbb A_\omega)=\operatorname{supp}(\check{\mathbb A})\).
The rationed exchange creates no link the sanction had not left in place.

The sector-block column sums follow from the column constraint of the flow-balancing program, which reads, for every buyer \(j\) and every sector \(r\) that retains a supplier for \(j\),
\[
\sum_{i\in\mathcal N_r}a^{(\omega)}_{ij}
=\mathcal F(\mathbf A_\omega)_{rj}
=t^{(\omega)}_{rj}
=\mathcal F(\mathbf A)_{rj}-\tau(1-\delta)\!\!\sum_{i\in\mathcal N_r:\,i\to j\text{ severed}}\!\!a_{ij}.
\]
This is \eqref{eq:m-blocksum}. If sector \(r\) retains no supplier for \(j\),
the target is zero and the sector-\(r\) block of column \(j\) vanishes. Since
\(0\le\mathcal F(\mathbf A)_{rj}-\tau(1-\delta)\!\!\sum_{i\in\mathcal N_r:\,i\to j\text{ severed}}\!\!a_{ij}\le\mathcal F(\mathbf A)_{rj}\),
the block sum lies in \([0,\mathcal F(\mathbf A)_{rj}]\).

Summing those block sums over sectors gives the column totals. Using
\(\sum_r\mathcal F(\mathbf A)_{rj}=\sum_i a_{ij}=1\),
\begin{align*}
\sum_{i=1}^N a^{(\omega)}_{ij}
&=\sum_{r}\Bigl(\mathcal F(\mathbf A)_{rj}-\tau(1-\delta)\!\!\sum_{i\in\mathcal N_r:\,i\to j\text{ severed}}\!\!a_{ij}\Bigr)\\
&=1-\tau(1-\delta)\sum_{i:\,i\to j\text{ severed}}a_{ij}
\le 1.
\end{align*}
Here the last sum gives the total share of column \(j\) that \(\omega\) severs.
An unaffected buyer has \(\sum_i a^{(\omega)}_{ij}=1\), so every column of
\(\mathbf A_\omega\) is nonnegative and sums to at most one.

The row totals are governed by the seller constraint, which binds sales, a flow quantity, over every buyer column of the block,
\[
\sum_b a^{(\omega)}_{ib}\,w_b
=g^{(\omega)}_i
=(s_i-f^{o}_i)-\tau(1-\delta)\!\!\sum_{b:\,i\to b\,\text{severed}}\!\!a_{ib}\,w_b
\le s_i-f^{o}_i. \] The rationed exchange therefore pushes no seller beyond its benchmark sales to producers and households. Splitting the sum into its producer and household parts gives $\sum_j a^{(\omega)}_{ij}w_j=[\mathbf A_\omega(\mathbf I-\boldsymbol\beta)\mathbf s]_i$ and, by the definition of the effective final demand, $\sum_c a^{(\omega)}_{ic}F^{c}=f_{\omega,i}-f^{o}_i$. Substituting $g^{(\omega)}_i=s_i-f^{o}_i-\xi_{\omega,i}$ from \eqref{eq:m-lost-sales} then gives the row identity of the statement. The coefficient
row sum $\sum_j a^{(\omega)}_{ij}$ over producer columns is read directly off the balanced matrix and
may lie slightly above its benchmark value, and it enters the analysis only through
the finite bound $\bar a_\omega$.

That leaves the propagation norm of the post-sanction kernel. Post-sanction propagation has entries
\(\bigl[\mathbf A_\omega(\mathbf I-\boldsymbol\beta)\bigr]_{ij}=(1-\beta_j)a^{(\omega)}_{ij}\ge0\),
with column \(j\) summing to
\((1-\beta_j)\sum_i a^{(\omega)}_{ij}\le1-\beta_j\). Taking the maximum over
columns,
\[
\|\mathbf A_\omega(\mathbf I-\boldsymbol\beta)\|_1
\le\max_j(1-\beta_j)
=\|\mathbf A(\mathbf I-\boldsymbol\beta)\|_1<1. 
\]
This proves the lemma.
\end{proof}

\subsection{Well-posedness of the rationed exchange}
\label{app:m-wellposed}

The budgets and capacities of Section~\ref{subsec:model_rerouting} are consistent block by block, $\sum_{i\in\mathcal N_r}g^{(\omega)}_i=\sum_j t^{(\omega)}_{rj}\,w_j$, each side equal to the block's benchmark sales to producers and households less the fraction $\tau(1-\delta)$ of its severed sales. Propositions~\ref{prop:m-balancing} and~\ref{prop:m-leak} together prove Lemma~\ref{lem:m-fixedpoint}, a classical result on matrix scaling \citep{sinkhorn1967concerning,bacharach1970}. Online Appendix~\ref{subsec:recomposition} records the accommodation multiplier that the iterative proportional fitting of \citet{stonebrown1962} and \citet{bacharach1970} assigns to each link.

\begin{proposition}[Well-posedness of the rationed exchange under positive attenuation]
\label{prop:m-leak}
Positive attenuation makes the rationed exchange well posed on every sector block and its computation convergent.
\begin{enumerate}
\item[\textnormal{(i)}] The disrupted matrix keeps the benchmark support, $\operatorname{supp}(\check{\mathbf A})=\operatorname{supp}(\mathbf A)$,
and for every sector block the prescribed-margin transportation polytope on the
support-reduced block contains a matrix that is strictly positive on its
support. Positive attenuation therefore supplies the support compatibility that
Proposition~\ref{prop:m-balancing} requires, and that proposition delivers the
unique balanced matrix, its strictly positive active-block scalings, and their
uniqueness up to one reciprocal gauge per connected component. No zero-retention
corner arises either, because no buyer loses a benchmark sector and no seller with
positive benchmark intermediate
sales loses every benchmark buyer, so $\kappa^{(\omega)}_j(\mathbf h)>0$ for
every producer $j$ and every strictly positive $\mathbf h$.
\item[\textnormal{(ii)}] On every
sector block whose benchmark pattern is strictly positive, a Sinkhorn
half-update is the composition of a strictly positive linear map with
componentwise reciprocation and multiplication by a positive target vector.
The latter two operations are isometries of the Hilbert projective metric, so
each half-update is a strict projective contraction and a full sweep converges
geometrically. Writing
\[
q_r:=\tanh\!\left(\frac{\operatorname{diam}
(\check{\mathbf A}^{(r)})}{4}\right)<1,
\]
the full-sweep modulus is bounded by $q_r^2$. Positive attenuation controls the
projective diameter through the bound
\begin{equation}
\operatorname{diam}(\check{\mathbf A}^{(r)})
\;\le\;
\operatorname{diam}(\mathbf A^{(r)})+2\log(1/\delta).
\label{eq:m-diam}
\end{equation}
The constant two falls to one when the severed set of the block is a single rectangle of links.
The diameter therefore grows at most logarithmically in $1/\delta$, while the upper bound on the contraction modulus obtained from \eqref{eq:m-diam} approaches one as
$\delta\to0$. On blocks with benchmark zeros, part \textnormal{(i)} still
delivers existence and uniqueness of the balanced matrix together with
convergence of iterative proportional fitting. At $\delta=0$, a nonempty restricted transportation polytope still gives
a unique optimizer and convergence of the flow iterates to it. Finite positive
scalings additionally require strict feasibility on every retained support edge.
\end{enumerate}
\end{proposition}

\begin{proof}
For part \textnormal{(i)}, an unbarred benchmark link satisfies
$\check a_{ij}=a_{ij}>0$, while a barred benchmark link satisfies
$\check a_{ij}=\delta a_{ij}>0$. Hence
$\operatorname{supp}(\check{\mathbf A})
=\operatorname{supp}(\mathbf A)$, and the matrix $X^*$ constructed in the proof
of Proposition~\ref{prop:m-balancing} is strictly positive on this support and
meets the prescribed margins. Support compatibility therefore holds on every
block, and the balanced matrix, its scalings, and their gauge follow from that
proposition. What remains is the zero-retention corner. Every benchmark edge
persists, so no used benchmark
buyer-sector block or positive-sales seller row loses all of its support, and
$\kappa^{(\omega)}_j(\mathbf h)>0$ for strictly positive $\mathbf h$.

For part \textnormal{(ii)}, fix a sector block whose benchmark coefficient
pattern is strictly positive, and work in flow units on the block. The kernel
is $\check{\mathbf Z}^{(r)}=\check{\mathbf A}^{(r)}\diag(\mathbf w)$, whose
projective diameter equals that of $\check{\mathbf A}^{(r)}$, since
right-multiplication by a positive diagonal matrix is a projective isometry.
Write $\mathbf g$ for the positive row targets and $\widetilde{\mathbf t}$,
with $\widetilde t_{rj}:=t^{(\omega)}_{rj}w_j$, for the positive flow column
targets. The row half-update is
$\mathbf v\mapsto \mathbf g/(\check{\mathbf Z}^{(r)}\mathbf v)$, and the
column half-update is
$\mathbf u\mapsto \widetilde{\mathbf t}/((\check{\mathbf Z}^{(r)})^{\!\top}\mathbf u)$. Each
half-update composes a strictly positive linear map with componentwise
reciprocation and multiplication by a positive vector. The latter two
operations are isometries of the Hilbert projective metric. Birkhoff's theorem
therefore bounds the contraction modulus of each half-update by
$q_r=\tanh(\operatorname{diam}(\check{\mathbf A}^{(r)})/4)$. The modulus of a
full sweep is at most $q_r^2<1$.

To bound the projective diameter of the disrupted coefficient block as
$\delta$ varies, write
$\check a_{ij}=\delta^{e_{ij}}a_{ij}$, with $e_{ij}=1$ on severed links and
$e_{ij}=0$ otherwise. A cross-ratio equals its benchmark value times
$\delta^{-[(e_{i'j}+e_{ij'})-(e_{ij}+e_{i'j'})]}$, so
\[
\operatorname{diam}(\check{\mathbf A}^{(r)})
\;\le\;
\operatorname{diam}(\mathbf A^{(r)})
+\Bigl[\max_{i,i',j,j'}
\bigl(e_{i'j}+e_{ij'}-e_{ij}-e_{i'j'}\bigr)\Bigr]\log(1/\delta).
\]
Each exponent lies in $\{0,1\}$, so the bracket never exceeds two, which proves \eqref{eq:m-diam}. For the bilateral sanction $c'\notnotleftrightarrow c$ the severed set is the union of the two rectangles $\{(i,j):\mathcal C(i)=c',\,\mathcal C(j)=c\}$ and $\{(i,j):\mathcal C(i)=c,\,\mathcal C(j)=c'\}$, with $\mathcal C(j)$ the country of buyer $j$, and the bracket reaches two, for instance at $\mathcal C(i)=\mathcal C(j)=c$ and $\mathcal C(i')=\mathcal C(j')=c'$. When the severed set is a single rectangle the bracket is at most one. The bracket is positive only if at least one of $e_{i'j}$ and $e_{ij'}$ equals one, and if both equal one then both rows lie in its seller set and both columns in its buyer set, placing $(i,j)$ and $(i',j')$ in the rectangle as well, so $e_{ij}=e_{i'j'}=1$ and the bracket is zero. The general bound \eqref{eq:m-diam} grows at most logarithmically in $1/\delta$, so the corresponding upper bound on $q_r$ approaches one as $\delta\to0$. The realized modulus need not. An attenuation that is uniform, or that acts as a scaling of rows and columns, leaves the cross-ratios that set the projective diameter unchanged. At $\delta=0$ the severed entries
leave the support. If the resulting prescribed-margin transportation polytope is nonempty,
strict convexity gives a unique optimizer, and alternating relative-entropy
projections onto the row and column constraints converge to it. A feasible
point positive on every retained edge is additionally required for a
representation by finite strictly positive scalings. Thus, the absence
of empty retained rows and columns alone does not guarantee existence of the
balanced flow block or convergence of the Sinkhorn iteration. This proves the proposition.
\end{proof}
A diameter that grows logarithmically in $1/\delta$ does not imply that the iteration count does, so we report the realized convergence rate computationally, and at $\delta=0$ the conclusions do not follow merely from the absence of empty rows or columns. The block sums \eqref{eq:m-blocksum} are exactly those of a column rescaling
on the buyer side alone, since the column target \eqref{eq:m-coltarget} prescribes
the buyer margin independently of the supplier margin, so the column-substochastic bound of Lemma~\ref{lem:m-kernel}
is independent of the supplier margin. The coefficient
row-sum bound $\bar a_\omega:=\max_i\sum_j a^{(\omega)}_{ij}$ is read directly off the
balanced matrix. On the 2022 trade table it stays near its benchmark value,
$\bar a_\omega\approx19$ in every scenario, so the conservative $1$-norm delivery-feedback condition of Online
Appendix~\ref{app:m-proofs} fails throughout and the empirical exercise relies
on the benchmark-path diagnostics that the same section reports. The logarithms of the scalings
are dual variables for the seller and buyer-sector margin constraints of the flow-balancing program, but only up to a componentwise additive gauge. Same-side
log differences and edge sums $\log u_i+\log v_{rj}$ survive this gauge, and
whether an unnormalized scaling lies above or below one does not.
Proposition~\ref{prop:m-decentralization} interprets these dual variables as the
pseudo-prices of the block's rationed exchange (Definition~\ref{def:oa-rationed-exchange}).

\subsection{The outer CES aggregator on the unit interval}

The intermediate output elasticity summarizes the local pass-through from the efficiency factor to output,
\begin{equation}
e_j(\kappa)
\;:=\;
\frac{d\log G_j}{d\log\kappa}
\;=\;
\frac{(1-\beta_j)\,\kappa^{\varrho}}{\beta_j+(1-\beta_j)\,\kappa^{\varrho}}
\;\in\;[1-\beta_j,\,1)
\qquad\text{for }\kappa\in(0,1].
\label{eq:m-shadow}
\end{equation}
At the benchmark, $e_j(1)=1-\beta_j$. Away from the benchmark the elasticity rises as $\kappa$ falls because $\varrho<0$. The elasticity stays below one at every positive bundle and approaches one as the bundle collapses, at every finite maintained curvature. The elasticity is identically one on the degraded region only in the excluded limit $\varrho\to-\infty$. At that boundary, a bilateral sanction with a positive unrecovered share of severed trade makes collapse the unique equilibrium under the directed reachability condition of Online Appendix~\ref{app:m-correction}.

\begin{lemma}[Properties of the outer CES aggregator]
\label{lem:m-outer}
Fix $\beta_j\in(0,1)$ and $\varrho\in(-\infty,0)$, and extend the outer
aggregator
$G_j(\kappa)=\bigl[\beta_j+(1-\beta_j)\,\kappa^{\varrho}\bigr]^{1/\varrho}$ to
$[0,\infty)$. Then
\textnormal{(i)} $G_j$ is continuous, strictly increasing, and smooth on
$(0,\infty)$, strictly concave, with $G_j(0)=0$ and $G_j(1)=1$.
\textnormal{(ii)} $\kappa\wedge1\le G_j(\kappa)\le\kappa^{1-\beta_j}$ for
$\kappa\le1$, with strict inequalities on $(0,1)$.
\textnormal{(iii)} $\log G_j$ is concave in $\log\kappa$, with slope the
output elasticity $e_j(\kappa)$, which is strictly
decreasing in $\kappa$ on $(0,\infty)$ with
$e_j(\kappa)\uparrow1$ as $\kappa\downarrow0$ and $e_j(1)=1-\beta_j$.
Consequently, for $0<t\le1\le T$ and $\kappa>0$,
\begin{equation}
t\,G_j(\kappa)\;\le\;t^{\,e_j(t\kappa)}\,G_j(\kappa)\;\le\;G_j(t\kappa)\;\le\;t^{\,e_j(\kappa)}\,G_j(\kappa),
\qquad
G_j(T\kappa)\;\le\;T^{\,e_j(\kappa)}\,G_j(\kappa),
\label{eq:m-outer-subhom}
\end{equation}
the first inequality strict for $t<1$.
\textnormal{(iv)} $G_j'(\kappa)=e_j(\kappa)\,G_j(\kappa)/\kappa$, and on any
interval $[\underline\kappa,1]$ with $\underline\kappa>0$,
$G_j'(\kappa)\le e_j(\underline\kappa)\,\underline\kappa^{-\beta_j}$.
\end{lemma}
\begin{proof}
Write \(b:=\beta_j\in(0,1)\) and \(G:=G_j\), so
\(G(\kappa)=[\,b+(1-b)\kappa^{\varrho}]^{1/\varrho}\) with \(\varrho<0\). We first prove part (i), that \(G\) is smooth, strictly increasing, and strictly concave with \(G(0)=0\) and \(G(1)=1\). On \((0,\infty)\) the bracket is a positive smooth function of
\(\kappa\), and \(1/\varrho\ne0\), so \(G\) is smooth. It increases strictly
because \(\kappa\mapsto\kappa^{\varrho}\) and \(t\mapsto t^{1/\varrho}\) both
decrease strictly. As \(\kappa\downarrow0\), \(\kappa^{\varrho}\uparrow+\infty\),
the bracket diverges, and its \(1/\varrho\)-th power tends to \(0\), matching the
convention \(G(0)=0\). Continuity at \(0\) follows, and normalization gives
\(G(1)=1\). Direct differentiation gives
\[
G''(\kappa)
=b(1-b)(\varrho-1)\,\kappa^{\varrho-2}
\bigl[b+(1-b)\kappa^{\varrho}\bigr]^{1/\varrho-2}<0.
\]
Hence $G$ is strictly concave.

We next prove part (ii), the bound between unit pass-through and constant share. Here \(G(\kappa)=M_{\varrho}(1,\kappa;\,b,1-b)\) is the weighted power mean of order \(\varrho\) of the pair \((1,\kappa)\). The power mean increases
strictly in its order when the entries differ, so for \(\kappa\in(0,1)\),
\[
\kappa
=\min\{1,\kappa\}
=M_{-\infty}
\;<\;
M_{\varrho}
\;<\;
M_{0}
=\kappa^{1-b}.
\]
This is \textnormal{(ii)}. At \(\kappa\in\{0,1\}\) all three coincide.

We now establish part (iii), the log-concavity of \(G\) and the subhomogeneity bounds. Substituting \(\kappa=\exp(v)\) gives
\(\log G=\tfrac1\varrho\log\bigl(b+(1-b)\exp(\varrho v)\bigr)\). Its
derivative in \(v\) is
\(e_j(\kappa)=(1-b)\kappa^{\varrho}/(b+(1-b)\kappa^{\varrho})\), the output
elasticity \eqref{eq:m-shadow}, and its second derivative is
\(\varrho\,e_j(1-e_j)<0\), so \(\log G\) is strictly concave in \(\log\kappa\).
Since \(e_j\) increases strictly in \(\kappa^{\varrho}\) and \(\varrho<0\),
\(e_j\) decreases strictly in \(\kappa\), with \(e_j\uparrow1\) as
\(\kappa\downarrow0\) and \(e_j(1)=1-b\). For \(0<t\le1\), integrate the slope
over \([\log t\kappa,\log\kappa]\). That slope lies in
\([e_j(\kappa),e_j(t\kappa)]\subset(0,1)\), so
\[
e_j(\kappa)\log(1/t)
\;\le\;
\log G(\kappa)-\log G(t\kappa)
\;\le\;
e_j(t\kappa)\log(1/t)
\;\le\;
\log(1/t).
\]
The lower bound on the gap rearranges to
\(G(t\kappa)\le t^{e_j(\kappa)}G(\kappa)\) and the upper bound to
\(G(t\kappa)\ge t^{e_j(t\kappa)}G(\kappa)\). The strict bound
\(e_j<1\) on \((0,\infty)\) gives
\(t^{e_j(t\kappa)}>t\) for \(t\in(0,1)\). Together these give
\eqref{eq:m-outer-subhom} and the strictness claim. For
\(T\ge1\) the slope on \([\log\kappa,\log T\kappa]\) is at most
\(e_j(\kappa)\), giving \(G(T\kappa)\le T^{e_j(\kappa)}G(\kappa)\).

We prove part (iv) last, the slope identity and the derivative bound. The identity \(G'(\kappa)=e_j(\kappa)G(\kappa)/\kappa\) restates \textnormal{(iii)} in levels. On \([\underline\kappa,1]\), \(e_j(\kappa)\le
e_j(\underline\kappa)\) by monotonicity, and
\(G(\kappa)/\kappa\le\kappa^{1-b}/\kappa=\kappa^{-b}\le\underline\kappa^{-b}\)
by \textnormal{(ii)}, whence
\(G'(\kappa)\le e_j(\underline\kappa)\,\underline\kappa^{-b}\). This proves the lemma.
\end{proof}

\subsection{Bounds on conditional economic activity}

\begin{lemma}[Capacity bounds on conditional economic activity]
\label{lem:m-capacity}
For every fulfillment vector $\mathbf h\in[0,1]^N$, the conditional economic activity vector $\mathbf s(\mathbf h)$ satisfies
\begin{equation*}
\mathbf 0
\;\le\;
\mathbf s(\mathbf h)
\;\le\;
\mathbf s,
\qquad
\bigl[\mathbf A_\omega(\mathbf I-\boldsymbol\beta)
\mathbf s(\mathbf h)\bigr]_i+f_{\omega,i}
\;\le\;
g^{(\omega)}_i+f^{o}_i
\;\le\;
s_i,
\end{equation*}
for every supplier $i$. The last inequality is an equality exactly at sellers with no severed sales or when $\tau(1-\delta)=0$. The sellers' capacities bound standing demand, intermediate and final together, while the buyers' budgets and the benchmark-normalized production representation bound the output ratio. Together, these restrictions imply that conditional economic activity, and therefore activity in every sanction equilibrium, cannot exceed its benchmark level.
\end{lemma}
\begin{proof}

Fix \(\omega\in\Omega\), \(\tau\in[0,1]\), \(\delta\in(0,\bar\delta]\), and
\(\mathbf h\in[0,1]^N\). Write
\(\mathbf B_\omega:=\mathbf A_\omega(\mathbf I-\boldsymbol\beta)\) and
\(\mathbf M_\omega(\mathbf h):=\diag(\mathbf h)\mathbf B_\omega\). The two
sides of the construction provide separate benchmark bounds, one on
production and one on flows.

On the buyer side, the prescribed sector budgets and \(\mathbf h\le\mathbf 1\)
give
\[
0\;\le\;\theta^{(\omega)}_{rj}(\mathbf h)\;\le\;1,
\]
for every buyer \(j\) and every used input sector \(r\in\mathcal R_j\)
(Lemma~\ref{lem:m-kernel}). Because the maintained nested-CES representation
is increasing and normalized at the benchmark input mix, the
\(\boldsymbol\alpha\)-weighted \(\rho\)-power mean and the outer aggregator
\(G_j\) carry the unit interval into itself (Lemma~\ref{lem:m-outer}), so
\[
0\;\le\;\kappa^{(\omega)}_j(\mathbf h)\;\le\;1,
\qquad
0\;\le\;\lambda^{(\omega)}_j(\mathbf h)\;\le\;1.
\]
Hence \(\mathbf 0\le\boldsymbol\lambda^{(\omega)}(\mathbf h)\le\mathbf 1\). On the seller side, the row identity of Lemma~\ref{lem:m-kernel} gives
\[
\bigl[\mathbf B_\omega\mathbf s\bigr]_i+f_{\omega,i}
=
g_i^{(\omega)}+f^{o}_i
\;\le\;
s_i
\qquad\text{for every }i.
\]
Hence $\mathbf s-\mathbf B_\omega\mathbf s-\mathbf f_\omega=\boldsymbol\xi_\omega\ge\mathbf 0$ by \eqref{eq:m-lost-sales}.
The activity comparison combines the production-side and the flow-side
bounds. Lemma~\ref{lem:m-kernel} gives
\(\spr(\mathbf M_\omega(\mathbf h))<1\), so
\((\mathbf I-\mathbf M_\omega(\mathbf h))^{-1}\ge\mathbf 0\). With
\(\mathbf 0\le\mathbf h\le\mathbf 1\) and the production-side bound
\(\mathbf 0\le\boldsymbol\lambda^{(\omega)}(\mathbf h)\le\mathbf 1\),
\[
\begin{aligned}
(\mathbf I-\mathbf M_\omega(\mathbf h))\mathbf s
-\boldsymbol\lambda^{(\omega)}(\mathbf h)\circ\mathbf f_\omega
&=
\mathbf s-\mathbf M_\omega(\mathbf h)\mathbf s
-\boldsymbol\lambda^{(\omega)}(\mathbf h)\circ\mathbf f_\omega\\
&\ge
\mathbf s-\mathbf B_\omega\mathbf s-\mathbf f_\omega
=\boldsymbol\xi_\omega\ge\mathbf 0.
\end{aligned}
\]
Multiplying by the nonnegative inverse and using $\mathbf s(\mathbf h)=[\mathbf I-\diag(\mathbf h)\mathbf B_\omega]^{-1}(\boldsymbol\lambda^{(\omega)}(\mathbf h)\circ\mathbf f_\omega)$ gives
\(\mathbf s-\mathbf s(\mathbf h)\ge\mathbf 0\). The same inverse applied to the
nonnegative source
\(\boldsymbol\lambda^{(\omega)}(\mathbf h)\circ\mathbf f_\omega\) gives
\(\mathbf s(\mathbf h)\ge\mathbf 0\). Finally, nonnegativity of
\(\mathbf B_\omega\) gives
\[
\bigl[\mathbf B_\omega\mathbf s(\mathbf h)\bigr]_i+f_{\omega,i}
\;\le\;
\bigl[\mathbf B_\omega\mathbf s\bigr]_i+f_{\omega,i}
=
g^{(\omega)}_i+f^{o}_i
\;\le\;
s_i.
\]
The equality condition for the last inequality follows from the capacity
\eqref{eq:m-rowtarget}. This proves the lemma.
\end{proof}

The no-activity-gain conclusion of Lemma~\ref{lem:m-capacity} is a property
of the maintained short-run production representation at fixed benchmark
prices. The conclusion need not hold if the change of suppliers alters effective input prices,
supplier quality, production technology, or the efficiency of the benchmark
input mix.

\subsection{The sanction-equilibrium map}

Sanction equilibria are the fixed points of the joint map on $\mathbb R_+^N\times[0,1]^N$
\begin{equation}
\Phi_\omega(\mathbf x,\mathbf h)
:=
\Bigl(
\underbrace{\boldsymbol\lambda^{(\omega)}(\mathbf h)\circ
\bigl[\mathbf f_\omega+\mathbf A_\omega(\mathbf I-\boldsymbol\beta)\,\mathbf x\bigr]}_{\text{activity update}},\;\;
\underbrace{\boldsymbol\lambda^{(\omega)}(\mathbf h)}_{\text{fulfillment update}}
\Bigr).
\label{eq:m-operator}
\end{equation}
The equivalence with Definition~\ref{def:m-sanction-equilibrium} follows from the fulfillment coordinate. At a fixed point, $\mathbf h^*=\boldsymbol\lambda^{(\omega)}(\mathbf h^*)$, so the activity coordinate of \eqref{eq:m-operator} reduces to \eqref{eq:m-equilibrium}. Conversely, any pair satisfying the two equilibrium conditions fixes $\Phi_\omega$. The map characterizes the equilibrium set rather than computing a point in it. Its activity coordinate is a one-step update at the given pair rather than the conditional solution $\mathbf s(\mathbf h)$, and the computation of the benchmark path uses $\mathbf s(\mathbf h)$ directly.

\begin{lemma}[The fulfillment rule]
\label{lem:m-rule}
The fulfillment rule $\boldsymbol\lambda^{(\omega)}$ is continuous and entrywise nondecreasing on $[0,1]^N$, and $\boldsymbol\lambda^{(\omega)}(\mathbf 1)\le\mathbf 1$.
\end{lemma}

\begin{proof}
Since \(\mathbf A_\omega\ge\mathbf 0\), each realized
sectoral input-retention ratio
\(\theta^{(\omega)}_{rj}(\mathbf h)=\sum_{i\in\mathcal N_r}h_i\,a^{(\omega)}_{ij}/\alpha_{rj}\)
rises entrywise in \(\mathbf h\). The
\(\boldsymbol\alpha\)-weighted \(\rho\)-power mean rises in each input-retention
ratio on \([0,1]^{\mathcal R_j}\), including at the zero-retention boundary, where our
convention assigns the minimal value zero. The outer CES aggregator \(G_j\)
increases on \([0,1]\) by Lemma~\ref{lem:m-outer}\textnormal{(i)}. Hence the
fulfillment rule \(\boldsymbol\lambda^{(\omega)}\) is entrywise nondecreasing.

The same components make the fulfillment rule continuous. The bundle
efficiency factor \(\kappa^{(\omega)}_j(\mathbf h)\) is smooth wherever every
realized sectoral input-retention ratio is positive. If some
\(\theta^{(\omega)}_{r_0,j}(\mathbf h)\) tends to zero, then
\((\theta^{(\omega)}_{r_0,j})^\rho\) diverges because \(\rho<0\), and the
\(1/\rho\)-th power tends to zero. This limit agrees with the zero-retention convention. Composing the resulting continuous efficiency factor with
the continuous outer CES aggregator gives a continuous fulfillment rule on
\([0,1]^N\).

At \(\mathbf h=\mathbf 1\), Lemma~\ref{lem:m-kernel} gives
\(\theta^{(\omega)}_{rj}(\mathbf 1)=
\mathcal F(\mathbf A_\omega)_{rj}/\alpha_{rj}\in[0,1]\). Hence \(\boldsymbol\lambda^{(\omega)}(\mathbf 1)\le\mathbf 1\). This proves the lemma.
\end{proof}

\subsection{Proof of Proposition~\ref{prop:m-monotone}}
\label{app:m-proof-monotone}
\begin{proof}

We prove parts (i)--(iii) in order, the monotone convergence of the benchmark path, the maximality of its limit, and the interiority condition that keeps that limit away from collapse. Part (i) also establishes existence and the benchmark bound on economic activity. Nothing in this proof uses Proposition~\ref{prop:m-interior} of the Online Appendix, which rests on these three parts.

For part (i), Lemma~\ref{lem:m-rule} gives \(\boldsymbol\lambda^{(\omega)}(\mathbf 1)\le\mathbf 1=\mathbf h^{(0)}\). Monotonicity then gives
\[
\mathbf h^{(n+1)}
=
\boldsymbol\lambda^{(\omega)}(\mathbf h^{(n)})
\;\le\;
\boldsymbol\lambda^{(\omega)}(\mathbf h^{(n-1)})
=
\mathbf h^{(n)}.
\]
The benchmark path is entrywise nonincreasing and bounded below by
\(\mathbf 0\). It therefore converges to some
\(\mathbf h_\omega\in[0,1]^N\), and the continuity of the fulfillment rule gives
\(\mathbf h_\omega=\boldsymbol\lambda^{(\omega)}(\mathbf h_\omega)\).

We next recover economic activity from this fulfillment fixed point. Write
\(\mathbf B_\omega:=\mathbf A_\omega(\mathbf I-\boldsymbol\beta)\), whose norm satisfies
\(\|\mathbf B_\omega\|_1<1\) by Lemma~\ref{lem:m-kernel}. Uniformly over \(\mathbf h\in[0,1]^N\),
\(\|\diag(\mathbf h)\mathbf B_\omega\|_1\le\|\mathbf B_\omega\|_1\), so
\[
\mathbf s(\mathbf h)
=
\sum_{t\ge0}
\bigl[\diag(\mathbf h)\mathbf B_\omega\bigr]^t
\bigl(\boldsymbol\lambda^{(\omega)}(\mathbf h)\circ\mathbf f_\omega\bigr).
\]
The series converges uniformly and consists of continuous, entrywise
nondecreasing terms. Hence \(\mathbf s(\mathbf h)\) is continuous and
entrywise nondecreasing in \(\mathbf h\). It follows that
\(\mathbf x^{(n)}=\mathbf s(\mathbf h^{(n)})\) converges to
\(\mathbf s_\omega:=\mathbf s(\mathbf h_\omega)\). The pair
\((\mathbf s_\omega,\mathbf h_\omega)\) satisfies the two fixed-point
equations and is therefore a fixed point of \(\Phi_\omega\).
Lemma~\ref{lem:m-capacity} gives
\(\mathbf 0\le\mathbf s_\omega\le\mathbf s\).

More generally, let \((\mathbf x^*,\mathbf h^*)\) be any fixed point of
\(\Phi_\omega\). The first coordinate and the unique solution of the linear economic-activity equation
give \(\mathbf x^*=\mathbf s(\mathbf h^*)\). Lemma~\ref{lem:m-capacity} then
gives \(\mathbf 0\le\mathbf x^*\le\mathbf s\). This proves existence, the
benchmark bound on every fixed point, and part (i).

We now prove part (ii), the maximality of the benchmark-selected fulfillment vector. Let \(\mathbf g=\boldsymbol\lambda^{(\omega)}(\mathbf g)\) with \(\mathbf g\in[0,1]^N\). From \(\mathbf g\le\mathbf 1\) and monotonicity,
\(\mathbf g=(\boldsymbol\lambda^{(\omega)})^n(\mathbf g)
\le(\boldsymbol\lambda^{(\omega)})^n(\mathbf 1)=\mathbf h^{(n)}\) for every
\(n\). Letting \(n\to\infty\), \(\mathbf g\le\mathbf h_\omega\). Now let
\((\mathbf x^*,\mathbf h^*)\) be a fixed point of \(\Phi_\omega\). Its second
coordinate gives
\(\mathbf h^*=\boldsymbol\lambda^{(\omega)}(\mathbf h^*)\le\mathbf h_\omega\), and
its first coordinate identifies \(\mathbf x^*\) as the unique solution
\(\mathbf s(\mathbf h^*)\) of the linear activity equation. Monotonicity of \(\mathbf s(\cdot)\)
gives
\(\mathbf x^*=\mathbf s(\mathbf h^*)\le\mathbf s(\mathbf h_\omega)=\mathbf s_\omega\).

We turn last to part (iii), interiority. Fix an active \(i\) and abbreviate \(\varkappa_i:=\kappa^{(\omega)}_i(\mathbf 1)\), positive by Proposition~\ref{prop:m-leak}\textnormal{(i)}. Consider the scalar gap
\[
\varsigma_i(t)
\;:=\;
G_i(t\,\varkappa_i)-t,
\qquad t\in[0,1].
\]
The aggregator \(G_i\) is a power mean of order \(\varrho<1\) of the pair
\((1,\kappa)\) and is strictly concave in \(\kappa\)
(Lemma~\ref{lem:m-outer}), so \(\varsigma_i\) is strictly concave with
\(\varsigma_i(0)=0\). Its right derivative at zero is
\(\varkappa_i\,G_i'(0^+)-1=\varkappa_i\,(1-\beta_i)^{1/\varrho}-1\), where one
reads the limit slope \(G_i'(0^+)=(1-\beta_i)^{1/\varrho}\) off the stable
form \(G_i(\kappa)=\kappa\,[(1-\beta_i)+\beta_i\kappa^{-\varrho}]^{1/\varrho}\).
The derivative turns strictly positive exactly when the interiority condition
\eqref{eq:m-cushion-cond} holds at \(i\),
\(\varkappa_i(1-\beta_i)^{1/\varrho}>1
\iff\varkappa_i>(1-\beta_i)^{-1/\varrho}\). A strictly concave function
vanishing at zero with strictly positive initial slope stays strictly positive
up to its unique positive root and turns negative beyond it. Solve
\(G_i(t\,\varkappa_i)=t\), that is
\(\beta_i+(1-\beta_i)t^{\varrho}\varkappa_i^{\varrho}=t^{\varrho}\), for the
root in closed form,
\begin{equation}
c_i=\Bigl[\beta_i\big/\bigl(1-(1-\beta_i)\varkappa_i^{\varrho}\bigr)\Bigr]^{1/\varrho},
\qquad
c_\omega:=\min_i c_i,
\label{eq:m-floor}
\end{equation}
the interiority floor of part (iii). The interiority condition is exactly what
makes the bracketed denominator positive, and
\(\varsigma_i(1)=G_i(\varkappa_i)-1\le0\) gives \(c_i\le1\). Hence \(\varsigma_i\ge0\)
on \([0,c_i]\), and, with \(c_\omega=\min_i c_i\le c_i\),
\(G_i(c_\omega\varkappa_i)\ge c_\omega\) for every \(i\). For
\(\mathbf h\ge c_\omega\mathbf 1\), monotonicity together with positive
\(1\)-homogeneity of \(\kappa^{(\omega)}_i\) gives
\(\kappa^{(\omega)}_i(\mathbf h)\ge\kappa^{(\omega)}_i(c_\omega\mathbf 1)
=c_\omega\,\varkappa_i\), hence
\[
\lambda^{(\omega)}_i(\mathbf h)
\;\ge\;
G_i\bigl(c_\omega\,\varkappa_i\bigr)
\;\ge\;
c_\omega.
\]
Since
\(\mathbf h^{(0)}=\mathbf 1\ge c_\omega\mathbf 1\), induction gives
\(\mathbf h^{(n)}\ge c_\omega\mathbf 1\) for every \(n\), and hence
\(\mathbf h_\omega\ge c_\omega\mathbf 1\). The same computation shows
\(\lambda^{(\omega)}_i(\mathbf h)\ge c_\omega\) for every
\(\mathbf h\in[c_\omega,1]^N\) and every \(i\). Thus the set
\([c_\omega,1]^N\) is forward invariant under
\(\boldsymbol\lambda^{(\omega)}\), a fact
Online Appendix~\ref{app:m-proofs} uses in its conditional contraction analysis.
If the interiority condition fails, no positive vector on the ray $t\mathbf 1$ is a subsolution. At any \(i\) with
\(\varkappa_i\le(1-\beta_i)^{-1/\varrho}\) we have \(\varsigma_i<0\) on all of
\((0,1]\), since concavity gives the result when the initial slope is negative
and strict concavity gives it when that slope is zero. This does not exclude an interior equilibrium with unequal fulfillment rates. Parts \textnormal{(i)} and
\textnormal{(ii)} are unaffected. This proves part \textnormal{(iii)} and the
proposition.
\end{proof}

\subsection{Why no activity is lost when displaced trade is fully recovered}

\begin{lemma}[The full-recovery limit]
\label{lem:m-frictionless}
At $\tau=0$ the benchmark-selected equilibrium is the benchmark pair, $(\mathbf s_\omega,\mathbf h_\omega)=(\mathbf s,\mathbf 1)$, and the loss vector vanishes.
\end{lemma}

\begin{proof}

We verify the full-recovery limit stated in Section~\ref{subsec:model_counterfactual}. Set $\tau=0$. The column
targets \eqref{eq:m-coltarget} and row targets \eqref{eq:m-rowtarget} then sit
at benchmark, $t^{(\omega)}_{rb}=\alpha_{rb}$ and
$g^{(\omega)}_i=s_i-f^{o}_i$, and the lost-sales vector \eqref{eq:m-lost-sales} vanishes. We first show that the fulfillment component stays at
$\mathbf h_\omega=\mathbf 1$. By Lemma~\ref{lem:m-kernel}, every buyer's
sector block sum returns to benchmark, $\mathcal F(\mathbf A_\omega)_{rj}
=\alpha_{rj}$, so the realized sectoral input-retention ratio at full fulfillment is
$\theta^{(\omega)}_{rj}(\mathbf 1)=1$ for every $r\in\mathcal R_j$. The bundle
efficiency factor is then $\kappa^{(\omega)}_j(\mathbf 1)=1$ at every buyer,
and $G_j(1)=1$ (Lemma~\ref{lem:m-outer}\textnormal{(i)}) gives
$\boldsymbol\lambda^{(\omega)}(\mathbf 1)=\mathbf 1$. The vector $\mathbf 1$ is
therefore a fixed point of $\boldsymbol\lambda^{(\omega)}$, and it is the
greatest by Proposition~\ref{prop:m-monotone}\textnormal{(ii)}. Hence
$\mathbf h_\omega=\mathbf 1$.

We next show that the activity component returns the benchmark vector. The
row identity of Lemma~\ref{lem:m-kernel} gives, at every supplier $i$,
\[
\bigl[\mathbf A_\omega(\mathbf I-\boldsymbol\beta)\,\mathbf s\bigr]_i+f_{\omega,i}
=\sum_b a^{(\omega)}_{ib}\,w_b+f^{o}_i
=g^{(\omega)}_i+f^{o}_i
=s_i.
\]
The benchmark activity vector therefore solves
$\mathbf s=\mathbf f_\omega+\mathbf A_\omega(\mathbf I-\boldsymbol\beta)\,\mathbf s$,
the post-sanction activity equation \eqref{eq:m-equilibrium} at
$\mathbf h=\mathbf 1$. Since
$\|\mathbf A_\omega(\mathbf I-\boldsymbol\beta)\|_1<1$
(Lemma~\ref{lem:m-kernel}), that equation has exactly one solution, so
$\mathbf s(\mathbf 1)=\mathbf s$. The selected equilibrium is therefore
$(\mathbf s_\omega,\mathbf h_\omega)=(\mathbf s,\mathbf 1)$, and
$\boldsymbol\Delta\mathbf s_\omega=\mathbf s-\mathbf s_\omega=\mathbf 0$. This proves the lemma.
\end{proof}

\subsection{Proof of Proposition~\ref{prop:m-anatomy}}
\label{app:m-proof-anatomy}

\begin{proof}
Write $\mathbf B_\omega:=\mathbf A_\omega(\mathbf I-\boldsymbol\beta)$ and let $(\mathbf x^*,\mathbf h^*)$ be a sanction equilibrium. The essence of the proof is that the row identity of Lemma~\ref{lem:m-kernel} and the activity equation carry the same two objects, the effective final demand and the effective intermediate demand, and the effective final demand cancels between them.

We first apply $\mathbf I-\diag(\mathbf h^*)\mathbf B_\omega$ to the loss. The activity equation \eqref{eq:m-equilibrium} gives $\diag(\mathbf h^*)\mathbf B_\omega\mathbf x^*=\mathbf x^*-\boldsymbol\lambda^{(\omega)}(\mathbf h^*)\circ\mathbf f_\omega$, so
\[
\bigl(\mathbf I-\diag(\mathbf h^*)\mathbf B_\omega\bigr)(\mathbf s-\mathbf x^*)
=\mathbf s-\diag(\mathbf h^*)\mathbf B_\omega\mathbf s-\boldsymbol\lambda^{(\omega)}(\mathbf h^*)\circ\mathbf f_\omega.
\]
The row identity of Lemma~\ref{lem:m-kernel} reads $\mathbf B_\omega\mathbf s=\mathbf s-\mathbf f_\omega-\boldsymbol\xi_\omega$. Substituting it on the right,
\[
\bigl(\mathbf I-\diag(\mathbf h^*)\mathbf B_\omega\bigr)(\mathbf s-\mathbf x^*)
=(\mathbf 1-\mathbf h^*)\circ\mathbf s+\mathbf h^*\circ\boldsymbol\xi_\omega
+\bigl(\mathbf h^*-\boldsymbol\lambda^{(\omega)}(\mathbf h^*)\bigr)\circ\mathbf f_\omega.
\]
At a sanction equilibrium $\mathbf h^*=\boldsymbol\lambda^{(\omega)}(\mathbf h^*)$, so the last term vanishes. The effective final demand drops out at this step.

We next invert. Lemma~\ref{lem:m-kernel} gives $\|\diag(\mathbf h^*)\mathbf B_\omega\|_1\le\|\mathbf B_\omega\|_1<1$, so $\mathbf I-\diag(\mathbf h^*)\mathbf B_\omega$ is invertible and its inverse is the nonnegative Neumann series $\sum_{t\ge0}[\diag(\mathbf h^*)\mathbf B_\omega]^t$. Multiplying through by that inverse gives \eqref{eq:m-anatomy}. The first bracketed vector is nonnegative because $\mathbf 0\le\mathbf h^*\le\mathbf 1$ and $\mathbf s\ge\mathbf 0$, and the second because $\boldsymbol\xi_\omega\ge\mathbf 0$ by construction and $\mathbf h^*\ge\mathbf 0$. This proves the proposition.
\end{proof}

\subsection{Bounded asymmetries}
\label{app:m-psi}

The normalized difference $\psi$ of \eqref{eq:m-psi} carries a positive ratio to a number in $(-1,1)$, and the asymmetry, the trade balance, and the size asymmetry of a pair are its values at three ratios. The size asymmetry is the value at the ratio of the two sizes,
\begin{equation}
\mathcal Q_{cc'}:=\psi\bigl(Q(c')/Q(c)\bigr),
\label{eq:m-size-asymmetry}
\end{equation}
positive when country $c$ is the smaller of the two. For $u,v\in(-1,1)$ define the bounded sum
\begin{equation}
u\oplus v:=\frac{u+v}{1+uv}\in(-1,1).
\label{eq:m-oplus}
\end{equation}

\begin{lemma}[Bounded asymmetries]
\label{lem:m-psi}
\begin{enumerate}
\item[\textnormal{(i)}] $\psi$ is a strictly increasing bijection of $(0,\infty)$ onto $(-1,1)$ with inverse $r=(1+\psi)/(1-\psi)$, and $\psi(1/r)=-\psi(r)$.
\item[\textnormal{(ii)}] $\psi(r_1r_2)=\psi(r_1)\oplus\psi(r_2)$ for all positive $r_1$ and $r_2$. The operation $\oplus$ is associative and commutative, has the neutral element $0$ and the inverse $-u$, and $\psi(r_1r_2\cdots r_n)=\psi(r_1)\oplus\cdots\oplus\psi(r_n)$.
\item[\textnormal{(iii)}] $\psi\bigl((1+x)/(1-x)\bigr)=x$ for $x\in(-1,1)$.
\end{enumerate}
\end{lemma}

\begin{proof}
We first prove \textnormal{(i)}. The derivative $\psi'(r)=2/(r+1)^{2}$ is positive, $\psi(r)\to-1$ as $r\to0$, and $\psi(r)\to1$ as $r\to\infty$, so $\psi$ is a strictly increasing bijection of $(0,\infty)$ onto $(-1,1)$. Solving $\psi=(r-1)/(r+1)$ for $r$ gives the inverse, and $\psi(1/r)=(1-r)/(1+r)=-\psi(r)$. This proves \textnormal{(i)}.

We next prove \textnormal{(ii)}. Write $u=\psi(r_1)$ and $v=\psi(r_2)$, so that $r_1=(1+u)/(1-u)$ and $r_2=(1+v)/(1-v)$ by \textnormal{(i)}. Then
\[
\psi(r_1r_2)
=\frac{(1+u)(1+v)-(1-u)(1-v)}{(1+u)(1+v)+(1-u)(1-v)}
=\frac{2(u+v)}{2(1+uv)}
=u\oplus v .
\]
Since $\psi$ is a bijection that carries products to bounded sums, the operation inherits associativity, commutativity, the neutral element $\psi(1)=0$, and the inverse $\psi(1/r)=-\psi(r)$ from the multiplicative group of the positive reals, and the $n$-fold statement follows by induction on $n$. This proves \textnormal{(ii)}.

Part \textnormal{(iii)} is the inverse of \textnormal{(i)} read in the other direction, since $(1+x)/(1-x)$ is the ratio whose normalized difference is $x$. This proves the lemma.
\end{proof}

\subsection{Sizes depend only on the row and column totals}
\label{app:m-proof-fiber}

Each producer leaks the share $\beta_j$ of its activity out of the intermediate network, and adding that share back, spent across the sellers in the proportions of final demand, closes the propagation matrix,
\begin{equation}
\mathbf P_0
:=\mathbf A(\mathbf I-\boldsymbol\beta)
+\frac{\mathbf f\,\mathbf 1^{\top}\boldsymbol\beta}{\mathbf 1^{\top}\mathbf f}.
\label{eq:m-size-invariant}
\end{equation}
Write $E$ for the set of trading links of the network, the links with a positive sourcing share, and $K_E$ for the number of connected components of the bipartite graph that $E$ draws on the sellers and the buyers. The lemma states that the benchmark sizes are fixed by the row and column totals of the trade table and are shared by every network with those totals.

\begin{lemma}[Sizes depend only on the row and column totals]
\label{lem:m-fiber}
Fix the leakages $\boldsymbol\beta$, the household matrix $\mathbb A$, the budgets $\mathbf F$, and the residual outside demand $\mathbf f^{o}$, and let the network $\mathbf A$ vary over the nonnegative matrices with unit column sums that are zero outside $E$.
\begin{enumerate}
\item[\textnormal{(i)}] The closed benchmark matrix $\mathbf P_0$ is column stochastic, and $\mathbf s/(\mathbf 1^{\top}\mathbf s)$ is its unique nonnegative invariant vector of unit sum.
\item[\textnormal{(ii)}] Two networks generate the same benchmark activity if and only if they generate the same seller totals of intermediate sales, $\mathbf A'(\mathbf I-\boldsymbol\beta)\,\mathbf s(\mathbf A)=\mathbf A(\mathbf I-\boldsymbol\beta)\,\mathbf s(\mathbf A)$.
\item[\textnormal{(iii)}] The networks that generate the activity $\mathbf s(\mathbf A)$ form a convex set, of dimension $|E|-2N+K_E$ when $\mathbf A$ is positive on $E$. On it the size of every country, every household flow, and the invariant distribution of $\mathbf P_0$ are constant, while the sourcing shares and the intermediate flows vary.
\end{enumerate}
\end{lemma}

Throughout this subsection the leakages $\boldsymbol\beta$, the household matrix $\mathbb A$, the budgets $\mathbf F$, and the residual outside demand $\mathbf f^{o}$ are fixed, so that the final demand $\mathbf f=\mathbb A\mathbf F+\mathbf f^{o}$ is fixed. The vector $\mathbf s(\mathbf A)$ denotes the benchmark activity of Lemma~\ref{lem:m-benchmark} at the network $\mathbf A$. A circulation at the activity $\mathbf s$ is a matrix $\mathbf D$ supported on $E$ with
\begin{equation}
\mathbf 1^{\!\top}\mathbf D=\mathbf 0^{\!\top}
\qquad\text{and}\qquad
\mathbf D\,(\mathbf I-\boldsymbol\beta)\,\mathbf s=\mathbf 0,
\label{eq:m-circulation}
\end{equation}
a rearrangement of who buys from whom that leaves every buyer's sourcing shares summing to one and every seller's intermediate sales where they were.

\begin{proof}
For part \textnormal{(i)}, summing the benchmark equation \eqref{eq:m-benchmark} over the country-sectors gives $\mathbf 1^{\top}\mathbf s=\mathbf 1^{\top}\mathbf f+\mathbf 1^{\top}(\mathbf I-\boldsymbol\beta)\mathbf s$, since the columns of $\mathbf A$ sum to one, so that $\mathbf 1^{\top}\boldsymbol\beta\mathbf s=\mathbf 1^{\top}\mathbf f>0$. Every entry of $\mathbf P_0$ is nonnegative, and column $j$ of $\mathbf P_0$ sums to $(1-\beta_j)+\beta_j=1$, so the matrix is column stochastic. Substituting $\mathbf 1^{\top}\boldsymbol\beta\mathbf s=\mathbf 1^{\top}\mathbf f$ into $\mathbf P_0\mathbf s=\mathbf A(\mathbf I-\boldsymbol\beta)\mathbf s+\mathbf f\,(\mathbf 1^{\top}\boldsymbol\beta\mathbf s)/(\mathbf 1^{\top}\mathbf f)$ gives $\mathbf P_0\mathbf s=\mathbf A(\mathbf I-\boldsymbol\beta)\mathbf s+\mathbf f=\mathbf s$. If $\mathbf v\ge\mathbf 0$ is any invariant vector with $\mathbf 1^{\top}\mathbf v=1$, then $[\mathbf I-\mathbf A(\mathbf I-\boldsymbol\beta)]\mathbf v=\mathbf f\,(\mathbf 1^{\top}\boldsymbol\beta\mathbf v)/(\mathbf 1^{\top}\mathbf f)$. The inverse of Lemma~\ref{lem:m-benchmark} then makes $\mathbf v$ the multiple $(\mathbf 1^{\top}\boldsymbol\beta\mathbf v)/(\mathbf 1^{\top}\mathbf f)$ of $\mathbf s$, which the normalization $\mathbf 1^{\top}\mathbf v=1$ fixes at $1/\mathbf 1^{\top}\mathbf s$. No irreducibility assumption is needed. This proves \textnormal{(i)}.

For part \textnormal{(ii)}, write $\mathbf s:=\mathbf s(\mathbf A)$ and let $\mathbf A'$ be admissible on $E$. If $\mathbf A'(\mathbf I-\boldsymbol\beta)\mathbf s=\mathbf A(\mathbf I-\boldsymbol\beta)\mathbf s$, then $\mathbf f+\mathbf A'(\mathbf I-\boldsymbol\beta)\mathbf s=\mathbf f+\mathbf A(\mathbf I-\boldsymbol\beta)\mathbf s=\mathbf s$, so $\mathbf s$ solves the benchmark equation \eqref{eq:m-benchmark} at the network $\mathbf A'$, and the uniqueness in Lemma~\ref{lem:m-benchmark} gives $\mathbf s(\mathbf A')=\mathbf s$. Conversely, if $\mathbf s(\mathbf A')=\mathbf s$, then $\mathbf A'(\mathbf I-\boldsymbol\beta)\mathbf s=\mathbf s-\mathbf f=\mathbf A(\mathbf I-\boldsymbol\beta)\mathbf s$. This proves \textnormal{(ii)}.

For part \textnormal{(iii)}, by \textnormal{(ii)} the network $\mathbf A'$ generates the activity $\mathbf s$ if and only if $\mathbf D:=\mathbf A'-\mathbf A$ is supported on $E$ and satisfies $\mathbf D(\mathbf I-\boldsymbol\beta)\mathbf s=\mathbf 0$. Since both matrices have unit column sums, $\mathbf D$ also satisfies $\mathbf 1^{\top}\mathbf D=\mathbf 0^{\top}$, so the two requirements are the conditions \eqref{eq:m-circulation}. Nonnegativity of $\mathbf A+\mathbf D$ is the only inequality, so the set of such networks is the intersection of an affine space with an orthant, a convex polytope. Multiplying an admissible $\mathbf A'$ on the right by $\diag\bigl((\mathbf 1-\boldsymbol\beta)\circ\mathbf s\bigr)$ carries the polytope bijectively onto the nonnegative matrices on $E$ with column totals $(1-\beta_j)s_j$ and row totals $s_i-f_i$. The column totals encode the unit column sums and the row totals encode the condition of \textnormal{(ii)}. In flow units the conditions \eqref{eq:m-circulation} say that the flow on $E$ has zero net total at every seller and at every buyer, which is the kernel of the incidence map of the bipartite graph of $E$ on its $2N$ vertices. That map has rank $2N-K_E$, one relation for each connected component, so the kernel has dimension $|E|-2N+K_E$. When $\mathbf A$ is positive on $E$ every direction of the kernel is feasible for a small step, so the polytope has the dimension of the kernel. The country sizes are sums of entries of the common activity vector, and the household flows are fixed by assumption. Part \textnormal{(i)} applied to $\mathbf A'$, whose benchmark activity is $\mathbf s$, gives $\mathbf P_0(\mathbf A')\mathbf s=\mathbf s$ together with the uniqueness of the normalized invariant vector, so the invariant distribution is common to the polytope. The matrix $\mathbf P_0(\mathbf A')$ contains $\mathbf A'(\mathbf I-\boldsymbol\beta)$ and moves with $\mathbf A'$. The sourcing shares are the entries of $\mathbf A'$, and the intermediate flows $\mathbf A'\diag\bigl((\mathbf 1-\boldsymbol\beta)\circ\mathbf s\bigr)$ are the columns of $\mathbf A'$ rescaled by fixed positive numbers. This proves the lemma.
\end{proof}

\subsection{Content rows and the direct losses}
\label{app:m-content}

Let $(\mathbf s_\omega,\mathbf h_\omega)$ be the benchmark-selected equilibrium of Proposition~\ref{prop:m-monotone}. Write $\mathbf B_\omega:=\mathbf A_\omega(\mathbf I-\boldsymbol\beta)$ for the sanctioned propagation matrix, $\mathbf R^{*}_\omega:=[\mathbf I-\diag(\mathbf h_\omega)\mathbf B_\omega]^{-1}$ for the inverse of Proposition~\ref{prop:m-anatomy}, and $\boldsymbol\nu_\omega:=(\mathbf 1-\mathbf h_\omega)\circ\mathbf s+\mathbf h_\omega\circ\boldsymbol\xi_\omega$ for the direct-loss vector, so that the anatomy \eqref{eq:m-anatomy} reads $\mathbf s-\mathbf s_\omega=\mathbf R^{*}_\omega\boldsymbol\nu_\omega$. The content row of a country $k$ and its dollar loss are
\begin{equation}
\begin{aligned}
\mathbf m^{*\top}_k&:=\mathbbm 1_k\,\mathbf R^{*}_\omega,\\
Q(k)-Q_\omega(k)&=\mathbf m^{*\top}_k\boldsymbol\nu_\omega.
\end{aligned}
\label{eq:m-content}
\end{equation}
Here $m^{*}_{k,i}=\sum_{l:\,\mathcal C(l)=k}[\mathbf R^{*}_\omega]_{li}$ is the content of country $k$ in a dollar of producer $i$'s output, and $\underline\beta:=\min_j\beta_j$. Every step of this subsection uses only the two conditions of Definition~\ref{def:m-sanction-equilibrium} and the bound $\mathbf h^{*}\le\mathbf h_\omega$ of Proposition~\ref{prop:m-monotone}\textnormal{(ii)}, so the statements hold at every sanction equilibrium $(\mathbf x^{*},\mathbf h^{*})$ with $\mathbf h^{*}$ in place of $\mathbf h_\omega$.

\begin{lemma}[Content rows]
\label{lem:m-content}
\begin{enumerate}
\item[\textnormal{(i)}] $\mathbf R^{*}_\omega=\sum_{t\ge0}[\diag(\mathbf h_\omega)\mathbf B_\omega]^{t}$, and for every country $k$ and country-sector $i$,
\begin{equation}
m^{*}_{k,i}
=\mathbbm 1[\mathcal C(i)=k]
+\sum_{t\ge1}\ \sum_{l:\,\mathcal C(l)=k}\ \sum_{l_1,\dots,l_{t-1}}
h_{\omega,l}\,B_{\omega,ll_1}\,h_{\omega,l_1}\,B_{\omega,l_1l_2}\cdots h_{\omega,l_{t-1}}\,B_{\omega,l_{t-1}i}.
\label{eq:m-content-chains}
\end{equation}
Here $B_{\omega,ab}=a^{(\omega)}_{ab}(1-\beta_b)$, and each summand is a chain of $t$ effective demands from country-sector $i$ upstream to a country-sector $l$ of $k$, every link weighted by the buyer's intermediate spending share and the seller's fulfillment rate.
\item[\textnormal{(ii)}] $m^{*}_{k,i}\ge1$ if $\mathcal C(i)=k$ and $m^{*}_{k,i}\ge0$ otherwise, and $\sum_km^{*}_{k,i}=\mathbf 1^{\top}\mathbf R^{*}_\omega\mathbf e_i\le1/\underline\beta$ for every country-sector $i$.
\item[\textnormal{(iii)}] $m^{*}_{k,i}$ is a function of the effective demand matrix $\mathbf A_\omega$, the leakages $\boldsymbol\beta$, and the fulfillment rates $\mathbf h_\omega$, and the level of economic activity of a country-sector does not enter $m^{*}_{k,i}$.
\end{enumerate}
\end{lemma}

\begin{proof}
We first prove \textnormal{(i)}. By Lemma~\ref{lem:m-kernel}, $\|\diag(\mathbf h_\omega)\mathbf B_\omega\|_1\le\|\mathbf B_\omega\|_1<1$, so the Neumann series converges to the inverse. Writing out the $(l,i)$ entry of the $t$-th power and summing over the country-sectors $l$ of $k$ gives \eqref{eq:m-content-chains}, the term $t=0$ contributing $\mathbbm 1[l=i]$. This proves \textnormal{(i)}.

We next prove \textnormal{(ii)}. Every summand of \eqref{eq:m-content-chains} is nonnegative, and the term $t=0$ equals one exactly when $i$ is a country-sector of $k$. For the sum over $k$, note that $\sum_km^{*}_{k,i}=\sum_l[\mathbf R^{*}_\omega]_{li}=\mathbf 1^{\top}\mathbf R^{*}_\omega\mathbf e_i$. The column sums of $\diag(\mathbf h_\omega)\mathbf B_\omega$ satisfy $\sum_lh_{\omega,l}B_{\omega,lb}\le(1-\beta_b)\sum_la^{(\omega)}_{lb}\le1-\underline\beta$, the last step by the column sums of Lemma~\ref{lem:m-kernel}. Hence $\mathbf 1^{\top}[\diag(\mathbf h_\omega)\mathbf B_\omega]^{t}\le(1-\underline\beta)^{t}\mathbf 1^{\top}$ entrywise, by induction on $t$, and summing the series gives $\mathbf 1^{\top}\mathbf R^{*}_\omega\le\underline\beta^{-1}\mathbf 1^{\top}$. This proves \textnormal{(ii)}.

Part \textnormal{(iii)} is read off \eqref{eq:m-content-chains}, whose every factor is a standing sourcing share, an intermediate spending share $1-\beta_b$, or a fulfillment rate. This proves the lemma.
\end{proof}

The severed sales and the severed purchases of a country in Section~\ref{subsec:power_size} are entries of the cross-border trade matrix. Summing the intermediate flows $Z_{ij}$ of \eqref{eq:m-flows} and the household purchases $f_i^c$ over the country-sectors of each country, with the buyer in the first index, gives the matrix
\begin{equation}
T^{\mathrm{cb}}_{cc'}
:=\sum_{i:\mathcal C(i)=c'}\Bigl[\sum_{j:\mathcal C(j)=c}Z_{ij}+f_i^c\Bigr]
\quad(c\ne c'),\qquad T^{\mathrm{cb}}_{cc}:=0.
\label{eq:country_flow_table_crossborder}
\end{equation}
The entry $T^{\mathrm{cb}}_{cc'}$ is what the producers and the household of country $c$ buy from the producers of country $c'$, the flows that the severance $c'\notnotleftrightarrow c$ bars in one direction, so that $M_{cc'}=T^{\mathrm{cb}}_{cc'}$ and $X_{cc'}=T^{\mathrm{cb}}_{c'c}$.

Write $X_i:=\sum_{b:\,i\to b\text{ severed}}Z_{ib}$ for the severed sales of country-sector $i$, to producers and households alike, so that $\xi_{\omega,i}=\tau(1-\delta)X_i$ by \eqref{eq:m-lost-sales}, and $M_j:=\sum_{i:\,i\to j\text{ severed}}Z_{ij}$ for the severed intermediate purchases of producer $j$. For a producer $j$ and an input sector $r\in\mathcal R_j$ write $\sigma^{(\omega)}_{rj}:=\alpha_{rj}^{-1}\sum_{i\in\mathcal N_r:\,i\to j\text{ severed}}a_{ij}$ for the severed suppliers' share of producer $j$'s benchmark purchases of input $r$. Write $\bar\sigma_{\omega,j}:=\sum_{r\in\mathcal R_j}\alpha_{rj}\sigma^{(\omega)}_{rj}=\sum_{i:\,i\to j\text{ severed}}a_{ij}$ for their share of its intermediate purchases in all, so that $M_j=w_j\bar\sigma_{\omega,j}$ by \eqref{eq:m-flows}.

\begin{lemma}[The production shortfalls]
\label{lem:m-shortfall}
For every producer $j$, $(1-h_{\omega,j})\,s_j\ge\tau(1-\delta)\,M_j$.
\end{lemma}

\begin{proof}
The proof takes four steps. First, the benchmark path of Proposition~\ref{prop:m-monotone}\textnormal{(i)} is nonincreasing from $\mathbf h^{(0)}=\mathbf 1$, so its limit satisfies $\mathbf h_\omega\le\mathbf h^{(1)}=\boldsymbol\lambda^{(\omega)}(\mathbf 1)$. Second, at full delivery the retention ratio \eqref{eq:m-theta} is the block sum \eqref{eq:m-blocksum} divided by $\alpha_{rj}$, which is $\theta^{(\omega)}_{rj}(\mathbf 1)=1-\tau(1-\delta)\sigma^{(\omega)}_{rj}$. The efficiency factor \eqref{eq:m-kappa} therefore satisfies
\[
\kappa^{(\omega)}_j(\mathbf 1)
=\Bigl(\sum_{r\in\mathcal R_j}\alpha_{rj}\bigl(1-\tau(1-\delta)\sigma^{(\omega)}_{rj}\bigr)^{\rho}\Bigr)^{1/\rho}
\le\sum_{r\in\mathcal R_j}\alpha_{rj}\bigl(1-\tau(1-\delta)\sigma^{(\omega)}_{rj}\bigr)
=1-\tau(1-\delta)\bar\sigma_{\omega,j},
\]
the inequality being the power-mean inequality between the order $\rho<1$ and the order one, the weights $\alpha_{rj}$ summing to one over $\mathcal R_j$. Third, Lemma~\ref{lem:m-outer}\textnormal{(ii)} gives $G_j(\kappa)\le\kappa^{1-\beta_j}$ on $[0,1]$, and $\kappa\mapsto\kappa^{1-\beta_j}$ is increasing, so
\[
\lambda^{(\omega)}_j(\mathbf 1)
=G_j\bigl(\kappa^{(\omega)}_j(\mathbf 1)\bigr)
\le\bigl(1-\tau(1-\delta)\bar\sigma_{\omega,j}\bigr)^{1-\beta_j}
\le1-(1-\beta_j)\,\tau(1-\delta)\,\bar\sigma_{\omega,j},
\]
the last step by Bernoulli's inequality $(1-x)^{a}\le1-ax$ for $x\in[0,1]$ and $a\in[0,1]$. Fourth, the three displays give $1-h_{\omega,j}\ge1-\lambda^{(\omega)}_j(\mathbf 1)\ge(1-\beta_j)\tau(1-\delta)\bar\sigma_{\omega,j}$, and multiplying by $s_j$ gives $(1-h_{\omega,j})s_j\ge\tau(1-\delta)\,w_j\bar\sigma_{\omega,j}=\tau(1-\delta)M_j$. This proves the lemma.
\end{proof}

\begin{lemma}[The balance in the lost sales]
\label{lem:m-injuries}
Suppose that the unrecovered share $\tau$ is positive and that countries $c$ and $c'$ trade in both directions. At the selected equilibrium of the severance $c'\notnotleftrightarrow c$, the lost sales \eqref{eq:m-lost-sales} of the two trading partners sum to $\sum_{i\in c}\xi_{\omega,i}=\tau(1-\delta)X_{cc'}$ and $\sum_{i\in c'}\xi_{\omega,i}=\tau(1-\delta)M_{cc'}$. Their normalized difference is $-\mathcal B_{cc'}$, the negative of the trade balance \eqref{eq:direct_trade_asymmetry}.
\end{lemma}

\begin{proof}
Under the severance $c'\notnotleftrightarrow c$ a link from a seller of country $c$ is severed exactly when its buyer, producer or household, sits in country $c'$. Summing $\xi_{\omega,i}=\tau(1-\delta)X_i$ over the country-sectors of country $c$ therefore gives $\tau(1-\delta)$ times the benchmark flow from the sellers of country $c$ to the producers and the household of country $c'$, which is $T^{\mathrm{cb}}_{c'c}=X_{cc'}$ by \eqref{eq:country_flow_table_crossborder}. The same argument on the sellers of country $c'$ gives $\tau(1-\delta)\,T^{\mathrm{cb}}_{cc'}=\tau(1-\delta)M_{cc'}$. The normalized difference of the two totals is $\psi(X_{cc'}/M_{cc'})=-\psi(M_{cc'}/X_{cc'})=-\mathcal B_{cc'}$ by \eqref{eq:direct_trade_asymmetry} and Lemma~\ref{lem:m-psi}\textnormal{(i)}. This proves the lemma.
\end{proof}

Under the bilateral severance $c'\notnotleftrightarrow c$ the lost-sales vector vanishes outside the two countries of the pair, so the dollar loss of country $c$ in \eqref{eq:m-content} has three parts,
\begin{equation}
Q(c)\,\gamma^{(c'\notnotleftrightarrow c)}_c
=\underbrace{\sum_{i\in c}m^{*}_{c,i}h_{\omega,i}\xi_{\omega,i}}_{\text{own lost sales}}
+\underbrace{\sum_{i\in c'}m^{*}_{c,i}h_{\omega,i}\xi_{\omega,i}}_{\text{partner's lost sales}}
+\underbrace{\sum_{i}m^{*}_{c,i}(1-h_{\omega,i})s_i}_{\text{shortfalls}} .
\label{eq:m-three-parts}
\end{equation}
The own lost sales are what country $c$ loses through its own producers' lost sales to the partner, each dollar carrying the content of country $c$ in that producer's output. The partner's lost sales are what country $c$ loses when the partner's producers lose their sales to it and order fewer inputs from it, each dollar carrying the content of country $c$ in the partner's output. The shortfalls are what country $c$ loses through the production shortfalls of the world, its own producers' degraded bundles first among them. The own lost sales are a cost of lost sales, and the partner's lost sales are a cost of lost purchases. The shortfalls have two causes, the severed purchases of country $c$, which starve its buyer sectors, and its severed sales, which starve the buyer sectors of the partner, and we split them by cause with two further fixed points.

Write $\boldsymbol\lambda^{(\omega,c)}$ for the one-sided fulfillment rule of country $c$, the output ratios \eqref{eq:m-lambda-rule} evaluated at the retention ratios
\begin{equation}
\theta^{(\omega,c)}_{rj}(\mathbf h):=
\begin{cases}
\theta^{(\omega)}_{rj}(\mathbf h)+\tau(1-\delta)\,\sigma^{(\omega)}_{rj}, & j\in c',\\
\theta^{(\omega)}_{rj}(\mathbf h), & j\notin c',
\end{cases}
\label{eq:m-onesided-theta}
\end{equation}
and $\boldsymbol\lambda^{(\omega,c')}$ for the one-sided fulfillment rule of country $c'$, defined by the same display with the roles of the two countries exchanged. By the second step of the proof of Lemma~\ref{lem:m-shortfall}, $\theta^{(\omega)}_{rj}(\mathbf 1)=1-\tau(1-\delta)\sigma^{(\omega)}_{rj}$, so the rule $\boldsymbol\lambda^{(\omega,c)}$ gives every buyer sector of the partner, at full delivery, the retention ratios of a producer with no severed link. It leaves every other producer, and every chain of deliveries, as the severance has them, since the effective demand matrix $\mathbf A_\omega$ stays where the rationed exchange put it. The rule $\boldsymbol\lambda^{(\omega,c)}$ therefore describes an economy in which only the buyer sectors of country $c$ have lost their inputs, and $\boldsymbol\lambda^{(\omega,c')}$ one in which only the buyer sectors of the partner have. Write $\mathbf h^{(c)}_\omega$ and $\mathbf h^{(c')}_\omega$ for the greatest fixed points of the two one-sided rules, the one-sided fulfillment rates, and write
\begin{equation}
\begin{aligned}
S_c&:=\sum_i m^{*}_{c,i}\,(1-h_{\omega,i})\,s_i,\\
S^{(c)}_c&:=\sum_i m^{*}_{c,i}\,\bigl(1-h^{(c)}_{\omega,i}\bigr)\,s_i,\qquad
S^{(c')}_c:=\sum_i m^{*}_{c,i}\,\bigl(1-h^{(c')}_{\omega,i}\bigr)\,s_i,\\
S^{\times}_c&:=S_c-S^{(c)}_c-S^{(c')}_c,
\end{aligned}
\label{eq:m-shortfall-split}
\end{equation}
for the shortfall loss of country $c$, the shortfall loss that its lost purchases cause on their own, the shortfall loss that its lost sales cause on their own, and the interaction of the two causes. The content row $\mathbf m^{*}_c$ is that of the selected equilibrium in all four sums. The shortfall loss $S_c$ is the third part of \eqref{eq:m-three-parts}, and the two sums of \eqref{eq:m-two-parts} in Section~\ref{subsec:power_size} that run over all producers are $S^{(c')}_c$ and $S^{(c)}_c$. The interaction $L^{\times}_c=L_c-L^{\downarrow}_c-L^{\uparrow}_c$ of that section is therefore $S^{\times}_c$.

\begin{lemma}[The one-sided fulfillment rates]
\label{lem:m-onesided}
Suppose that the unrecovered share $\tau$ is positive and that countries $c$ and $c'$ trade in both directions, and let $k\in\{c,c'\}$.
\begin{enumerate}
\item[\textnormal{(i)}] The one-sided rule $\boldsymbol\lambda^{(\omega,k)}$ is continuous and entrywise nondecreasing on $[0,1]^N$, and it satisfies $\boldsymbol\lambda^{(\omega)}\le\boldsymbol\lambda^{(\omega,k)}$ and $\boldsymbol\lambda^{(\omega,k)}(\mathbf 1)\le\mathbf 1$. Its benchmark path converges to its greatest fixed point $\mathbf h^{(k)}_\omega$, and $\mathbf h_\omega\le\mathbf h^{(k)}_\omega\le\mathbf 1$.
\item[\textnormal{(ii)}] $0\le S^{(c)}_c\le S_c$, $0\le S^{(c')}_c\le S_c$, and $S^{\times}_c\ge-\min\{S^{(c)}_c,S^{(c')}_c\}$. Write $T_c:=X_{cc'}+M_{cc'}$ for the pair's trade. The export multiplier, the import multiplier, and the interaction rate \eqref{eq:m-rates} satisfy
\begin{equation}
\begin{aligned}
\tau(1-\delta)X_{cc'}\,(\ell^{\downarrow}_c+\ell^{\times}_c)&=\sum_{i\in c}m^{*}_{c,i}h_{\omega,i}\xi_{\omega,i}
+\sum_i m^{*}_{c,i}s_i\Bigl[\bigl(1-h^{(c')}_{\omega,i}\bigr)\frac{M_{cc'}}{T_c}+\bigl(h^{(c)}_{\omega,i}-h_{\omega,i}\bigr)\frac{X_{cc'}}{T_c}\Bigr],\\
\tau(1-\delta)M_{cc'}\,(\ell^{\uparrow}_c+\ell^{\times}_c)&=\sum_{i\in c'}m^{*}_{c,i}h_{\omega,i}\xi_{\omega,i}
+\sum_i m^{*}_{c,i}s_i\Bigl[\bigl(1-h^{(c)}_{\omega,i}\bigr)\frac{X_{cc'}}{T_c}+\bigl(h^{(c')}_{\omega,i}-h_{\omega,i}\bigr)\frac{M_{cc'}}{T_c}\Bigr],
\end{aligned}
\label{eq:m-rates-nonneg}
\end{equation}
so that each multiplier with its interaction rate added is a nonnegative combination of the entries of the content row, and $\ell^{\times}_c\ge-\min\{\ell^{\downarrow}_c,\ell^{\uparrow}_c\}$.
\item[\textnormal{(iii)}] For every producer $j$ of country $c$, $(1-h^{(c)}_{\omega,j})\,s_j\ge\tau(1-\delta)\,M_j$, and for every producer $j$ of country $c'$, $(1-h^{(c')}_{\omega,j})\,s_j\ge\tau(1-\delta)\,M_j$.
\end{enumerate}
\end{lemma}

\begin{proof}
We first prove \textnormal{(i)} for $k=c$, the case $k=c'$ being the same with the roles of the two countries exchanged. The rule $\boldsymbol\lambda^{(\omega,c)}$ differs from $\boldsymbol\lambda^{(\omega)}$ only in the retention ratios of the producers of country $c'$, each raised by a constant in $[0,1]$. For such a producer $j$ and $\mathbf h\le\mathbf 1$, Lemma~\ref{lem:m-rule} gives $\theta^{(\omega)}_{rj}(\mathbf h)\le\theta^{(\omega)}_{rj}(\mathbf 1)=1-\tau(1-\delta)\sigma^{(\omega)}_{rj}$, so $\theta^{(\omega,c)}_{rj}(\mathbf h)\in[0,1]$ and $\theta^{(\omega,c)}_{rj}(\mathbf 1)=1$. The argument of Lemma~\ref{lem:m-rule} therefore applies to the one-sided rule without change. It is continuous and entrywise nondecreasing, and $\boldsymbol\lambda^{(\omega,c)}(\mathbf 1)\le\mathbf 1$ because every retention ratio at full delivery lies in $[0,1]$. Since the efficiency factor \eqref{eq:m-kappa} and the outer aggregator $G_j$ are nondecreasing in the retention ratios, $\boldsymbol\lambda^{(\omega)}\le\boldsymbol\lambda^{(\omega,c)}$ entrywise. The benchmark path of the one-sided rule, $\mathbf h^{(0)}=\mathbf 1$ and $\mathbf h^{(n+1)}=\boldsymbol\lambda^{(\omega,c)}(\mathbf h^{(n)})$, is nonincreasing and bounded below by $\mathbf 0$, by the argument of part (i) of Proposition~\ref{prop:m-monotone}, so it converges to a fixed point $\mathbf h^{(c)}_\omega$, and every fixed point $\mathbf h^{*}\le\mathbf 1$ of the rule satisfies $\mathbf h^{*}=\boldsymbol\lambda^{(\omega,c)}(\mathbf h^{*})\le\boldsymbol\lambda^{(\omega,c)}(\mathbf h^{(n)})=\mathbf h^{(n+1)}$ for every $n$ by induction, hence $\mathbf h^{*}\le\mathbf h^{(c)}_\omega$. For the comparison with $\mathbf h_\omega$, note that $\mathbf h_\omega\le\mathbf 1=\mathbf h^{(0)}$, and that $\mathbf h_\omega\le\mathbf h^{(n)}$ gives $\mathbf h_\omega=\boldsymbol\lambda^{(\omega)}(\mathbf h_\omega)\le\boldsymbol\lambda^{(\omega,c)}(\mathbf h_\omega)\le\boldsymbol\lambda^{(\omega,c)}(\mathbf h^{(n)})=\mathbf h^{(n+1)}$. Taking limits gives $\mathbf h_\omega\le\mathbf h^{(c)}_\omega$. This proves \textnormal{(i)}.

We next prove \textnormal{(ii)}. The content row is nonnegative by Lemma~\ref{lem:m-content}\textnormal{(ii)}, and by \textnormal{(i)} every factor $1-h^{(k)}_{\omega,i}$ lies between $0$ and $1-h_{\omega,i}$, which gives $0\le S^{(k)}_c\le S_c$ for both $k$. The lower bound on $S^{\times}_c$ follows from $S^{\times}_c=(S_c-S^{(c)}_c)-S^{(c')}_c\ge-S^{(c')}_c$ and its counterpart with the two countries exchanged. For the display, add the interaction rate to the export multiplier of \eqref{eq:m-rates} and substitute $S^{\times}_c=S_c-S^{(c)}_c-S^{(c')}_c$, which gives $\tau(1-\delta)X_{cc'}(\ell^{\downarrow}_c+\ell^{\times}_c)=\sum_{i\in c}m^{*}_{c,i}h_{\omega,i}\xi_{\omega,i}+S^{(c')}_c\,M_{cc'}/(X_{cc'}+M_{cc'})+(S_c-S^{(c)}_c)\,X_{cc'}/(X_{cc'}+M_{cc'})$, and write out $S^{(c')}_c$ and $S_c-S^{(c)}_c$ as sums over producers. The import multiplier is treated in the same way. Every bracket in \eqref{eq:m-rates-nonneg} is nonnegative by \textnormal{(i)}, and so is every other factor. For the last inequality, $S^{\times}_c\ge-S^{(c')}_c$ and $S^{(c')}_c\le\tau(1-\delta)X_{cc'}\ell^{\downarrow}_c$ by \eqref{eq:m-rates} give $\tau(1-\delta)(X_{cc'}+M_{cc'})\ell^{\times}_c\ge-\tau(1-\delta)X_{cc'}\ell^{\downarrow}_c$, hence $\ell^{\times}_c\ge-\ell^{\downarrow}_c$, and $\ell^{\times}_c\ge-\ell^{\uparrow}_c$ follows in the same way from $S^{\times}_c\ge-S^{(c)}_c$. This proves \textnormal{(ii)}.

We finally prove \textnormal{(iii)}. For a producer $j$ of country $c$ the retention ratios of the one-sided rule $\boldsymbol\lambda^{(\omega,c)}$ are those of the severance, so $\theta^{(\omega,c)}_{rj}(\mathbf 1)=1-\tau(1-\delta)\sigma^{(\omega)}_{rj}$, and the benchmark path of the one-sided rule is nonincreasing from $\mathbf 1$ by \textnormal{(i)}. The four steps of the proof of Lemma~\ref{lem:m-shortfall} therefore apply to $\mathbf h^{(c)}_\omega$ at producer $j$ without change. The statement for the producers of country $c'$ follows with the roles of the two countries exchanged. This proves the lemma.
\end{proof}

The interaction $S^{\times}_c$ is a mixed second difference. For $(t,t')\in[0,1]^2$ let $\boldsymbol\lambda^{(\omega)}[t,t']$ be the fulfillment rule in which the retention ratio of every buyer sector $j$ of country $c$ is raised by $(1-t)\tau(1-\delta)\sigma^{(\omega)}_{rj}$ and that of every buyer sector of the partner by $(1-t')\tau(1-\delta)\sigma^{(\omega)}_{rj}$. The rule scales the two input shortfalls that the severance plants, at the buyer sectors of country $c$ and at the buyer sectors of the partner, by $t$ and $t'$. Let $\mathbf h_\omega[t,t']$ be its greatest fixed point and $S_c(t,t'):=\sum_im^{*}_{c,i}(1-h_{\omega,i}[t,t'])s_i$ the shortfall loss at the equilibrium content row. The rule at $(1,1)$ is $\boldsymbol\lambda^{(\omega)}$, the rules at $(1,0)$ and $(0,1)$ are the two one-sided rules, and at $(0,0)$ every retention ratio at full delivery is one, so that $\mathbf h_\omega[0,0]=\mathbf 1$ and $S_c(0,0)=0$. Hence $S^{(c)}_c=S_c(1,0)-S_c(0,0)$ and $S^{(c')}_c=S_c(0,1)-S_c(0,0)$ are first differences, and $S^{\times}_c=S_c(1,1)-S_c(1,0)-S_c(0,1)+S_c(0,0)$ is the mixed second difference. Lemma~\ref{lem:m-interaction} shows that, when the two input shortfalls are small, the first differences are of first order in them and the mixed second difference is of second order.

\begin{lemma}[The interaction is of second order]
\label{lem:m-interaction}
Suppose that the unrecovered share $\tau$ is positive and that countries $c$ and $c'$ trade in both directions. Then there are constants $\bar t\in(0,1]$ and $K<\infty$ such that, for every $(t,t')\in[0,\bar t]^2$,
\begin{equation}
\begin{aligned}
S_c(t,0)&\le K\,t,\\
S_c(0,t')&\le K\,t',\\
\bigl|S_c(t,t')-S_c(t,0)-S_c(0,t')\bigr|&\le K\,t\,t'.
\end{aligned}
\label{eq:m-interaction-order}
\end{equation}
\end{lemma}

\begin{proof}
The retention ratios of the rule $\boldsymbol\lambda^{(\omega)}[t,t']$ are those of \eqref{eq:m-theta} raised by amounts that are affine in $t$ and $t'$, so they are affine in $(\mathbf h,t,t')$, and they equal one at $(\mathbf 1,0,0)$. The efficiency factor \eqref{eq:m-kappa} is smooth where every retention ratio is positive, and the outer aggregator $G_j$ is smooth on $(0,\infty)$ by Lemma~\ref{lem:m-outer}\textnormal{(i)}. The rule is therefore a smooth function of $(\mathbf h,t,t')$ on a neighborhood of $(\mathbf 1,0,0)$ in $\mathbb R^{N}\times\mathbb R^{2}$.

We first compute the Jacobian of the rule with respect to $\mathbf h$ at $(\mathbf 1,0,0)$. There every retention ratio and every efficiency factor equals one, the derivative of the efficiency factor of producer $j$ with respect to its sector-$r$ retention ratio is $\alpha_{rj}$, and $G_j'(1)=1-\beta_j$. By \eqref{eq:m-theta} and the chain rule, the derivative of the output ratio of producer $j$ with respect to $h_i$ is $(1-\beta_j)\,a^{(\omega)}_{ij}$, which is the $(i,j)$ entry of $\mathbf B_\omega$. The Jacobian is therefore $\mathbf B_\omega^{\top}$. Since $\|\mathbf B_\omega\|_1<1$ by Lemma~\ref{lem:m-kernel}, the matrix $\mathbf I-\mathbf B_\omega^{\top}$ is nonsingular, with the nonnegative inverse $\sum_{n\ge0}(\mathbf B_\omega^{\top})^{n}$. By the implicit function theorem there are a neighborhood $U$ of $(0,0)$, a neighborhood $V$ of $\mathbf 1$, and a twice continuously differentiable map $\mathbf z:U\to V$ with $\mathbf z(0,0)=\mathbf 1$ such that, for every $(t,t')\in U$, the point $\mathbf z(t,t')$ is the only fixed point of $\boldsymbol\lambda^{(\omega)}[t,t']$ in $V$.

We next show that the greatest fixed point $\mathbf h_\omega[t,t']$ lies in $V$ when $t$ and $t'$ are small. Write $\mathbf u:=(\mathbf I-\mathbf B_\omega^{\top})^{-1}\mathbf 1$, so that $\mathbf u-\mathbf B_\omega^{\top}\mathbf u=\mathbf 1$. Since the rule is smooth near $(\mathbf 1,0,0)$ and equals $\mathbf 1$ there, Taylor's theorem gives constants $\eta_0>0$ and $K_0<\infty$, with $\mathbf 1-\eta_0\mathbf u\ge\mathbf 0$, such that
\[
\boldsymbol\lambda^{(\omega)}[t,t'](\mathbf 1-\eta\mathbf u)
\ge\mathbf 1-\eta\,\mathbf B_\omega^{\top}\mathbf u-K_0\,(t+t'+\eta^{2})\,\mathbf 1
\]
whenever $0\le t,t',\eta\le\eta_0$. Take $\eta:=2K_0(t+t')$. When $t+t'$ is small enough that $\eta\le\eta_0$ and $4K_0^{2}(t+t')\le1$, the right-hand side is at least $\mathbf 1-\eta\mathbf u$, so the vector $\mathbf 1-\eta\mathbf u$ lies below its own image under the rule. The rule is entrywise nondecreasing and maps $\mathbf 1$ below $\mathbf 1$, by the argument of Lemma~\ref{lem:m-onesided}\textnormal{(i)}, so every point of its benchmark path from $\mathbf 1$ lies above $\mathbf 1-\eta\mathbf u$, and so does the limit of the path, which is $\mathbf h_\omega[t,t']$. Hence $\mathbf 1-2K_0(t+t')\mathbf u\le\mathbf h_\omega[t,t']\le\mathbf 1$, and $\mathbf h_\omega[t,t']$ lies in $V$ once $t+t'$ is small. Being a fixed point of the rule in $V$, it is then $\mathbf z(t,t')$.

Choose $\bar t\in(0,1]$ so small that $[0,\bar t]^2\subset U$ and $\mathbf h_\omega[t,t']=\mathbf z(t,t')$ on $[0,\bar t]^2$. On this square $S_c(t,t')=\sum_im^{*}_{c,i}\bigl(1-z_i(t,t')\bigr)s_i$ is twice continuously differentiable, and $S_c(0,0)=0$. Let $K$ bound the absolute values of $\partial S_c/\partial t$, $\partial S_c/\partial t'$, and $\partial^{2}S_c/\partial t\,\partial t'$ on the square. Then $S_c(t,0)=\int_0^{t}\partial S_c/\partial t\,(x,0)\,dx\le Kt$, and $S_c(0,t')\le Kt'$ in the same way. The mixed second difference is $S_c(t,t')-S_c(t,0)-S_c(0,t')=\int_0^{t}\!\!\int_0^{t'}\partial^{2}S_c/\partial t\,\partial t'\,(x,y)\,dy\,dx$, which is at most $K\,t\,t'$ in absolute value. This proves the lemma.
\end{proof}

At the full size of the severance the same integral gives $|S^{\times}_c|\le\max_{[0,1]^2}|\partial^2S_c/\partial t\,\partial t'|$ whenever $S_c$ is twice continuously differentiable on the whole unit square. That differentiability holds when the greatest fixed point is the unique interior fixed point of the rule at every $(t,t')$ and the Jacobian $\mathbf I-D\boldsymbol\lambda^{(\omega)}[t,t']$ is nonsingular there.

For $\tau>0$ and trade in both directions, measuring each part per dollar of the lost trade that produces it gives the export multiplier, the import multiplier, and the interaction rate of Section~\ref{subsec:power_size},
\begin{equation}
\begin{aligned}
\ell^{\downarrow}_c&:=\frac{\sum_{i\in c}m^{*}_{c,i}h_{\omega,i}\xi_{\omega,i}+S^{(c')}_c}{\tau(1-\delta)X_{cc'}},
\qquad
\ell^{\uparrow}_c:=\frac{\sum_{i\in c'}m^{*}_{c,i}h_{\omega,i}\xi_{\omega,i}+S^{(c)}_c}{\tau(1-\delta)M_{cc'}},\\
\ell^{\times}_c&:=\frac{S^{\times}_c}{\tau(1-\delta)(X_{cc'}+M_{cc'})},
\end{aligned}
\label{eq:m-rates}
\end{equation}
the activity country $c$ loses per dollar of lost sales, per dollar of lost purchases, and, through the interaction, per dollar of the pair's lost trade. The numerator of $\ell^{\downarrow}_c$ is $L^{\downarrow}_c$ of \eqref{eq:m-two-parts}, that of $\ell^{\uparrow}_c$ is $L^{\uparrow}_c$, and $S^{\times}_c$ is $L^{\times}_c$.

\subsection{Proof of Proposition~\ref{prop:m-exposure}}
\label{app:m-proof-exposure}

Throughout this subsection and the two that follow, the pair $\{c,c'\}$ trades in both directions, the unrecovered share $\tau$ is positive, and the severance is $c'\notnotleftrightarrow c$. The equilibrium is its selected equilibrium, as in Section~\ref{sec:power_size}, so that the export multiplier $\ell^{\downarrow}_c$ and the import multiplier $\ell^{\uparrow}_c$ of \eqref{eq:m-rates} are well defined. We first record the scale invariance, and we then record the dependence of the export multiplier and the import multiplier on shares. The text of Section~\ref{subsec:power_size} uses both.

\begin{lemma}[Scale invariance]
\label{lem:m-scale}
Hold the sourcing shares $\mathbf A$ and $\mathbb A$ and the leakages $\boldsymbol\beta$ fixed. Carrying $(\mathbf s,\mathbf F,\mathbf f^{o})$ to $\varsigma(\mathbf s,\mathbf F,\mathbf f^{o})$ for a positive $\varsigma$ leaves the effective demand matrices $\mathbf A_\omega$ and $\mathbb A_\omega$, the fulfillment fixed point $\mathbf h_\omega$, the content rows, and every vulnerability where they were. The same change carries $\mathbf s_\omega$, the lost sales $\boldsymbol\xi_\omega$, and the direct-loss vector $\boldsymbol\nu_\omega$ to $\varsigma$ times themselves.
\end{lemma}

\begin{proof}
Under the rescaling the spending $w_b$ of every buyer and every flow $Z_{ib}$ of \eqref{eq:m-flows} are multiplied by $\varsigma$, and so are $\mathbf f=\mathbb A\mathbf F+\mathbf f^{o}$ and the benchmark activity, which solves \eqref{eq:m-benchmark} at the rescaled $\mathbf f$. The disrupted coefficients \eqref{eq:m-disrupted} and the column targets \eqref{eq:m-coltarget} are built from sourcing shares alone and do not move, and the row targets \eqref{eq:m-rowtarget} are multiplied by $\varsigma$. If a flow block is feasible for the program \eqref{eq:m-balprog} at the original data, then $\varsigma$ times it is feasible at the rescaled data. Its objective at the rescaled data is $\varsigma$ times the objective of the original block, since $\varsigma X_{ij}\log\bigl(\varsigma X_{ij}/(\varsigma\check a_{ij}w_j)\bigr)=\varsigma X_{ij}\log\bigl(X_{ij}/(\check a_{ij}w_j)\bigr)$. The unique optimizer of Proposition~\ref{prop:m-balancing} is therefore multiplied by $\varsigma$, and the coefficients \eqref{eq:m-balanced-coeff}, flows divided by spending, do not move. The retention ratios \eqref{eq:m-theta}, the efficiency factors \eqref{eq:m-kappa}, and the output ratios \eqref{eq:m-lambda-rule} are built from the effective demand, the sector shares, the leakages, and the exponents, so the fulfillment rule and its greatest fixed point $\mathbf h_\omega$ do not move. The one-sided rules \eqref{eq:m-onesided-theta} add severed sourcing shares to the retention ratios, so they and their greatest fixed points do not move either. Neither do the content rows, which Lemma~\ref{lem:m-content}\textnormal{(iii)} builds from $\mathbf A_\omega$, $\boldsymbol\beta$, and $\mathbf h_\omega$. The lost sales \eqref{eq:m-lost-sales} are sums of flows and scale by $\varsigma$, and so does the effective final demand $\mathbf f_\omega=\mathbb A_\omega\mathbf F+\mathbf f^{o}$. The unique solution $\mathbf s_\omega$ of the activity equation \eqref{eq:m-equilibrium} at $\mathbf h_\omega$ therefore scales by $\varsigma$, and with it the direct-loss vector and the loss vector. The own-size normalization of Definition~\ref{def:country_vulnerability} removes the factor from every vulnerability. This proves the lemma.
\end{proof}

\begin{lemma}[The export multiplier and the import multiplier depend on levels only through shares]
\label{lem:m-rates-shares}
The export multiplier $\ell^{\downarrow}_c$ and the import multiplier $\ell^{\uparrow}_c$ depend on the sourcing shares $a_{ib}$, the shares $Z_{ib}/s_i$ of each seller's activity that its customers take, the leakages, and the parameters. A level of activity therefore reaches either multiplier only through the shares $Z_{ib}/s_i$.
\end{lemma}

\begin{proof}
Write $a^{\downarrow}_{ib}:=Z_{ib}/s_i$ for the share of seller $i$'s activity that buyer $b$ takes, so that $a^{\downarrow}_{ij}=a_{ij}(1-\beta_j)s_j/s_i$ on a producer link and $a^{\downarrow}_{ic}=a_{ic}F^{c}/s_i$ on a household link by \eqref{eq:m-flows}. On a link with $Z_{ib}>0$ these identities determine the ratio $s_j/s_i$, or $F^{c}/s_i$, from the two shares and the leakage. The support graph of the benchmark flows is connected, a standing condition of Section~\ref{subsec:model_benchmark}, so a path from a fixed seller $i_0$ to any seller or household determines every ratio $s_i/s_{i_0}$ and $F^{c}/s_{i_0}$ from the shares and the leakages. The benchmark row identity $s_i=f^{o}_i+\sum_bZ_{ib}$ gives $f^{o}_i=s_i\bigl(1-\sum_ba^{\downarrow}_{ib}\bigr)$. The shares and the leakages thus determine $(\mathbf s,\mathbf F,\mathbf f^{o})$ up to one positive factor. By Lemma~\ref{lem:m-scale} that factor reaches neither the effective demand, nor $\mathbf h_\omega$, nor the one-sided fulfillment rates, nor the content rows, while it multiplies the direct losses, the flows, and the activities alike. The numerator and the denominator of each fraction in \eqref{eq:m-rates} are sums of products of contents, fulfillment rates, and one direct loss, flow, or activity each, so the factor cancels from every multiplier. Each multiplier is a function of the sourcing shares, the shares $a^{\downarrow}_{ib}$, the leakages, and the parameters. This proves the lemma.
\end{proof}

\begin{proof}[Proof of Proposition~\ref{prop:m-exposure}]
We first prove \eqref{eq:m-exact-exposure}. By Definition~\ref{def:country_vulnerability} and the anatomy \eqref{eq:m-anatomy}, $Q(c)\gamma^{(c'\notnotleftrightarrow c)}_c=\mathbbm 1_c(\mathbf s-\mathbf s_\omega)=\mathbbm 1_c\mathbf R^{*}_\omega\boldsymbol\nu_\omega=\mathbf m^{*\top}_c\boldsymbol\nu_\omega$, which is $\sum_im^{*}_{c,i}\bigl[(1-h_{\omega,i})s_i+h_{\omega,i}\xi_{\omega,i}\bigr]$. A seller outside the pair has no severed link, so $\xi_{\omega,i}$ vanishes outside the country-sectors of countries $c$ and $c'$, and the lost-sales part of the sum splits into the sum over country $c$ and the sum over country $c'$, which gives the three parts of \eqref{eq:m-three-parts}. The third part is the shortfall loss $S_c=S^{(c)}_c+S^{(c')}_c+S^{\times}_c$ of \eqref{eq:m-shortfall-split}. By \eqref{eq:m-rates}, $\tau(1-\delta)X_{cc'}\ell^{\downarrow}_c$ is the first part together with $S^{(c')}_c$, $\tau(1-\delta)M_{cc'}\ell^{\uparrow}_c$ is the second part together with $S^{(c)}_c$, and $\tau(1-\delta)(X_{cc'}+M_{cc'})\ell^{\times}_c$ is $S^{\times}_c$, so the three sum to the three parts. Dividing by $Q(c)$ gives \eqref{eq:m-exact-exposure}. For country $c'$ the severed sales are $M_{cc'}$ and the severed purchases $X_{cc'}$, and the same argument applies. This proves \eqref{eq:m-exact-exposure}.

We next prove \eqref{eq:m-size-bound}. Since $\ell^{\times}_c(X_{cc'}+M_{cc'})=\ell^{\times}_cX_{cc'}+\ell^{\times}_cM_{cc'}$, \eqref{eq:m-exact-exposure} gives the dollar loss of country $c$ as $L_c=Q(c)\gamma^{(c'\notnotleftrightarrow c)}_c=\tau(1-\delta)[(\ell^{\downarrow}_c+\ell^{\times}_c)X_{cc'}+(\ell^{\uparrow}_c+\ell^{\times}_c)M_{cc'}]$. For country $c'$ the severed sales are $M_{cc'}$ and the severed purchases $X_{cc'}$, so the dollar loss of country $c'$ is $L_{c'}=Q(c')\gamma^{(c'\notnotleftrightarrow c)}_{c'}=\tau(1-\delta)[(\ell^{\uparrow}_{c'}+\ell^{\times}_{c'})X_{cc'}+(\ell^{\downarrow}_{c'}+\ell^{\times}_{c'})M_{cc'}]$. Dividing the dollar loss of country $c$ by the dollar loss of country $c'$ gives
\[
\frac{L_c}{L_{c'}}
=\frac{(\ell^{\downarrow}_c+\ell^{\times}_c)X_{cc'}+(\ell^{\uparrow}_c+\ell^{\times}_c)M_{cc'}}{(\ell^{\uparrow}_{c'}+\ell^{\times}_{c'})X_{cc'}+(\ell^{\downarrow}_{c'}+\ell^{\times}_{c'})M_{cc'}},
\]
which is a weighted average of the two ratios of \eqref{eq:m-size-bound}, with the weights $(\ell^{\uparrow}_{c'}+\ell^{\times}_{c'})X_{cc'}$ and $(\ell^{\downarrow}_{c'}+\ell^{\times}_{c'})M_{cc'}$ divided by their sum. The two weights are positive, since the partner's two multipliers with their interaction rate added, $X_{cc'}$, and $M_{cc'}$ are, and they sum to one, so the right side lies between the two ratios. This proves the proposition.
\end{proof}

\begin{lemma}[Bounds on the export multiplier and the import multiplier]
\label{lem:m-rate-bounds}
Write $\underline h_c$ for the lowest fulfillment rate among the producers of country $c$ that sell to the partner and $\iota_c$ for the intermediate share of country $c$'s severed purchases, the fraction of $M_{cc'}$ that its producers rather than its household buy. Then $\ell^{\downarrow}_c\ge\underline h_c$ and $\ell^{\uparrow}_c\ge\iota_c$.
\end{lemma}

\begin{proof}
By \eqref{eq:m-rates} and Lemma~\ref{lem:m-onesided}\textnormal{(ii)}, $\tau(1-\delta)X_{cc'}\ell^{\downarrow}_c=\sum_{i\in c}m^{*}_{c,i}h_{\omega,i}\xi_{\omega,i}+S^{(c')}_c\ge\sum_{i\in c}m^{*}_{c,i}h_{\omega,i}\xi_{\omega,i}$. Every term of this sum with $\xi_{\omega,i}>0$ has $h_{\omega,i}\ge\underline h_c$ and $m^{*}_{c,i}\ge1$, the latter by Lemma~\ref{lem:m-content}\textnormal{(ii)}, and $\sum_{i\in c}\xi_{\omega,i}=\tau(1-\delta)X_{cc'}$ by Lemma~\ref{lem:m-injuries}, so the sum is at least $\underline h_c\tau(1-\delta)X_{cc'}$. For the bound on $\ell^{\uparrow}_c$, the numerator of $\ell^{\uparrow}_c$ in \eqref{eq:m-rates} has a nonnegative first term, and in $S^{(c)}_c$ keep only the producers $j$ of country $c$, use $m^{*}_{c,j}\ge1$, and apply Lemma~\ref{lem:m-onesided}\textnormal{(iii)}, which gives $S^{(c)}_c\ge\tau(1-\delta)\sum_{j\in c}M_j$. Hence $\ell^{\uparrow}_c\ge\sum_{j\in c}M_j/M_{cc'}=\iota_c$, since the severed purchases of country $c$ are the severed intermediate purchases of its producers together with the severed purchases of its household. This proves the lemma.
\end{proof}

Equation~\eqref{eq:m-exact-exposure} takes a closed form when every producer sources every intermediate input at home and every producer of country $c$ has the leakage $\beta_c$. Within each input sector a producer then has one incumbent supplier, its own country's producer of that sector, no such link is severed under $c'\notnotleftrightarrow c$, and the column constraint of the program \eqref{eq:m-balprog} fixes that coefficient at its benchmark value, so that $\mathbf A_\omega=\mathbf A$. Every retention ratio at full delivery is then one, $\boldsymbol\lambda^{(\omega)}(\mathbf 1)=\mathbf 1$, and the benchmark path selects $\mathbf h_\omega=\mathbf 1$, so the shortfalls vanish, the one-sided fulfillment rates are one as well by Lemma~\ref{lem:m-onesided}\textnormal{(i)}, and $\boldsymbol\nu_\omega=\boldsymbol\xi_\omega$. Domestic sourcing and unit column sums give $\mathbbm 1_c\mathbf A=\mathbbm 1_c$, and $\mathbf B_\omega=\mathbf A(\mathbf I-\boldsymbol\beta)$ carries no entry between countries, so that $\mathbbm 1_c\mathbf B_\omega=(1-\beta_c)\mathbbm 1_c$ and $\mathbbm 1_c\mathbf R^{*}_\omega=\sum_{t\ge0}(1-\beta_c)^{t}\mathbbm 1_c=\beta_c^{-1}\mathbbm 1_c$. The content row of country $c$ is therefore $\beta_c^{-1}$ on the country-sectors of country $c$ and zero elsewhere, the partner's lost sales contribute nothing to the loss of country $c$, and \eqref{eq:m-exact-exposure} reads
\begin{equation}
\ell^{\downarrow}_c=\frac{1}{\beta_c},
\qquad
\ell^{\uparrow}_c=\ell^{\times}_c=0,
\qquad
\gamma^{(c'\notnotleftrightarrow c)}_c=\frac{\tau(1-\delta)}{\beta_c}\,\frac{X_{cc'}}{Q(c)}.
\label{eq:m-home-sourcing}
\end{equation}
Here $X_{cc'}$ is the final sales of country $c$ to the household of country $c'$, since no intermediate link crosses the border. Suppose in addition that the leakage is a common $\beta$, that no household buys at home, that the residual outside demand vanishes, and that every household relationship of the network carries the same flow $t$ in each direction. A country with $n_c$ trading partners then sells $t$ to each of their households and has the budget $F^{c}=tn_c$ and the size $Q(c)=tn_c/\beta$, so \eqref{eq:m-home-sourcing} gives $\gamma^{(c'\notnotleftrightarrow c)}_c=\tau(1-\delta)/n_c$ for every pair of partners. The asymmetry $\mathcal H_{cc'}=\psi\bigl(n_{c'}/n_c\bigr)$ then compares only the numbers of trading relationships of the two countries.

\subsection{Home sourcing and the cost of a lost dollar}
\label{app:m-proof-homechain}

We say that the orders of a country-sector $l$ reach a producer $j$ if a chain of effective demands with positive fulfillment leads from $l$ upstream to $j$, so that a dollar of orders at $l$ generates orders at $j$. The lemma states that the content of a country in its producers, and with it the export multiplier and the import multiplier, rise as its producers buy more of their inputs at home, and it bounds the content from below in closed form when home sourcing is uniform.

\begin{lemma}[Home sourcing and the cost of a lost dollar]
\label{lem:m-homechain}
Fix a sanction, its effective demand matrix $\mathbf A_\omega$, and fulfillment rates $\mathbf h\in(0,1]^N$, and let $\mathbf m^{*}_c$ be the content row of country $c$ at those orders and rates. For the multipliers, fix as well the one-sided fulfillment rates of the severance.
\begin{enumerate}
\item[\textnormal{(i)}] Let a producer $j$ move a share of its effective demand in one input sector from a supplier $k'$ outside country $c$ to a supplier $k$ inside it. If $h_k m^{*}_{c,k}\ge h_{k'}m^{*}_{c,k'}$, then no entry of $\mathbf m^{*}_c$ falls, and if the inequality is strict the entry at producer $j$ and at every country-sector whose orders reach it rises. The export multiplier $\ell^{\downarrow}_c$ and the import multiplier $\ell^{\uparrow}_c$ of country $c$ are nondecreasing in the content row.
\item[\textnormal{(ii)}] Suppose the producers of country $c$ share the leakage $\beta_c$ and each buys the fraction $\varphi_c$ of its intermediate inputs from producers of country $c$. At the benchmark orders and full delivery, the content of country $c$ in each of its producers is at least $1/(1-\varphi_c(1-\beta_c))$, with equality when no producer abroad buys from country $c$.
\end{enumerate}
\end{lemma}

\begin{proof}
Write $\mathbf M:=\diag(\mathbf h)\mathbf A_\omega(\mathbf I-\boldsymbol\beta)$ and $\mathbf R:=(\mathbf I-\mathbf M)^{-1}$, so that $\mathbf m^{*}_c=\mathbbm 1_c\mathbf R$. Since $\mathbf h\le\mathbf 1$, Lemma~\ref{lem:m-kernel} gives $\|\mathbf M\|_1\le\max_j(1-\beta_j)<1$, so $\mathbf I-\mathbf M$ is a nonsingular M-matrix, $\mathbf R=\sum_{t\ge0}\mathbf M^{t}\ge\mathbf 0$, and $\det(\mathbf I-\mathbf M)>0$. The entry $R_{jl}$ is positive exactly when $(\mathbf M^{t})_{jl}>0$ for some $t\ge0$, that is, when $l=j$ or the orders of $l$ reach $j$.

We first prove \textnormal{(i)}. The move replaces $a^{(\omega)}_{k'j}$ by $a^{(\omega)}_{k'j}-\Delta$ and $a^{(\omega)}_{kj}$ by $a^{(\omega)}_{kj}+\Delta$ for some $\Delta\in(0,a^{(\omega)}_{k'j}]$, with $k$ and $k'$ in the same input sector, and it leaves every other entry of the effective demand matrix and the fulfillment rates where they were. The new propagation matrix is $\mathbf M':=\mathbf M+\mathbf u\mathbf e_j^{\top}$ with $\mathbf u:=\Delta(1-\beta_j)(h_k\mathbf e_k-h_{k'}\mathbf e_{k'})$. Column $j$ of the new effective demand matrix still sums to at most one and $\mathbf h\le\mathbf 1$, so $\|\mathbf M'\|_1\le\max_j(1-\beta_j)<1$ as well, and $\mathbf R':=(\mathbf I-\mathbf M')^{-1}$ is a nonnegative matrix with $\det(\mathbf I-\mathbf M')>0$. The matrix determinant lemma gives $\det(\mathbf I-\mathbf M')=\det(\mathbf I-\mathbf M)\,(1-\mathbf e_j^{\top}\mathbf R\mathbf u)$, so the scalar $1-\mathbf e_j^{\top}\mathbf R\mathbf u$ is positive, and the Sherman--Morrison formula gives
\[
\mathbf R'
=\mathbf R+\frac{\mathbf R\mathbf u\,\mathbf e_j^{\top}\mathbf R}{1-\mathbf e_j^{\top}\mathbf R\mathbf u}.
\]
Multiplying on the left by $\mathbbm 1_c$,
\[
\mathbbm 1_c\mathbf R'-\mathbbm 1_c\mathbf R
=\frac{\Delta(1-\beta_j)\bigl[h_k m^{*}_{c,k}-h_{k'}m^{*}_{c,k'}\bigr]}{1-\mathbf e_j^{\top}\mathbf R\mathbf u}\;\mathbf e_j^{\top}\mathbf R,
\]
since $\mathbbm 1_c\mathbf R\mathbf u=\Delta(1-\beta_j)(h_km^{*}_{c,k}-h_{k'}m^{*}_{c,k'})$. The row $\mathbf e_j^{\top}\mathbf R$ is nonnegative, and its entry at $l$ is positive exactly when $l=j$ or the orders of $l$ reach $j$. Under the condition $h_km^{*}_{c,k}\ge h_{k'}m^{*}_{c,k'}$ the fraction is nonnegative, so no entry of the content row falls, and when the condition is strict the fraction is positive and every entry with $R_{jl}>0$ rises. By \eqref{eq:m-rates} and Lemma~\ref{lem:m-onesided}\textnormal{(i)} the export multiplier and the import multiplier of country $c$ are nonnegative combinations of the entries of the content row, with weights built from the lost sales, the fulfillment rates, the one-sided fulfillment rates, the benchmark activities, and the severed flows. None of these moves, so both the export multiplier and the import multiplier are nondecreasing in the entries of the content row. This proves \textnormal{(i)}.

We next prove \textnormal{(ii)}. At the benchmark orders and full delivery the propagation matrix is $\mathbf B:=\mathbf A(\mathbf I-\boldsymbol\beta)$ and the content row of country $c$ solves $\mathbf m_c=\mathbbm 1_c+\mathbf m_c\mathbf B$. Write $\mathbf m$ for the restriction of $\mathbf m_c$ to the country-sectors of country $c$ and $\mathbf A_{cc}$ for the block of $\mathbf A$ whose rows and columns are the country-sectors of country $c$. By hypothesis every column of $\mathbf A_{cc}$ sums to $\varphi_c$, so $\mathbf 1^{\top}\mathbf A_{cc}=\varphi_c\mathbf 1^{\top}$, and every producer of country $c$ has the leakage $\beta_c$. For a country-sector $i$ of country $c$,
\[
m_{c,i}
=1+(1-\beta_c)\sum_{k\in c}m_{c,k}a_{ki}+(1-\beta_c)\sum_{k\notin c}m_{c,k}a_{ki}
\;\ge\;1+(1-\beta_c)\sum_{k\in c}m_{c,k}a_{ki},
\]
since the content row is nonnegative by Lemma~\ref{lem:m-content}. In matrix form, $\mathbf m\bigl(\mathbf I-(1-\beta_c)\mathbf A_{cc}\bigr)\ge\mathbf 1^{\top}$. The columns of $(1-\beta_c)\mathbf A_{cc}$ sum to $(1-\beta_c)\varphi_c<1$, so $\bigl(\mathbf I-(1-\beta_c)\mathbf A_{cc}\bigr)^{-1}=\sum_{t\ge0}\bigl((1-\beta_c)\mathbf A_{cc}\bigr)^{t}$ is nonnegative, and multiplying the inequality by it on the right preserves it. Since $\mathbf 1^{\top}\mathbf A_{cc}^{t}=\varphi_c^{t}\mathbf 1^{\top}$ for every $t$, by induction on $t$,
\[
\mathbf m
\;\ge\;\mathbf 1^{\top}\sum_{t\ge0}\bigl((1-\beta_c)\mathbf A_{cc}\bigr)^{t}
=\sum_{t\ge0}\bigl((1-\beta_c)\varphi_c\bigr)^{t}\,\mathbf 1^{\top}
=\frac{\mathbf 1^{\top}}{1-\varphi_c(1-\beta_c)} .
\]
If no producer abroad buys from country $c$, then $a_{ki}=0$ whenever $k$ is a country-sector of country $c$ and $i$ is not, so every chain of sales that leaves country $c$ never returns to it, $(\mathbf B^{t})_{lk'}=0$ for every $l$ in country $c$, every $k'$ outside it, and every $t$, and $m_{c,k'}=0$ for every $k'$ outside country $c$. The inequality above is then an equality at every step. This proves the lemma.
\end{proof}

\subsection{Lost purchases and lost sales}
\label{app:m-ordering}

The anatomy \eqref{eq:m-anatomy} measures each direct loss at the producer that suffers it, and the same content row carries a direct loss to the country whichever cause produced it. The two causes differ in the size of the direct loss a lost dollar produces. Write $X_i:=\sum_{b:\,i\to b\text{ severed}}Z_{ib}$ for the severed sales of country-sector $i$ and $M_j:=\sum_{i:\,i\to j\text{ severed}}Z_{ij}$ for the severed intermediate purchases of producer $j$, as in Appendix~\ref{app:m-content}.

\begin{lemma}[A lost input costs more than a lost order]
\label{lem:m-ordering}
Let $(\mathbf s_\omega,\mathbf h_\omega)$ be the selected equilibrium of a sanction and let $j$ be a producer of country $c$ with severed sales $X_j>0$ and severed intermediate purchases $M_j>0$. In the anatomy \eqref{eq:m-anatomy}, the direct loss at producer $j$ from its own lost sales costs country $c$ the amount $m^{*}_{c,j}h_{\omega,j}\,\tau(1-\delta)X_j$, at most $m^{*}_{c,j}$ per unrecovered dollar of its sales. The direct loss from its shortfall costs country $c$ the amount $m^{*}_{c,j}(1-h_{\omega,j})s_j$, at least $m^{*}_{c,j}$ per unrecovered dollar of its intermediate purchases.
\end{lemma}

\begin{proof}
The loss of country $c$ is $\mathbf m^{*\top}_c\boldsymbol\nu_\omega$ with $\boldsymbol\nu_\omega=(\mathbf 1-\mathbf h_\omega)\circ\mathbf s+\mathbf h_\omega\circ\boldsymbol\xi_\omega$, by Proposition~\ref{prop:m-anatomy}, so the two direct losses at producer $j$ enter it as $m^{*}_{c,j}h_{\omega,j}\xi_{\omega,j}$ and $m^{*}_{c,j}(1-h_{\omega,j})s_j$. The lost sales of producer $j$ are $\xi_{\omega,j}=\tau(1-\delta)X_j$ by \eqref{eq:m-lost-sales}, which gives the first amount, and $h_{\omega,j}\le1$ gives its bound. Lemma~\ref{lem:m-shortfall} gives $(1-h_{\omega,j})s_j\ge\tau(1-\delta)M_j$, which divided by the unrecovered intermediate purchases $\tau(1-\delta)M_j$ gives the second bound. This proves the lemma.
\end{proof}

The second amount is the contribution of producer $j$'s own shortfall. Its shortfall lowers what it delivers to each of its customers, whose bundles degrade in turn and whose own shortfalls enter the anatomy at their own country-sectors. With its input bundle held fixed, its lost sales do not travel that way, for the fulfillment rule of Lemma~\ref{lem:m-rule} depends on the producer's input bundle alone, so the customers it keeps are served at the same rate as before.

\subsection{Proof of Proposition~\ref{prop:m-decomposition}}
\label{app:m-proof-decomposition}

\begin{proof}
The gross volume of the pair's trade is $M_{cc'}+X_{cc'}$, so that $M_{cc'}=\tfrac12(M_{cc'}+X_{cc'})(1+\mathcal B_{cc'})$ and $X_{cc'}=\tfrac12(M_{cc'}+X_{cc'})(1-\mathcal B_{cc'})$ by \eqref{eq:direct_trade_asymmetry}. Each trading partner sells to the other and the unrecovered share is positive, so each has a producer with positive lost sales. The dollar loss \eqref{eq:m-content} of a country is a sum of nonnegative terms. At such a producer $i$ of country $c$ the term is $m^{*}_{c,i}\bigl[(1-h_{\omega,i})s_i+h_{\omega,i}\xi_{\omega,i}\bigr]\ge\xi_{\omega,i}>0$, since $m^{*}_{c,i}\ge1$ by Lemma~\ref{lem:m-content}\textnormal{(ii)} and $\xi_{\omega,i}\le s_i$ by \eqref{eq:m-lost-sales}. Both vulnerabilities are therefore positive, so $\Sigma_c$ and $\Sigma_{c'}$ are positive, and the two import premia of \eqref{eq:m-gap} lie in $[-1,1]$ because, by Lemma~\ref{lem:m-onesided}\textnormal{(ii)}, each interaction rate is at least minus the smaller of the two multipliers of the same country, so that $\Sigma_k\ge|\ell^{\uparrow}_k-\ell^{\downarrow}_k|$. By \eqref{eq:m-exact-exposure} and the two expressions of $M_{cc'}$ and $X_{cc'}$ in the trade balance,
\begin{align*}
Q(c)\gamma^{(c'\notnotleftrightarrow c)}_c
&=\tau(1-\delta)\bigl[\ell^{\downarrow}_cX_{cc'}+\ell^{\uparrow}_cM_{cc'}+\ell^{\times}_c(X_{cc'}+M_{cc'})\bigr]\\
&=\tfrac12\tau(1-\delta)(M_{cc'}+X_{cc'})\bigl[\ell^{\downarrow}_c(1-\mathcal B_{cc'})+\ell^{\uparrow}_c(1+\mathcal B_{cc'})+2\ell^{\times}_c\bigr]\\
&=\tfrac12\tau(1-\delta)(M_{cc'}+X_{cc'})\,\Sigma_c\bigl(1+\mathcal B_{cc'}\mathcal G_c\bigr),
\end{align*}
the last step because $\ell^{\downarrow}_c(1-\mathcal B_{cc'})+\ell^{\uparrow}_c(1+\mathcal B_{cc'})+2\ell^{\times}_c=(\ell^{\downarrow}_c+\ell^{\uparrow}_c+2\ell^{\times}_c)+\mathcal B_{cc'}(\ell^{\uparrow}_c-\ell^{\downarrow}_c)=\Sigma_c(1+\mathcal B_{cc'}\mathcal G_c)$ by \eqref{eq:m-gap}. Since $L_c=Q(c)\gamma^{(c'\notnotleftrightarrow c)}_c$, this display is \eqref{eq:m-balanced-loss}. For country $c'$ the severed sales are $M_{cc'}$ and the severed purchases $X_{cc'}$, so
\begin{align*}
Q(c')\gamma^{(c'\notnotleftrightarrow c)}_{c'}
&=\tau(1-\delta)\bigl[\ell^{\downarrow}_{c'}M_{cc'}+\ell^{\uparrow}_{c'}X_{cc'}+\ell^{\times}_{c'}(X_{cc'}+M_{cc'})\bigr]\\
&=\tfrac12\tau(1-\delta)(M_{cc'}+X_{cc'})\bigl[\ell^{\downarrow}_{c'}(1+\mathcal B_{cc'})+\ell^{\uparrow}_{c'}(1-\mathcal B_{cc'})+2\ell^{\times}_{c'}\bigr]\\
&=\tfrac12\tau(1-\delta)(M_{cc'}+X_{cc'})\,\Sigma_{c'}\bigl(1-\mathcal B_{cc'}\mathcal G_{c'}\bigr),
\end{align*}
which is \eqref{eq:m-balanced-loss} for the partner, the sign of the trade balance reversed since $\mathcal B_{c'c}=-\mathcal B_{cc'}$. Both final brackets are positive, being positive dollar losses divided by positive constants. Dividing the first display by the second and then by the ratio of the two sizes gives \eqref{eq:m-exact-decomposition}, the common factor $\tfrac12\tau(1-\delta)(M_{cc'}+X_{cc'})$ cancelling. If each country sold the other the same $t>0$ dollars, with the multipliers and the interaction rates unchanged, \eqref{eq:m-exact-exposure} would give the dollar losses $\tau(1-\delta)\,t\,\Sigma_c$ and $\tau(1-\delta)\,t\,\Sigma_{c'}$. The loss ratio of the pair at balanced trade is therefore the product of the first two ratios of \eqref{eq:m-exact-decomposition}, whatever the volume $t$.
\end{proof}

We next derive \eqref{eq:m-balance-transform}. By \eqref{eq:m-exact-decomposition} the ratio of the loss ratio $r_{cc'}$ to the loss ratio $r^{\mathrm b}_{cc'}$ at balanced trade is the trade-imbalance effect, so by \eqref{eq:m-psi} the normalized difference of the two loss ratios is the normalized difference of the trade-imbalance effect. Write $x:=\mathcal B_{cc'}\mathcal G_c$ and $y:=\mathcal B_{cc'}\mathcal G_{c'}$. Both flows are positive, so $|\mathcal B_{cc'}|<1$, and by the proof the two import premia lie in $[-1,1]$, so $|x|<1$ and $|y|<1$. Evaluating $\psi$ by \eqref{eq:m-psi} gives
\[
\psi\Bigl(\frac{1+x}{1-y}\Bigr)=\frac{(1+x)-(1-y)}{(1+x)+(1-y)}=\frac{x+y}{2+x-y},
\]
whose denominator is positive, and substituting for $x$ and $y$ gives \eqref{eq:m-balance-transform}.

The bound \eqref{eq:m-balance-bound} of the text and its equality case follow from \eqref{eq:m-balance-transform}. With the same $x$ and $y$, write $u:=\max\{|x|,|y|\}=|\mathcal B_{cc'}|\max\{|\mathcal G_c|,|\mathcal G_{c'}|\}<1$. If $x+y\ge0$, then
\[
u(2+x-y)-(x+y)=(u-x)(1-u)+(u-y)(1+u)\ge0,
\]
since $x\le u$, $y\le u$, and $u<1$, with equality if and only if $x=y=u$. If $x+y<0$, then
\[
u(2+x-y)+(x+y)=(u+x)(1+u)+(u+y)(1-u)\ge0,
\]
since $x\ge-u$ and $y\ge-u$, with equality if and only if $x=y=-u$. In either case $|x+y|\le u(2+x-y)$, which is \eqref{eq:m-balance-bound} by \eqref{eq:m-balance-transform}, and equality holds if and only if $x=y$, that is, if and only if $\mathcal B_{cc'}=0$ or $\mathcal G_c=\mathcal G_{c'}$. The denominator $2+x-y$ is positive, so the normalized difference, and with it the logarithm of the trade-imbalance effect, has the sign of $x+y=\mathcal B_{cc'}(\mathcal G_c+\mathcal G_{c'})$. The normalized difference vanishes if and only if $\mathcal B_{cc'}=0$ or $\mathcal G_c=-\mathcal G_{c'}$, and when $\mathcal G_c=\mathcal G_{c'}=\mathcal G$ it equals $\mathcal B_{cc'}\mathcal G$. To first order in the trade balance $\mathcal B_{cc'}$, the normalized difference is $\mathcal B_{cc'}$ times the mean of the two import premia, since its numerator is linear in $\mathcal B_{cc'}$ and its denominator equals two at $\mathcal B_{cc'}=0$. Once the export multipliers and the import multipliers are given, the gross volume $M_{cc'}+X_{cc'}$ enters the two dollar losses only through the common factor $\tfrac12\tau(1-\delta)(M_{cc'}+X_{cc'})$, which cancels from the loss ratio.

We now show that the three ratios of \eqref{eq:m-exact-decomposition} do not depend on the level at which the two flows of the pair are made to meet. Suppose that each country sold the other $t_\zeta:=\zeta M_{cc'}+(1-\zeta)X_{cc'}$ dollars for some weight $\zeta\in(0,1)$, with the export multipliers, the import multipliers, and the interaction rates held at their values at the selected equilibrium. By the last step of the proof with $t=t_\zeta$, the dollar losses of the two countries would be $\tau(1-\delta)\,t_\zeta\,\Sigma_c$ and $\tau(1-\delta)\,t_\zeta\,\Sigma_{c'}$, so the linkage effect at balanced trade is $\Sigma_c/\Sigma_{c'}$ at every $\zeta$. And since the loss ratio itself does not involve $\zeta$, neither does its ratio to the loss ratio at balanced trade, which is the trade-imbalance effect. The weight enters only the comparison of a country's own loss with its loss at balanced trade. Since $t_\zeta=\tfrac12(M_{cc'}+X_{cc'})\bigl(1+(2\zeta-1)\mathcal B_{cc'}\bigr)$, the dollar loss \eqref{eq:m-balanced-loss} of country $c$ is its loss at balanced trade times $(1+\mathcal B_{cc'}\mathcal G_c)/\bigl(1+(2\zeta-1)\mathcal B_{cc'}\bigr)$, and that of country $c'$ is its loss at balanced trade times $(1-\mathcal B_{cc'}\mathcal G_{c'})/\bigl(1+(2\zeta-1)\mathcal B_{cc'}\bigr)$. The common denominator cancels from the ratio of the two factors. At $\zeta=1/2$, which is the balanced trade of Section~\ref{subsec:power_surplus}, the common denominator is one, and the first factors of each line of \eqref{eq:m-balanced-loss} are the loss at balanced trade.

\subsection{The rationed exchange for a sector block}
\label{app:m-decentralization}

This subsection states the fixed-price equilibrium with rationing behind the effective demand of Section~\ref{subsec:model_rerouting}, together with the result that section quotes, and the proof that follows works on the support-reduced block. Fix a sanction $\omega\in\Omega$, an unrecovered share $\tau\in[0,1]$, a barred-link attenuation $\delta$ as in \eqref{eq:m-leak}, and a sector $r$ whose block contains a severed link. All objects are defined on the support-reduced block $\mathcal I_r^{\omega}\times\mathcal J_r^{\omega}$ of active sellers and active buyers, the column index $j$ runs over producers and households alike, and $\check Z_{ij}=\check a_{ij}w_j$ is the disrupted flow kernel. Flows are valued at fixed benchmark prices, so a flow is also a quantity. The buyers' budgets $E_{rj}=t^{(\omega)}_{rj}w_j$ and the sellers' capacities $g^{(\omega)}_i$ are the column and row targets of the balancing program \eqref{eq:m-balprog}, and the proofs use the two names interchangeably. Normalizing the incumbency weights $\pi_{ij}$ in the log-linear criterion of Definition~\ref{def:oa-rationed-exchange} rescales the criterion built on $\check Z_{ij}$ itself by the positive constant $1/\sum_k\check Z_{kj}$, so it leaves every demand, pseudo-price, and equilibrium allocation unchanged.

\begin{definition}[Fixed-price equilibrium with rationing]
\label{def:oa-rationed-exchange}
Fix a sector $r$ whose block contains a severed link. Posted prices are one on every link. Each active buyer $b$, producer or household, holds the sector budget $E_{rb}:=t^{(\omega)}_{rb}w_b>0$ and ranks bundles $X_{\cdot b}\ge0$ on its incumbent links $\{i:\check Z_{ib}>0\}$ by the log-linear criterion
\[
U_b(X_{\cdot b})
:=
\sum_{i:\,\check Z_{ib}>0}\pi_{ib}\log X_{ib},
\qquad
\pi_{ib}:=\check Z_{ib}\Big/\sum_k\check Z_{kb}.
\]
Here $U_b:=-\infty$ whenever an incumbent purchase is zero. Each active seller $i$ holds the capacity $g^{(\omega)}_i>0$. A fixed-price equilibrium with rationing on the block is an allocation $X\ge0$ on the incumbent links together with strictly positive pseudo-prices $\tilde{\mathbf p}$ that satisfy two conditions.
\begin{enumerate}
\item[\textnormal{(i)}] For every active buyer $b$ there is a pseudo-income $\tilde E_{rb}>0$ such that $X_{\cdot b}$ maximizes $U_b$ over $\{Y_{\cdot b}\ge0:\sum_i\tilde p_iY_{ib}\le\tilde E_{rb}\}$. The bundle also exhausts the posted-price budget, $\sum_iX_{ib}=E_{rb}$.
\item[\textnormal{(ii)}] Every active seller sells its capacity, $\sum_bX_{ib}=g^{(\omega)}_i$.
\end{enumerate}
\end{definition}

\begin{proposition}[Existence, uniqueness, and decentralization of the rationed exchange]
\label{prop:m-decentralization}
Fix a sector $r$ whose block contains a severed link, and consider the fixed-price equilibrium with rationing of Definition~\ref{def:oa-rationed-exchange} on that block. Then:
\begin{enumerate}
\item[\textnormal{(i)}] An equilibrium exists, and its allocation is unique. The equilibrium coefficients are supported on $\operatorname{supp}(\check{\mathbf A})$, and on the block they take the biproportional form \eqref{eq:m-reroute} of Section~\ref{subsec:model_rerouting}, with positive seller scalings $u_i$ and buyer scalings $v_{rb}$ on the active rows and columns. Since the attenuation is positive, the active rows and columns are the rows and columns with a positive entry in the benchmark table.
\item[\textnormal{(ii)}] The pseudo-prices are the reciprocals of the seller scalings of \eqref{eq:m-reroute}, $\tilde p_i=1/u_i$, and they are determined up to one positive scalar on each connected component of the block's support graph, the choice of num\'eraire.
\end{enumerate}
\end{proposition}

\subsection{Proof of Proposition~\ref{prop:m-decentralization}}

\begin{proof}
We first prove part (i) by identifying the equilibrium allocation with the
optimizer of the flow-balancing program \eqref{eq:m-balprog}, which shows that an equilibrium
with pseudo-prices exists and that no other allocation is an equilibrium. Throughout we hold the sector index $r$ fixed and write
$v_j:=v_{rj}$ for the column scaling of \eqref{eq:m-reroute}.

We first show that the balanced block is an equilibrium allocation. By
Proposition~\ref{prop:m-balancing}, the balanced flow block
$\mathbf Z^{(r)}_\omega$ is strictly positive on the retained support, meets
the row targets $g^{(\omega)}_i$ and the column flow targets
$t^{(\omega)}_{rj}w_j$, and has the biproportional form
$X_{ij}=u_i\check Z_{ij}v_j$ with strictly positive active-block scalings
$(u_i,v_j)$. Define $\tilde p_i:=u_i^{-1}$ and
$\tilde E_{rj}:=\sum_i\tilde p_iX_{ij}=v_j\sum_k\check Z_{kj}$. The log-linear criterion is strictly concave on the positive orthant and tends to $-\infty$ as any incumbent purchase tends to zero, so its maximizer on the pseudo-budget set is interior and unique, and the first-order conditions give the constant-share demand
\[
X^{d}_{ij}
=
\frac{\pi_{ij}\,\tilde E_{rj}}{\tilde p_i}
=
\frac{\check Z_{ij}\,\tilde E_{rj}}{\tilde p_i\sum_k\check Z_{kj}}.
\]
Substituting $\tilde p_i^{-1}=u_i$ and $\tilde E_{rj}=v_j\sum_k\check Z_{kj}$ gives $X^{d}_{ij}=u_i\check Z_{ij}v_j=X_{ij}$. The column margin
gives $\sum_iX_{ij}=t^{(\omega)}_{rj}w_j=E_{rj}$, which is posted-price budget
exhaustion, and the row margin gives capacity clearing. The pair
$(\mathbf Z^{(r)}_\omega,\tilde{\mathbf p})$ is therefore an equilibrium with
pseudo-prices.

We next show that every equilibrium allocation equals the balanced block. Let
$(X,\tilde{\mathbf p})$ be an equilibrium with pseudo-prices. The criterion
is strictly increasing, so the pseudo-budget binds, and the
first-order conditions of buyer optimality give
$X_{ij}=\check Z_{ij}\tilde E_{rj}/(\tilde p_i\sum_k\check Z_{kj})$ on the incumbent links, strictly
positive. Setting $u_i:=\tilde p_i^{-1}$ and
$v_j:=\tilde E_{rj}/\sum_k\check Z_{kj}$ puts the
allocation in the biproportional form $X_{ij}=u_i\check Z_{ij}v_j$ with
strictly positive scalings. Posted-price budget exhaustion delivers the
column margins and capacity clearing the row margins. Any matrix of this form
that meets both prescribed margins is stationary for the strictly convex
flow-balancing program \eqref{eq:m-balprog}, because
$\log\bigl(X_{ij}/\check Z_{ij}\bigr)=\log u_i+\log v_j$ is a sum of row and
column potentials, and stationarity on the transportation polytope identifies
the unique optimizer. Hence $X=\mathbf Z^{(r)}_\omega$, so the equilibrium
flows minimize relative entropy against the disrupted flows among the matrices
meeting both prescribed margins, which is the characterization stated in
Section~\ref{subsec:model_rerouting}. This proves part \textnormal{(i)}.

We last prove part (ii), on the pseudo-prices and the gauge. The
identification $u_i=\tilde p_i^{-1}$ in either direction of part \textnormal{(i)}
gives $\tilde p_i=u_i^{-1}$. By
Proposition~\ref{prop:m-balancing}, the scalings are unique up to
$(u,v)\mapsto(\chi u,\chi^{-1}v)$ on each connected component of the support
graph. Under $\tilde p_i=u_i^{-1}$ this freedom is the rescaling
$\tilde{\mathbf p}\mapsto\chi^{-1}\tilde{\mathbf p}$ with the virtual
incomes rescaled in proportion, and the constant-share demand is homogeneous
of degree zero in the price-income pair, so the allocation is unchanged. Fixing
one pseudo-price on each component removes the freedom, which is the choice of
num\'eraire. This proves part (ii).
\end{proof}

The log-linear criterion of Definition~\ref{def:oa-rationed-exchange} is the Cobb--Douglas member of the CES family, and the allocation is the same for every member. We record the statement, since the flexible-price economy of Online Appendix~\ref{app:flex} makes the within-sector curvature explicit.

\begin{lemma}[Curvature invariance of the rationed exchange]
\label{lem:m-curvature}
Fix the sector block of Proposition~\ref{prop:m-decentralization}, and replace the log-linear criterion of Definition~\ref{def:oa-rationed-exchange} by the CES criterion with within-sector curvature $\rho_r\in(-\infty,0)\cup(0,1)$,
\[
U_b(X_{\cdot b})
:=
\Bigl(\sum_{i:\,\check Z_{ib}>0}\pi_{ib}^{\,1-\rho_r}\,X_{ib}^{\,\rho_r}\Bigr)^{1/\rho_r}.
\]
Here $U_b:=0$ whenever an incumbent purchase is zero and $\rho_r<0$. The equilibrium allocation is the balanced block $\mathbf Z^{(r)}_\omega$ for every such curvature, and the supporting pseudo-prices are $\tilde p_i=u_i^{\,\rho_r-1}$ for the seller scalings $u_i$ of \eqref{eq:m-reroute}.
\end{lemma}

\begin{proof}
Hold the sector index $r$ fixed and write $v_j:=v_{rj}$. At prices $\tilde{\mathbf p}$ and income $\tilde E_{rj}$, the CES demand of buyer $j$ is
\[
X^{d}_{ij}
=
\frac{\check Z_{ij}\,\tilde p_i^{\,1/(\rho_r-1)}\,\tilde E_{rj}}
{\sum_{k}\check Z_{kj}\,\tilde p_k^{\,\rho_r/(\rho_r-1)}}.
\]
The display uses the disrupted flows in place of the normalized weights $\pi_{ij}$, because the normalizing constant $\sum_k\check Z_{kj}$ cancels between the numerator and the denominator of a CES share. The maximizer is interior. For $\rho_r<0$ the marginal valuation at a zero purchase is finite, so the ordinary Inada argument does not apply, but a bundle that drops any incumbent link has zero utility by the boundary convention while every strictly positive feasible bundle has positive utility, which excludes the boundary. For $0<\rho_r<1$ the marginal valuation diverges at zero and the Inada argument applies directly. The criterion is strictly quasiconcave, so the maximizer is also unique.

For the balanced block $X_{ij}=u_i\check Z_{ij}v_j$ of Proposition~\ref{prop:m-balancing}, define $\tilde p_i:=u_i^{\,\rho_r-1}$ and $\tilde E_{rj}:=\sum_i\tilde p_iX_{ij}$. Substituting $\tilde p_i^{\,1/(\rho_r-1)}=u_i$ gives $\tilde E_{rj}=v_j\sum_k\check Z_{kj}\tilde p_k^{\,\rho_r/(\rho_r-1)}$, hence $X^{d}_{ij}=u_i\check Z_{ij}v_j=X_{ij}$, and the two margins deliver budget exhaustion and capacity clearing as in the proof of Proposition~\ref{prop:m-decentralization}. Conversely, at any equilibrium the pseudo-budget binds and the first-order conditions give the display. Setting $u_i:=\tilde p_i^{\,1/(\rho_r-1)}$ and $v_j:=\tilde E_{rj}/\sum_k\check Z_{kj}\tilde p_k^{\,\rho_r/(\rho_r-1)}$ then puts the allocation in biproportional form with both margins met, and the stationarity argument of that proof identifies it with $\mathbf Z^{(r)}_\omega$. As the curvature varies, the bijection $u\mapsto u^{\,\rho_r-1}$ of the positive orthant relabels the supporting pseudo-prices and leaves the allocation fixed. The gauge freedom $(u,v)\mapsto(\chi u,\chi^{-1}v)$ becomes $\tilde{\mathbf p}\mapsto\chi^{\,\rho_r-1}\tilde{\mathbf p}$ with the pseudo-incomes rescaled in proportion, which leaves CES demand unchanged. This proves the lemma.
\end{proof}

\clearpage

\setcounter{section}{0}
\setcounter{footnote}{0}
\renewcommand{\thesection}{\arabic{section}}
\makeatletter
\renewcommand{\theequation}{\thesection.\arabic{equation}}
\@addtoreset{equation}{section}
\renewcommand{\thetable}{\thesection.\arabic{table}}
\@addtoreset{table}{section}
\renewcommand{\thefigure}{\thesection.\arabic{figure}}
\@addtoreset{figure}{section}
\renewcommand{\thetheorem}{\thesection.\arabic{theorem}}
\@addtoreset{theorem}{section}
\renewcommand{\theproposition}{\thesection.\arabic{proposition}}
\@addtoreset{proposition}{section}
\renewcommand{\thelemma}{\thesection.\arabic{lemma}}
\@addtoreset{lemma}{section}
\renewcommand{\thecorollary}{\thesection.\arabic{corollary}}
\@addtoreset{corollary}{section}
\renewcommand{\theassumption}{\thesection.\arabic{assumption}}
\@addtoreset{assumption}{section}
\renewcommand{\thedefinition}{\thesection.\arabic{definition}}
\@addtoreset{definition}{section}
\renewcommand{\theremark}{\thesection.\arabic{remark}}
\@addtoreset{remark}{section}
\makeatother
\setcounter{equation}{0}
\setcounter{table}{0}
\setcounter{figure}{0}
\setcounter{theorem}{0}
\setcounter{proposition}{0}
\setcounter{lemma}{0}
\setcounter{corollary}{0}
\setcounter{assumption}{0}
\setcounter{definition}{0}
\setcounter{remark}{0}

\renewcommand{\theHsection}{OA.\arabic{section}}
\makeatletter
\@addtoreset{equation}{section}
\@addtoreset{table}{section}
\@addtoreset{figure}{section}
\@addtoreset{theorem}{section}
\@addtoreset{proposition}{section}
\@addtoreset{lemma}{section}
\@addtoreset{corollary}{section}
\@addtoreset{assumption}{section}
\@addtoreset{definition}{section}
\@addtoreset{remark}{section}
\makeatother

\graphicspath{{./Images/}}
\hypersetup{pdfborder={0 0 0}}
\setlength{\emergencystretch}{1em}
\onehalfspacing

\pdfbookmark[0]{Online Appendix}{OnlineAppendix}
\null
\vskip 2em
\begin{center}
{\LARGE \textbf{Online Appendix\\ \large ``Economic Power in International Trade''}\par}
\vskip 1.5em
{\large \lineskip .5em
\begin{tabular}[t]{c}Ashwin Bhattathiripad\end{tabular}\hskip 1em plus .17fil
\begin{tabular}[t]{c}Vipin P. Veetil\end{tabular}\par}
\vskip 1em
{\large \today}
\end{center}
\par
\vskip 1.5em
\noindent This appendix carries the constructions, the calibration, and the empirical checks on which the paper rests, beyond the proofs of its main results, which are in Appendix~\ref{app:m-coreproofs} of the paper. Sections~\ref{app:m-proofs} and~\ref{app:m-correction} support the model of Section~\ref{sec:model} of the paper. Section~\ref{app:m-proofs} establishes the uniqueness and the convergence of the sanction equilibrium and bounds the error that the local CES representation of the technology can carry into the reported vulnerabilities. Section~\ref{app:m-correction} shows that the model degenerates at the Leontief limit of the outer aggregator. Section~\ref{oa:diversification} holds size fixed and compares diversification on the two sides of trade, and it bounds the vulnerabilities by the shares that a partner holds in a country's sales and purchases, in support of the comparisons of power with size of Section~\ref{sec:power_size} of the paper. Section~\ref{app:m-tau} calibrates the two disruption parameters from disruption episodes. Section~\ref{app:m-empirics} records the construction of the benchmark network and the computations behind Sections~\ref{sec:statistics} to~\ref{sec:power_balance} of the paper, together with their sensitivity to the parameters, the specification, the year of the trade table, and the construction of the table. Section~\ref{oa:balanced-trade} erases the trade imbalance of each pair of countries from the trade table, one pair at a time, and solves the trade severance again to measure how much power the trade imbalance confers. Section~\ref{app:flex} solves the same severances in an economy whose prices clear its markets and compares the distribution of power under the two adjustment closures. Section~\ref{oa:datasets} defines the coercion decisions behind the empirical test of the paper and explains how the Global Sanctions Data Base is made comparable with them. Section~\ref{app:notation} collects the notation and the conventions that govern it.

\section{Uniqueness, Convergence, and Local Sensitivity}
\label{app:m-proofs}

We establish the uniqueness of the sanction equilibrium along two routes and record the convergence diagnostics of the empirical exercise. The first route makes the fulfillment rule a contraction on a forward-invariant positive domain, which gives a unique fulfillment fixed point on that domain and geometric convergence to it from every point of the domain. Activity is then recovered from the fulfillment vector by a linear solve. The second route establishes uniqueness among the interior sanction equilibria without a contraction condition. Both routes rest on the existence and selection results of Proposition~\ref{prop:m-monotone} of the paper. The section closes with the input-level representation of the CES-shaped fulfillment response and a bound on the error that the representation carries into the reported vulnerabilities.

\subsection{Benchmark selection, contraction, and convergence diagnostics}

Both routes start from the benchmark path, so we record its definition first. The selection matters because complete collapse is always a sanction equilibrium (Section~\ref{subsec:model_counterfactual} of the paper).

\begin{definition}[Benchmark path]
\label{def:m-selected-equilibrium}
Starting from $\mathbf h^{(0)}=\mathbf 1$, iterate
\begin{equation}
\mathbf x^{(n)}=\mathbf s(\mathbf h^{(n)}),
\qquad
\mathbf h^{(n+1)}
=\boldsymbol\lambda^{(\omega)}(\mathbf h^{(n)}).
\label{eq:m-selected-iteration}
\end{equation}
This benchmark-initialized sequence is the benchmark path, and its limit is the benchmark-selected equilibrium $(\mathbf s_\omega,\mathbf h_\omega)$.
\end{definition}

Proposition~\ref{prop:m-monotone} of the paper establishes that the benchmark path converges and that its limit is the greatest fixed point of the fulfillment rule and hence the equilibrium of least loss. Under the interiority condition \eqref{eq:m-cushion-cond} the limit stays away from collapse, with the floor $c_\omega=\min_ic_i\in(0,1]$ of \eqref{eq:m-floor} fixed by the benchmark network and the maintained parameters alone. The argument behind the proposition is a monotone fixed-point argument and not a contraction, so it returns no uniqueness. Nor could it, since collapse is a second fixed point whenever the selected equilibrium is nonzero. Without \eqref{eq:m-cushion-cond}, parts \textnormal{(i)} and \textnormal{(ii)} of the proposition still hold, and the positivity of the equilibrium has to be verified ex post.

Throughout, all objects are restricted to the active country-sectors of Section~\ref{subsec:model_benchmark} of the paper. For a maintained fulfillment floor $\underline h_\omega\in(0,1]$, define the bundle-feedback domain
\begin{equation}
\mathcal D_\omega(\underline h_\omega)
:=
[\underline h_\omega,1]^N
\;\subseteq\;
[0,1]^N.
\label{eq:m-Domega}
\end{equation}
The domain contains fulfillment vectors alone, since both uniqueness arguments reduce the equilibrium problem to the fulfillment vector. On $\mathcal D_\omega$ every used sector satisfies $\theta^{(\omega)}_{rj}(\mathbf h)\ge\underline h_\omega\,\mathcal F(\mathbf A_\omega)_{rj}/\alpha_{rj}$ for $r\in\mathcal R_j$, so the retention ratios on the domain have the floor
\begin{equation}
\underline\theta_\omega
:=
\underline h_\omega\,
\min_{j}\min_{r\in\mathcal R_j}
\frac{\mathcal F(\mathbf A_\omega)_{rj}}{\alpha_{rj}}
\;>\;0.
\label{eq:m-underbarf}
\end{equation}
The benchmark matrix $\mathbf A$, the sanction $\omega$, the parameters $\tau$ and $\delta$, and the floor $\underline h_\omega$ fix $\underline\theta_\omega$ without an equilibrium solve, and they fix the supplier row-sum bound $\bar a_\omega:=\max_i\sum_j a^{(\omega)}_{ij}$ of Appendix~\ref{app:m-wellposed} of the paper as well.

The reduction to the fulfillment vector runs through the activity equation. Write $\mathbf B_\omega:=\mathbf A_\omega(\mathbf I-\boldsymbol\beta)$ for the sanctioned propagation matrix. Every $\mathbf h\in[\underline h_\omega,1]^N$ satisfies
\[
\|\diag(\mathbf h)\mathbf B_\omega\|_1\le\|\mathbf B_\omega\|_1<1.
\]
The activity equation is therefore linear in activity, and its solution has the Neumann representation
\begin{equation}
\mathbf s(\mathbf h)
\;:=\;
\bigl[\mathbf I-\diag(\mathbf h)\mathbf B_\omega\bigr]^{-1}\bigl(\boldsymbol\lambda^{(\omega)}(\mathbf h)\circ\mathbf f_\omega\bigr)
\;=\;
\sum_{t\ge 0}\bigl[\diag(\mathbf h)\mathbf B_\omega\bigr]^t\bigl(\boldsymbol\lambda^{(\omega)}(\mathbf h)\circ\mathbf f_\omega\bigr).
\label{eq:m-s-of-h}
\end{equation}
The output-ratio vector $\boldsymbol\lambda^{(\omega)}(\mathbf h)$ scales the effective final demand $\mathbf f_\omega$ because the productivity penalty falls on all output, final sales included, and the effective final demand itself is fixed by the rationed exchange and does not depend on $\mathbf h$. The fulfillment coordinate of a sanction equilibrium solves
\begin{equation*}
\mathbf h
\;=\;
\boldsymbol\lambda^{(\omega)}(\mathbf h).
\end{equation*}
The reduction closes the nonlinear problem in $\mathbf h$ alone and halves the effective state space. Definition~\ref{def:m-selected-equilibrium} and the empirical computation iterate on fulfillment and recover activity from the linear solve $\mathbf s(\mathbf h)$. The problem stays nonlinear, since the fulfillment vector is endogenous.

The fulfillment rule $\boldsymbol\lambda^{(\omega)}$ collects the output ratios $\lambda^{(\omega)}_i$ of \eqref{eq:m-lambda-rule} and depends on $\mathbf h\in[0,1]^N$ alone. Two Lipschitz constants bound its response to fulfillment. The first bounds the rule itself in the $1$-norm,
\begin{equation*}
\|\boldsymbol\lambda^{(\omega)}(\mathbf h)-\boldsymbol\lambda^{(\omega)}(\mathbf g)\|_1
\le
\mathrm{Lip}_h^\omega\,\|\mathbf h-\mathbf g\|_1.
\end{equation*}
The second bounds the source term $\boldsymbol\lambda^{(\omega)}(\mathbf h)\circ\mathbf f_\omega$ of the activity equation,
\begin{equation*}
\|\boldsymbol\lambda^{(\omega)}(\mathbf h)\circ\mathbf f_\omega-\boldsymbol\lambda^{(\omega)}(\mathbf g)\circ\mathbf f_\omega\|_1
\;\le\;
\|\mathbf f_\omega\|_\infty\,\mathrm{Lip}_\lambda^\omega\,\|\mathbf h-\mathbf g\|_1.
\end{equation*}
Both inequalities hold for all $\mathbf h,\mathbf g\in\mathcal D_\omega$, and both constants are finite on $\mathcal D_\omega$ because $\underline\theta_\omega>0$ there and the maintained CES-exponent intervals are compact subsets of $(-\infty,0)$. The bounds \eqref{eq:m-Lh-bound} on the two constants carry the outer-aggregator factor $\bar e_\omega\,\underline\kappa_\omega^{-\bar\beta}$. Here $\underline\kappa_\omega:=\underline h_\omega\min_i\kappa^{(\omega)}_i(\mathbf 1)$ is the efficiency floor on the domain, and $\bar e_\omega:=\max_i e_i(\underline\kappa_\omega)<1$ is an upper bound on the intermediate output elasticities on the domain. The leakage bounds are $\bar\beta:=\max_i\beta_i$ and $\underline\beta:=\min_i\beta_i$, the largest and the smallest leakage over the active country-sectors.

Positive attenuation keeps the retention floor positive without an equilibrium solve. By Lemma~\ref{lem:m-kernel} of the paper, $\mathcal F(\mathbf A_\omega)_{rj}/\alpha_{rj}\ge1-\tau(1-\delta)\ge\delta$ on every affected $(j,r)$, so \eqref{eq:m-underbarf} gives
\begin{equation*}
\underline\theta_\omega\;\ge\;\underline h_\omega\,\delta\;>\;0.
\end{equation*}
Hence $\mathrm{Lip}_\lambda^\omega\le\bar e_\omega\,\underline\kappa_\omega^{-\bar\beta}\,(\underline h_\omega\delta)^{\rho-1}\bar a_\omega<\infty$. The floor excludes the singularity at $\underline\theta_\omega\to0$ that full severance, $\delta=0$, can create, and it keeps $\bar e_\omega$ below the unit gain reached as $\underline\kappa_\omega\to0$.

The contraction route rests on two conditions on $\mathcal D_\omega(\underline h_\omega)$, both computable from the benchmark network and the maintained parameters without solving an equilibrium. The first is the contraction condition
\begin{equation}
\mathrm{Lip}_h^\omega\;<\;1.
\label{eq:m-suff-scalar}
\end{equation}
The condition involves fulfillment alone, since activity does not feed back into the fulfillment rule. The second is forward invariance, $\inf_{\mathbf h\in\mathcal D_\omega}\lambda^{(\omega)}_i(\mathbf h)\ge\underline h_\omega$ for every $i$, which keeps every iterate that starts in the domain inside it, so that the Lipschitz bound applies at every later iterate. Too high a floor makes the domain noninvariant and too low a floor weakens the contraction bound. The canonical floor $\underline h_\omega=c_\omega$ of Proposition~\ref{prop:m-monotone} is invariant whenever the interiority condition that defines it holds. At a fixed attenuation $\delta$, write $\mathcal P_\omega$ for the set of triples $(\tau,\rho,\varrho)$ that meet both conditions. On that set the fulfillment iteration converges geometrically from every point of the domain.

The bound \eqref{eq:m-Lh-bound} on the fulfillment derivative turns the contraction condition into the computable sufficient condition
\begin{equation}
\bar e_\omega\,\bar a_\omega
\;<\;
\underline\kappa_\omega^{\,\bar\beta}\,
\underline\theta_\omega^{\,1-\rho}.
\label{eq:m-suff}
\end{equation}
Here $\underline\theta_\omega$ is the retention floor \eqref{eq:m-underbarf} and $\underline\kappa_\omega$ the efficiency floor, and at the constant-share anchor $\varrho\to0$ the elasticity bound $\bar e_\omega$ reduces to $1-\underline\beta$. Every quantity in \eqref{eq:m-suff} is fixed by $\mathbf A$, $\omega$, $\tau$, $\rho$, $\varrho$, $\delta$, and $\underline h_\omega$, so the condition can be checked sanction by sanction.

\begin{proposition}[Conditional uniqueness and convergence on $\mathcal P_\omega$]
\label{prop:m-uniqueness}
Maintain a positive barred-link attenuation $\delta\in(0,\bar\delta]$. Fix a
sanction $\omega$ and a parameter triple
$(\tau,\rho,\varrho)\in\mathcal P_\omega$. The fulfillment rule
$\boldsymbol\lambda^{(\omega)}$ is a contraction self-map of
$\mathcal D_\omega(\underline h_\omega)$ in the $1$-norm. It has a unique
fixed point $\mathbf h_\omega$ on that domain, and
$\mathbf h^{(n+1)}=\boldsymbol\lambda^{(\omega)}(\mathbf h^{(n)})$
converges to it geometrically from every
$\mathbf h^{(0)}\in\mathcal D_\omega$. The corresponding activity vectors
$\mathbf s(\mathbf h^{(n)})$ converge to
$\mathbf s_\omega=\mathbf s(\mathbf h_\omega)$.
\end{proposition}

\begin{proof}
Fix \((\tau,\rho,\varrho)\in\mathcal P_\omega\) and work on the maintained domain \(\mathcal D_\omega(\underline h_\omega)=[\underline h_\omega,1]^N\), which forward invariance keeps every fulfillment iterate inside. We first bound the derivatives of the fulfillment rule, then apply Banach's theorem, and last recover activity from the conditional solve.

By \eqref{eq:m-lambda-rule}, \(\lambda^{(\omega)}_i(\mathbf h)=G_i(\kappa^{(\omega)}_i(\mathbf h))\). Positive attenuation in \eqref{eq:m-leak} leaves each buyer a positive-weight supplier in every sector it uses, so the efficiency factor \(\kappa^{(\omega)}_i\) is monotone and positively 1-homogeneous in fulfillment and attains its minimum on the domain at \(\mathbf h=\underline h_\omega\mathbf 1\), so the efficiency floor is
\[
\underline\kappa_\omega
\;=\;
\inf_{\mathbf h\in\mathcal D_\omega}\min_i\kappa^{(\omega)}_i(\mathbf h)
\;=\;
\underline h_\omega\,\min_i\kappa^{(\omega)}_i(\mathbf 1)\;>\;0.
\]
Lemma~\ref{lem:m-outer}\textnormal{(ii)} of the paper then gives \(\lambda^{(\omega)}_i\ge G_i(\underline\kappa_\omega)\ge\underline\kappa_\omega>0\). To bound the Lipschitz constants, differentiate \(\kappa^{(\omega)}_i(\mathbf h)=\bigl(\sum_{r\in\mathcal R_i}\alpha_{ri}\theta^{(\omega)}_{ri}(\mathbf h)^\rho\bigr)^{1/\rho}\) where every retention is positive, and compose with \(\lambda^{(\omega)}_i=G_i(\kappa^{(\omega)}_i)\) through the slope identity of Lemma~\ref{lem:m-outer}\textnormal{(iv)}. The composition gives
\begin{equation}
\begin{aligned}
\frac{\partial\kappa^{(\omega)}_i}{\partial h_j}
&=
\bigl(\kappa^{(\omega)}_i\bigr)^{1-\rho}
\bigl(\theta^{(\omega)}_{\mathcal S(j),i}\bigr)^{\rho-1}\,a^{(\omega)}_{ji},\\
\frac{\partial\lambda^{(\omega)}_i}{\partial h_j}
&=
e_i\bigl(\kappa^{(\omega)}_i\bigr)\,
\frac{G_i(\kappa^{(\omega)}_i)}{\kappa^{(\omega)}_i}\,
\frac{\partial\kappa^{(\omega)}_i}{\partial h_j}.
\end{aligned}
\label{eq:m-kappa-deriv}
\end{equation}
Here \(\mathcal S(j)\) is the sector of supplier \(j\), as in Definition~\ref{def:m-country-sector}, and \(\partial\lambda^{(\omega)}_i/\partial h_j=0\) when buyer \(i\) does not use that sector.

On \(\mathcal D_\omega\) the bounds \(\kappa^{(\omega)}_i\le1\) and \(1-\rho>0\) give \(\bigl(\kappa^{(\omega)}_i\bigr)^{1-\rho}\le1\), and the retention floor \(\theta^{(\omega)}_{\mathcal S(j),i}\ge\underline\theta_\omega\) gives \(\bigl(\theta^{(\omega)}_{\mathcal S(j),i}\bigr)^{\rho-1}\le\underline\theta_\omega^{\rho-1}\). Since \(\kappa^{(\omega)}_i\ge\underline\kappa_\omega\), Lemma~\ref{lem:m-outer}\textnormal{(iii,iv)} also gives \(e_i(\kappa^{(\omega)}_i)\,G_i(\kappa^{(\omega)}_i)/\kappa^{(\omega)}_i\le e_i(\underline\kappa_\omega)\,\underline\kappa_\omega^{-\beta_i}\le\bar e_\omega\,\underline\kappa_\omega^{-\bar\beta}\). Combining these bounds yields
\[
\left|\frac{\partial\lambda^{(\omega)}_i}{\partial h_j}\right|
\;\le\;
\bar e_\omega\,\underline\kappa_\omega^{-\bar\beta}\,
\underline\theta_\omega^{\rho-1}\,a^{(\omega)}_{ji}.
\]
The same derivatives enter the fulfillment Jacobian and the source term, so the two constants share the induced \(1\)-norm bound
\begin{equation}
\mathrm{Lip}_h^\omega,\ \mathrm{Lip}_\lambda^\omega
\;\le\;
\bar e_\omega\,\underline\kappa_\omega^{-\bar\beta}\,
\underline\theta_\omega^{\rho-1}\,\bar a_\omega.
\label{eq:m-Lh-bound}
\end{equation}
The maximum seller row sum $\bar a_\omega$ appears as the worst column sum of the Jacobian over the supplier index $j$, the factor $\bar e_\omega\,\underline\kappa_\omega^{-\bar\beta}$ is the contribution of the outer CES aggregator, and the bound rises as the domain floor falls.

The closed-form condition \eqref{eq:m-suff}, $\bar e_\omega\,\bar a_\omega<\underline\kappa_\omega^{\bar\beta}\,\underline\theta_\omega^{1-\rho}$, is equivalent to $\bar e_\omega\,\underline\kappa_\omega^{-\bar\beta}\,\underline\theta_\omega^{\rho-1}\bar a_\omega<1$, and by \eqref{eq:m-Lh-bound} it implies $\mathrm{Lip}_h^\omega<1$, the condition \eqref{eq:m-suff-scalar}. The closed-form condition is sufficient and not necessary, and the proposition assumes only the contraction bound itself. Membership in $\mathcal P_\omega$ supplies that bound together with the forward invariance of $\mathcal D_\omega$, so Banach's theorem gives a unique fulfillment fixed point on $\mathcal D_\omega$ and geometric convergence of the iteration.

We conclude by recovering activity, for which Lemma~\ref{lem:m-kernel} gives $\spr(\diag(\mathbf h)\mathbf B_\omega)<1$ for every $\mathbf h\in[0,1]^N$, so the conditional solve \eqref{eq:m-s-of-h} is continuous in fulfillment and $\mathbf s(\mathbf h^{(n)})\to\mathbf s(\mathbf h_\omega)=\mathbf s_\omega$. Lemma~\ref{lem:m-capacity} bounds that solve by the benchmark activity vector, $\mathbf0\le\mathbf s(\mathbf h)\le\mathbf s$, a bound that follows from the seller flow targets and not from a restriction on the state space. This proves the proposition.
\end{proof}

Alongside \eqref{eq:m-suff}, the empirical exercise records a convergence diagnostic for each scenario.

\begin{definition}[Benchmark-path convergence diagnostic]
\label{def:m-convergence-diagnostic}
Fix a sanction $\omega\in\Omega$, maintained parameters
$(\tau,\rho,\varrho,\delta)$, a floor $\underline h_\omega\in(0,1]$, and a
numerical tolerance $\varepsilon_{\mathrm{tol}}>0$.
For the reduced fulfillment iteration initiated at the benchmark fulfillment
vector $\mathbf h^{(0)}=\mathbf 1$, record these diagnostics at its
terminal pair
$(\mathbf s_\omega,\mathbf h_\omega)$.
\begin{enumerate}
\item[\textnormal{(i)}] The terminal ratio of successive fulfillment-iterate differences
$\|\mathbf h^{(n+1)}-\mathbf h^{(n)}\|_1
\big/\|\mathbf h^{(n)}-\mathbf h^{(n-1)}\|_1$
is strictly below one. If the denominator is zero, the deterministic reduced
fulfillment iteration also makes the numerator zero, and we record the ratio
as zero.
\item[\textnormal{(ii)}] The terminal fixed-point residual
$\|\Phi_\omega(\mathbf s_\omega,\mathbf h_\omega)-(\mathbf s_\omega,\mathbf h_\omega)\|_1$
is below $\varepsilon_{\mathrm{tol}}$.
\item[\textnormal{(iii)}] As a basin check, reduced fulfillment iterations
launched from randomized
$\mathbf h^{(0)}\in\mathcal D_\omega(\underline h_\omega)$ return the
same fixed point within $\varepsilon_{\mathrm{tol}}$ in the $1$-norm.
\item[\textnormal{(iv)}] The computed fulfillment fixed point lies in the maintained
domain, $\mathbf h_\omega\in\mathcal D_\omega(\underline h_\omega)$.
\end{enumerate}
\end{definition}

Checks~\textnormal{(i)}, \textnormal{(ii)}, and \textnormal{(iv)} record the terminal decay, the residual accuracy, and the membership of the terminal iterate in the maintained positive domain, and check~\textnormal{(iii)} probes a wider basin. The ratio in \textnormal{(i)} is an observed ratio of successive iterate differences and neither a Jacobian spectral radius nor a contraction modulus, so the diagnostics support the computed solution without establishing contraction on a neighborhood or global uniqueness.

Condition~\eqref{eq:m-suff} holds in no reported scenario on the observed trade table. The largest row total $\bar a_\omega$ of $\mathbf A_\omega$ is about $19$ on the 2022 trade table of Section~\ref{subsec:vulnerability_estimates} of the paper, while the right side of the condition cannot exceed one, so the $1$-norm bound fails for every sanction at every reported $(\tau,\rho,\varrho)$. The failure shows neither that the actual Lipschitz constant of $\boldsymbol\lambda^{(\omega)}$ exceeds one nor that $\mathcal P_\omega$ is empty, since the bound is governed by the worst column of the fulfillment Jacobian and is conservative. We therefore report the benchmark-path diagnostics of Definition~\ref{def:m-convergence-diagnostic} sanction by sanction. At the baseline the terminal iterate-difference ratio is at most $0.694$ and has a median of $0.595$ across the $3{,}160$ scenarios, and every scenario meets checks \textnormal{(i)}, \textnormal{(ii)}, and \textnormal{(iv)}.\footnote{We carry out the basin check \textnormal{(iii)} on the headline scenario, where solutions from randomized starting vectors agree to $10^{-10}$. The per-scenario diagnostics, like the scenario-level values behind every summary in this appendix, are in the Replication Package.} The ratios vary little in the inner exponent $\rho$ because the factor $(\theta^{(\omega)}_{\mathcal S(j),i}/\kappa^{(\omega)}_i)^{\rho-1}$ in the fulfillment derivative saturates. The power-mean identity gives $\theta^{(\omega)}_{ri}/\kappa^{(\omega)}_i\ge\alpha_{ri}^{1/|\rho|}$, so that factor is bounded by $\alpha_{\mathcal S(j),i}^{-(1+1/|\rho|)}$, which is nonincreasing in $|\rho|$ and tends to the finite Leontief bound $\alpha_{\mathcal S(j),i}^{-1}$. At the baseline the target retention floor $1-\tau(1-\delta)=0.73$ keeps the retention-to-efficiency ratios near one. Extending the United States--China severance beyond the maintained interval to $\rho\in\{-8,-16,-32\}$ moves the terminal ratio only from $0.59$ to $0.61$.

\subsection{Interior uniqueness and convergence}

Proposition~\ref{prop:m-uniqueness} reaches only the fixed points in $\mathcal D_\omega(\underline h_\omega)$, and its sufficient inequality fails throughout the 2022 trade table, so the benchmark-selected equilibria of the empirical exercise rest on a second argument. The argument combines the entrywise monotonicity of the fulfillment rule $\boldsymbol\lambda^{(\omega)}$ (Proposition~\ref{prop:m-monotone}) with strict subhomogeneity, $\lambda^{(\omega)}_i(t\,\mathbf h)>t\,\lambda^{(\omega)}_i(\mathbf h)$ for $t\in(0,1)$, which the degree-one homogeneity of $\kappa^{(\omega)}_i$ and Lemma~\ref{lem:m-outer}\textnormal{(iii)} supply. Together the two properties rule out multiple strictly positive equilibria, and no contraction condition enters. The convergence is stated in two forms, as a two-sided envelope around the fixed point and as a contraction estimate in Thompson's part metric, $d_T(\mathbf g,\mathbf h):=\|\log\mathbf g-\log\mathbf h\|_\infty$ on the strictly positive orthant, with the logarithm acting entrywise.

\begin{proposition}[Interior uniqueness and convergence without a contraction condition]
\label{prop:m-interior}
Maintain a positive barred-link attenuation $\delta\in(0,\bar\delta]$. Fix
$\omega\in\Omega$,
$\tau\in[0,1]$,
$\rho\in[\underline\rho,\overline\rho]\subset(-\infty,0)$, and
$\varrho\in[\underline\varrho,\overline\varrho]\subset(-\infty,0)$. Suppose the selected equilibrium is interior, $\mathbf h_\omega>\mathbf 0$.
\begin{enumerate}
\item[\textnormal{(i)}] The fulfillment rule
$\boldsymbol\lambda^{(\omega)}$ has exactly one fixed point $\mathbf h>\mathbf 0$
in $[0,1]^N$, the benchmark-selected $\mathbf h_\omega$. Consequently
$(\mathbf s_\omega,\mathbf h_\omega)$ is the unique fixed point of
$\Phi_\omega$ whose fulfillment coordinate is strictly positive.
\item[\textnormal{(ii)}] For every
$\mathbf h^{(0)}\in[0,1]^N$ with $\mathbf h^{(0)}>\mathbf 0$, the reduced
iteration $\mathbf h^{(n+1)}=\boldsymbol\lambda^{(\omega)}(\mathbf h^{(n)})$
converges to $\mathbf h_\omega$, with the two-sided envelope
\begin{equation}
\begin{aligned}
t_0^{\,\bar e_*^{\,n}}\,\mathbf h_\omega
&\;\le\;
\mathbf h^{(n)}
\;\le\;
\bar t_0^{\,\bar e_*^{\,n}}\,\mathbf h_\omega,
\qquad n\ge0,\\
\bar e_*
&:=
\max_i e_i\bigl(t_0\,\kappa^{(\omega)}_i(\mathbf h_\omega)\bigr)
\;<\;1.
\end{aligned}
\label{eq:m-envelope}
\end{equation}
Here
$t_0:=\min\{1,\min_i h^{(0)}_i/h_{\omega,i}\}$ and
$\bar t_0:=\max\{1,\max_i h^{(0)}_i/h_{\omega,i}\}$. The fulfillment rule
$\boldsymbol\lambda^{(\omega)}$ is nonexpansive in Thompson's part metric on the strictly
positive orthant and satisfies the
contraction estimate
$d_T\bigl(\boldsymbol\lambda^{(\omega)}(\mathbf h),\mathbf h_\omega\bigr)
\le\bar e_*\,d_T(\mathbf h,\mathbf h_\omega)$ for every
$\mathbf h\in[t_0\,\mathbf h_\omega,\,t_0^{-1}\,\mathbf h_\omega]$. For the same fulfillment sequence, the conditional
activity vectors $\mathbf x^{(n)}=\mathbf s(\mathbf h^{(n)})$ converge to
$\mathbf s_\omega$.
\item[\textnormal{(iii)}] Every fixed point
$(\mathbf x^*,\mathbf h^*)$ of $\Phi_\omega$ other than
$(\mathbf s_\omega,\mathbf h_\omega)$ has a nonempty zero set
$Z:=\{i:h^*_i=0\}$, and $Z$ is starvation-supported. Specifically, for every
$i\in Z$, there is a used input sector $r\in\mathcal R_i$ whose suppliers all
lie in $Z$:
$\{k\in\mathcal N_r:a_{ki}>0\}\subseteq Z$. If no proper nonempty subset of the active country-sectors is
starvation-supported, the
fixed points of $\Phi_\omega$ are exactly the collapse
$(\mathbf 0,\mathbf 0)$ and the benchmark-selected
$(\mathbf s_\omega,\mathbf h_\omega)$.
\end{enumerate}
\end{proposition}

The interiority condition \eqref{eq:m-cushion-cond} is sufficient for the hypothesis of Proposition~\ref{prop:m-interior}, which imposes no contraction condition on the maintained parameter ranges. The modulus $\bar e_*$ of part~\textnormal{(ii)} is uniform on every order interval around the fixed point but tends to one as $t_0\downarrow0$, so the estimate is a contraction on each such interval and not on the whole orthant. Because the input sectors are complements, losing every supplier in one used sector makes zero output self-consistent for each member of $Z$. The implication from a nonempty zero set to starvation support uses neither the interiority hypothesis nor any restriction on $\rho$ beyond $\rho<0$, while the claim that every nonselected fixed point has a nonempty zero set, and the two-point classification, use the interiority hypothesis and part~\textnormal{(i)}.

\begin{proof}
Fix $\omega$, $\tau$, $\rho$, $\varrho$, and $\delta$ as in the proposition. We first extend the fulfillment rule to the nonnegative cone and derive two scaling inequalities, then prove interior uniqueness and convergence, and last characterize the starvation boundary. The formulas \eqref{eq:m-theta}, \eqref{eq:m-kappa}, and \eqref{eq:m-lambda-rule} define $\boldsymbol\theta^{(\omega)}$, $\boldsymbol\kappa^{(\omega)}$, and $\boldsymbol\lambda^{(\omega)}$ for every $\mathbf h\ge\mathbf0$ and not only on $[0,1]^N$. The proof of entrywise monotonicity uses neither the upper bound $\mathbf h\le\mathbf1$ nor differentiability, and the same holds for the exact scaling $\kappa^{(\omega)}_i(t\,\mathbf h)=t\,\kappa^{(\omega)}_i(\mathbf h)$ for $t>0$, which the proof of Proposition~\ref{prop:m-monotone} in Appendix~\ref{app:m-coreproofs} of the paper establishes. Both properties therefore persist on the cone, and Lemma~\ref{lem:m-outer} gives the corresponding properties of $G_i$ on $(0,\infty)$. The cone extension serves only to bound the iterates from above, and every fixed point considered lies in $[0,1]^N$. Composing the exact scaling with \eqref{eq:m-outer-subhom} turns homogeneity into the entrywise inequalities
\begin{equation}
\begin{aligned}
\lambda^{(\omega)}_i(t\,\mathbf h)
&\;\ge\;
t^{\,e_i(t\kappa^{(\omega)}_i(\mathbf h))}\,\lambda^{(\omega)}_i(\mathbf h)
\;>\;
t\,\lambda^{(\omega)}_i(\mathbf h)
\quad(0<t<1),\\
\lambda^{(\omega)}_i(\bar t\,\mathbf h)
&\;\le\;
\bar t^{\,e_i(\kappa^{(\omega)}_i(\mathbf h))}\,\lambda^{(\omega)}_i(\mathbf h)
\quad(\bar t\ge1).
\end{aligned}
\label{eq:m-subhom}
\end{equation}
These inequalities hold at every $\mathbf h\ge\mathbf0$ with $\kappa^{(\omega)}_i(\mathbf h)>0$. They use the pointwise gap $1-e_i>0$ and do not require that gap to be uniform on the positive cone.

We next establish uniqueness among the strictly positive fixed points of the fulfillment rule. Let $\mathbf g,\mathbf h\in[0,1]^N$ be strictly positive fixed points of $\boldsymbol\lambda^{(\omega)}$, and define the largest scalar with $t^*\mathbf g\le\mathbf h$ by
\[
t^*:=\min_i\,h_i/g_i\;\in\;(0,\infty).
\]
Strict positivity makes $t^*$ positive and finite, and the fixed-point identity gives $\kappa^{(\omega)}_i(\mathbf g)>0$ at every $i$, since $g_i=G_i(\kappa^{(\omega)}_i(\mathbf g))>0$ holds only for a positive argument. Suppose $t^*<1$. Monotonicity and the strict inequality of \eqref{eq:m-subhom} give, at every coordinate,
\[
h_i
\;=\;
\lambda^{(\omega)}_i(\mathbf h)
\;\ge\;
\lambda^{(\omega)}_i(t^*\mathbf g)
\;>\;
t^*\,\lambda^{(\omega)}_i(\mathbf g)
\;=\;
t^*\,g_i.
\]
Because the index set is finite, $\min_i h_i/g_i>t^*$, which contradicts the definition of $t^*$. Hence $t^*\ge1$ and $\mathbf g\le\mathbf h$. Exchanging the roles of $\mathbf g$ and $\mathbf h$ gives $\mathbf h\le\mathbf g$, so the two fixed points coincide. This comparison is the cone argument of \citet{oa-krasnoselskii1964} for monotone strictly subhomogeneous operators, and it does not extend to the starvation boundary, where zero coordinates can make the ratios that define $t^*$ vanish or become undefined.

By the maintained interiority hypothesis, $\mathbf h_\omega>\mathbf0$ is the unique strictly positive fixed point of $\boldsymbol\lambda^{(\omega)}$. Proposition~\ref{prop:m-monotone}\textnormal{(iii)} gives this interiority under the interiority condition, and the empirical exercise verifies $\mathbf h_\omega>\mathbf0$ ex post in the remaining cases. If $(\mathbf x^*,\mathbf h^*)$ is a fixed point of $\Phi_\omega$ with $\mathbf h^*>\mathbf0$, its fulfillment coordinate satisfies $\mathbf h^*=\boldsymbol\lambda^{(\omega)}(\mathbf h^*)$, hence $\mathbf h^*=\mathbf h_\omega$, and its activity coordinate solves the linear activity equation at $\mathbf h_\omega$, whose unique solution is $\mathbf s(\mathbf h_\omega)=\mathbf s_\omega$ (Lemma~\ref{lem:m-kernel}). This proves \textnormal{(i)}.

We next prove the two-sided envelope and the convergence of the reduced
iteration from a strictly positive starting vector. Fix
$\mathbf h^{(0)}\in[0,1]^N$ with $\mathbf h^{(0)}>\mathbf 0$ and let
$t_0\in(0,1]$ and $\bar t_0\ge1$ be as in the statement. Here $\bar t_0$ is finite because
$\mathbf h_\omega>\mathbf 0$. Then
$t_0\,\mathbf h_\omega\le\mathbf h^{(0)}\le \bar t_0\,\mathbf h_\omega$. Set
$\bar e_*:=\max_i e_i\bigl(t_0\,\kappa^{(\omega)}_i(\mathbf h_\omega)\bigr)<1$,
the worst intermediate output elasticity on the initial order interval. By
Lemma~\ref{lem:m-outer}\textnormal{(iii)}, $e_i$ is decreasing, so for every
$t\in[t_0,1]$ and $\bar t\ge1$,
\begin{equation}
\begin{aligned}
e_i\bigl(t\,\kappa^{(\omega)}_i(\mathbf h_\omega)\bigr)&\;\le\;\bar e_*,\\
e_i\bigl(\kappa^{(\omega)}_i(\mathbf h_\omega)\bigr)&\;\le\;\bar e_*.
\end{aligned}
\label{eq:m-estar-dom}
\end{equation}
We claim that, for every $n\ge0$,
\[
t_0^{\,\bar e_*^{\,n}}\,\mathbf h_\omega
\;\le\;
\mathbf h^{(n)}
\;\le\;
\bar t_0^{\,\bar e_*^{\,n}}\,\mathbf h_\omega.
\]
The bounds at $n=0$ follow from the definitions of $t_0$ and $\bar t_0$. For
the inductive step, apply the monotonicity of
$\boldsymbol\lambda^{(\omega)}$ to the three points in the order interval, and
combine the resulting inequalities with \eqref{eq:m-subhom} and the fixed-point
identity
$\boldsymbol\lambda^{(\omega)}(\mathbf h_\omega)=\mathbf h_\omega$. With
$t:=t_0^{\bar e_*^{\,n}}\in[t_0,1]$ and
$\bar t:=\bar t_0^{\bar e_*^{\,n}}\ge1$,
\[
h^{(n+1)}_i
\;\ge\;
\lambda^{(\omega)}_i(t\,\mathbf h_\omega)
\;\ge\;
t^{\,e_i(t\kappa^{(\omega)}_i(\mathbf h_\omega))}\,h_{\omega,i}
\;\ge\;
t^{\,\bar e_*}\,h_{\omega,i}
\;=\;
t_0^{\,\bar e_*^{\,n+1}}\,h_{\omega,i}.
\]
The third inequality uses \eqref{eq:m-estar-dom} and $t\le1$. For the upper
bound,
$\lambda^{(\omega)}_i(\bar t\mathbf h_\omega)\le
\bar t^{e_i(\kappa^{(\omega)}_i(\mathbf h_\omega))}h_{\omega,i}\le
\bar t^{\bar e_*}h_{\omega,i}$ because $\bar t\ge1$. Only this upper bound
uses the cone extension, since $\bar t\,\mathbf h_\omega$ may leave
$[0,1]^N$. The two inequalities prove \eqref{eq:m-envelope}. Because
$\bar e_*<1$ and $t_0,\bar t_0$ are
positive and finite, $t_0^{\bar e_*^{\,n}}\to1$ and
$\bar t_0^{\bar e_*^{\,n}}\to1$, so the bounds in
\eqref{eq:m-envelope} give
$\mathbf h^{(n)}\to\mathbf h_\omega$. For the metric formulation, applying
the weak inequality in \eqref{eq:m-subhom} in both directions gives
nonexpansiveness on the positive orthant in Thompson's part metric
\citep{oa-thompson1963,oa-lemmensnussbaum2012}. For the contraction estimate, take
$\mathbf h\in[t_0\mathbf h_\omega,\,t_0^{-1}\mathbf h_\omega]$ with
$d:=d_T(\mathbf h,\mathbf h_\omega)\le\log(1/t_0)$, so that
$\exp(-d)\mathbf h_\omega\le\mathbf h\le \exp(d)\mathbf h_\omega$ with
$\exp(-d)\ge t_0$. Monotonicity,
\eqref{eq:m-subhom} anchored at $\mathbf h_\omega$, and the domination
\eqref{eq:m-estar-dom} give
\[
\exp(-\bar e_*d)\,\mathbf h_\omega
\;\le\;
\boldsymbol\lambda^{(\omega)}(\exp(-d)\mathbf h_\omega)
\;\le\;
\boldsymbol\lambda^{(\omega)}(\mathbf h)
\;\le\;
\boldsymbol\lambda^{(\omega)}(\exp(d)\mathbf h_\omega)
\;\le\;
\exp(\bar e_*d)\,\mathbf h_\omega.
\]
Equivalently,
$d_T\bigl(\boldsymbol\lambda^{(\omega)}(\mathbf h),\mathbf h_\omega\bigr)
\le\bar e_*\,d_T(\mathbf h,\mathbf h_\omega)$. The envelope
\eqref{eq:m-envelope} is the order-interval counterpart of this metric
estimate, and both give geometric decay of the part-distance
$d_T(\mathbf h^{(n)},\mathbf h_\omega)$ at rate $\bar e_*$. The envelope
retains the separate factors $t_0$ and $\bar t_0$, whereas the metric combines
them as $\max\{\log \bar t_0,\log(1/t_0)\}$.

Convergence of fulfillment carries over to the conditional activity. Lemma~\ref{lem:m-kernel} makes the conditional solve
\[
\mathbf s(\mathbf h)
=
\bigl[\mathbf I-\diag(\mathbf h)\mathbf B_\omega\bigr]^{-1}
\bigl(\boldsymbol\lambda^{(\omega)}(\mathbf h)\circ\mathbf f_\omega\bigr),
\]
continuous on $[0,1]^N$, so $\mathbf s(\mathbf h^{(n)})\to\mathbf s(\mathbf h_\omega)=\mathbf s_\omega$. This proves \textnormal{(ii)}.

We conclude by characterizing the fixed points on the starvation boundary. Let
$(\mathbf x^*,\mathbf h^*)$ be a fixed
point of $\Phi_\omega$ other than $(\mathbf s_\omega,\mathbf h_\omega)$. By
\textnormal{(i)}, $\mathbf h^*>\mathbf 0$ fails, so
$Z=\{i:h^*_i=0\}\neq\emptyset$. Fix $i\in Z$. From
$0=h^*_i=G_i(\kappa^{(\omega)}_i(\mathbf h^*))$ and
Lemma~\ref{lem:m-outer}\textnormal{(i)} we get
$\kappa^{(\omega)}_i(\mathbf h^*)=0$. With $\rho<0$ and weights
$\alpha_{ri}>0$ on $\mathcal R_i$, the power mean \eqref{eq:m-kappa}
vanishes if and only if $\theta^{(\omega)}_{ri}(\mathbf h^*)=0$ for some
$r\in\mathcal R_i$, that is, $h^*_k=0$ for every $k\in\mathcal N_r$ with
$a^{(\omega)}_{ki}>0$. On the active country-sectors the disrupted coefficient matrix
$\check{\mathbf A}$ retains the benchmark support,
$\operatorname{supp}(\check{\mathbf A})
=\operatorname{supp}(\mathbf A)$, and each balanced coefficient block of
$\mathbf A_\omega$ is strictly positive on its support-reduced active set
(Proposition~\ref{prop:m-leak}\textnormal{(i)}). Moreover,
$\tau(1-\delta)\le1-\delta<1$ keeps every relevant target positive:
\[
g^{(\omega)}_k\ge\bigl(1-\tau(1-\delta)\bigr)(s_k-f^{o}_k)>0
\quad\text{whenever }s_k-f^{o}_k>0,
\qquad
t^{(\omega)}_{ri}\ge\delta\,\alpha_{ri}>0
\quad\text{whenever }\alpha_{ri}>0.
\]
A benchmark link $a_{ki}>0$ entails both $s_k-f^{o}_k\ge a_{ki}w_i>0$ and
$\alpha_{\mathcal S(k),i}\ge a_{ki}>0$. Hence
$\{k\in\mathcal N_r:a^{(\omega)}_{ki}>0\}
=\{k\in\mathcal N_r:a_{ki}>0\}$, and $Z$ is starvation-supported as claimed.
Suppose the only nonempty starvation-supported subset of the active
country-sectors is the full set. Then $\mathbf h^*=\mathbf 0$, so
$\boldsymbol\lambda^{(\omega)}(\mathbf h^*)=\mathbf 0$ by the boundary
convention of \eqref{eq:m-lambda-rule} that a producer stripped of its intermediate composite produces nothing, and the activity equation returns $\mathbf x^*=\mathbf 0$. The
collapse pair is always a fixed point
(Section~\ref{subsec:model_counterfactual} of the paper). Part \textnormal{(i)} and
$\mathbf h_\omega>\mathbf 0$ therefore imply that the fixed point set is
exactly $\{(\mathbf 0,\mathbf 0),(\mathbf s_\omega,\mathbf h_\omega)\}$.
This proves \textnormal{(iii)}.
\end{proof}

On the 2022 trade table the hypothesis of Proposition~\ref{prop:m-interior} holds in every reported scenario. The interiority condition \eqref{eq:m-cushion-cond} holds in $3{,}151$ of the $3{,}160$ scenarios, and in the other nine we verify interiority ex post, with $\min_i h_{\omega,i}\ge0.868$ across them. Part~\textnormal{(iii)} permits fixed points on the starvation boundary, for which starvation support is necessary and not sufficient, since the restricted fulfillment rule on the complement of the zero set must also admit a strictly positive fixed point. Every such equilibrium lies below the selected one by Proposition~\ref{prop:m-monotone}\textnormal{(ii)}. Neither Proposition~\ref{prop:m-monotone} of the paper, which proves the convergence of the fulfillment iteration started at the benchmark, nor Proposition~\ref{prop:m-interior} supplies the global $1$-norm sufficient condition \eqref{eq:m-suff} for contraction, and neither rests on the numerical checks of Definition~\ref{def:m-convergence-diagnostic}.

\subsection{CES-shaped fulfillment and the input-level representation}
\label{app:m-local-efficiency}\label{app:m-local-expansion}

The trade table's cost shares supply the inner weights of the efficiency factor, and a representation in input levels shows why. For the sector-$r$ input quantity $q_{rj}$ that producer $j$ receives, define
\begin{equation}
Y_j(\mathbf q_{\cdot j})
:=
\biggl(\sum_{r\in\mathcal R_j}\alpha_{rj}
\Bigl(\tfrac{q_{rj}}{\alpha_{rj}}\Bigr)^{\rho}\biggr)^{1/\rho}.
\label{eq:m-ces}
\end{equation}
The restriction to the used sectors keeps the expression from dividing by zero. The benchmark bundle has the quantities $q^0_{rj}=\alpha_{rj}(1-\beta_j)s_j$, collected in $\mathbf q^0_{\cdot j}$, at which direct substitution gives $Y_j(\mathbf q^0_{\cdot j})=(1-\beta_j)s_j$. Under a sanction, at the activity $x_j$ and the fulfillment vector $\mathbf h$, producer $j$ receives the sector-$r$ quantity $q^{(\omega)}_{rj}(x_j,\mathbf h)=\alpha_{rj}(1-\beta_j)x_j\,\theta^{(\omega)}_{rj}(\mathbf h)$. Substituting these quantities and factoring out the intermediate budget gives
\[
Y^{(\omega)}_j(x_j,\mathbf h)
=
(1-\beta_j)x_j\kappa^{(\omega)}_j(\mathbf h).
\]
The identity rests on the homogeneity of the inner composite, which is why the ratio vector rather than the level of activity enters the efficiency factor. The fulfillment rule $h_j=G_j(\kappa_j(\mathbf h))$ remains a separate assumption on the operating response, calibrated to the benchmark shares, and substituting the realized inputs into an outer technology with the specific factor fixed at $\beta_js_j$ does not derive it. That comparator would instead require $x_j/s_j\le G_j\bigl((x_j/s_j)\,\kappa_j(\mathbf h)\bigr)$ at the activity $x_j$, a feasibility inequality that neither the fulfillment fixed point nor the transaction accounts impose. Reading the fulfillment rule as an operating response keeps the maintained equations without claiming that they are equivalent to a technology.

The share parameters $\alpha_{rj}$ need no estimation, for in benchmark units the benchmark bundle $\mathbf q^0_{\cdot j}$ is the unique interior minimizer of $\sum_rq_{rj}$ subject to $Y_j(\mathbf q_{\cdot j})=(1-\beta_j)s_j$, so the calibration rationalizes the trade table's sectoral proportions at every maintained $\rho$. Differentiating \eqref{eq:m-ces} gives $\partial Y_j/\partial q_{rj}=\bigl((q_{rj}/\alpha_{rj})/Y_j\bigr)^{\rho-1}$, which equals one in every used sector at $\mathbf q_{\cdot j}=\mathbf q^0_{\cdot j}$, so every used input has the same marginal contribution at the benchmark. A sanction moves the delivered quantities apart, and the concavity of the $\rho$-mean turns that dispersion into a lower input index and, through the assumed response, into lower fulfillment than a proportional bundle of the same mean would give.

We now derive the local loss expansion of the output ratio in the bundle shortfalls. Fix a producer $j$ and suppress the sanction and fulfillment arguments. Write $d_{rj}:=1-\theta_{rj}$ and $\bar d_j:=\sum_{r\in\mathcal R_j}\alpha_{rj}d_{rj}$. A third-order expansion of the inner power sum at $\mathbf d_j=\mathbf0$ gives
\[
\sum_{r\in\mathcal R_j}\alpha_{rj}(1-d_{rj})^\rho
=
1-\rho\bar d_j
+\frac{\rho(\rho-1)}{2}
\sum_{r\in\mathcal R_j}\alpha_{rj}d_{rj}^2
+O\bigl(\|\mathbf d_j\|_\infty^3\bigr).
\]
Applying $(1+u)^{1/\rho}=1+u/\rho+\tfrac12(1/\rho)(1/\rho-1)u^2+O(u^3)$, the efficiency factor satisfies
\begin{align}
\kappa_j
&=
1-\bar d_j
+\frac{\rho-1}{2}
\left(
\sum_{r\in\mathcal R_j}\alpha_{rj}d_{rj}^2-\bar d_j^2
\right)
+O\bigl(\|\mathbf d_j\|_\infty^3\bigr)\nonumber\\
&=
1-\bar d_j-\frac{1-\rho}{2}V_j
+O\bigl(\|\mathbf d_j\|_\infty^3\bigr).
\label{eq:m-kappa-local-expansion}
\end{align}
Here $V_j=\sum_{r\in\mathcal R_j}\alpha_{rj}(d_{rj}-\bar d_j)^2$ is the weighted variance of the sectoral losses.

Direct differentiation of the outer CES aggregator at $\kappa=1$ gives
\[
G_j(1)=1,\qquad
G_j'(1)=1-\beta_j,\qquad
G_j''(1)=\beta_j(1-\beta_j)(\varrho-1).
\]
Set $\Delta_j:=\kappa_j-1$. A Taylor expansion gives
\[
G_j(1+\Delta_j)
=
1+(1-\beta_j)\Delta_j
+\frac{\beta_j(1-\beta_j)(\varrho-1)}{2}\Delta_j^2
+O(|\Delta_j|^3).
\]
By \eqref{eq:m-kappa-local-expansion},
\[
\Delta_j
=-\bar d_j-\frac{1-\rho}{2}V_j
+O\bigl(\|\mathbf d_j\|_\infty^3\bigr),
\qquad
\Delta_j^2
=\bar d_j^2+O\bigl(\|\mathbf d_j\|_\infty^3\bigr).
\]
Substitution yields the local loss expansion
\begin{equation}
1-G_j(\kappa_j)
=
(1-\beta_j)\bar d_j
+\frac{(1-\beta_j)(1-\rho)}{2}V_j
+\frac{\beta_j(1-\beta_j)(1-\varrho)}{2}\bar d_j^2
+O\bigl(\|\mathbf d_j\|_\infty^3\bigr).
\label{eq:m-local-loss}
\end{equation}
The linear term is the cost-share-weighted mean loss of intermediate input intensity. The two quadratic terms are the composition cost of unequal sectoral losses, governed by $\rho$, and the curvature cost of the mean contraction, governed by $\varrho$ and $\beta_j$. Only the mean loss contributes at first order. For a fixed support $\mathcal R_j$ the remainder is uniform over the maintained compact curvature intervals for $\|\mathbf d_j\|_\infty$ in a sufficiently small neighborhood of zero. The expansion is one-sided on $d_{rj}\ge0$ and does not cover a sector falling to zero retention.

\subsection{Local functional-form error and its network amplification}
\label{app:m-form-robustness}

The fixed-price economy assumes a CES-shaped fulfillment response, and the world input--output table does not identify the response away from the benchmark. We bound the local derivative of a country's vulnerability under a perturbation of the response map. The bound joins the displacement of the input bundles and the mismatch of the derivatives to the network operator that transmits the mismatch through the fulfillment fixed point.

We first write the output ratio as a function of the sectoral retention ratios. For producer $i$, collect the ratios in $\boldsymbol\theta_i=(\theta_{ri})_{r\in\mathcal R_i}$ and define
\begin{equation}
\mathcal L^{\mathrm{ces}}_i(\boldsymbol\theta_i)
:=
G_i\Bigl[\bigl({\textstyle\sum_{r\in\mathcal R_i}}
\alpha_{ri}\theta_{ri}^{\rho}\bigr)^{1/\rho}\Bigr].
\label{eq:m-form-ces-map}
\end{equation}
The fulfillment rule of Section~\ref{subsec:model_ces} of the paper then reads $\lambda^{(\omega)}_i(\mathbf h)=\mathcal L^{\mathrm{ces}}_i\bigl(\boldsymbol\theta^{(\omega)}_i(\mathbf h)\bigr)$. Let $\mathcal L^{\mathrm{alt}}_i$ be an alternative map in the same coordinates, normalized at the benchmark bundle as the CES representation is, $\mathcal L^{\mathrm{alt}}_i(\mathbf1)=1$, and write its deviation from the CES response as
\begin{equation}
\upsilon_i(\boldsymbol\theta_i)
:=
\mathcal L^{\mathrm{alt}}_i(\boldsymbol\theta_i)
-\mathcal L^{\mathrm{ces}}_i(\boldsymbol\theta_i).
\label{eq:m-form-deviation}
\end{equation}
Beyond that normalization and the stated smoothness, \eqref{eq:m-form-deviation} places no restriction on the global shape of the alternative.

Fix a sanction $\omega$ and interpolate between the two maps. For $t\in[0,1]$, set $\mathcal L^t_i:=\mathcal L^{\mathrm{ces}}_i+t\upsilon_i$ and
\[
\bigl[\boldsymbol\lambda^{(\omega)}_t(\mathbf h)\bigr]_i
:=
\mathcal L^{t}_i\bigl(\boldsymbol\theta^{(\omega)}_i(\mathbf h)\bigr).
\]
The maintained economy sits at $t=0$, where $\boldsymbol\lambda^{(\omega)}_0=\boldsymbol\lambda^{(\omega)}$, and the alternative at $t=1$. The sanction, the unrecovered share, and the effective demand matrix $\mathbf A_\omega$ stay fixed along the interpolation, since Section~\ref{subsec:model_rerouting} of the paper builds the effective demand from benchmark quantities and the two disruption parameters, with no curvature entering.

The argument differentiates a country's activity along a branch of fulfillment fixed points, so we record equilibrium activity as a function of fulfillment alone. Write
\begin{equation}
\mathcal X_\omega(\mathbf h)
:=
\bigl[\mathbf I-\diag(\mathbf h)\mathbf B_\omega\bigr]^{-1}
\diag(\mathbf h)\,\mathbf f_\omega,
\qquad
Q_\omega(c;\mathbf h):=\mathbbm{1}_c\mathcal X_\omega(\mathbf h).
\label{eq:m-form-activity-map}
\end{equation}
The map $\mathcal X_\omega$ scales the effective final demand by $\mathbf h$, whereas the conditional solve $\mathbf s(\mathbf h)$ of \eqref{eq:m-s-of-h} scales it by $\boldsymbol\lambda^{(\omega)}(\mathbf h)$, so the two agree along a branch of fixed points and not at arbitrary fulfillment vectors. If $\mathbf h(t)$ satisfies $\boldsymbol\lambda^{(\omega)}_t(\mathbf h(t))=\mathbf h(t)$ at every $t$, then $\diag(\mathbf h(t))=\diag\bigl(\boldsymbol\lambda^{(\omega)}_t(\mathbf h(t))\bigr)$ holds identically in $t$, and the equilibrium activity of the economy indexed by $t$ is $\mathcal X_\omega(\mathbf h(t))$. Two functions of $t$ that agree on an interval have the same derivative there, so the chain rule in the proof applies to that composition and not to the two maps separately. At the selected fulfillment vector the country aggregate of \eqref{eq:m-form-activity-map} is the post-sanction size the paper reports, $Q_\omega(c;\mathbf h_\omega)=Q_\omega(c)$ of \eqref{eq:counterfactual_country_size} of the paper.

\begin{lemma}[Local functional-form error and its network amplification]
\label{lem:m-local-form}
Fix a sanction $\omega\in\Omega$, an unrecovered share $\tau\in[0,1]$, a
barred-link attenuation $\delta\in(0,\bar\delta]$, and CES exponents
$\rho\in[\underline\rho,\overline\rho]\subset(-\infty,0)$ and
$\varrho\in[\underline\varrho,\overline\varrho]\subset(-\infty,0)$. Suppose the
benchmark-selected equilibrium is interior,
$\mathbf h_\omega>\mathbf 0$, and suppose
$\boldsymbol\lambda^{(\omega)}_t(\mathbf h)$ is continuously differentiable in
$(\mathbf h,t)$ on a neighborhood of $(\mathbf h_\omega,0)$. Write the
fulfillment Jacobian of the maintained economy as
\begin{equation*}
\mathbf J_\omega
:=
D_{\mathbf h}\boldsymbol\lambda^{(\omega)}(\mathbf h_\omega).
\end{equation*}
The fulfillment Jacobian is then nonnegative and satisfies
\begin{equation}
\spr(\mathbf J_\omega)
\;\le\;
\max_i e_i\bigl(\kappa^{(\omega)}_i(\mathbf h_\omega)\bigr)
\;<\;1.
\label{eq:m-form-spectral}
\end{equation}
Hence $\mathbf I-\mathbf J_\omega$ is nonsingular and the fulfillment resolvent
\[
\mathbf R_\omega
:=
(\mathbf I-\mathbf J_\omega)^{-1}
=
\sum_{n\ge0}\mathbf J_\omega^{\,n},
\]
exists and is nonnegative. Moreover:
\begin{enumerate}
\item[\textnormal{(i)}] On a neighborhood of
$t=0$ there is a unique continuously differentiable branch of fulfillment
fixed points $\mathbf h(t)$ through $\mathbf h(0)=\mathbf h_\omega$. Collect
the functional-form deviations evaluated at the selected equilibrium bundles
into the local functional-form residual
\begin{equation}
\upsilon_{\omega,i}
:=
\upsilon_i\bigl(\boldsymbol\theta^{(\omega)}_i(\mathbf h_\omega)\bigr),
\qquad
\boldsymbol\upsilon_\omega
:=(\upsilon_{\omega,1},\dots,\upsilon_{\omega,N})^{\top}.
\label{eq:m-form-residual}
\end{equation}
The fulfillment response to the change of response map is then
\begin{equation}
\frac{d\mathbf h}{dt}\Big|_{t=0}
=
\mathbf R_\omega\,\boldsymbol\upsilon_\omega.
\label{eq:m-form-response}
\end{equation}

\item[\textnormal{(ii)}] Write the activity gradient of
country $c$ as
$\mathbf y_{c,\omega}^{\top}:=D_{\mathbf h}Q_\omega(c;\mathbf h_\omega)$, and
let $\gamma^{(\omega)}_c(t):=\bigl[Q(c)-Q_\omega(c;\mathbf h(t))\bigr]/Q(c)$
be the country vulnerability of Definition~\ref{def:country_vulnerability}
along the branch. Then
\begin{equation}
\frac{dQ_\omega(c;\mathbf h(t))}{dt}\Big|_{t=0}
=
\mathbf y_{c,\omega}^{\top}\mathbf R_\omega\boldsymbol\upsilon_\omega,
\qquad
\frac{d\gamma^{(\omega)}_c(t)}{dt}\Big|_{t=0}
=
-\frac{1}{Q(c)}\,
\mathbf y_{c,\omega}^{\top}\mathbf R_\omega\boldsymbol\upsilon_\omega.
\label{eq:m-form-country}
\end{equation}
Define the country-specific network amplification factor as the own-size
normalization
\begin{equation}
\mathcal A_{c,\omega}
:=
\frac{\bigl\|\mathbf y_{c,\omega}^{\top}\mathbf R_\omega\bigr\|_{*}}{Q(c)}.
\label{eq:m-form-amplification}
\end{equation}
Here $\|\cdot\|_{*}$ is the norm dual to a chosen norm on the residual. Then
\begin{equation}
\left|\,\frac{d\gamma^{(\omega)}_c}{dt}\Big|_{t=0}\right|
\;\le\;
\mathcal A_{c,\omega}\,\bigl\|\boldsymbol\upsilon_\omega\bigr\|.
\label{eq:m-form-bound}
\end{equation}
Taking the supremum norm on the residual and reading
\eqref{eq:m-form-country} entrywise gives the sharper bound
\begin{equation}
\left|\,\frac{d\gamma^{(\omega)}_c}{dt}\Big|_{t=0}\right|
\;\le\;
\frac{1}{Q(c)}\sum_i
\bigl|[\mathbf y_{c,\omega}^{\top}\mathbf R_\omega]_i\bigr|\,
|\upsilon_{\omega,i}|.
\label{eq:m-form-bound-entrywise}
\end{equation}

\item[\textnormal{(iii)}] Fix a norm on
the retention coordinates and give Hessians and third derivatives the induced bilinear and trilinear operator norms. Suppose $\upsilon_i$ is three times continuously differentiable on the
segment joining $\mathbf 1$ to
$\boldsymbol\theta^{(\omega)}_i(\mathbf h_\omega)$, and suppose the
alternative response map matches the CES representation at the benchmark in
value and in gradient,
\begin{equation}
\upsilon_i(\mathbf 1)=0,
\qquad
D\upsilon_i(\mathbf 1)=\mathbf 0^{\top}.
\label{eq:m-form-match}
\end{equation}
Write
$\mathbf d_{\omega,i}
:=\mathbf 1-\boldsymbol\theta^{(\omega)}_i(\mathbf h_\omega)$ for the bundle
displacement, $\Delta\mathbf H_i:=D^2\upsilon_i(\mathbf 1)$ for the curvature
mismatch, and $\Upsilon_i$ for an upper bound on the norm of $D^3\upsilon_i$
along that segment. Then
\begin{equation}
|\upsilon_{\omega,i}|
\;\le\;
\tfrac12\|\Delta\mathbf H_i\|\,\|\mathbf d_{\omega,i}\|^2
+\tfrac16\,\Upsilon_i\,\|\mathbf d_{\omega,i}\|^3.
\label{eq:m-form-taylor}
\end{equation}
Matching the benchmark value and the cost-share gradient therefore makes the
local functional-form error second order in the bundle displacement. If the
curvature also matches, so that $\Delta\mathbf H_i=\mathbf 0$, the error is
third order.
\end{enumerate}
\end{lemma}

\begin{proof}
We first bound the fulfillment Jacobian, since parts \textnormal{(i)} and \textnormal{(ii)} rest on the resolvent it delivers. We then differentiate the fixed point and the country activity, and we last bound the residual of the response map. The maintained fulfillment rule is entrywise nondecreasing (Lemma~\ref{lem:m-rule} of the paper), so $\mathbf J_\omega\ge\mathbf0$. Bundle efficiency is positively homogeneous of degree one in fulfillment (Section~\ref{subsec:model_ces} of the paper), so Euler's identity gives $\sum_j h_{\omega,j}\,\partial\kappa^{(\omega)}_i(\mathbf h_\omega)/\partial h_j=\kappa^{(\omega)}_i(\mathbf h_\omega)$. Write $\kappa_i:=\kappa^{(\omega)}_i(\mathbf h_\omega)$. With the slope identity $G_i'(\kappa)\kappa=e_i(\kappa)G_i(\kappa)$ of Lemma~\ref{lem:m-outer}\textnormal{(iv)},
\[
[\mathbf J_\omega\mathbf h_\omega]_i
=
G_i'(\kappa_i)\sum_j
\frac{\partial\kappa^{(\omega)}_i}{\partial h_j}(\mathbf h_\omega)\,
h_{\omega,j}
=
G_i'(\kappa_i)\,\kappa_i
=
e_i(\kappa_i)\,G_i(\kappa_i)
=
e_i(\kappa_i)\,h_{\omega,i}.
\]
The last equality is the fixed-point condition $h_{\omega,i}=G_i(\kappa_i)$. Since $\mathbf h_\omega>\mathbf0$, the Collatz--Wielandt bound for a nonnegative matrix gives
\[
\spr(\mathbf J_\omega)
\;\le\;
\max_i\frac{[\mathbf J_\omega\mathbf h_\omega]_i}{h_{\omega,i}}
\;=\;
\max_i e_i(\kappa_i)
\;<\;1.
\]
The strict inequality is the bound $e_i(\kappa)<1$ on $(0,1]$ of \eqref{eq:m-shadow}, which applies because $\mathbf h_\omega>\mathbf0$ puts every $\kappa_i$ strictly above zero. This proves \eqref{eq:m-form-spectral} and delivers the nonnegative Neumann series for $\mathbf R_\omega$.

We now prove \textnormal{(i)}. Define $\mathbf F(\mathbf h,t):=\mathbf h-\boldsymbol\lambda^{(\omega)}_t(\mathbf h)$, so that $\mathbf F(\mathbf h_\omega,0)=\mathbf0$ and $D_{\mathbf h}\mathbf F(\mathbf h_\omega,0)=\mathbf I-\mathbf J_\omega$, which the spectral bound makes nonsingular. The implicit function theorem therefore gives a neighborhood of $t=0$ and a unique continuously differentiable branch $\mathbf h(t)$ of fulfillment fixed points through $\mathbf h_\omega$. The interpolation is affine in $t$, so $\partial\boldsymbol\lambda^{(\omega)}_t(\mathbf h)/\partial t=\boldsymbol\upsilon_\omega$ at $(\mathbf h_\omega,0)$, and differentiating $\mathbf F(\mathbf h(t),t)=\mathbf0$ at $t=0$ gives $(\mathbf I-\mathbf J_\omega)\mathbf h'(0)=\boldsymbol\upsilon_\omega$, which inverts to \eqref{eq:m-form-response}. This proves \textnormal{(i)}.

We next prove \textnormal{(ii)}. Along the branch the economy indexed by $t$ has the equilibrium activity $\mathcal X_\omega(\mathbf h(t))$ by \eqref{eq:m-form-activity-map}, so the chain rule and \eqref{eq:m-form-response} give
\[
\frac{dQ_\omega(c;\mathbf h(t))}{dt}\Big|_{t=0}
=
D_{\mathbf h}Q_\omega(c;\mathbf h_\omega)\,\mathbf h'(0)
=
\mathbf y_{c,\omega}^{\top}\mathbf R_\omega\boldsymbol\upsilon_\omega.
\]
The benchmark size $Q(c)$ does not vary with $t$, so differentiating $\gamma^{(\omega)}_c(t)=[Q(c)-Q_\omega(c;\mathbf h(t))]/Q(c)$ returns the second identity of \eqref{eq:m-form-country}. The dual-norm inequality applied to that scalar product gives \eqref{eq:m-form-bound}, and bounding the same product term by term gives \eqref{eq:m-form-bound-entrywise}. This proves \textnormal{(ii)}.

We prove \textnormal{(iii)} last. Apply Taylor's theorem to $\upsilon_i$ at $\mathbf1$ in the direction $-\mathbf d_{\omega,i}$. The two equalities of \eqref{eq:m-form-match} remove the constant and the linear term, and the trilinear bound along the segment gives
\[
\Bigl|
\upsilon_{\omega,i}
-\tfrac12\,\mathbf d_{\omega,i}^{\top}
\Delta\mathbf H_i\,\mathbf d_{\omega,i}
\Bigr|
\;\le\;
\tfrac16\,\Upsilon_i\,\|\mathbf d_{\omega,i}\|^3.
\]
Together with $|\mathbf d^{\top}\Delta\mathbf H_i\mathbf d|\le\|\Delta\mathbf H_i\|\,\|\mathbf d\|^2$ this gives \eqref{eq:m-form-taylor}. This proves \textnormal{(iii)}.
\end{proof}

The branch of part~\textnormal{(i)} is a local continuation of the selected CES equilibrium. The branch remains the selected equilibrium of the economy indexed by $t$ when the alternative fulfillment rules are admissible continuous monotone self-maps of $[0,1]^N$ with a positive greatest fixed point and a unique positive fixed point. The CES-specific uniqueness results do not establish those properties for an arbitrary alternative.

The CES derivatives make the matching condition of part~\textnormal{(iii)} explicit. Write $\boldsymbol\alpha_i=(\alpha_{ri})_{r\in\mathcal R_i}$ and $\mathbf D_{\alpha_i}:=\diag(\boldsymbol\alpha_i)$. In retention coordinates, \eqref{eq:m-local-loss} gives
\begin{align}
D\mathcal L^{\mathrm{ces}}_i(\mathbf 1)
&=(1-\beta_i)\boldsymbol\alpha_i^{\top}
\label{eq:m-form-ces-gradient}\\
D^2\mathcal L^{\mathrm{ces}}_i(\mathbf 1)
&=(1-\beta_i)(\rho-1)
\bigl(\mathbf D_{\alpha_i}
-\boldsymbol\alpha_i\boldsymbol\alpha_i^{\top}\bigr)
+\beta_i(1-\beta_i)(\varrho-1)
\boldsymbol\alpha_i\boldsymbol\alpha_i^{\top}.
\label{eq:m-form-ces-hessian}
\end{align}
The first component of the Hessian governs changes in composition at a fixed mean retention, and the second governs the mean contraction. An alternative may take any smooth shape away from the benchmark and still satisfy the third-order conclusion of the lemma, provided its value at $\mathbf1$ is one and its gradient and Hessian there agree with \eqref{eq:m-form-ces-gradient} and \eqref{eq:m-form-ces-hessian}.

Lemma~\ref{lem:m-local-form} is a local statement, for the implicit function theorem gives the branch only near $t=0$, so that a finite comparison of the two economies would need a resolvent bounded uniformly along the whole path from the CES response to the alternative. The Taylor order is local in bundle space as well, since the third-derivative bound has to hold on each producer's segment from the benchmark bundle to the realized one, and a deeper sanction lengthens that segment. The curvature matching is a restriction the calibration does not verify, because the paper takes $\rho$ and $\varrho$ from the disruption literature rather than from an estimated Hessian. A true Hessian outside the two-direction form \eqref{eq:m-form-ces-hessian} leaves the quadratic mismatch term $\Delta\mathbf H_i$ in place, and even the second-order statement of part~\textnormal{(iii)} requires the value and gradient matching of \eqref{eq:m-form-match}, which restricts the alternative beyond what the trade table supplies. The lemma perturbs the response map alone, moreover, and an alternative that also moves the effective demand, the prices, the cost shares, or the prescribed margins needs residual terms of its own.

We evaluate the amplification factor and the quadratic curvature term of the lemma at the eight headline bilateral severances of Table~\ref{tab:s4_headline}, and Table~\ref{tab:form-robustness} reports them. The spectral radius of the fulfillment Jacobian is $0.596$ to three decimals at all eight severances, against the Collatz--Wielandt bound $\max_ie_i(\kappa^{(\omega)}_i)=0.980$ of \eqref{eq:m-form-spectral}. The bound is loose for a reason its proof shows, since $e_i$ falls in $\kappa$ and $e_i(1)=1-\beta_i$, so the bound cannot fall below the largest benchmark intermediate share in the trade table, the $0.98$ of the two capped buyer columns. The spectral radius is also the local rate at which the linearized benchmark path converges, and $0.596$ is close to the median terminal ratio of $0.595$ that the benchmark-path diagnostics record over the $3{,}160$ scenarios.

With the supremum norm on the residual, the amplification factor $\mathcal A_{c,\omega}$ runs from $3.1$ to $6.7$, so a residual bounded by $\epsilon$ at every producer bounds $|d\gamma_c/dt|_{t=0}$ by $\mathcal A_{c,\omega}\epsilon$. A small finite change multiplies that derivative bound by $|t|$ and adds the local remainder $o(|t|)$. The factor varies far more across countries than across severances. That of the United States lies between $3.4$ and $3.5$ across its three headline severances, and that of China is $6.7$ across all five of its own. The factor measures the exposure of a country's own activity to a disturbance anywhere in the fulfillment feedback, and in these comparisons the country's position in the network fixes that exposure far more than the identity of the severed link does.

Parts~\textnormal{(ii)} and~\textnormal{(iii)} bound the vulnerability derivative with a quadratic and a cubic term. In the Euclidean norm on the retention coordinates the largest bundle displacement $\max_i\|\mathbf d_{\omega,i}\|$ runs from $0.095$ at the United States--Russia severance to $0.664$ at the United States--Mexico severance. Define the quadratic curvature-sensitivity component by
\[
\mathcal D^{(2)}_{c,\omega}
:=\tfrac12 Q(c)^{-1}\sum_i\bigl|[\mathbf y_{c,\omega}^{\top}\mathbf R_\omega]_i\bigr|
\,\|\Delta\mathbf H_i\|\,\|\mathbf d_{\omega,i}\|^2.
\]
The full bound also contains the cubic term
\[
\tfrac16 Q(c)^{-1}\sum_i\bigl|[\mathbf y_{c,\omega}^{\top}\mathbf R_\omega]_i\bigr|
\,\Upsilon_i\,\|\mathbf d_{\omega,i}\|^3.
\]
Table~\ref{tab:form-robustness} sets each mismatch norm at one tenth of that producer's maintained CES Hessian norm in \eqref{eq:m-form-ces-hessian}, whose median across producers is $0.24$. As a share of the reported vulnerability the quadratic component has a median of $3.1\%$, a minimum of $0.8\%$, and a maximum of $8.8\%$. These numbers evaluate the quadratic term alone and neither evaluate nor bound the third-derivative term.

\begin{table}[!htbp]
\centering
\small
\begin{adjustbox}{max width=\textwidth}
\begin{tabular}{llrrrr}
\toprule
Severance & Country & $\gamma^{(\omega)}_c$ (\%) & $\|\mathbf y_{c,\omega}^{\top}\mathbf R_\omega\|_{*}/Q(c)$ & $\max_i\|\mathbf d_{\omega,i}\|$ & $\mathcal D^{(2)}_{c,\omega}/\gamma^{(\omega)}_c$ (\%) \\
\midrule
United States--China & United States & 0.56 & 3.42 & 0.136 & 5.9 \\
 & China & 1.37 & 6.66 & 0.136 & 2.5 \\
China--Chinese Taipei & China & 0.63 & 6.69 & 0.201 & 8.8 \\
 & Chinese Taipei & 7.43 & 4.90 & 0.201 & 1.7 \\
United States--Russia & United States & 0.05 & 3.47 & 0.095 & 0.8 \\
 & Russia & 0.62 & 5.50 & 0.095 & 1.7 \\
Germany--Russia & Germany & 0.57 & 4.76 & 0.191 & 7.3 \\
 & Russia & 0.95 & 5.47 & 0.191 & 2.0 \\
Japan--China & Japan & 2.57 & 4.17 & 0.187 & 3.9 \\
 & China & 0.89 & 6.67 & 0.187 & 0.9 \\
India--China & India & 1.18 & 4.05 & 0.154 & 5.3 \\
 & China & 0.23 & 6.70 & 0.154 & 0.8 \\
United States--Mexico & United States & 0.62 & 3.44 & 0.664 & 3.9 \\
 & Mexico & 10.88 & 3.11 & 0.664 & 7.6 \\
Germany--China & Germany & 1.51 & 4.64 & 0.154 & 3.6 \\
 & China & 0.56 & 6.69 & 0.154 & 1.2 \\
\bottomrule
\end{tabular}
\end{adjustbox}
\caption{The network amplification of a local response-map error at the eight headline bilateral severances. The third column repeats the country vulnerability of Table~\ref{tab:s4_headline}. The fourth is the amplification factor $\mathcal A_{c,\omega}$ of \eqref{eq:m-form-amplification} at the supremum norm on the residual. The fifth is the largest bundle displacement the severance produces at any producer, in the Euclidean norm. The last column is $\mathcal D^{(2)}_{c,\omega}$ as a percentage of the vulnerability, with each mismatch norm set at one tenth of the producer's maintained CES Hessian norm. The quadratic component omits the higher-order remainder. Baseline calibration $\tau=0.30$, $\rho=-1$, $\varrho=-1$, $\delta=0.10$.}
\label{tab:form-robustness}
\end{table}

\section{The Degeneracy at the Leontief Limit of the Outer CES Aggregator}
\label{app:m-correction}

We establish what becomes of the sanction equilibrium at the Leontief limit of the outer CES aggregator, which the maintained interval for $\varrho$ excludes. At that limit the output ratio equals the efficiency factor, $\lambda_i^{(\omega)}=\kappa_i^{(\omega)}$, and since the efficiency factor is homogeneous of degree one, unit pass-through makes every uniform fulfillment vector a fixed point of the benchmark fulfillment rule. The benchmark therefore ceases to be an isolated fixed point, a degeneracy that is structural and not numerical.

\begin{lemma}[Degeneracy of the unit pass-through rule]
\label{lem:m-degenerate}
Fix $\rho\in[\underline\rho,\overline\rho]\subset(-\infty,0)$ and maintain
the active-set convention. At the excluded outer limit, set
$\lambda_i^{(\omega)}=\kappa_i^{(\omega)}$ for every sanction $\omega$.
\begin{enumerate}
\item[\textnormal{(i)}] Under the no-sanction benchmark map
$\Phi_{\varnothing}$, the pairs
\[
\bigl(\mathbf x(t),\,t\,\mathbf 1\bigr),
\qquad
\mathbf x(t):=t\,(\mathbf I-t\,\mathbf B)^{-1}\mathbf f,
\qquad t\in[0,1],
\]
are all fixed points. Here
$\mathbf B=\mathbf A(\mathbf I-\boldsymbol\beta)$. The equilibrium set is a continuum
containing the benchmark $(\mathbf x(1),\mathbf 1)=(\mathbf s,\mathbf 1)$.
\item[\textnormal{(ii)}] Maintain a positive barred-link attenuation
$\delta\in(0,\bar\delta]$ and fix a bilateral sanction $\omega$ with
$\tau(1-\delta)>0$. Suppose that every country-sector can be reached from some producer of the sanctioned pair that buys intermediate inputs from its partner at the benchmark, along a chain of positive input shares running downstream from seller to buyer. Under $\omega$, the unique fixed point of
$\Phi_\omega$ on $\mathbb R_+^N\times[0,1]^N$ is the
collapse equilibrium $(\mathbf 0,\mathbf 0)$.
\end{enumerate}
\end{lemma}

\begin{proof}
We first prove part (i). Fix $t\in[0,1]$ and $\mathbf h=t\mathbf1$. Homogeneity and unit pass-through give $\boldsymbol\lambda^{(\varnothing)}(t\mathbf1)=\boldsymbol\kappa^{(\varnothing)}(t\mathbf1)=t\mathbf1$ on the active set. The activity equation $\mathbf x=t\,\mathbf f+t\,\mathbf B\mathbf x$ has the unique solution $\mathbf x(t)$, which is entrywise increasing in $t$ and bounded by $\mathbf s$. The activity and the fulfillment identities together make every pair a fixed point. This proves part (i).

We next prove part (ii). Fix the stated sanction and attenuation, let $(\mathbf x^*,\mathbf h^*)$ be a fixed point, and let $m:=\max_i h^*_i$. Suppose $m>0$ and define $\mathcal W:=\{i:h^*_i=m\}$. For $i_0\in\mathcal W$, monotonicity, homogeneity, and $\mathbf h^*\le m\mathbf1$ give
\[
m=h^*_{i_0}
=\kappa^{(\omega)}_{i_0}(\mathbf h^*)
\;\le\;\kappa^{(\omega)}_{i_0}(m\,\mathbf 1)
\;=\;m\,\kappa^{(\omega)}_{i_0}(\mathbf 1)\;\le\;m.
\]
Every inequality is then tight, so $\kappa^{(\omega)}_{i_0}(\mathbf1)=1$. Since $m>0$, the retention ratio of $i_0$ is positive in every input sector it uses, the $\rho$-mean is strictly increasing in these positive ratios, and each ratio is a weighted sum of its suppliers' fulfillment rates, so the equality requires $h^*_{i'}=m$ for every retained supplier $i'$ of $i_0$. The set $\mathcal W$ is therefore closed under passage to suppliers, and every $i\in\mathcal W$ has $\kappa^{(\omega)}_i(\mathbf1)=1$. Repeating the argument along the input paths places every country-sector upstream of $i_0$ in $\mathcal W$.

Positive attenuation and positive balancing scalings preserve the benchmark input support. By the reachability hypothesis of part (ii), some producer $j$ of the sanctioned pair buys intermediate inputs from its partner at the benchmark and lies upstream of $i_0$, so $j\in\mathcal W$ and $\kappa^{(\omega)}_j(\mathbf1)=1$. Yet Lemma~\ref{lem:m-kernel} and $\tau(1-\delta)>0$ put the retention ratio of $j$ strictly below one in every used sector it sources from the severed partner, even at $\mathbf h=\mathbf1$. At least one such sector carries positive weight and every retention ratio is at most one, so the $\rho$-mean is strictly below one, which contradicts $\kappa^{(\omega)}_j(\mathbf1)=1$. Hence $m=0$ and $\mathbf h^*=\mathbf0$. The active-set convention gives $\boldsymbol\lambda^{(\omega)}(\mathbf0)=\mathbf0$, and the activity equation gives $\mathbf x^*=\mathbf0$. This proves the lemma.
\end{proof}

The collapse at the limit comes from unit pass-through and not from the rationed exchange of Definition~\ref{def:oa-rationed-exchange} or from the fulfillment condition it enters. The maintained band excludes the limit, and the interiority condition \eqref{eq:m-cushion-cond} of the paper is sufficient for an interior equilibrium away from it.

\section{Diversification on the Two Sides of Trade}
\label{oa:diversification}

Two countries alike in everything but their trade with a partner need not be equally vulnerable to it, and which of them fares better depends on the side of trade on which they differ. We compare such a pair at fixed size and ask when the lower trade share improves the outcome of the severance. Write $\sigma^{(\omega)}_{rj}$ for the severed partner's share of producer $j$'s benchmark purchases of input $r$. These shares keep the sectoral detail that the trade shares of Proposition~\ref{prop:m-exposure} of the paper add up over the producers and the household of a country. On the selling side, Proposition~\ref{prop:m-diversification} orders bilateral power at the equilibrium under symmetry and home dominance. On the buying side it orders the output ratios $\lambda^{(\omega)}_j(\mathbf 1)$ at full delivery and states a further condition under which that ordering survives the fixed point. Its proof rests on the anatomy of the loss and on the uniqueness of the balanced allocation.

To compare the two sides of a pair we hold everything else alike. A bijection $\pi$ carries the country-sectors of country $c$ onto those of country $c'$, fixes every third country, and exchanges the two households. The pair is otherwise alike under $\pi$ if the bijection preserves every producer's sector shares $\alpha_{rj}$ and leakage $\beta_j$, the benchmark activity, the households' budgets and sector shares, and every relation to a third country, and it leaves the trade inside the pair free. Write $\mathbf R^{*}_\omega$ for the operator of \eqref{eq:m-content} of the paper at the selected equilibrium and $\mathbf e_m$ for the unit vector of country-sector $m$. We call the pair home-dominant if a dollar of activity lost at any producer of the pair costs the producer's own country more activity than it costs the partner,
\begin{equation}
\mathbbm{1}_c\mathbf R^{*}_\omega\mathbf e_m\;>\;\mathbbm{1}_{c'}\mathbf R^{*}_\omega\mathbf e_m
\qquad\text{for every country-sector $m$ of country $c$}.
\label{eq:m-home-dominance}
\end{equation}
The mirror inequality must hold at every country-sector of country $c'$ as well. The proof of Proposition~\ref{prop:m-diversification} gives a sufficient condition on the effective demand, that every producer of the pair source each input sector at least as heavily from its own country as from the partner's counterpart, and the attenuation of the severed links makes that condition mild. We call a sector a consumer-goods sector if no producer uses its output, so that it sells to households and to the residual outside demand alone.

\begin{proposition}[Diversification on the two sides of trade]
\label{prop:m-diversification}
Fix a bilateral sanction $c'\notnotleftrightarrow c$ and the maintained parameters with $\tau>0$, and let the pair be otherwise alike under $\pi$.
\begin{enumerate}
\item[\textnormal{(i)}] On the selling side, suppose the producers of the pair buy alike, so that $\pi$ preserves every producer column of $\mathbf A$, and that the two households agree outside the consumer-goods sectors. In the consumer-goods sectors the household of country $c'$ takes a weakly smaller share of every sector's sales in country $c$ than the household of country $c$ takes of the corresponding sector's sales in country $c'$, strictly for some sector. The residual outside demand $\mathbf f^{o}$ takes up the difference in sales, so that the two sides keep the same benchmark activity. If the pair is home-dominant and the selected equilibrium is interior, then
\begin{equation}
\gamma^{(c'\notnotleftrightarrow c)}_c\;<\;\gamma^{(c'\notnotleftrightarrow c)}_{c'},
\qquad\text{so}\qquad
\mathcal H_{cc'}
=
\frac{\gamma^{(c'\notnotleftrightarrow c)}_c-\gamma^{(c'\notnotleftrightarrow c)}_{c'}}{\gamma^{(c'\notnotleftrightarrow c)}_c+\gamma^{(c'\notnotleftrightarrow c)}_{c'}}
\;<\;0.
\label{eq:m-tilt}
\end{equation}
The pair then tilts in favor of country $c$. The tilt holds at every admissible $(\rho,\varrho)$ at which the home-dominance and interiority hypotheses hold, since the exponents enter it only through the fulfillment rates and the output ratios, which the two sides share.
\item[\textnormal{(ii)}] On the buying side, suppose the producers of the pair differ only in the partner's share of each input sector they use, with the third-country coefficients preserved by $\pi$ and the own-country share absorbing the difference in the partner's share. The residual outside demand $\mathbf f^{o}$ takes up the difference in sales, so that the two sides keep the same benchmark activity. If $\sigma^{(\omega)}_{rj}\le\sigma^{(\omega)}_{r\pi(j)}$ at every input sector, then at full delivery the retention ratios are ordered sector by sector,
\begin{equation}
\theta^{(\omega)}_{rj}(\mathbf 1)-\theta^{(\omega)}_{r\pi(j)}(\mathbf 1)
=\tau(1-\delta)\bigl(\sigma^{(\omega)}_{r\pi(j)}-\sigma^{(\omega)}_{rj}\bigr)
\;\ge\;0.
\label{eq:m-level-full}
\end{equation}
The output ratios satisfy $\lambda^{(\omega)}_j(\mathbf 1)\ge\lambda^{(\omega)}_{\pi(j)}(\mathbf 1)$ as well, strictly wherever the share ordering is strict, at every admissible $\rho$ and $\varrho$. Suppose further that the interiority condition \eqref{eq:m-cushion-cond} holds, let $c_\omega$ be its floor \eqref{eq:m-floor}, and write $k_r$ for the seller of country $c$ in input sector $r$, so that $\pi(k_r)$ is the seller of country $c'$. If at every input sector and every producer $j$ of country $c$
\begin{equation}
a^{(\omega)}_{k_rj}\;\ge\;a^{(\omega)}_{k_r\pi(j)}
\qquad\text{and}\qquad
(1-c_\omega)\!\!\sum_{i\in\mathcal N_r:\,\mathcal C(i)\notin\{c,c'\}}\!\!\bigl|a^{(\omega)}_{ij}-a^{(\omega)}_{i\pi(j)}\bigr|
\;\le\;
c_\omega\,\alpha_{rj}\,\tau(1-\delta)\bigl(\sigma^{(\omega)}_{r\pi(j)}-\sigma^{(\omega)}_{rj}\bigr),
\label{eq:m-buying-cond}
\end{equation}
then the ordering of the output ratios survives the fixed point. Every producer of country $c$ delivers at least the fraction of the effective demand it faces that its counterpart delivers, $h_{\omega,j}\ge h_{\omega,\pi(j)}$, strictly wherever the second condition is strict. The unpropagated fulfillment shortfall $(1-h_{\omega,j})s_j$ of Proposition~\ref{prop:m-anatomy} is then weakly smaller at every producer of country $c$ than at its counterpart.
\item[\textnormal{(iii)}] At full delivery the retention profile is the benchmark object $1-\tau(1-\delta)\sigma^{(\omega)}_{rj}$. If its distribution over the sectors producer $j$ uses, weighted by the cost shares $\alpha_{rj}$, dominates its counterpart's in the increasing-concave (second-order) order, then $\lambda^{(\omega)}_j(\mathbf 1)\ge\lambda^{(\omega)}_{\pi(j)}(\mathbf 1)$, strictly under a nontrivial mean-preserving contraction. When a nonconstant positive retention profile is compared with the constant profile at its own mean, with the outer aggregator fixed, the ratio of the constant profile's output ratio to the dispersed profile's increases as $\rho$ falls. When two constant profiles at $0<\theta_{c'}<\theta_c\le1$ are compared, the ratio of their output ratios does not depend on $\rho$ and increases as $\varrho$ falls. The proposition asserts no monotonicity in $\rho$ for the advantage of an arbitrary mean-preserving contraction or of a uniform improvement of a heterogeneous profile.
\end{enumerate}
\end{proposition}

\noindent Proof in Section~\ref{oa:proof-diversification}.

The two conditions of \eqref{eq:m-buying-cond} are read off the effective demand and the floor. The first asks that a seller of the pair weigh at least as much in its own country's producer as in the foreign one, and the second bounds the third-country orders by which the two producers' effective demands differ. Both are evaluated at the stated curvatures, since the floor $c_\omega$ depends on $\rho$ and $\varrho$, and a statement uniform over a set of curvatures requires a common positive floor and the same inequalities imposed with that floor. The ordering of the fulfillment shortfalls that the conditions deliver does not by itself order the productivity channels of the two countries, which carry the shortfalls through $\mathbf R^{*}_\omega$.

The two buying-side parts are first-round statements, and neither carries over to the equilibrium without the further hypotheses of part~(ii). Part~(ii) orders the levels of the two producers' retention profiles at full delivery and part~(iii) their shapes, and both orderings can reverse at the selected equilibrium. The equilibrium reweights each input sector by what its suppliers actually ship, and the rationed exchange rescales the two producers' orders differently. The producer that loses more of the partner's supply reroutes more of its budget to third-country suppliers, whose deliveries stay close to full. The sellers of the two countries carry different scalings as well, since the country whose producers buy less from the partner is the country the partner buys more from. A benchmark profile that dominates its counterpart at $\mathbf 1$ therefore need not dominate it at $\mathbf h_\omega$.\footnote{Three-country benchmarks in the maintained band show both reversals, and the second condition of \eqref{eq:m-buying-cond} fails in the benchmarks that reverse the level ordering, as it must. The code that builds and solves these benchmarks is in the Replication Package.}

Diversifying purchases lowers both the level of a producer's purchases from the partner and, when its partner shares become more even across inputs, the composition cost of those purchases. Diversifying sales lowers only the orders at risk. In the two comparisons of part~(iii) the outer exponent $\varrho$ governs what an even shortfall of inputs costs and the inner exponent $\rho$ what unevenness across inputs costs. Neither exponent enters the direct source of lost sales, though both can move its propagation, since inputs must be assembled and orders need not be. The buying-side ordering of the output ratios reaches the two countries' activity through the anatomy of Proposition~\ref{prop:m-anatomy}. On its own it does not order the asymmetry of the pair, since the two countries also differ in their lost markets and in the paths along which both losses propagate. The empirical comparisons of Section~\ref{sec:power_size} of the paper describe the observed network, and they do not impose these conditional orderings on every pair.

\subsection{Proof of Proposition~\ref{prop:m-diversification}}
\label{oa:proof-diversification}

Throughout, $\mathbf P$ is the permutation matrix of the bijection $\pi$, extended by the identity on third countries and exchanging the two households, so that $\mathbf P^{2}=\mathbf I$. A vector $\mathbf x$ or a matrix $\mathbf X$ is symmetric if $\mathbf P\mathbf x=\mathbf x$ or $\mathbf P\mathbf X\mathbf P=\mathbf X$. A row vector $\boldsymbol\phi$ is antisymmetric if $\boldsymbol\phi\mathbf P=-\boldsymbol\phi$, which is to say $\phi_{\pi(k)}=-\phi_k$ on the pair and $\phi_k=0$ on third countries. The hypothesis that the pair is otherwise alike makes the cost shares, the leakages, the benchmark activity, the household budgets, and every third-country coefficient symmetric.

\begin{proof}
We first prove (i). The essence of the argument is that the hypotheses make everything symmetric under $\pi$ except the lost sales, so an antisymmetric decomposition of the anatomy \eqref{eq:m-anatomy} isolates what the lost sales do to the difference between the two countries' losses. We take the argument in five steps, the effective demand, the fulfillment and propagation operators, the lost sales, the decomposition, and a sufficient condition for home dominance.

We begin with the effective demand, which is symmetric off the consumer-goods sectors. The bijection $\pi$ carries the severed set into itself, since a link joins the two countries of the pair exactly when its image does. Fix an input sector $r$ that some producer uses. Its block of the rationed exchange has a symmetric disrupted kernel, since the producer columns are symmetric by hypothesis, the household columns agree outside the consumer-goods sectors, and $\pi$ fixes the third-country columns. Its column targets \eqref{eq:m-coltarget} are symmetric for the same reason, and its row targets \eqref{eq:m-rowtarget} are symmetric because a seller's sales to producers and households and its severed sales in such a sector are sums of symmetric flows. The balanced block is the unique optimizer of the program \eqref{eq:m-balprog} by Proposition~\ref{prop:m-balancing}. The reflection of the optimizer is feasible for the reflected program and attains the same value, and the reflected program is the same program, so uniqueness makes the optimizer symmetric. A consumer-goods sector has no producer column. Hence the producer block of the effective demand is symmetric, $\mathbf P\mathbf A_\omega\mathbf P=\mathbf A_\omega$, and so is the household block outside the consumer-goods sectors.

Fulfillment and propagation inherit the symmetry. The retention ratios \eqref{eq:m-theta} are built from the producer block and the symmetric cost shares, so $\theta^{(\omega)}_{r,\pi(j)}(\mathbf P\mathbf h)=\theta^{(\omega)}_{rj}(\mathbf h)$ for every $\mathbf h$, and with $\beta_{\pi(j)}=\beta_j$ the fulfillment rule is equivariant, $\boldsymbol\lambda^{(\omega)}(\mathbf P\mathbf h)=\mathbf P\boldsymbol\lambda^{(\omega)}(\mathbf h)$. The benchmark path starts at the symmetric vector $\mathbf 1$ and stays symmetric, so its limit $\mathbf h_\omega$ is symmetric. The propagation matrix $\mathbf M:=\diag(\mathbf h_\omega)\mathbf A_\omega(\mathbf I-\boldsymbol\beta)$ and the operator $\mathbf R^{*}_\omega=\sum_{t\ge0}\mathbf M^{t}$ are therefore symmetric as well.

The lost sales are where the two sides differ. A seller in a sector that producers use has symmetric severed sales, so $\xi_{\omega,m}=\xi_{\omega,\pi(m)}$ there. A seller $m$ in a consumer-goods sector of country $c$ severs only its sales to the household of country $c'$. The hypothesis orders the two households' shares of those sales, and $\pi$ preserves the benchmark activity the shares are taken of, so it orders the lost sales themselves,
\[
\xi_{\omega,m}=\tau(1-\delta)f^{c'}_m\;\le\;\tau(1-\delta)f^{c}_{\pi(m)}=\xi_{\omega,\pi(m)}.
\]
The inequality is strict for some $m$ by hypothesis. With $\mathbf s$ and $\mathbf h_\omega$ symmetric, the direct-loss vector $\boldsymbol\nu_\omega:=(\mathbf 1-\mathbf h_\omega)\circ\mathbf s+\mathbf h_\omega\circ\boldsymbol\xi_\omega$ of Proposition~\ref{prop:m-anatomy} satisfies $\nu_{\omega,m}-\nu_{\omega,\pi(m)}=h_{\omega,m}(\xi_{\omega,m}-\xi_{\omega,\pi(m)})$, which vanishes except on the consumer-goods sectors of the pair, where it is nonpositive.

We now decompose the difference between the two losses. Let $\mathbbm{1}_{\pm}:=\mathbbm{1}_c-\mathbbm{1}_{c'}$, an antisymmetric row vector. Since $\mathbf P\mathbf R^{*}_\omega\mathbf P=\mathbf R^{*}_\omega$, the row vector $\boldsymbol\Delta:=\mathbbm{1}_{\pm}\mathbf R^{*}_\omega$ satisfies $\boldsymbol\Delta\mathbf P=\mathbbm{1}_{\pm}\mathbf P\mathbf R^{*}_\omega=-\boldsymbol\Delta$, so it is antisymmetric too, and its entry at a country-sector $m$ of country $c$ is the gap $\mathbbm{1}_c\mathbf R^{*}_\omega\mathbf e_m-\mathbbm{1}_{c'}\mathbf R^{*}_\omega\mathbf e_m$ of \eqref{eq:m-home-dominance}. By Proposition~\ref{prop:m-anatomy} the two countries' dollar losses are $\mathbbm{1}_c\mathbf R^{*}_\omega\boldsymbol\nu_\omega$ and $\mathbbm{1}_{c'}\mathbf R^{*}_\omega\boldsymbol\nu_\omega$. Pairing each $m$ of country $c$ with $\pi(m)$ and using the antisymmetry of $\boldsymbol\Delta$ gives
\[
\bigl[Q(c)-Q_\omega(c)\bigr]-\bigl[Q(c')-Q_\omega(c')\bigr]
=\boldsymbol\Delta\boldsymbol\nu_\omega
=\sum_{m\in c}\Delta_m\bigl(\nu_{\omega,m}-\nu_{\omega,\pi(m)}\bigr)
=\sum_{m\in c}\Delta_m\,h_{\omega,m}\bigl(\xi_{\omega,m}-\xi_{\omega,\pi(m)}\bigr).
\]
Home dominance is $\Delta_m>0$ at every $m$ of country $c$, interiority is $h_{\omega,m}>0$, and the ordering of the lost sales makes every summand nonpositive and one of them negative, so the difference is negative. The two sizes agree, $Q(c)=Q(c')$, which gives $\gamma^{(c'\notnotleftrightarrow c)}_c<\gamma^{(c'\notnotleftrightarrow c)}_{c'}$, and \eqref{eq:bounded-Hirschman} then gives $\mathcal H_{cc'}<0$, which is \eqref{eq:m-tilt}. The exponents $\rho$ and $\varrho$ enter the display only through $\mathbf h_\omega$ and $\mathbf R^{*}_\omega$, which are symmetric and weight both sides alike, so under the stated hypotheses the sign rests on the ordering of the lost sales. This proves (i).

We next give the sufficient condition for home dominance. Index the country-sectors of country $c$ and write, for $k$ and $m$ in country $c$, $D_{km}:=M_{km}$ for the propagation from the home seller $k$ to the buyer $m$ and $X_{km}:=M_{\pi(k)m}$ for the propagation from the partner's counterpart $\pi(k)$ to the same buyer, and set $\mathbf N:=\mathbf D-\mathbf X$. Let $\boldsymbol\phi$ be an antisymmetric row vector. For a buyer $m$ of country $c$, $(\boldsymbol\phi\mathbf M)_m=\sum_{k\in c}\phi_k(M_{km}-M_{\pi(k)m})=(\boldsymbol\phi|_c\mathbf N)_m$. For a third-country buyer the same bracket vanishes, since such a buyer sources alike from $k$ and $\pi(k)$ and $\mathbf h_\omega$ is symmetric, and for $m=\pi(m_0)$ the symmetry of $\mathbf M$ turns the bracket into its negative at $m_0$. Hence $\boldsymbol\phi\mathbf M$ is antisymmetric with $(\boldsymbol\phi\mathbf M)|_c=\boldsymbol\phi|_c\mathbf N$, and by induction $(\mathbbm{1}_{\pm}\mathbf M^{t})|_c=\mathbf 1^{\top}\mathbf N^{t}$ for every $t$, so
\[
\boldsymbol\Delta|_c=\mathbf 1^{\top}\sum_{t\ge0}\mathbf N^{t}.
\]
Suppose every producer $m$ of country $c$ sources each input sector at least as heavily from its own country as from the partner's counterpart in the effective demand, $a^{(\omega)}_{km}\ge a^{(\omega)}_{\pi(k)m}$ for all $k$ and $m$ in country $c$, with the mirror image in country $c'$. Since $h_{\omega,k}=h_{\omega,\pi(k)}$, this is $\mathbf N\ge\mathbf 0$. Then $\mathbf 0\le\mathbf N\le\mathbf D$, and $\mathbf D$ is a principal submatrix of the nonnegative matrix $\mathbf M$, so $\spr(\mathbf N)\le\spr(\mathbf D)\le\spr(\mathbf M)<1$ by Lemma~\ref{lem:m-kernel}. The series converges to $(\mathbf I-\mathbf N)^{-1}\ge\mathbf I$, and $\boldsymbol\Delta|_c\ge\mathbf 1^{\top}>\mathbf 0$. The mirror argument gives the condition at country $c'$.

The condition can be read off the benchmark table. When the home and the partner links are positive in the same buyer column, their common support component is carried into itself by $\pi$, so the reflected scalings $(\mathbf u\circ\pi,\mathbf v\circ\pi)$ balance the same symmetric optimizer. Proposition~\ref{prop:m-balancing} then makes them a gauge of the original, $\mathbf u\circ\pi=\chi\mathbf u$ on the component. Applying $\pi$ twice gives $\chi^{2}=1$ and positivity gives $\chi=1$, so the seller scalings are symmetric, and the common buyer scaling cancels from the comparison of $a^{(\omega)}_{km}$ with $a^{(\omega)}_{\pi(k)m}$. Where a link is zero, the preservation of the support settles the inequality directly. The condition is then $a_{km}\ge\delta\,a_{\pi(k)m}$ on the benchmark coefficients, since the sanction attenuates the partner's link and leaves the home link alone, and at the calibrated $\delta=0.10$ it asks only that a producer's home sourcing of an input reach a tenth of its sourcing from the partner's counterpart. Proposition~\ref{prop:m-balancing} makes $\mathbf A_\omega$ independent of both CES exponents, and the sign of $\mathbf N$ turns on $\mathbf A_\omega$ alone. The sufficient condition therefore holds at every admissible $\rho$ and $\varrho$, and part~(i) holds throughout any range of curvatures on which the selected equilibrium stays interior.

We now prove (ii). Fix a producer $j$ of country $c$ and an input sector $r$, and write $D_i:=a^{(\omega)}_{ij}-a^{(\omega)}_{i\pi(j)}$ for $i\in\mathcal N_r$, the difference between the two producers' effective demands from seller $i$. The two producers share the cost share $\alpha_{rj}$, so at every fulfillment vector
\[
\alpha_{rj}\bigl[\theta^{(\omega)}_{rj}(\mathbf h)-\theta^{(\omega)}_{r\pi(j)}(\mathbf h)\bigr]
=\sum_{i\in\mathcal N_r}h_iD_i.
\]
At $\mathbf h=\mathbf 1$ the sum is the gap between the two sector budgets \eqref{eq:m-coltarget}, which Lemma~\ref{lem:m-kernel} makes the column totals of the effective demand. Each budget is $\alpha_{rj}$ less $\tau(1-\delta)$ times the producer's benchmark purchases from the partner in the sector, so
\[
\sum_{i\in\mathcal N_r}D_i
=t^{(\omega)}_{rj}-t^{(\omega)}_{r\pi(j)}
=\alpha_{rj}\,\tau(1-\delta)\bigl(\sigma^{(\omega)}_{r\pi(j)}-\sigma^{(\omega)}_{rj}\bigr).
\]
This proves \eqref{eq:m-level-full}. Differentiating \eqref{eq:m-kappa} gives $\partial\kappa^{(\omega)}_j/\partial\theta^{(\omega)}_{rj}=\alpha_{rj}(\theta^{(\omega)}_{rj})^{\rho-1}(\kappa^{(\omega)}_j)^{1-\rho}>0$ at every used sector, so the efficiency factor is strictly increasing in each retention ratio that carries positive weight. The two producers share their cost shares and their leakage, and the outer aggregator is increasing by Lemma~\ref{lem:m-outer}\textnormal{(i)}, so $\lambda^{(\omega)}_j(\mathbf 1)\ge\lambda^{(\omega)}_{\pi(j)}(\mathbf 1)$, strictly where the share ordering is strict. Proposition~\ref{prop:m-balancing} makes $\mathbf A_\omega$ independent of both CES exponents, so the ordering at full delivery holds at every admissible pair of exponents.

We next carry the ordering to the fixed point under \eqref{eq:m-buying-cond}, writing $\Gamma:=\alpha_{rj}\tau(1-\delta)(\sigma^{(\omega)}_{r\pi(j)}-\sigma^{(\omega)}_{rj})\ge0$ for the budget gap. Write $D^{\mathrm h}:=D_{k_r}$ and $D^{\mathrm f}:=D_{\pi(k_r)}$ for the differences at the home and the foreign seller of the pair. Write $T^{+}$ and $T^{-}$ for the sums of the positive and of the negative parts of $D_i$ over the third-country sellers of the sector. These four quantities satisfy $D^{\mathrm h}+D^{\mathrm f}+T^{+}-T^{-}=\Gamma$. Let $\mathcal K:=\{\mathbf h\in[c_\omega,1]^N:h_k\ge h_{\pi(k)}\text{ for every }k\text{ in }c\}$ be the ordering cone above the floor, and fix $\mathbf h\in\mathcal K$. The first condition of \eqref{eq:m-buying-cond} is $D^{\mathrm h}\ge0$, and with $h_{k_r}\ge h_{\pi(k_r)}\ge c_\omega$ it gives
\[
h_{k_r}D^{\mathrm h}+h_{\pi(k_r)}D^{\mathrm f}
\;\ge\;h_{\pi(k_r)}\bigl(D^{\mathrm h}+D^{\mathrm f}\bigr)
\;\ge\;c_\omega\bigl(D^{\mathrm h}+D^{\mathrm f}\bigr)-(1-c_\omega)\bigl(D^{\mathrm h}+D^{\mathrm f}\bigr)^{-}.
\]
Here $x^{-}$ is the negative part of $x$, and the second inequality uses $h_{\pi(k_r)}\ge c_\omega$ when the bracket is nonnegative and $h_{\pi(k_r)}\le1$ when it is negative. On the third-country sellers, $h_i\in[c_\omega,1]$ gives $\sum h_iD_i\ge c_\omega T^{+}-T^{-}$. Adding the two bounds and substituting $D^{\mathrm h}+D^{\mathrm f}=\Gamma-T^{+}+T^{-}$ gives
\[
\sum_{i\in\mathcal N_r}h_iD_i
\;\ge\;
c_\omega\,\Gamma-(1-c_\omega)\,T^{-}-(1-c_\omega)\bigl(D^{\mathrm h}+D^{\mathrm f}\bigr)^{-}
\;\ge\;
c_\omega\,\Gamma-(1-c_\omega)\bigl(T^{+}+T^{-}\bigr).
\]
The last step uses $(D^{\mathrm h}+D^{\mathrm f})^{-}=(T^{+}-T^{-}-\Gamma)^{+}\le T^{+}$. The second condition of \eqref{eq:m-buying-cond} makes the right side nonnegative, so $\theta^{(\omega)}_{rj}(\mathbf h)\ge\theta^{(\omega)}_{r\pi(j)}(\mathbf h)$ at every used sector and, by the monotonicity of the efficiency factor, $\lambda^{(\omega)}_j(\mathbf h)\ge\lambda^{(\omega)}_{\pi(j)}(\mathbf h)$. The fulfillment rule therefore carries $\mathcal K$ into the ordering cone, and the proof of Proposition~\ref{prop:m-monotone}\textnormal{(iii)} keeps every iterate of the benchmark path above the floor. The path starts at $\mathbf 1\in\mathcal K$, so every iterate and the limit lie in $\mathcal K$, which is $h_{\omega,j}\ge h_{\omega,\pi(j)}$. Where the second condition is strict at some used sector, the bound is strict at $\mathbf h_\omega$, so $\kappa^{(\omega)}_j(\mathbf h_\omega)>\kappa^{(\omega)}_{\pi(j)}(\mathbf h_\omega)$, and the fixed-point identity $h_{\omega,j}=\lambda^{(\omega)}_j(\mathbf h_\omega)$ carries the strict inequality to the deliveries. With $s_j=s_{\pi(j)}$ the shortfall $(1-h_{\omega,j})s_j$ is weakly smaller at producer $j$ than at $\pi(j)$. This proves (ii).

We last prove (iii). At $\mathbf h=\mathbf 1$ Lemma~\ref{lem:m-kernel} gives $\theta^{(\omega)}_{rj}(\mathbf 1)=\mathcal F(\mathbf A_\omega)_{rj}/\alpha_{rj}=1-\tau(1-\delta)\sigma^{(\omega)}_{rj}$, so the profile is the benchmark object of the statement. The cost shares sum to one over $\mathcal R_j$, so the efficiency factor \eqref{eq:m-kappa} is the $\boldsymbol\alpha_j$-weighted power mean of order $\rho$ of that profile, and it depends on the profile only through its weighted distribution. Since $\rho<0$, the map $\theta\mapsto\theta^{\rho}$ is strictly decreasing and strictly convex on $(0,\infty)$, so $-\theta^{\rho}$ is increasing and concave. Second-order dominance and the common cost shares therefore give $\sum_{r}\alpha_{rj}\theta^{(\omega)}_{rj}(\mathbf 1)^{\rho}\le\sum_{r}\alpha_{rj}\theta^{(\omega)}_{r\pi(j)}(\mathbf 1)^{\rho}$. Raising both sides to the power $1/\rho<0$ reverses the inequality, so $\kappa^{(\omega)}_j(\mathbf 1)\ge\kappa^{(\omega)}_{\pi(j)}(\mathbf 1)$, strictly under a nontrivial mean-preserving contraction of a non-degenerate profile by strict convexity. The outer aggregator is common to the two producers, since $\beta_j=\beta_{\pi(j)}$, and strictly increasing by Lemma~\ref{lem:m-outer}\textnormal{(i)}, so the output ratios are ordered at full delivery.

For the first curvature comparison, let producer $j$ have the constant profile $\bar\theta$ and its counterpart a non-degenerate profile with the same mean, of which the constant profile is a mean-preserving contraction. Then $\kappa^{(\omega)}_j(\mathbf 1)=\bar\theta$ at every $\rho$, while $\kappa^{(\omega)}_{\pi(j)}(\mathbf 1)$ is the power mean of order $\rho$ of the dispersed profile, which lies below $\bar\theta$, is nondecreasing in its order, and rises to $\bar\theta$ as $\rho$ rises to one. Since $G_j$ is increasing, the ratio $\lambda^{(\omega)}_j(\mathbf 1)/\lambda^{(\omega)}_{\pi(j)}(\mathbf 1)$ is nonincreasing in $\rho$ and tends to one as $\rho$ rises to one, so the advantage widens as $\rho$ falls. Its disappearance at the arithmetic mean is a limit of the formula and not a case of the model, since the maintained band holds $\rho$ below zero. For the second comparison, take constant profiles at $\theta_c>\theta_{c'}$. Their efficiency factors equal $\theta_c$ and $\theta_{c'}$ at every $\rho$, so the ratio of the two output ratios does not move with $\rho$, and the slope identity of Lemma~\ref{lem:m-outer}\textnormal{(iii)} gives
\[
\log\frac{\lambda^{(\omega)}_j(\mathbf 1)}{\lambda^{(\omega)}_{\pi(j)}(\mathbf 1)}
=\int_{\log\theta_{c'}}^{\log\theta_c}e_j(\kappa)\,d\log\kappa,
\qquad
e_j(\kappa)=\frac{1-\beta_j}{1-\beta_j+\beta_j\,\kappa^{-\varrho}}.
\]
On $\kappa\in(0,1)$ the term $\kappa^{-\varrho}$ falls as $\varrho$ falls, so the integrand rises at every $\kappa$ and the ratio widens as $\varrho$ falls. This proves (iii) and the proposition.
\end{proof}

\subsection{Shares, links, and the sign of power}
\label{app:m-shares}

This subsection bounds the vulnerabilities by the shares that a partner holds in a country's sales and purchases, and it turns those shares into counts of links. Throughout, $\tau>0$ and the pair trades in both directions. Section~\ref{subsec:power_size} of the paper defines the severed sales $X_{cc'}$ and the severed purchases $M_{cc'}$ of country $c$, its export multiplier $\ell^{\downarrow}_c$, its import multiplier $\ell^{\uparrow}_c$, and its interaction rate $\ell^{\times}_c$. Appendix~\ref{app:m-content} of the paper defines the severed sales $X_i$ of seller $i$, the severed intermediate purchases $M_j$ of producer $j$, and the partner's share $\bar\sigma_{\omega,j}$ of producer $j$'s intermediate purchases. Lemma~\ref{lem:m-rate-bounds} of the paper defines $\underline h_c$, the lowest fulfillment rate among the producers of country $c$ that sell to the partner. Write $\zeta_{\omega,i}:=\sum_{b:\,i\to b\text{ severed}}Z_{ib}/s_i=X_i/s_i$ for the severed-sales share of seller $i$, the fraction of its activity sold on severed links, so that $\xi_{\omega,i}=\tau(1-\delta)\zeta_{\omega,i}s_i$. Write $\sigma^{\mathrm H}_c:=\sum_{i\in c'}a_{ic}$ for the partner's share of the household of country $c$'s spending. Write
\begin{equation}
\bar\zeta_c:=\max_{i\in c}\zeta_{\omega,i},
\qquad
\bar\sigma_c:=\max\Bigl\{\max_{j\in c}\bar\sigma_{\omega,j},\ \sigma^{\mathrm H}_c\Bigr\},
\qquad
Q^{\uparrow}(c):=\sum_{j\in c}(1-\beta_j)s_j+F^{c},
\label{eq:m-maxshares}
\end{equation}
for the largest fraction of activity that any producer of country $c$ sells to the partner, the largest fraction of intermediate spending, or of household spending, that any buyer of country $c$ directs to the partner, and the total spending of its buyers.

\begin{lemma}[The two trade shares as weighted averages]
\label{lem:m-exposures}
\begin{equation}
\frac{X_{cc'}}{Q(c)}=\sum_{i\in c}\frac{s_i}{Q(c)}\,\zeta_{\omega,i},
\qquad
\frac{M_{cc'}}{Q(c)}=\sum_{j\in c}\frac{(1-\beta_j)s_j}{Q(c)}\,\bar\sigma_{\omega,j}+\frac{F^{c}}{Q(c)}\,\sigma^{\mathrm H}_c .
\label{eq:m-exposures}
\end{equation}
\end{lemma}

\begin{proof}
The severed sales of country $c$ are $X_{cc'}=\sum_{i\in c}X_i=\sum_{i\in c}\zeta_{\omega,i}s_i$, by the proof of Lemma~\ref{lem:m-injuries} of the paper. The severed purchases of country $c$ are the flows from the producers of country $c'$ to the producers and the household of country $c$, which by \eqref{eq:m-flows} are $\sum_{j\in c}M_j+\sum_{i\in c'}f^{c}_i=\sum_{j\in c}w_j\bar\sigma_{\omega,j}+F^{c}\sigma^{\mathrm H}_c$ with $w_j=(1-\beta_j)s_j$. Dividing by $Q(c)$ proves the lemma.
\end{proof}

\begin{proposition}[Shares bound the vulnerabilities]
\label{prop:m-shares}
\begin{enumerate}
\item[\textnormal{(i)}] The vulnerability of country $c$ satisfies
\[
\tau(1-\delta)\,\underline h_c\,\frac{X_{cc'}}{Q(c)}
\;\le\;\gamma^{(c'\notnotleftrightarrow c)}_c
\;\le\;\tau(1-\delta)\Bigl[(\ell^{\downarrow}_c+\ell^{\times}_c)\,\bar\zeta_c+(\ell^{\uparrow}_c+\ell^{\times}_c)\,\bar\sigma_c\,\frac{Q^{\uparrow}(c)}{Q(c)}\Bigr].
\]
\item[\textnormal{(ii)}] If $\ell^{\downarrow}_c+\ell^{\times}_c\le\bar r$ and $\ell^{\uparrow}_c+\ell^{\times}_c\le\bar r$, then $\mathcal H_{cc'}<0$ whenever
\begin{equation}
\bar r\,\bigl(\bar\zeta_c+\bar\sigma_c\,Q^{\uparrow}(c)/Q(c)\bigr)<\underline h_{c'}\,\frac{M_{cc'}}{Q(c')}.
\label{eq:m-sign-condition}
\end{equation}
If $\ell^{\downarrow}_{c'}+\ell^{\times}_{c'}\le\bar r$ and $\ell^{\uparrow}_{c'}+\ell^{\times}_{c'}\le\bar r$, then $\mathcal H_{cc'}>0$ whenever $\bar r(\bar\zeta_{c'}+\bar\sigma_{c'}\,Q^{\uparrow}(c')/Q(c'))<\underline h_c\,X_{cc'}/Q(c)$.
\end{enumerate}
\end{proposition}

\begin{proof}
For \textnormal{(i)}, the three parts \eqref{eq:m-three-parts} of the dollar loss of country $c$ are nonnegative, so the dollar loss is at least the first of them, the lost sales of its own producers, $Q(c)\gamma^{(c'\notnotleftrightarrow c)}_c\ge\sum_{i\in c}m^{*}_{c,i}h_{\omega,i}\xi_{\omega,i}$. The proof of Lemma~\ref{lem:m-rate-bounds} shows that this sum is at least $\underline h_c\,\tau(1-\delta)X_{cc'}$, which gives the lower bound. For the upper bound, the proof of Proposition~\ref{prop:m-exposure} of the paper writes \eqref{eq:m-exact-exposure} as $Q(c)\gamma^{(c'\notnotleftrightarrow c)}_c=\tau(1-\delta)[(\ell^{\downarrow}_c+\ell^{\times}_c)X_{cc'}+(\ell^{\uparrow}_c+\ell^{\times}_c)M_{cc'}]$. Each multiplier with its interaction rate added is nonnegative by Lemma~\ref{lem:m-onesided}\textnormal{(ii)} of the paper. Lemma~\ref{lem:m-exposures} gives $X_{cc'}/Q(c)\le\bar\zeta_c$, since the weights $s_i/Q(c)$ sum to one over the country-sectors of country $c$, and $M_{cc'}/Q(c)\le\bar\sigma_c\bigl[\sum_{j\in c}(1-\beta_j)s_j+F^{c}\bigr]/Q(c)=\bar\sigma_c\,Q^{\uparrow}(c)/Q(c)$. For \textnormal{(ii)}, the upper bound of \textnormal{(i)} for country $c$ with $\ell^{\downarrow}_c+\ell^{\times}_c\le\bar r$ and $\ell^{\uparrow}_c+\ell^{\times}_c\le\bar r$ and the lower bound of \textnormal{(i)} for country $c'$, whose severed sales are $M_{cc'}$, give $\gamma^{(c'\notnotleftrightarrow c)}_c<\gamma^{(c'\notnotleftrightarrow c)}_{c'}$ under \eqref{eq:m-sign-condition}. The inequality gives $\mathcal H_{cc'}<0$ by Lemma~\ref{lem:m-rigidity}\textnormal{(i)} and Lemma~\ref{lem:m-psi}\textnormal{(i)} of the paper. The symmetric statement exchanges the roles of the two countries. This proves the proposition.
\end{proof}

The number of nonzero entries of the network reaches these bounds through the shares, and only under a condition on how evenly the entries of a sector block are spread. Call a buyer column $\Theta$-even in sector $r$ if the positive sourcing shares $a_{ib}$, $i\in\mathcal N_r$, of buyer $b$ lie within the ratio $\Theta\ge1$ of one another. Call a seller row $\Theta$-even if the positive shares $a^{\downarrow}_{ib}=Z_{ib}/s_i$ of seller $i$ over its producer and household customers lie within the ratio $\Theta$. Write $n^{\uparrow}_{rb}$ for the number of suppliers of sector $r$ that buyer $b$ uses and $n^{\uparrow}_{rb}(c')$ for the number of them in the partner, and $n^{\downarrow}_i$ for the number of producer and household customers of seller $i$ and $n^{\downarrow}_i(k)$ for the number of them in country $k$. Write $\sigma^{(\omega)}_{rb}$ for the partner's share of buyer $b$'s purchases of sector $r$, defined for the household of country $c$ as for a producer, and $\chi_i:=(s_i-f^{o}_i)/s_i$ for the share of seller $i$'s activity sold to producers and households.

\begin{corollary}[Links under evenness]
\label{cor:m-links}
If every buyer column of country $c$ is $\Theta$-even in every sector it uses, then $\sigma^{(\omega)}_{rb}\le\Theta\,n^{\uparrow}_{rb}(c')/n^{\uparrow}_{rb}$ for every buyer $b$ of country $c$ and sector $r$, hence $\bar\sigma_c\le\Theta\max_{b,r}n^{\uparrow}_{rb}(c')/n^{\uparrow}_{rb}$. If every seller row of country $c$ is $\Theta$-even, then $\zeta_{\omega,i}\le\Theta\,\chi_i\,n^{\downarrow}_i(c')/n^{\downarrow}_i$ for every seller $i$ of country $c$, hence $\bar\zeta_c\le\Theta\max_{i\in c}n^{\downarrow}_i(c')/n^{\downarrow}_i$. The reverse inequalities for the individual shares hold with $\Theta$ replaced by $1/\Theta$. Under $\Theta$-evenness on both sides of both countries, the sign condition \eqref{eq:m-sign-condition} holds whenever
\begin{equation}
\Theta^{2}\,\bar r\,\Bigl[\max_{i\in c}\frac{n^{\downarrow}_i(c')}{n^{\downarrow}_i}+\frac{Q^{\uparrow}(c)}{Q(c)}\max_{b\in c,\,r}\frac{n^{\uparrow}_{rb}(c')}{n^{\uparrow}_{rb}}\Bigr]
<\underline h_{c'}\sum_{i\in c'}\frac{s_i}{Q(c')}\,\chi_i\,\frac{n^{\downarrow}_i(c)}{n^{\downarrow}_i} .
\label{eq:m-link-condition}
\end{equation}
\end{corollary}

\begin{proof}
For a $\Theta$-even column, the partner's suppliers contribute at most $n^{\uparrow}_{rb}(c')$ times the largest positive share and all suppliers at least $n^{\uparrow}_{rb}$ times the smallest, so $\sigma^{(\omega)}_{rb}\le\Theta n^{\uparrow}_{rb}(c')/n^{\uparrow}_{rb}$. The shares $\bar\sigma_{\omega,j}=\sum_r\alpha_{rj}\sigma^{(\omega)}_{rj}$ and $\sigma^{\mathrm H}_c=\sum_r\alpha_{rc}\sigma^{(\omega)}_{rc}$ are weighted averages of the sector shares, so the same bound holds for them. For a $\Theta$-even row the same argument applies to the shares $a^{\downarrow}_{ib}$, whose sum over its producer and household customers is $\chi_i$ by the benchmark row identity $s_i=f^{o}_i+\sum_bZ_{ib}$. The reverse inequalities exchange the roles of the largest and the smallest share. Inserting the upper bounds for country $c$ into the left side of \eqref{eq:m-sign-condition} and the lower bounds for sales from country $c'$ to country $c$ into $M_{cc'}/Q(c')=\sum_{i\in c'}(s_i/Q(c'))\zeta_{\omega,i}$, which is Lemma~\ref{lem:m-exposures} applied to the partner, gives \eqref{eq:m-link-condition}. This proves the corollary.
\end{proof}

The corollary shows how the support of the network enters power. By Proposition~\ref{prop:m-leak}\textnormal{(i)} of the paper the sanctioned network keeps the support of the benchmark network, so every buyer and every seller has the same number of links in the two networks, while the shares on those links move. The corollary bounds the partner's share of a buyer's spending, or of a seller's activity, by the fraction of its links that point to the partner at the cost of the factor $\Theta$. On the network of \eqref{eq:m-home-sourcing} of the paper with equal flows, in which country $c$ has $n_c$ trading partners, the factor is one on every household link, and the corollary returns the comparison $\mathcal H_{cc'}=\psi(n_{c'}/n_c)$ of the two numbers of relationships. On a general network the count of links enters the sign of power through the share it bounds, and the bound contains the evenness factor twice, once on each side, and the range of the export multipliers and the import multipliers once.

\section{Calibrating the Unrecovered Share \texorpdfstring{$\tau$}{tau} from Disruption Episodes}
\label{app:m-tau}

Two disruption parameters stand outside the benchmark input--output table, the unrecovered share $\tau\in[0,1]$ and the barred-link attenuation $\delta$, and observed disruption episodes bracket them. The two CES exponents $\rho$ and $\varrho$ take the short-run complementarity values of Section~\ref{subsec:vulnerability_estimates} of the paper, whose sources Table~\ref{tab:calibration} lists.\footnote{A single post-sanction bilateral flow does not structurally identify either disruption parameter. Identification would require many sanctions episodes and an econometric procedure that estimates the distributions of $\widehat\tau$ and $\widehat\delta$, since observed flows reflect stochastic adjustment and are not deterministic functions of the two parameters. The episodes therefore discipline plausible magnitudes without identifying the parameters.}

The unrecovered share $\tau$ is the fraction of displaced mass removed from the row and column targets, and it is not observed directly. We proxy it by one minus the share of displaced trade that reappears with unbarred partners, measured after residual barred trade and demand destruction have been separated out. Consider a sanctions episode that bars the sector-$r$ flows from country $c'$ to country $c$, and write $X^{0}>0$ for the benchmark flow on the barred links. For a single buyer $j$ this flow is
\[
X^{0}_j
=
w_j\!\!\sum_{i\in\mathcal N_r:\,i\to j\text{ severed}}\!\!a_{ij}.
\]
Here the benchmark severed share is scaled by the buyer's intermediate spending, and an episode covering several buyers sums them, $X^{0}=\sum_j X^{0}_j$. The targets of Section~\ref{subsec:model_rerouting} of the paper divide $X^{0}$ into three accounting components,
\[
X^0
=
\underbrace{\delta X^0}_{\text{pre-balancing retained flow}}
+
\underbrace{(1-\tau)(1-\delta)X^0}_{\text{recovered}}
+
\underbrace{\tau(1-\delta)X^0}_{\text{unrecovered}}.
\]
The recovered mass is therefore $M^{\mathrm{rec}}=(1-\tau)(1-\delta)X^{0}$, and solving the target identity for the unrecovered share gives
\begin{equation}
\tau
\;=\;
\frac{(1-\delta)X^{0}-M^{\mathrm{rec}}}{(1-\delta)X^{0}}
\;=\;
1-\frac{M^{\mathrm{rec}}}{(1-\delta)\,X^{0}}.
\label{eq:tau-estimand}
\end{equation}
The denominator carries the factor $1-\delta$ because only that fraction of the flow is displaced in the disrupted matrix. The identity concerns the construction of the targets and not the allocation across links, so the observed proxy does not invert $\tau$ exactly. The identity is the same on the two sides of the market, so buyer-side and seller-side estimates should be comparable, and their difference is a calibration-consistency diagnostic rather than an overidentifying test.

An observed bilateral change cannot be inserted into \eqref{eq:tau-estimand} directly, because supplier switching is only one of the adjustments it combines.\footnote{Trade can also change its physical route while the partners and the goods stay the same. That logistical adjustment leaves every entry of $\mathbf A$ unchanged and does not affect $\tau$, and it matters for $\widehat\delta$, which we measure with mirror statistics on re-exports. We exclude the episodes dominated by transshipment from the calibration of $\tau$.} Supplier switching raises the purchases from unbarred suppliers and the barred seller's sales to unbarred buyers, and that increase is the observable proxy for the recovered target mass. Residual barred trade or evasion keeps the barred relationship active at $X^{1}>0$, directly or through transshipment, re-invoicing, or third-country blending. The residual-flow share $X^{1}/X^{0}$ disciplines the attenuation without being the primitive $\delta$, so we treat $\widehat\delta$ as a reduced-form calibration input and measure recovery against the displaced flow $(1-\widehat\delta)X^{0}$. Demand destruction lowers the buyer's total sector-$r$ intake rather than shifting it between suppliers, which is not a failure of supplier switching and must not enter $\tau$. We therefore measure recovery against the no-disruption intake rather than the lower realized intake.

\begin{definition}[Calibration of $\tau$ from a sanctions episode]
\label{def:tau-estimator}
Fix a sanctions episode that bars the sector-$r$ flows from country $c'$ to country $c$. Let $X^{0}>0$ be the benchmark flow on the barred links, and let $\widehat\delta\in[0,1)$ be the episode's residual-flow share used to discipline the barred-link attenuation. Write $X^{t}_{rc''c}$ for the sector-$r$ flow from seller country $c''$ to buyer country $c$, at the benchmark for $t=0$ and after the severance for $t=1$. Let $\widetilde X^{1}_{rc''c}$ denote the post-severance flow from an unbarred seller $c''$ after the buyer's total sector-$r$ intake $\nu_{rc}$ is adjusted to its estimated no-disruption level $\widetilde\nu_{rc}$. A proportional adjustment does this by $\widetilde X^{1}_{rc''c}=(\widetilde\nu_{rc}/\nu_{rc})X^{1}_{rc''c}$ when $\nu_{rc}>0$. The demand-adjusted increase in purchases from unbarred suppliers is $\widehat{M}^{\mathrm{rec}}=\sum_{c''\neq c'}\bigl(\widetilde X^{1}_{rc''c}-X^{0}_{rc''c}\bigr)$, and the projected calibration proxy is
\begin{equation}
\begin{aligned}
\widehat\tau^{\,\mathrm{raw}}
&:=
1-\frac{\widehat{M}^{\mathrm{rec}}}{(1-\widehat\delta)\,X^{0}},\\
\widehat\tau
&:=
\Pi_{[0,1]}\!\left(\widehat\tau^{\,\mathrm{raw}}\right)
=
\min\!\left\{1,\max\!\left\{0,\widehat\tau^{\,\mathrm{raw}}\right\}\right\}.
\end{aligned}
\label{eq:tau-hat}
\end{equation}
The seller-side estimate is defined symmetrically.
\end{definition}

A panel design may estimate the no-disruption intake directly instead of adjusting proportionally, and recovery is measured at that counterfactual intake in either case. The projection keeps $\widehat\tau$ in $[0,1]$, and whenever it binds we also report $\widehat\tau^{\,\mathrm{raw}}$, since a raw value outside $[0,1]$ means that the episode's accounting proxy is incompatible with the model identity.

Separating residual barred trade matters most where evasion is large. Roughly $90\%$ of Iran's crude and condensate exports reach China through ship-to-ship transfers and relabeling as Malaysian-origin crude, and China's recorded imports from Malaysia rose from about $0.2$ to $1.1$ million barrels per day between 2019 and 2023, beyond Malaysia's own production \citep{oa-crs2024iran}. The Iranian episode points to weak attenuation and hence a high $\delta$, not to a high unrecovered share, which is why the denominator of \eqref{eq:tau-hat} is the displaced flow $(1-\widehat\delta)X^{0}$ rather than the raw severed flow.

Table~\ref{tab:tau-calibration} orders the commodity classes by how readily their trade can be redirected, and the fungible goods receive the lower $\tau$. Crude oil is the only class for which we compute the unrecovered share directly. In the year after the EU embargo and the G7 price cap, Russia's seaborne crude exports to coalition countries fell by about $91\%$, while exports to non-coalition buyers rose enough to leave total volume almost unchanged. Output stayed within roughly $0.2$ million barrels per day of its pre-invasion level \citep{oa-crea2023tracker,oa-eia2025russia,oa-iea2023oil}. Coalition purchases fell by roughly $2.0$ million barrels per day and about $1.8$ million reappeared in Asia and Oceania, so the recovered share $\widehat{M}^{\mathrm{rec}}/[(1-\widehat\delta)X^{0}]$ is about $0.9$ and $\widehat\tau\approx0.05$--$0.10$. Gas takes an intermediate bracket because regasification and pipeline capacity limited Europe's substitution toward liquefied natural gas \citep{oa-iea2023gas}. Differentiated inputs take the highest bracket, and the war-related disruptions to neon and palladium supply and the export-control evidence illustrate their short-run sourcing constraints \citep{oa-foreignpolicy2022neon,oa-chupilkin2023eurasian}.

\begin{table}[!htbp]
\centering
\small
\begin{adjustbox}{max width=\textwidth}
\begin{tabular}{@{}p{0.30\textwidth}p{0.26\textwidth}cc@{}}
\toprule
\textbf{Commodity class} & \textbf{Calibration evidence} &
\textbf{Evidence on the recovered share} & \textbf{Calibrated $\tau$ range} \\
\midrule
Fungible: crude oil & Russia 2022 (seaborne crude) & $\approx 0.90$--$0.95$ & $0.05$--$0.10$ \\
Fungible: gas (LNG-substitutable) & Russia 2022 (EU gas) & Partial, not directly estimated & $0.20$--$0.40$ \\
Specific: machinery, specialized inputs & Russia/Ukraine 2022 (neon, palladium, dual-use) & Minority, not directly estimated & $0.50$--$0.80$ \\
\bottomrule
\end{tabular}
\end{adjustbox}
\caption{Calibration of the unrecovered share $\tau$ from disruption episodes. The crude-oil row applies Definition~\ref{def:tau-estimator} to the displaced and recovered flows. The gas and specific-input rows are ordered calibration brackets, since their sources do not separately identify the demand-adjusted recovery term and the displaced base that the estimator requires. The sensitivity analysis varies $\tau$ over the interval that spans these ranges.}
\label{tab:tau-calibration}
\end{table}

The baseline applies one unrecovered share, $\tau=0.30$, to every sector, a value inside the gas bracket of Table~\ref{tab:tau-calibration}, above crude oil and below the specific inputs. The choice is not a formal mixture over sectors, since an input--output sector can combine fungible and specific intermediates, and the sensitivity range $\tau\in[0.05,0.80]$ spans both ends of the commodity brackets.

The same episodes bracket the barred-link attenuation by commodity class. Iranian oil illustrates weak attenuation for fungible goods and the export controls stronger attenuation for specific manufactures, while the Russian crude calculation treats evasion back into coalition markets as negligible and therefore calibrates $\tau$ rather than the attenuation. The grouped robustness exercise of Section~\ref{sec:robustness} assigns the fungible class $\delta=1/3$ and the specific class $\delta=0.05$, the baseline $\delta=0.10$ lies between them, and the sensitivity analysis covers $\delta\in[0.05,0.60]$.

\begin{table}[!htbp]
\centering
\small
\begin{adjustbox}{max width=\textwidth}
\begin{tabular}{@{}lcp{0.62\textwidth}@{}}
\toprule
Parameter & Baseline & Evidence \\
\midrule
Unrecovered share $\tau$ & $0.30$ &
Intermediate common calibration from the commodity brackets in
Table~\ref{tab:tau-calibration} \\
CES exponents $\rho,\varrho$ & $-1,-1$ &
Short-run complementarity of intermediate inputs
\citep{oa-atkesonkehoe1999,oa-boehmflaaen2019,oa-barrotsauvagnat2016,oa-bachmann2022whatif} \\
Barred-link attenuation $\delta$ & $0.10$ &
Residual barred trade and evasion evidence discussed in this section \\
\bottomrule
\end{tabular}
\end{adjustbox}
\caption{Baseline calibration. These values apply unless a robustness exercise states otherwise.}
\label{tab:calibration}
\end{table}

\section{Sensitivity, Computation, and Empirical Robustness}
\label{app:m-empirics}

The benchmark network behind Sections~\ref{sec:statistics}, \ref{sec:power_size}, and~\ref{sec:power_balance} of the paper is the 2022 OECD inter-country input--output table read onto the active set of the paper's Section~\ref{subsec:model_benchmark}. We record its construction first. We read the purchases, the leakages, the sourcing shares, and the seller residuals off the retained intermediate block $\mathbf Z$ in that order, so the benchmark identity \eqref{eq:m-benchmark} holds by construction and every residual is nonnegative. We read the final-demand block by purchasing country into the household matrix $\mathbb A$ and the budgets $\mathbf F$, counting household consumption, the consumption of nonprofit institutions, government consumption, and gross fixed capital formation as the household's purchases. The rest-of-world aggregate is an accounting closure and not a separate equilibrium. The purchases of the rest of the world sector, inventory changes, residents' purchases abroad, and the statistical discrepancy stay in the residual final demand $\mathbf f^{o}$ rather than becoming network links, and the residual is clipped at zero where the accounts make an entry negative.

No entry of the household block is negative in the 2022 trade table. In $91$ of the $4{,}000$ rows, however, the four household categories sum to slightly more than total final demand, because the residual categories are negative there, and we scale the household purchases back to that total in those rows, which moves $206$ million dollars in all. The household budgets so constructed account for $96.3\%$ of world final demand and the residual final demand for the remaining $3.7\%$. Of the household purchases $9.1\%$ cross a border, and these cross-border final purchases are $36\%$ of all cross-border flows in the trade table, the rest being intermediate trade.

Seventy active country-sectors buy no intermediate inputs at all, so that their leakage is one. These seventy country-sectors are the industry of households as employers in the seventy economies where that industry is active, and they account for $0.1\%$ of world activity. They sell only to the household of their own country, so no trade severance reaches them and their output ratio is one in every scenario. The conditions of Section~\ref{subsec:model_benchmark} of the paper, under which every producer buys intermediate inputs, describe the other $3{,}879$ of the $3{,}949$ active country-sectors.

Two buyer columns record raw intermediate spending above their own activity, which taken literally would imply $\beta_j<0$. They are Icelandic pharmaceuticals, at a raw share of $1.19$, and Cypriot air transport, at $1.05$, whose intermediate purchases of $67$ and $75$ million dollars are each below $0.0001\%$ of world activity. We cap both columns at $1-\beta_j=0.98$. The cap reaches only accounts that are internally inconsistent, leaves the sourcing shares unchanged, and moves no statistic at the precision the paper and this appendix report. Through the two capped columns it sets the $1$-norm of the sanctioned propagation matrix, $\|\mathbf A_\omega(\mathbf I-\boldsymbol\beta)\|_1$, at $0.98$ in every scenario. Unless a subsection states otherwise, the calibration throughout this section is the baseline $\tau=0.30$, $\rho=\varrho=-1$, and $\delta=0.10$ of Table~\ref{tab:calibration}, and every calculation takes the output ratio \eqref{eq:m-lambda-rule} as the fulfillment rule.

In the order of the paper, Sections~\ref{oa:sensitivity} to~\ref{oa:asymmetry-distribution} stand behind the vulnerabilities and the Hirschman matrix of Section~\ref{sec:statistics} of the paper, and Section~\ref{sec:robustness} tests the distribution of power across specifications, years, and table constructions. Sections~\ref{oa:fiber} to~\ref{oa:rates-content} stand behind the comparisons of power with size of Section~\ref{sec:power_size}, and Sections~\ref{oa:exact-decomposition} to~\ref{app:structural_scores_repro} behind the exact decomposition, its three ratios on the twenty largest bilateral deficits, its form when the unrecovered share differs across the severed links, and the trade-balance regressions of Section~\ref{sec:power_balance}.

\subsection{Sensitivity to the unrecovered share and the CES exponents}
\label{oa:sensitivity}

\begin{figure}[!htbp]
\centering
\includegraphics[width=0.98\linewidth]{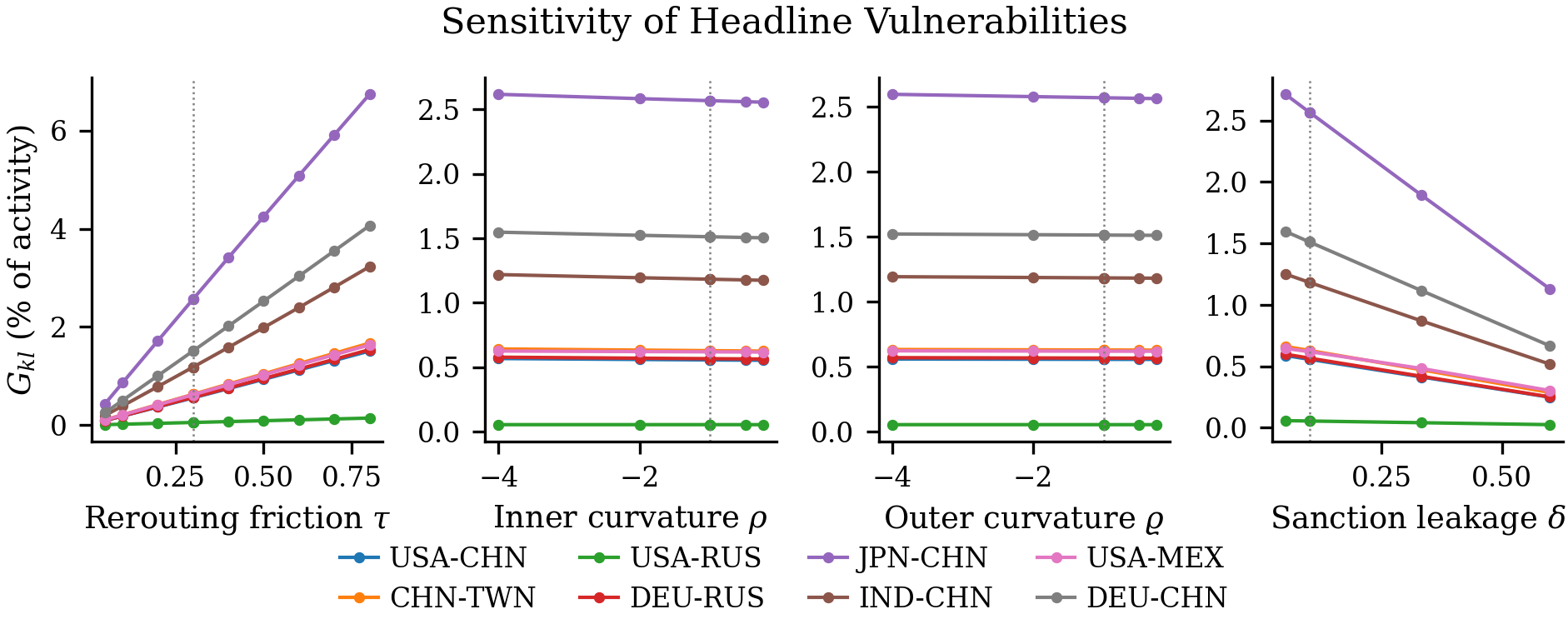}
\caption{Sensitivity of the headline bilateral vulnerabilities. Each panel varies one parameter and holds the others at $\tau=0.30$, $\rho=-1$, $\varrho=-1$, and $\delta=0.10$. From left to right, the panels vary the unrecovered share $\tau$, the inner CES exponent $\rho$, the outer CES exponent $\varrho$, and the barred-link attenuation $\delta$. Dotted vertical lines mark the baseline values.}
\label{fig:s4_sensitivity}
\end{figure}

Figure~\ref{fig:s4_sensitivity} varies one parameter at a time and traces the vulnerabilities of the headline pairs of Table~\ref{tab:s4_headline} of the paper. Its first two panels cover the calibration range of the unrecovered share, $\tau\in[0.05,0.80]$, and the maintained range of the inner CES exponent, $\rho\in[-4,-0.25]$. The unrecovered share moves the losses at first order, since it is the fraction of displaced trade removed from the budgets and the capacities. Raising $\tau$ from $0.10$ to $0.60$ raises the American loss from the United States--China severance about sixfold, from $0.18$ to $1.12\%$ of activity, and the other headline losses rise almost linearly as well. Every loss vanishes at $\tau=0$ (Lemma~\ref{lem:m-frictionless} of the paper). Before propagation a higher $\tau$ lowers every affected budget and capacity in proportion to the displaced mass, so the slope of each curve is, to a first approximation, the pair's severed mass times its network multiplier. The ordering of the curves does not change and no headline asymmetry reverses, so the comparisons of power in Section~\ref{subsec:hirschman_matrix} of the paper do not rest on the baseline choice of $\tau$.

The headline losses move little with the inner exponent $\rho$ over $[-4,-0.25]$, a property of the calibration and not of the model. At the baseline every affected buyer-sector keeps at least $1-\tau(1-\delta)=0.73$ of its benchmark budget, though that first-round floor does not bound the retention realized after supplier shortfalls and does not by itself account for the weak response to curvature. A larger displaced fraction $\tau(1-\delta)$ deepens the disruption, moves the losses toward their values at the Leontief limit, and makes them more sensitive to $\rho$. Carried beyond the maintained interval to $\rho\in\{-8,-16,-32\}$, the losses of the United States--China severance rise toward their values at that limit while the fulfillment iteration still converges (Section~\ref{app:m-proofs}).

The third panel varies the outer exponent $\varrho$, to which the headline losses are less sensitive still. Over $\varrho\in[-4,-0.25]$ the American loss from the United States--China severance stays between $0.555$ and $0.558\%$ of activity. At bundle degradations this small the intermediate output elasticity $e_j(\kappa)$ of \eqref{eq:m-shadow} remains close to its benchmark value $1-\beta_j$, so the outer aggregator agrees with its constant-share approximation to first order. The outer exponent matters where a bundle degrades severely. Belarus loses $13.9\%$ of its activity to the severance with Russia at $\varrho=-0.25$, $14.2\%$ at $-1$, $15.0\%$ at $-3$, and $17.6\%$ at $-9$, so at an outer elasticity of one tenth its loss is about a quarter above the baseline. Pass-through steepens there because the intermediate output elasticity approaches one, while the convergence of the iteration changes little, with a terminal ratio near $0.60$ throughout. Table~\ref{tab:rob-master} reports the full-system calculations at $\varrho\in\{-9,-3\}$. The last panel varies the barred-link attenuation, and the losses scale almost in proportion to the displaced share $1-\delta$.

\subsection{The two channels of the loss across countries}
\label{oa:loss-anatomy}

The lost-market share of a country's loss has a median of $47\%$ across the eighty economies and runs from $26$ to $82\%$, and what moves it is what a country sells. The share comes from carrying the anatomy of Proposition~\ref{prop:m-anatomy} of the paper to the country vulnerabilities. At the selected equilibrium the proposition writes the loss vector as two nonnegative direct losses carried through the operator $\mathbf R^{*}_\omega$ of \eqref{eq:m-content} of the paper. The direct losses are the fulfillment shortfalls $(\mathbf 1-\mathbf h_\omega)\circ\mathbf s$ of the productivity channel and the unrecovered sales $\mathbf h_\omega\circ\boldsymbol\xi_\omega$ of the lost-market channel. Summing each channel over a country's sectors and dividing by its benchmark size decomposes the vulnerability of Definition~\ref{def:country_vulnerability},
\begin{equation}
\gamma^{(\omega)}_c
=\underbrace{\frac{\mathbbm 1_c\mathbf R^{*}_\omega\bigl[(\mathbf 1-\mathbf h_\omega)\circ\mathbf s\bigr]}{Q(c)}}_{\text{productivity channel}}
+\underbrace{\frac{\mathbbm 1_c\mathbf R^{*}_\omega\bigl[\mathbf h_\omega\circ\boldsymbol\xi_\omega\bigr]}{Q(c)}}_{\text{lost-market channel}}.
\label{eq:oa-channels}
\end{equation}
The lost-sales vector \eqref{eq:m-lost-sales} splits by the kind of customer lost, $\boldsymbol\xi_\omega=\boldsymbol\xi^{I}_\omega+\boldsymbol\xi^{F}_\omega$, into the unrecovered sales to the partner's producers and the unrecovered sales to its household,
\[
\xi^{I}_{\omega,i}:=\tau(1-\delta)\!\!\sum_{j:\,i\to j\text{ severed}}\!\!Z_{ij},
\qquad
\xi^{F}_{\omega,i}:=\tau(1-\delta)\!\!\sum_{c:\,i\to c\text{ severed}}\!\!f^{c}_i.
\]
Substituting the split into \eqref{eq:oa-channels} divides the lost-market channel into an intermediate part and a final part. The two channels, and the final part within the second, are accounting components of the loss at the solved equilibrium, and none is the change in the loss that would follow from removing its direct loss and solving the economy again. Neither channel is confined to the two countries the sanction names. A third country's producers fill a smaller fraction of the effective demand they face when their own suppliers under-deliver, which enters the productivity channel at their own country-sectors. Its sellers lose the orders of the pair's export sectors without any severed sales of their own, since the operator carries the export sectors' lost sales upstream to every supplier behind them.

\begin{figure}[!htbp]
\centering
\includegraphics[width=0.62\linewidth]{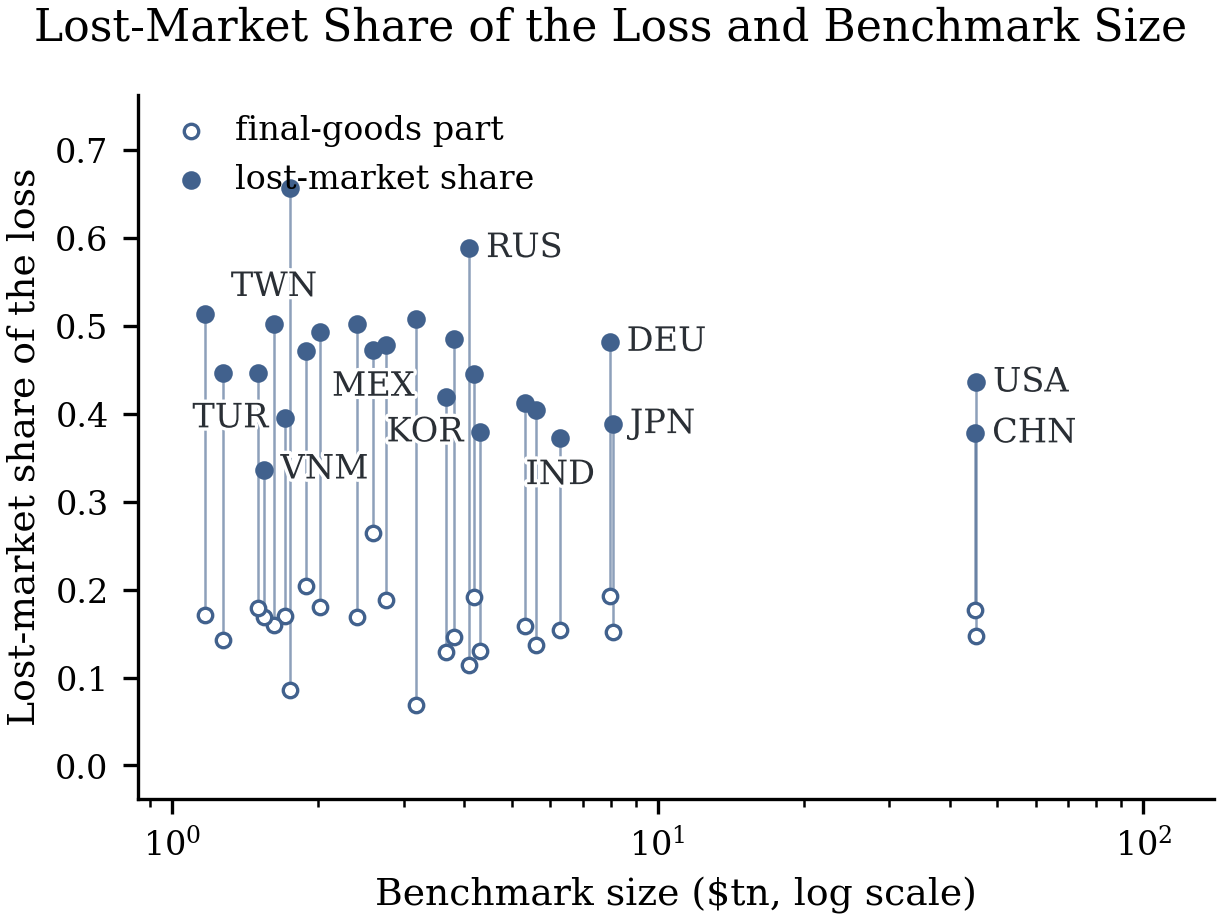}
\caption{The lost-market share of each country's loss, averaged across its bilateral severances, against its benchmark size. For country $c$ the share at a severance is the lost-market channel of \eqref{eq:oa-channels} divided by the country's vulnerability, and the marked part is the share due to severed final sales. The remainder is the productivity channel of Proposition~\ref{prop:m-anatomy}. Baseline calibration $\tau=0.30$, $\rho=-1$, $\varrho=-1$, and $\delta=0.10$.}
\label{fig:s6_decomposition}
\end{figure}

Figure~\ref{fig:s6_decomposition} reports the division country by country. For each economy the figure averages the lost-market share of its loss across its bilateral severances and marks the part due to severed final sales. The share is nearly unrelated to size, with a correlation of $-0.06$ against log size, and the two largest economies lie near the middle, the United States at $44\%$ and China at $38\%$. The commodity-exporting countries lose through their customers, Saudi Arabia at $66\%$ and Russia at $59\%$, with a small final-goods part, since what they sell abroad is inputs. Mexico's share is near the median at $47\%$, but $26$ points of it come from severed final sales, the largest final-goods part among the large economies, because much of what Mexico sells abroad is consumer goods bound for American households. Vietnam has the smallest share among the large economies, at $34\%$, and loses mainly through the input bundles its own producers can no longer complete. Weighted by dollars over all $3{,}160$ severances, $44\%$ of the world's activity loss runs through the lost-market channel, $16$ points of it through severed final sales, and $56\%$ through the productivity channel. Equal vulnerabilities can therefore hide losses of different kinds, and the difference matters for policy because the two channels reach different parts of an economy.

\subsection{The recomposition of the trade matrix under a severance}
\label{subsec:recomposition}

\begin{figure}[!htbp]
\centering
\includegraphics[width=0.58\linewidth]{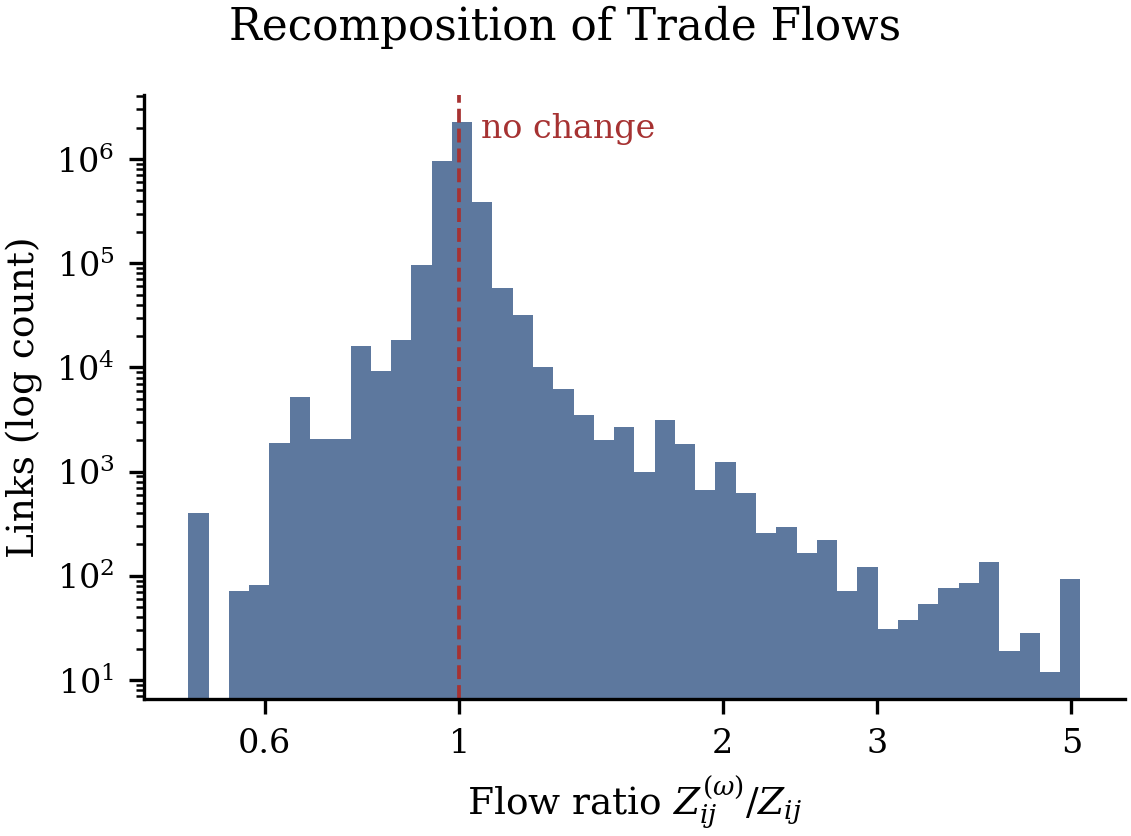}
\caption{The ratio $Z^{*}_{\omega,ij}/Z_{ij}$ of realized to benchmark flow over the unbarred links whose flow moves, pooled across forty randomly sampled bilateral severances. Link counts and the ratio are on log scales, and the dashed vertical line marks a ratio of one. Baseline calibration $\tau=0.30$, $\rho=-1$, $\varrho=-1$, $\delta=0.10$.}
\label{fig:s4_recomposition}
\end{figure}

A severance recomposes the benchmark flow matrix $\mathbf Z$ of \eqref{eq:m-flows} rather than scaling it down, since buyers redirect their effective demand to their retained suppliers and barred sellers place their capacity with their retained buyers. Under the United States--China severance $29\%$ of the benchmark links carry more than their benchmark flow. Among the links carrying at least a million dollars the largest rise is $1.6$ times, on the link that supplies American households with Russian textiles in place of Chinese ones. Figure~\ref{fig:s4_recomposition} pools forty randomly sampled bilateral severances and plots the ratio $Z^{*}_{\omega,ij}/Z_{ij}$ of realized to benchmark flow over the unbarred links whose flow moves.

The flow that unbarred intermediate links gain is concentrated on a few of them. Across the same forty severances the largest thousandth of the gaining links carries $58\%$ of the added flow and the largest hundredth $85\%$. By selling country, the five countries that gain most take $66\%$ of what a severance adds, and a country's mean share of the added flow correlates with its log benchmark size at $+0.42$, the United States and China each taking $5.1\%$ on average. Displaced trade therefore goes to the few relationships already large enough to take it rather than spreading across the many partners that could take it in principle.

The rationed exchange first cuts and then partly restores the flow on a barred link, and its two margins bound what it restores. The sector aggregation operator $\mathcal F$ of Definition~\ref{def:m-aggregate} of the paper, with $\mathbf A$ read as $\mathbb A$ on a household column, splits the mass of buyer $j$'s sector-$r$ column into the links the sanction severs and the links it leaves alone, and substituting the split into the column target \eqref{eq:m-coltarget} gives
\begin{equation}
t^{(\omega)}_{rj}=\underbrace{\sum_{i\in\mathcal N_r:\,i\to j\text{ not severed}}a_{ij}}_{\text{unbarred}}
+\,\big[\delta+(1-\tau)(1-\delta)\big]\!\!\sum_{i\in\mathcal N_r:\,i\to j\text{ severed}}\!\!a_{ij}.
\label{eq:m-coltarget-split}
\end{equation}
Of a severed link's coefficient the sanction leaves $\delta$ in place and displaces the remaining $1-\delta$, of which the exchange brings back the share $1-\tau$, so the bracket is $1-\tau(1-\delta)$, one at $\tau=0$ and $\delta$ at $\tau=1$.\footnote{Equation~\eqref{eq:m-coltarget-split} is an identity for the target total and not a rule for where the spending goes. Subject to the row targets, the rationed exchange spreads the retained total across the whole column, and it may leave a barred link carrying more than the bracket suggests or an unbarred link carrying less.} The biproportional form \eqref{eq:m-reroute} meets this target and the seller's capacity with one scaling per seller and one per buyer and sector. On each connected component of the support graph the rescaling $(u,v)\mapsto(\chi u,\chi^{-1}v)$ leaves $\mathbf A_\omega$ unchanged, so only the products $u_iv_{rj}$ are invariant, and whether an individual scaling lies above or below one is a property of the gauge.\footnote{Imposing one margin alone freezes the other side. Setting every seller scaling to one leaves the demand-oriented Leontief-style column rescaling, and setting every buyer scaling to one leaves the supply-oriented Ghosh-style row rescaling. Both are exact one-sided rescalings of $\check{\mathbf A}$ and not nests of the full Leontief or Ghosh quantity systems.} We call the product $u_iv_{rj}$ the accommodation multiplier of the link. On a block with no severed link the exchange is the identity and both scalings are one, so the multiplier is defined on every benchmark link, and by \eqref{eq:m-reroute} of the paper it is the single factor by which the exchange scales the disrupted coefficient \eqref{eq:m-disrupted}, $a^{(\omega)}_{ij}=u_i\,v_{rj}\,\check a_{ij}$. On an unsevered link $\check a_{ij}=a_{ij}$, so the multiplier is the whole of what the exchange does to it, and on a severed link $\check a_{ij}=\delta a_{ij}$, so the multiplier acts on top of the attenuation the sanction imposed.

A buyer that retains other suppliers of the sector meets its budget off the barred link, which therefore stays near the attenuation the sanction imposed. A buyer whose sector block holds no other supplier must spend its retained budget on the barred link, and a seller that keeps a single customer must likewise place its retained capacity there. On a barred link the accommodation multiplier gives $a^{(\omega)}_{ij}=\delta\,u_iv_{rj}\,a_{ij}$, so the realized attenuation of the link, the fraction of its benchmark coefficient that the effective demand keeps, is
\begin{equation}
\delta^{(\omega)}_{ij}
\;:=\;
\frac{a^{(\omega)}_{ij}}{a_{ij}}
\;=\;
\delta\,u_iv_{rj}.
\label{eq:m-realized-attenuation}
\end{equation}
The ratio is taken before fulfillment and activity scale the flow, so it records what the rationed exchange alone does to the attenuation the sanction imposed. Across the computed scenarios the realized attenuation stays within $1\%$ of the imposed one on $64\%$ of the barred links and within $10\%$ on $96\%$. The few links that depart from it by a sizable margin carry most of the trade that survives on barred links.

We set this dispersion against a sanction whose exceptions are on record. For the 2022 Russia measures the scope of the sanction, the exemptions written into it, and the trade that followed are recorded relationship by relationship, and the oil-import derogations of the European regulation accommodated markets with limited alternatives to Russian supply \citep{oa-eu2022reg879}. We represent the measures as a sector-partial severance between Russia and the thirty-seven sanctioning economies on the pre-invasion 2021 trade table. The four relationships holding the regulation's principal crude-oil derogations then rank first, fifth, sixth, and seventh of twenty-six European and aligned energy markets by benchmark-flow-weighted realized attenuation. Under a uniform random ranking of the twenty-six markets, the probability that all four fall within the top seven is $\binom{7}{4}/\binom{26}{4}=35/14{,}950\approx0.0023$. The comparison is descriptive and not causal, since the derogations were neither randomly assigned nor independent of pre-war volumes.

\subsection{Flow size and the first-order incidence of a severance}
\label{oa:flowsize}

The local loss expansion \eqref{eq:m-local-loss} ties the direct incidence of a severance to the size of the severed flow. Its leading term is the mean shortfall in input intensity, and when a single link is barred that shortfall is the part of the link's benchmark share that the rationed exchange does not restore.

\begin{lemma}[First-order incidence of a severed link]
\label{lem:m-flowsize}
Fix a buyer $j$ and a sanction $\omega$ that bars the single link $(i,j)$, with the seller $i$ in sector $r$. The benchmark coefficient $a_{ij}$ is the link's share of buyer $j$'s intermediate purchases. As $a_{ij}\to0$ with the benchmark sector share $\alpha_{rj}$ bounded away from zero, the direct loss of buyer $j$'s output ratio at full supplier fulfillment satisfies
\begin{equation}
1-G_j\bigl(\kappa^{(\omega)}_j(\mathbf 1)\bigr)
\;=\;
(1-\beta_j)\,\tau(1-\delta)\,a_{ij}
\;+\;
O\bigl(a_{ij}^{2}\bigr).
\label{eq:m-flowsize}
\end{equation}
\end{lemma}

\begin{proof}
Fix the buyer $j$ and the barred link $(i,j)$, with $i\in\mathcal N_r$, and evaluate at full supplier fulfillment $\mathbf h=\mathbf 1$. We first compute the mean shortfall $\bar d_j$ and then substitute it into the local loss expansion.

By Lemma~\ref{lem:m-kernel} the retention ratio at full fulfillment is $\theta^{(\omega)}_{r'j}(\mathbf 1)=\mathcal F(\mathbf A_\omega)_{r'j}/\alpha_{r'j}$ in every used sector $r'$, so, since $\sum_{r'}\alpha_{r'j}=1$,
\[
\bar d_j
=\sum_{r'\in\mathcal R_j}\alpha_{r'j}
\bigl(1-\theta^{(\omega)}_{r'j}(\mathbf 1)\bigr)
=\sum_{r'\in\mathcal R_j}
\bigl(\alpha_{r'j}-\mathcal F(\mathbf A_\omega)_{r'j}\bigr)
=1-\sum_{r'\in\mathcal R_j}\mathcal F(\mathbf A_\omega)_{r'j}.
\]
The mean shortfall is therefore the share of buyer $j$'s benchmark intermediate purchases that the rationed exchange does not restore, whatever the sectoral pattern of the restoration. The sanction leaves $\delta Z_{ij}$ on the barred link and displaces $(1-\delta)Z_{ij}$, of which the fraction $\tau$ is not recovered. No other link is barred and the budgets bind at $\mathbf h=\mathbf 1$, so
\[
\sum_{r'\in\mathcal R_j}\mathcal F(\mathbf A_\omega)_{r'j}
=1-\frac{\tau(1-\delta)Z_{ij}}{\sum_{i'}Z_{i'j}},
\qquad\text{hence}\qquad
\bar d_j=\tau(1-\delta)\,a_{ij}.
\]

Substituting $\bar d_j=\tau(1-\delta)a_{ij}$ into \eqref{eq:m-local-loss} gives
\[
1-G_j\bigl(\kappa^{(\omega)}_j(\mathbf 1)\bigr)
=(1-\beta_j)\tau(1-\delta)a_{ij}
+\frac{(1-\beta_j)(1-\rho)}{2}V_j
+\frac{\beta_j(1-\beta_j)(1-\varrho)}{2}\bar d_j^{2}
+O\bigl(\|\mathbf d_j\|_\infty^{3}\bigr).
\]
The single barred link places the whole shortfall in sector $r$, so $d_{r'j}=0$ for $r'\neq r$ and $\alpha_{rj}d_{rj}=\bar d_j$. The composition term is then $V_j=\alpha_{rj}d_{rj}^{2}-\bar d_j^{2}=\bar d_j^{2}\bigl(\alpha_{rj}^{-1}-1\bigr)$, which is $O(a_{ij}^{2})$ with $\alpha_{rj}$ bounded away from zero. The outer-curvature term is $O(\bar d_j^{2})=O(a_{ij}^{2})$, and $\|\mathbf d_j\|_\infty=d_{rj}=\bar d_j/\alpha_{rj}$ is $O(a_{ij})$, so the remainder is $O(a_{ij}^{3})$. Collecting the three terms gives \eqref{eq:m-flowsize}. This proves the lemma.
\end{proof}

Lemma~\ref{lem:m-flowsize} concerns the buyer's direct response and not the country equilibrium. Since $Z_{ij}=(1-\beta_j)s_ja_{ij}$, the leading term of \eqref{eq:m-flowsize} is $\tau(1-\delta)Z_{ij}/s_j$, so at fixed disruption parameters two small links of equal dollar value lower their buyers' output ratios in inverse proportion to the buyers' benchmark activities. The coefficient $(1-\beta_j)\tau(1-\delta)$ depends on the buyer's leakage and the two disruption parameters and on nothing in the network. The network enters through the composition term, which carries the factor $\alpha_{rj}^{-1}$, through the feedback of supplier shortfalls at the equilibrium, and through the propagation of both to the buyer's customers. A link that supplies its buyer's whole sector, $a_{ij}=\alpha_{rj}$, gives $V_j=\tau^{2}(1-\delta)^{2}\alpha_{rj}(1-\alpha_{rj})$, which enters at first order, and it also leaves the neighborhood in which the remainder of \eqref{eq:m-local-loss} is uniform, so the lemma claims no order there.

\begin{figure}[!htbp]
\centering
\includegraphics[width=\linewidth]{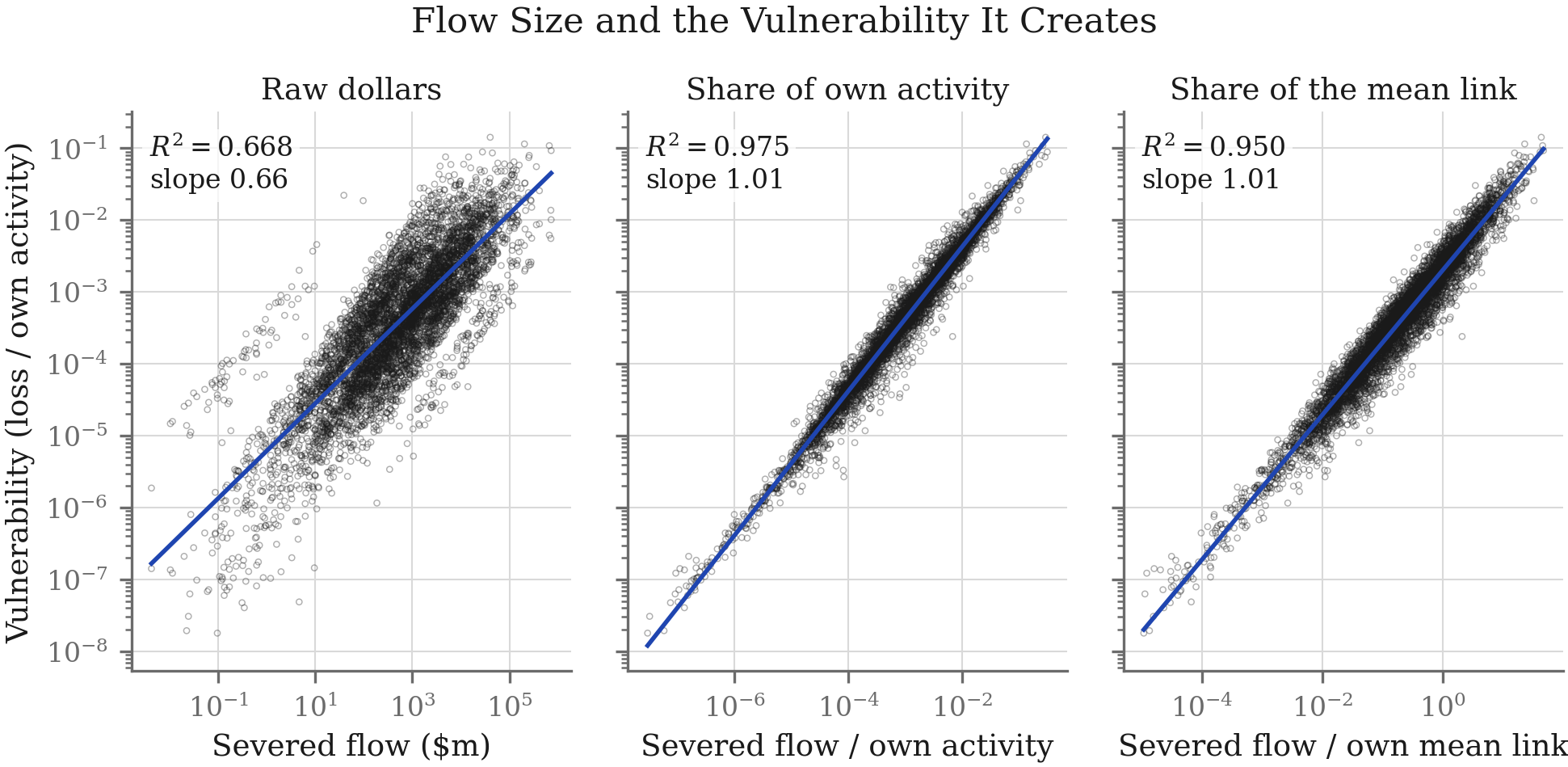}
\caption{Vulnerability against the severed flow for both sides of all $3{,}160$ pairs, on log scales. The three panels measure the severed flow in benchmark dollars, as a share of own gross activity, and as a share of the country's mean bilateral flow. The lines are log-log least-squares fits. Baseline calibration $\tau=0.30$, $\rho=-1$, $\varrho=-1$, $\delta=0.10$.}
\label{fig:flow_vs_vulnerability}
\end{figure}

The country losses add what the lemma leaves out, final trade and network propagation among them, and Figure~\ref{fig:flow_vs_vulnerability} sets each of the $6{,}320$ computed vulnerabilities against the severed flow behind it, intermediate and final together, under three normalizations. The lemma motivates the normalization by own activity, though its buyer-level law does not predict these finite country losses. Measured in dollars the severed flow orders the losses poorly. The log fit has an $R^2$ of $0.668$ and a slope of $0.66$, and the middle ninety percent of its residuals span a factor of sixty-seven, so a prediction from dollars alone can miss by more than an order of magnitude either way. Divided by the country's own gross activity the same flow fits with an $R^2$ of $0.975$ and a slope of $1.01$. The fit is an association consistent with the normalization, since own activity measures a dollar of severed trade against the activity that bears it, which the dollar flow alone does not. The third normalization, by the country's mean bilateral flow, measures the severed relationship against the country's typical relationship rather than against the activity that bears the loss.

\subsection{The distribution of bilateral asymmetry}
\label{oa:asymmetry-distribution}

\begin{figure}[!htbp]
\centering
\includegraphics[width=0.82\linewidth]{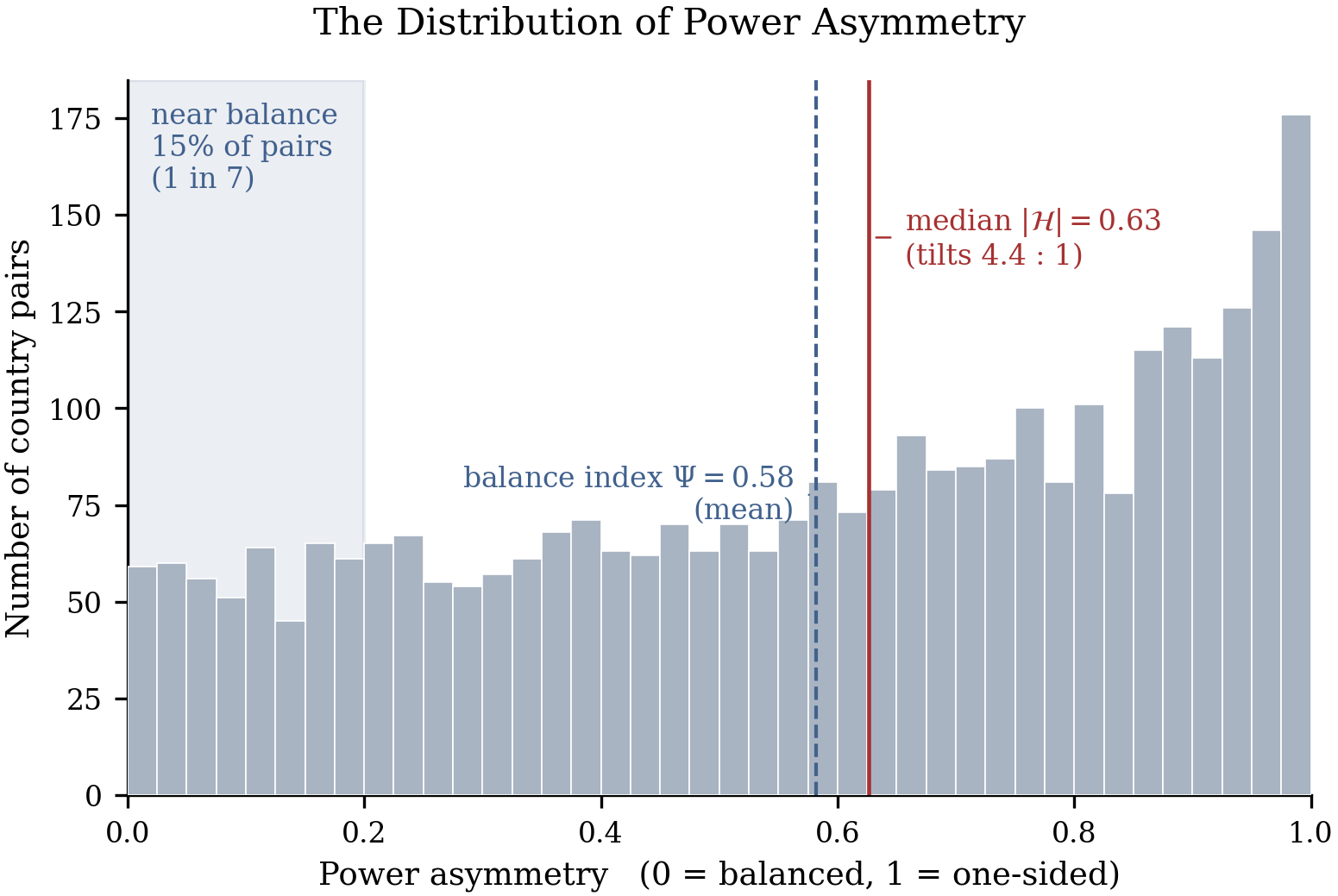}
\caption{The distribution of the absolute bilateral asymmetry $|\mathcal H_{cc'}|$ across the $3{,}160$ country pairs. The shaded bands separate the near-balanced pairs, at $|\mathcal H_{cc'}|<0.2$, from progressively more unequal relationships. Baseline calibration $\tau=0.30$, $\rho=-1$, $\varrho=-1$, and $\delta=0.10$.}
\label{fig:s6_tilt}
\end{figure}

Figure~\ref{fig:s6_tilt} draws the distribution of absolute asymmetry over the $3{,}160$ pairs that Section~\ref{subsec:hirschman_matrix} of the paper describes. The distribution rises toward the one-sided end rather than gathering near balance, its mean is the asymmetry index $\Psi=0.582$, and its median is $0.63$, a loss ratio of about $4.4$ to one.

\subsection{Robustness of the power structure across specifications, years, and table constructions}
\label{sec:robustness}

Figure~\ref{fig:s4_sensitivity} traces the headline vulnerabilities one parameter at a time. Whether the distribution of power survives a change of specification is a different question, and Table~\ref{tab:rob-master} collects every test of it in one place. Each row reports the asymmetry index $\Psi$, the rank correlation of the trade-weighted country ranking with the baseline ranking, the share of pairs that keep their baseline asymmetry sign, and the three highest-ranked economies. The country set and the materiality floor of a row determine which pairs enter it.

The parameter grid crosses $\tau\in\{0.10,0.30,0.60\}$ with $\rho\in\{-4,-1,-0.25\}$ at the baseline outer exponent, and further rows vary $\varrho$ and $\delta$ one at a time. The grouped exercise lets the unrecovered share and the attenuation vary by sector. Coal B05, crude oil and gas B06, and refined petroleum C19 are fungible goods under imperfect embargoes, and they receive $\tau=0.10$ and $\delta=1/3$. Electronics C26, electrical equipment C27, machinery C28, motor vehicles C29, and other transport C301/C302T309 are export-controlled manufactures, harder to re-source and more tightly enforced, and they receive $\tau=0.65$ and $\delta=0.05$. All other sectors keep the baseline $\tau=0.30$ and $\delta=0.10$. The joint configuration $(\tau,\rho,\varrho,\delta)=(0.60,-4,-9,0.05)$ combines a high unrecovered share and strong attenuation with strong complementarity in both nests, and two further rows relax it one parameter at a time. The joint configuration is a demanding sensitivity case and not a demonstrated worst case for reversals or rank changes. The last block varies the trade table rather than the parameters, across the annual world tables from 2016, the OECD's 2023 release at $45$ sectors against the 2025 release at $50$, an extended table that splits China into two accounts, and materiality floors on benchmark bilateral trade.

The distribution of power changes little across the rows of Table~\ref{tab:rob-master}. The asymmetry index $\Psi$ stays between $0.581$ and $0.584$ over the thirteen rows of the parameter grid and between $0.547$ and $0.590$ over every row of the table, the two floors that delete three fifths and four fifths of all pairs included. The rank correlation with the baseline is at least $0.999$ on every parameter specification and at least $0.985$ across the annual tables and the table constructions, and it falls to $0.952$ only on the floor that deletes four fifths of the pairs. Sign agreement is $99\%$ or better wherever it is defined, and the United States, China, and Germany are the three highest-ranked economies in every row. A change in the attenuation parameter $\delta$ scales the two sides' losses almost in proportion to the displaced share $1-\delta$, so the asymmetry, a ratio of the two losses, changes little with it. Across the parameter grid the cosine alignment of the asymmetry with the core-score gap, which Section~\ref{app:structural_scores_repro} defines, stays within $0.02$ of its baseline value of $0.69$.

\begin{table}[!htbp]
\centering\footnotesize
\begin{adjustbox}{max width=\textwidth}
\begin{tabular}{cccc ccc l}
\toprule
$\tau$ & $\rho$ & $\varrho$ & $\delta$ & $\Psi$ & rank corr. & sign agr. & top three \\
\midrule
0.10 & -4 & -1 & 0.10 & 0.582 & 1.000 & 100\% & USA, CHN, DEU \\
0.10 & -1 & -1 & 0.10 & 0.582 & 1.000 & 100\% & USA, CHN, DEU \\
0.10 & -0.25 & -1 & 0.10 & 0.581 & 1.000 & 100\% & USA, CHN, DEU \\
0.30 & -4 & -1 & 0.10 & 0.582 & 1.000 & 100\% & USA, CHN, DEU \\
0.30 & -1 & -1 & 0.10$^{\dagger}$ & 0.582 & --- & --- & USA, CHN, DEU \\
0.30 & -0.25 & -1 & 0.10 & 0.581 & 1.000 & 100\% & USA, CHN, DEU \\
0.60 & -4 & -1 & 0.10 & 0.584 & 0.999 & 99\% & USA, CHN, DEU \\
0.60 & -1 & -1 & 0.10 & 0.582 & 1.000 & 100\% & USA, CHN, DEU \\
0.60 & -0.25 & -1 & 0.10 & 0.582 & 1.000 & 100\% & USA, CHN, DEU \\
0.30 & -1 & -9 & 0.10 & 0.582 & 1.000 & 100\% & USA, CHN, DEU \\
0.30 & -1 & -3 & 0.10 & 0.582 & 1.000 & 100\% & USA, CHN, DEU \\
0.30 & -1 & -1 & 0.05 & 0.582 & 1.000 & 100\% & USA, CHN, DEU \\
0.30 & -1 & -1 & 0.33 & 0.581 & 1.000 & 100\% & USA, CHN, DEU \\
\addlinespace[0.35em]
\multicolumn{4}{l}{Grouped sector-specific $\tau_r$ and $\delta_r$} & 0.577 & 0.999 & 99\% & USA, CHN, DEU \\
0.60 & -4 & -9 & 0.05 & 0.586 & 0.999 & 99\% & USA, CHN, DEU \\
0.60 & -4 & -3 & 0.05 & 0.585 & 0.999 & 99\% & USA, CHN, DEU \\
0.60 & -4 & -9 & 0.10 & 0.586 & 0.999 & 99\% & USA, CHN, DEU \\
\addlinespace[0.35em]
\multicolumn{4}{l}{2016 world table} & 0.590 & 0.985 & --- & USA, CHN, DEU \\
\multicolumn{4}{l}{2017 world table} & 0.586 & 0.986 & --- & USA, CHN, DEU \\
\multicolumn{4}{l}{2018 world table} & 0.581 & 0.993 & --- & USA, CHN, DEU \\
\multicolumn{4}{l}{2019 world table} & 0.583 & 0.995 & --- & USA, CHN, DEU \\
\multicolumn{4}{l}{2020 world table} & 0.580 & 0.994 & --- & USA, CHN, DEU \\
\multicolumn{4}{l}{2021 world table} & 0.582 & 0.997 & --- & USA, CHN, DEU \\
\addlinespace[0.35em]
\multicolumn{4}{l}{2023 release, 45 sectors, 2020$^{\ddagger}$} & 0.573 & 0.990 & --- & USA, CHN, DEU \\
\multicolumn{4}{l}{Extended table splitting China, baseline} & 0.575 & 0.998 & 100\% & USA, CHN, DEU \\
\multicolumn{4}{l}{Extended table splitting China, joint configuration} & 0.580 & 0.998 & 99\% & USA, CHN, DEU \\
\addlinespace[0.35em]
\multicolumn{4}{l}{Materiality floor, 100 million dollars of pair trade} & 0.573 & 1.000 & --- & USA, CHN, DEU \\
\multicolumn{4}{l}{Materiality floor, 1 billion dollars} & 0.548 & 0.994 & --- & USA, CHN, DEU \\
\multicolumn{4}{l}{Materiality floor, 5 billion dollars} & 0.547 & 0.952 & --- & USA, CHN, DEU \\
\bottomrule
\end{tabular}
\end{adjustbox}
\caption{Robustness of the distribution of power. Each parameter and table row solves the severances under the stated specification, and each materiality row filters the baseline results without solving them again. The columns report the asymmetry index $\Psi$, the Spearman correlation of the trade-weighted country ranking with the baseline ranking, the share of pairs that keep their baseline asymmetry sign, and the three highest-ranked economies. $^{\dagger}$ marks the baseline against which the rank correlation and the sign agreement of every other row are computed. $^{\ddagger}$ The release comparison is computed against the 2025-edition table for the same year rather than against the 2022 baseline. A materiality floor keeps the baseline asymmetries of the pairs above the stated level of trade, so its sign agreement is one by construction and is not reported.}
\label{tab:rob-master}
\end{table}

The joint configuration tests the paper's claims at the higher unrecovered shares where input specificity is thought to concentrate, and it does not correct for the aggregation of inputs within industries. Each of the fifty industries pools varieties that have no substitute with varieties that have many, and the pooling can conceal specificity without implying a general direction of bias in the bilateral asymmetry. Raising $\tau$ on selected sectors raises their average rigidity, and it cannot reproduce the differences in rigidity among the varieties inside them.

Splitting China reveals a heterogeneity that the regular table hides. On the extended table China's trade-weighted advantage of $0.58$, against $0.59$ on the regular table (Table~\ref{tab:s6_power_ranking} of the paper), is the trade-weighted mean of $+0.61$ in the account that excludes processing trade and $-0.52$ in the processing-export account. Chinese leverage is therefore concentrated in the first account, which locates that leverage within China and does not qualify its size.

\subsection{Flattening the 2022 trade table at fixed sizes}
\label{oa:fiber}

The flattening experiment of Section~\ref{subsec:power_size_base} of the paper moves the 2022 trade table among the networks that share its seller totals of intermediate sales, holding the household matrix, the budgets, the leakages, and the residual outside demand fixed. Write $\mathbf Z^{(r)}$ for the intermediate flow block of input sector $r$, the flows from the producers of sector $r$ to every producer, with seller totals $\sum_jZ^{(r)}_{ij}$ and buyer totals $\sum_{i\in\mathcal N_r}Z^{(r)}_{ij}=\alpha_{rj}(1-\beta_j)s_j$. For a $\vartheta\in[0,1]$ we replace each block by the relative-entropy projection of the flattened kernel $\bigl(Z^{(r)}_{ij}\bigr)^{1-\vartheta}$, taken on the block's incumbent support, onto the block's own seller and buyer totals, which is the RAS rebalancing of the flattened block. Iterative proportional fitting computes the projection to a relative margin error of $10^{-13}$.\footnote{Alternating row-and-column rescaling is called iterative proportional fitting or raking in statistics, RAS updating in input--output economics, and Sinkhorn scaling in matrix analysis. The same procedure appears in probability as the static Schr\"odinger bridge, and it is the computational core of entropic optimal transport \citep{oa-demingstephan1940,oa-sinkhorn1964,oa-sinkhorn1967concerning,oa-stonebrown1962,oa-bacharach1970,oa-leonard2014,oa-cuturi2013}.} At $\vartheta=0$ the projection returns the trade table, and at $\vartheta=1$ it balances a flat kernel on the support, the pattern of maximum entropy given the margins. Dividing the flattened and rebalanced flows by the buyers' unchanged spending gives a network $\mathbf A'$ with unit column sums, the support of $\mathbf A$, and the seller totals $\mathbf A'(\mathbf I-\boldsymbol\beta)\mathbf s=\mathbf A(\mathbf I-\boldsymbol\beta)\mathbf s$. Only the pattern of who buys from whom therefore moves (Lemma~\ref{lem:m-fiber} of the paper).

At $\vartheta=1/4$ the flattening moves $10.7\%$ of the world's intermediate flows, measured as half the sum of the absolute cell changes divided by the total. It preserves the support cell by cell, meets the margins to $10^{-13}$, and leaves the activity vector unchanged to $2\times10^{-14}$ in relative terms. The world's domestic share of intermediate purchases falls from $0.852$ to $0.750$, and the effective number of foreign supplier countries of a buyer, the inverse Herfindahl index of its foreign partner shares averaged over the sectors it uses, rises from $7.19$ to $9.53$.

Re-solving all $3{,}160$ severances on the flattened table at the baseline calibration moves the vulnerabilities far more than the asymmetries. The median vulnerability rises by $57\%$, since the export share $X_{cc'}/Q(c)$ and the import share $M_{cc'}/Q(c)$ of every side rise by the same median of $73\%$. At fixed sizes each share moves with its flow, and every flow is the sales of one side and the purchases of the other. The export multiplier falls on $99\%$ of the sides, from a median of $1.63$ to $1.40$, in keeping with Lemma~\ref{lem:m-homechain} of the paper. The import multiplier per dollar of intermediate purchases falls with it, from $2.72$ to $2.03$ at the median and on $99.9\%$ of the sides with intermediate purchases. The import multiplier itself falls on only $52\%$ of the sides, from a median of $1.61$ to $1.58$. The rising intermediate share of the severed purchases, a median of $0.79$ against $0.61$, holds it up. The trade shares contribute a median of $+0.52$ log points to each side's change in the log vulnerability, evaluated at the old export multipliers and import multipliers, and the two multipliers contribute $-0.08$, evaluated at the new trade shares. The asymmetries move by a mean of $0.047$ and a median of $0.028$, with a ninetieth percentile of $0.118$ and a maximum of $0.633$, and they correlate with the baseline asymmetries at $0.994$. The sign of the asymmetry changes on $101$ pairs, $7$ of them among the $2{,}700$ pairs whose asymmetry exceeds $0.2$ in absolute value, and the asymmetry index $\Psi$ moves from $0.5819$ to $0.5820$. With the sizes fixed, the change in a pair's log loss ratio has no base effect in the decomposition \eqref{eq:m-exact-decomposition}, and over both orientations of every pair its variance splits into the change in the linkage effect at balanced trade, $53\%$, the change in the trade-imbalance effect, $53\%$, and twice their covariance, $-6\%$.

The second exercise freezes the pair's own trade. For each of the eight headline pairs of Table~\ref{tab:s4_headline} of the paper, we hold the pair's cross-border intermediate cells in both directions fixed, remove them from the margins, and flatten the remaining cells of every sector block at $\vartheta\in\{1/4,1/2,1\}$, with every household flow fixed as before. Both sizes, both trade shares, and the trade balance of the pair then stay where they are, so every movement of the asymmetry runs through the export multipliers and the import multipliers of the two sides. Table~\ref{tab:oa_fiber_frozen} reports the eight pairs on the benchmark table and at the three values of $\vartheta$. The domestic share of intermediate purchases falls on both sides as $\vartheta$ rises, to between $0.02$ and $0.46$ at $\vartheta=1$, and the export multiplier and the import multiplier of each side fall with it. That fall is a finding of the re-solved counterfactuals rather than a consequence of Lemma~\ref{lem:m-homechain} of the paper alone, and it holds for every multiplier of every pair at every step. The asymmetries move by a median of $0.015$ at $\vartheta=1/4$, $0.043$ at $\vartheta=1/2$, and $0.064$ at $\vartheta=1$. The largest movement is $0.16$, for India against China at $\vartheta=1$, and none of the eight asymmetries changes sign, although by $\vartheta=1$ more than two thirds of the world's intermediate flows have moved while every row and column total of the trade table stayed fixed. The linkage effect at balanced trade moves by a median absolute $0.05$, $0.13$, and $0.26$ log points at the three values of $\vartheta$, and the trade-imbalance effect, through the import premia alone, by $0.02$, $0.05$, and $0.10$.

\begin{table}[!htbp]
\centering
\small
\begin{adjustbox}{max width=\linewidth}
\begin{tabular}{@{}llrrrrrrrrrrr@{}}
\toprule
$c$ & $c'$ & $\vartheta$ & Flow moved & Dom.\ share $c$ & Dom.\ share $c'$ & $\gamma_c$ (\%) & $\gamma_{c'}$ (\%) & $\ell^{\downarrow}_c$ & $\ell^{\uparrow}_c$ & $\ell^{\downarrow}_{c'}$ & $\ell^{\uparrow}_{c'}$ & $\mathcal H_{cc'}$ \\
\midrule
USA & CHN & 0 & 0.00 & 0.92 & 0.93 & 0.556 & 1.373 & 1.81 & 1.11 & 2.67 & 4.57 & $-0.4236$ \\
USA & CHN & 0.25 & 0.10 & 0.86 & 0.89 & 0.525 & 1.275 & 1.76 & 1.02 & 2.53 & 4.12 & $-0.4170$ \\
USA & CHN & 0.5 & 0.28 & 0.75 & 0.80 & 0.487 & 1.136 & 1.71 & 0.92 & 2.33 & 3.47 & $-0.3999$ \\
USA & CHN & 1 & 0.67 & 0.39 & 0.46 & 0.446 & 0.903 & 1.69 & 0.80 & 1.98 & 2.41 & $-0.3388$ \\
CHN & TWN & 0 & 0.00 & 0.93 & 0.71 & 0.628 & 7.433 & 2.91 & 4.64 & 1.69 & 2.01 & $-0.8443$ \\
CHN & TWN & 0.25 & 0.11 & 0.88 & 0.56 & 0.564 & 6.242 & 2.81 & 4.10 & 1.48 & 1.52 & $-0.8342$ \\
CHN & TWN & 0.5 & 0.29 & 0.78 & 0.33 & 0.488 & 5.093 & 2.65 & 3.46 & 1.27 & 1.07 & $-0.8252$ \\
CHN & TWN & 1 & 0.71 & 0.38 & 0.04 & 0.373 & 4.154 & 2.31 & 2.54 & 1.08 & 0.75 & $-0.8352$ \\
USA & RUS & 0 & 0.00 & 0.92 & 0.92 & 0.054 & 0.621 & 1.77 & 2.42 & 2.08 & 2.45 & $-0.8412$ \\
USA & RUS & 0.25 & 0.11 & 0.86 & 0.82 & 0.049 & 0.540 & 1.71 & 2.16 & 1.89 & 2.01 & $-0.8326$ \\
USA & RUS & 0.5 & 0.29 & 0.72 & 0.60 & 0.044 & 0.413 & 1.65 & 1.83 & 1.57 & 1.39 & $-0.8079$ \\
USA & RUS & 1 & 0.71 & 0.29 & 0.07 & 0.037 & 0.262 & 1.57 & 1.42 & 1.12 & 0.72 & $-0.7514$ \\
DEU & RUS & 0 & 0.00 & 0.78 & 0.92 & 0.566 & 0.952 & 1.78 & 2.69 & 1.90 & 2.20 & $-0.2546$ \\
DEU & RUS & 0.25 & 0.11 & 0.64 & 0.82 & 0.457 & 0.838 & 1.57 & 2.09 & 1.77 & 1.78 & $-0.2936$ \\
DEU & RUS & 0.5 & 0.29 & 0.43 & 0.58 & 0.348 & 0.657 & 1.36 & 1.48 & 1.52 & 1.18 & $-0.3077$ \\
DEU & RUS & 1 & 0.71 & 0.07 & 0.05 & 0.249 & 0.440 & 1.13 & 0.96 & 1.13 & 0.62 & $-0.2778$ \\
JPN & CHN & 0 & 0.00 & 0.83 & 0.93 & 2.567 & 0.889 & 1.92 & 1.79 & 2.68 & 4.61 & $+0.4856$ \\
JPN & CHN & 0.25 & 0.10 & 0.74 & 0.89 & 2.245 & 0.818 & 1.75 & 1.50 & 2.57 & 4.13 & $+0.4661$ \\
JPN & CHN & 0.5 & 0.29 & 0.55 & 0.79 & 1.823 & 0.720 & 1.52 & 1.14 & 2.40 & 3.46 & $+0.4339$ \\
JPN & CHN & 1 & 0.70 & 0.10 & 0.40 & 1.293 & 0.559 & 1.21 & 0.70 & 2.07 & 2.45 & $+0.3959$ \\
IND & CHN & 0 & 0.00 & 0.84 & 0.93 & 1.182 & 0.235 & 1.69 & 1.99 & 2.64 & 2.93 & $+0.6687$ \\
IND & CHN & 0.25 & 0.11 & 0.74 & 0.88 & 1.010 & 0.220 & 1.56 & 1.66 & 2.52 & 2.62 & $+0.6429$ \\
IND & CHN & 0.5 & 0.29 & 0.55 & 0.78 & 0.780 & 0.200 & 1.36 & 1.22 & 2.37 & 2.17 & $+0.5913$ \\
IND & CHN & 1 & 0.70 & 0.10 & 0.40 & 0.520 & 0.169 & 1.11 & 0.74 & 2.12 & 1.50 & $+0.5100$ \\
USA & MEX & 0 & 0.00 & 0.92 & 0.67 & 0.619 & 10.882 & 1.98 & 1.27 & 1.52 & 1.64 & $-0.8924$ \\
USA & MEX & 0.25 & 0.11 & 0.86 & 0.58 & 0.574 & 9.684 & 1.88 & 1.15 & 1.39 & 1.40 & $-0.8881$ \\
USA & MEX & 0.5 & 0.29 & 0.72 & 0.37 & 0.516 & 7.912 & 1.75 & 0.99 & 1.21 & 1.05 & $-0.8775$ \\
USA & MEX & 1 & 0.71 & 0.28 & 0.02 & 0.441 & 6.009 & 1.56 & 0.80 & 1.01 & 0.66 & $-0.8631$ \\
DEU & CHN & 0 & 0.00 & 0.78 & 0.93 & 1.511 & 0.560 & 1.76 & 1.37 & 2.72 & 3.89 & $+0.4592$ \\
DEU & CHN & 0.25 & 0.11 & 0.65 & 0.88 & 1.290 & 0.516 & 1.57 & 1.12 & 2.60 & 3.47 & $+0.4285$ \\
DEU & CHN & 0.5 & 0.29 & 0.45 & 0.78 & 1.074 & 0.455 & 1.38 & 0.87 & 2.42 & 2.90 & $+0.4046$ \\
DEU & CHN & 1 & 0.71 & 0.10 & 0.39 & 0.858 & 0.354 & 1.18 & 0.64 & 2.06 & 2.03 & $+0.4159$ \\
\bottomrule
\end{tabular}

\end{adjustbox}
\caption{The eight headline severances on the network flattened with every row and column total of the trade table held fixed, and with the pair's own cross-border intermediate cells and every household flow frozen. Flow moved is half the sum of the absolute cell changes over the world's intermediate flows, and the domestic shares are each country's domestic share of intermediate purchases. Losses are percentages of unchanged benchmark activity, $\ell^{\downarrow}$ and $\ell^{\uparrow}$ are each side's export multiplier and import multiplier, and $\mathcal H_{cc'}$ is the asymmetry \eqref{eq:bounded-Hirschman} of the paper, positive when country $c$ is the weaker side. Baseline calibration $\tau=0.30$, $\rho=-1$, $\varrho=-1$, $\delta=0.10$.}
\label{tab:oa_fiber_frozen}
\end{table}

Corollary~\ref{cor:m-links} bounds the partner's shares by fractions of links, at the cost of the evenness factor $\Theta$, the ratio of the largest to the smallest positive share within a buyer's sector column. On the 2022 trade table $63.2\%$ of the cells of a sector block are positive over the active buyers, so a buyer of an input sources from $50.6$ of the eighty economies on average, and a single partner's link fraction varies across buyers rather than standing at one in seventy-nine. The median active block sources from $57$ of the eighty economies, the blocks with at least forty source countries account for more than $99\%$ of benchmark intermediate mass, and a single source country supplies the entire block in $1{,}821$ of roughly $1.9\times10^{5}$ active blocks. The evenness factor is far from one, with a median of about $56{,}000$ across the buyers' sector columns, a tenth percentile of $354$, and a ninetieth percentile of $2.7\times10^{6}$. Across the sides of the $3{,}160$ pairs, the largest share of a producer's activity that the partner takes has a median of $0.0068$ and a ninetieth percentile of $0.080$, and the largest share of a buyer's spending that the partner holds has a median of $0.0038$ and a ninetieth percentile of $0.049$. Against the link fraction $1/79\simeq0.0127$ of a complete network, the actual shares have median ratios of $0.5$ on the selling side and $0.3$ on the buying side. A count of links therefore misses the variation in the weighted trade shares, and evenness factors this large leave the bounds of the corollary loose.

We set the bound $\bar r=3.27$ at the ninety-fifth percentile of the export multipliers and the import multipliers, each with its interaction rate added. The sign condition of Proposition~\ref{prop:m-shares} then selects a negative asymmetry on $109$ pairs and a positive one on $68$, and all $177$ selected signs agree with the solved signs. A percentile bound does not establish these signs under the proposition, which requires the inequalities on the two multipliers of a side, each with its interaction rate added, to hold pair by pair. Substituting each pair's own export multipliers, import multipliers, and interaction rates into \eqref{eq:m-exact-exposure} recovers every sign, but that is an accounting reconstruction and not an independent prediction.

\subsection{Size and the trade-weighted advantage}
\label{oa:size-advantage}

Section~\ref{sec:power_size} of the paper compares size and power pair by pair. Country by country, the comparison runs through the trade-weighted advantage of Table~\ref{tab:s6_power_ranking} of the paper, which weights each of country $c$'s comparisons by the share of its cross-border trade that runs with country $c'$. Write $\mu_{cc'}$ for that share, so that the advantage is $\mathcal V_c:=-\sum_{c'\ne c}\mu_{cc'}\mathcal H_{cc'}$. The advantage therefore sets a country against all of its partners at once.

\begin{figure}[!htbp]
\centering
\includegraphics[width=0.86\linewidth]{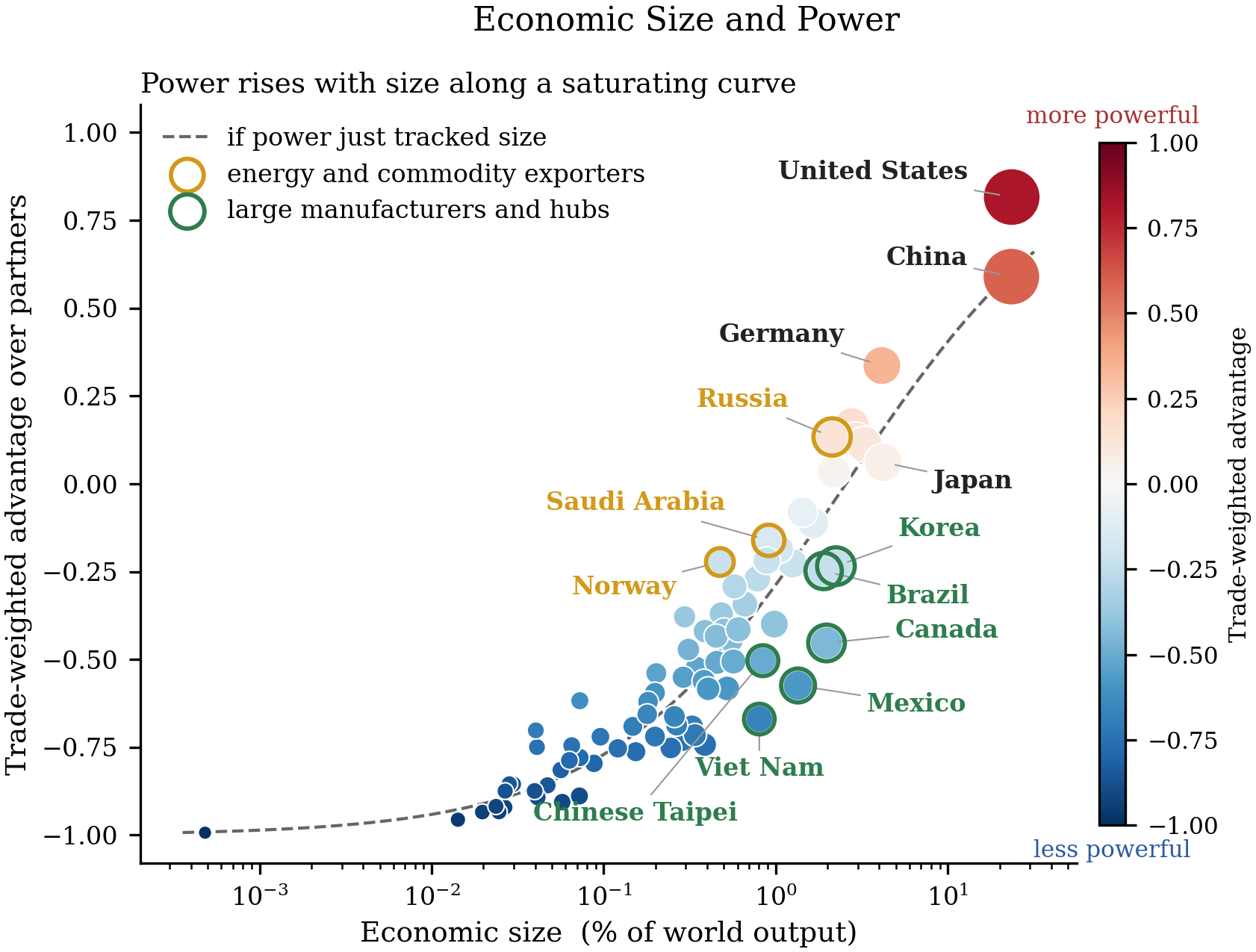}
\caption{The trade-weighted advantage against benchmark size on a log scale. The dashed curve is the fitted logistic in log size, $\psi(0.55\,q^{0.63})$ for an economy producing the percentage $q$ of world gross output, with $\psi$ the normalized difference \eqref{eq:m-psi} of the paper. Gold rings mark energy- and commodity-exporting countries and green rings large manufacturers and hubs. Marker area scales with size and color with the trade-weighted advantage. Baseline calibration $\tau=0.30$, $\rho=-1$, $\varrho=-1$, $\delta=0.10$.}
\label{fig:size_panel}
\end{figure}

Figure~\ref{fig:size_panel} plots the advantage of each of the eighty economies against its gross output in the 2022 trade table, in percent of world output on a log scale. The relation is strong and positive, and a logistic curve in log size fits it with an $R^{2}$ of $0.89$, leaving a residual standard deviation of $0.12$ on an advantage that runs from $-0.99$ to $0.82$.

The size statistic that matches this advantage is the same trade-weighted average of the size asymmetries \eqref{eq:m-size-asymmetry} of the paper, the size advantage
\begin{equation}
\mathcal V^{Q}_c:=-\sum_{c'\ne c}\mu_{cc'}\,\mathcal Q_{cc'}.
\label{eq:m-size-advantage}
\end{equation}
The size advantage is the advantage that the sizes alone predict. Table~\ref{tab:s6_size_advantage} reports standardized least-squares fits of the advantage on log size and on the size advantage, one observation per economy. On its own the size advantage takes a coefficient of $0.98$ with an $R^{2}$ of $0.964$, against $0.776$ for log size, and entering log size beside it adds one thousandth to the $R^{2}$.\footnote{Weighting every partner equally, running the regression on ranks, dropping the United States and China, or dropping the ten largest economies leaves the coefficient between $0.96$ and $0.99$ and the $R^{2}$ between $0.914$ and $0.971$. Entered beside the size advantage on the last two samples, log size takes coefficients of $0.08$ and $0.10$.} Own size fits less well because the advantage weights each comparison by the trade between the two countries, so two economies of the same size differ in power when their trade runs to partners of different sizes. Korea and Canada hold the weaker side at Russia's size because their heaviest weights fall on China and the United States, and weighting by trade lowers Japan's advantage by more than six tenths (Table~\ref{tab:s6_power_ranking} of the paper).

\begin{table}[!htbp]
\centering
\small
\begin{tabular}{lccc}
\toprule
 & (1) & (2) & (3) \\
\midrule
Log economic size & $0.881$ &  & $0.076$ \\
 & $(0.109)$ &  & $(0.036)$ \\[2pt]
Size advantage $\mathcal V^{Q}_c$ &  & $0.982$ & $0.915$ \\
 &  & $(0.023)$ & $(0.043)$ \\[2pt]
\midrule
$R^{2}$ & $0.776$ & $0.964$ & $0.965$ \\
Countries & $80$ & $80$ & $80$ \\
\bottomrule
\end{tabular}

\caption{Standardized least-squares fits of the trade-weighted advantage $\mathcal V_c$ on log size and on the size advantage $\mathcal V^{Q}_c$ of \eqref{eq:m-size-advantage}, one observation per economy, with heteroskedasticity-robust standard errors in parentheses. Both regressors are built from the benchmark table alone. Baseline calibration $\tau=0.30$, $\rho=-1$, $\varrho=-1$, $\delta=0.10$.}
\label{tab:s6_size_advantage}
\end{table}

\subsection{Power among economies of comparable size}
\label{oa:power_size_local}

Power varies widely among economies of comparable size, as the model permits, since networks with identical sizes can divide the losses of a severance differently between the two sides of a pair (Lemma~\ref{lem:m-fiber} of the paper). Among the ten economies above $2\%$ of world output the trade-weighted vulnerability, which weights partners as the advantage does and here exceeds the equally weighted mean of Table~\ref{tab:s6_power_ranking} of the paper, runs from $0.4\%$ of gross activity for the United States to $2.3\%$ for Korea. Among the thirty economies below a fifth of a percent of world output it runs from $0.4\%$ for Cameroon to $8.0\%$ for Belarus, so at the bottom of the size distribution the measure spans almost its whole range, and the two ranges overlap over most of their length. The countries differ in their foreign exposure, the concentration of their partners, their sectoral composition, and the propagation of losses through their networks, and the comparisons of this subsection describe the joint association of these differences with power. If scale alone determined power, the Hirschman matrix would add nothing to a table of national accounts, and the ranking of that table would be a list of the largest economies.

Section~\ref{subsec:comparable_size} of the paper measures how much of the variation in power remains among economies of comparable size by the within-size variation in power $\mathrm{sd}_x(c)/\mathrm{SD}$ of \eqref{eq:m-within-size} of the paper, and Figure~\ref{fig:size_window} of the paper plots it. Table~\ref{tab:size_window} reports the median of this ratio over the size bands of a given width, at four widths.\footnote{A fixed multiplicative width leaves a band thin at the ends of the size distribution and full in the middle, so we repeat the exercise on the $k$ economies nearest in log size, which fixes the number of economies in a band and lets its width float. At $k=12$, with a median width of a factor of $2.4$, the median within-size variation in power is $0.37$ against $0.42$ for the factor-of-two bands. The bracketed figures in Table~\ref{tab:size_window} are the fifth and ninety-fifth percentiles of a bootstrap over economies.} At a width of two the median is $0.42$, and the interval that holds $90\%$ of the advantages narrows from $1.02$ across all the economies to $0.43$ within a band, on the index's $[-1,+1]$ scale. The restriction to a band changes the comparison set, and it does not identify the effect of changing any one country's size.

The median of $0.42$ combines bands that are not alike. One band holds Brunei, Senegal, Malta, Iceland, Cambodia, Cyprus, and Cameroon, all within a hundredth of a percent of world output of one another, and every one of them holds the weaker side of almost every relationship it has, so their advantages run only from $-0.93$ to $-0.85$. The band that starts at Brazil spans a range of sizes no wider, and its advantages run from $-0.45$ to $0.17$, a spread eight times as wide (Section~\ref{subsec:comparable_size} of the paper). Part of this difference follows from the construction of the measure. A country's advantage is a trade-weighted mean of its pairwise asymmetries, so an economy that concentrates its trade on much larger partners places its heaviest weights on unfavorable entries, while one whose trade spreads across partners near or below its own size places them on favorable ones (Section~\ref{oa:size-advantage}).

\begin{table}[!htbp]
\centering
\small
\begin{tabular}{lrrcc}
\toprule
Comparison set & Windows & Median $n$ & Median sd\,/\,SD & $90\%$ band \\
\midrule
Within $1.5\times$ & 61 & 7 & 0.38\,[0.25,\,0.45] & 0.39 \\
Within $2\times$ & 72 & 10 & 0.42\,[0.31,\,0.49] & 0.43 \\
Within $3\times$ & 73 & 15 & 0.46\,[0.35,\,0.55] & 0.47 \\
Within $5\times$ & 73 & 21 & 0.55\,[0.43,\,0.62] & 0.55 \\
\midrule
All 77 economies & 1 & 77 & 1.00 & 1.02 \\
\bottomrule
\end{tabular}

\caption{The within-size variation in power, by the width of the size band. The windows column counts the bands of each width and the median $n$ column the economies a band holds, and the brackets give the bootstrap interval. Bands holding fewer than five economies are dropped, and S\~{a}o Tom\'{e}, China, and the United States are excluded throughout. Baseline calibration $\tau=0.30$, $\rho=-1$, $\varrho=-1$, $\delta=0.10$.}
\label{tab:size_window}
\end{table}

Table~\ref{tab:size_window} measures how much variation in power a restriction to comparable size leaves, and not what that variation is made of, so we relate it to the two margins of a country's exposure. Table~\ref{tab:s6_size_diversification} relates the trade-weighted advantage to log size and to the buying and selling margins of the exposures of Proposition~\ref{prop:m-exposure} of the paper, each a partner-specific flow divided by own activity and averaged over partners with the advantage's trade weights. Panel~B of that table divides the same flows by all intermediate purchases and all in-table sales instead, so its measures combine the foreign share of a country's trade with the allocation of that share across partners. Inside a narrow band log size barely varies, which holds down any within-band fit on size through the restriction of range alone, while the exposure margins keep their range, since economies of a common size differ widely in how much of their trade runs to particular partners. Table~\ref{tab:size_window_regress} runs size and exposure on the same bands and samples, and its fit on size carries that qualification.

\begin{table}[!htbp]
\centering
\small
\begin{adjustbox}{max width=\textwidth}
\begin{tabular}{l ccccc}
\toprule
 & (1) & (2) & (3) & (4) & (5) \\
\midrule
\multicolumn{6}{@{}l}{Panel A. The severed exposure, per dollar of own size} \\
\midrule
Log benchmark size & $0.88$ & $0.19$ & $0.33$ & $0.09$ & $0.05$ \\
 & ($0.11$) & ($0.06$) & ($0.09$) & ($0.04$) & ($0.03$) \\
Buying-side exposure &  & $0.82$ &  & $0.64$ &  \\
 &  & ($0.06$) &  & ($0.03$) &  \\
Selling-side exposure &  &  & $0.66$ & $0.30$ &  \\
 &  &  & ($0.08$) & ($0.03$) &  \\
Both margins summed &  &  &  &  & $0.94$ \\
 &  &  &  &  & ($0.03$) \\
$R^2$ & $0.776$ & $0.968$ & $0.906$ & $0.986$ & $0.975$ \\
\addlinespace[0.6em]
\multicolumn{6}{@{}l}{Panel B. Partner shares of all purchases and sales} \\
\midrule
Log benchmark size & $0.88$ & $0.18$ & $0.33$ & $0.10$ &  \\
 & ($0.11$) & ($0.06$) & ($0.09$) & ($0.04$) &  \\
Buying-side partner share &  & $0.83$ &  & $0.66$ &  \\
 &  & ($0.06$) &  & ($0.04$) &  \\
Selling-side partner share &  &  & $0.66$ & $0.27$ &  \\
 &  &  & ($0.08$) & ($0.04$) &  \\
$R^2$ & $0.776$ & $0.966$ & $0.906$ & $0.979$ &  \\
\midrule
Observations & $80$ & $80$ & $80$ & $80$ & $80$ \\
\bottomrule
\end{tabular}

\end{adjustbox}
\caption{Standardized least-squares fits of the trade-weighted advantage, one observation per country. Column~(1) uses log size alone, columns~(2) and~(3) add one margin of exposure each, column~(4) adds both separately, and column~(5) of Panel~A adds their sum, a descriptive control with unit weights. Panel~A divides the intermediate purchases from the partner and the total sales to it by own benchmark activity. Panel~B divides the same flows by all intermediate purchases and all in-table sales, domestic transactions included, so it retains differences in foreign exposure and does not measure concentration conditional on foreign trade. Every control is built from the benchmark table alone and averaged with the advantage's partner weights. Heteroskedasticity-robust standard errors in parentheses. Baseline calibration $\tau=0.30$, $\rho=-1$, $\varrho=-1$, $\delta=0.10$.}
\label{tab:s6_size_diversification}
\end{table}

\begin{table}[!htbp]
\centering
\small
\begin{adjustbox}{max width=\textwidth}
\begin{tabular}{l ccccc}
\toprule
\multicolumn{6}{@{}l}{Panel A. Pooled, with fixed effects on non-overlapping size bins} \\
\midrule
 & No bins & Within $5\times$ & Within $3\times$ & Within $2\times$ & Within $1.5\times$ \\
\midrule
Log benchmark size & $0.11$ & $0.12$ & $0.06$ & $0.04$ & $-0.08$ \\
 & ($0.05$) & ($0.05$) & ($0.05$) & ($0.05$) & ($0.05$) \\
Buying-side exposure & $0.65$ & $0.74$ & $0.76$ & $0.77$ & $0.87$ \\
 & ($0.03$) & ($0.04$) & ($0.04$) & ($0.04$) & ($0.05$) \\
Selling-side exposure & $0.28$ & $0.29$ & $0.29$ & $0.31$ & $0.26$ \\
 & ($0.03$) & ($0.04$) & ($0.05$) & ($0.05$) & ($0.05$) \\
\addlinespace[0.2em]
Within-bin $R^2$, all three & $0.982$ & $0.926$ & $0.917$ & $0.897$ & $0.909$ \\
Within-bin $R^2$, size alone & $0.793$ & $0.279$ & $0.235$ & $0.096$ & $0.072$ \\
Bins & $1$ & $4$ & $6$ & $9$ & $15$ \\
\addlinespace[0.6em]
\multicolumn{6}{@{}l}{Panel B. Band by band, medians over the moving windows of Table~\ref{tab:size_window}} \\
\midrule
Comparison set & Median $n$ & Median sd\,/\,SD & Adj.\ $R^2$, size & Adj.\ $R^2$, margins & Residual sd\,/\,SD \\
\midrule
Within $1.5\times$ & 7 & 0.38 & $-0.04$ & $0.87$ & 0.12\,[0.08,\,0.13] \\
Within $2\times$ & 10 & 0.42 & $0.03$ & $0.89$ & 0.12\,[0.08,\,0.14] \\
Within $3\times$ & 15 & 0.46 & $0.16$ & $0.90$ & 0.13\,[0.09,\,0.15] \\
Within $5\times$ & 21 & 0.55 & $0.33$ & $0.93$ & 0.13\,[0.10,\,0.14] \\
All 77 economies & 77 & 1.00 & $0.79$ & $0.98$ & 0.14 \\
\bottomrule
\end{tabular}

\end{adjustbox}
\caption{The trade-weighted advantage regressed on log size and on the exposure margins of Table~\ref{tab:s6_size_diversification}, over the seventy-seven economies of the comparison set. Panel~A absorbs scale with fixed effects on non-overlapping size bins, each holding the economies within a factor of the stated width, and reports the within-bin fit and coefficients standardized after the bin means are swept out. Each coefficient is therefore the within-bin standard deviations of the advantage that one within-bin standard deviation of the regressor moves. Heteroskedasticity-robust standard errors are in parentheses. The column with no bins repeats Table~\ref{tab:s6_size_diversification} on this sample, and the last fit row uses log size alone. Panel~B anchors a band at each economy, as Table~\ref{tab:size_window} does, and reports medians across bands of the degrees-of-freedom-adjusted fit. Its last column is the within-band standard deviation of the residual from the two margins divided by the standard deviation of the advantage across all the economies, on the same scale as the within-size variation in power beside it. Bands holding fewer than five economies are dropped, and S\~{a}o Tom\'{e}, China, and the United States are excluded throughout. Baseline calibration $\tau=0.30$, $\rho=-1$, $\varrho=-1$, $\delta=0.10$.}
\label{tab:size_window_regress}
\end{table}

Panel~A pools the seventy-seven economies and absorbs scale with fixed effects on non-overlapping size bins, which keeps the degrees of freedom that a band-by-band fit gives up while every comparison still runs between economies within a factor $x$ of one another. Its column with no bins differs slightly from Table~\ref{tab:s6_size_diversification} because it uses seventy-seven economies rather than eighty. As the bins narrow, the coefficient on the buying margin rises relative to that on the selling margin. The buying coefficient rises from $0.65$ with no bins to $0.87$ within factor-of-$1.5$ bins while the selling coefficient stays between $0.26$ and $0.31$, so the ratio of the two rises from a little over two to a little over three, and the joint within-bin fit stays near $0.90$ throughout. The association with size weakens on both measures. Its coefficient in the joint regression falls from $0.11$ with no bins to $0.04$ within factor-of-two bins and $-0.08$ within factor-of-$1.5$ bins, indistinguishable from zero at every width narrower than five, and its fit on its own falls from an $R^2$ of $0.793$ across the whole comparison set to $0.072$ within factor-of-$1.5$ bins. These comparisons condition on bilateral exposure while restricting the variation in size.

Panel~B repeats the moving bands of Table~\ref{tab:size_window} and takes medians across windows, so it follows the construction of the within-size variation in power. Within a factor of two the median adjusted fit is $0.03$ for log size and $0.89$ for the two exposure margins. Measured on the scale of the within-size variation in power, the within-band standard deviation of the residual from the two margins is $0.12$ of the standard deviation of the advantage across all the economies, against $0.42$ for the advantage itself. The two margins therefore account for most of the two fifths of the variation in power that a factor-of-two band leaves, and what neither size nor the margins reach is about an eighth of it.\footnote{The bracketed figures are interquartile ranges across the $61$ to $73$ windows and not bootstrap intervals over economies. A bootstrap that redraws economies with replacement places duplicates inside a band, which changes the weights of the observations and distorts the fits, at the narrow widths enough to put the point estimate below the fifth percentile of its own interval. The fits of Panel~B are adjusted for degrees of freedom, since the median band holds ten economies at a width of two and seven at a width of $1.5$.}

The buying and selling exposures thus keep most of their predictive content within narrow size bands, since their denominators and partner-specific flows together measure a country's foreign exposure and its allocation across relationships. The larger buying-side coefficients describe these standardized regressions, and they establish neither that concentration among suppliers has a larger causal effect on power than concentration among customers nor that the two margins exhaust the network features behind power. Section~\ref{oa:fiber} moves the allocation of exposure directly, at fixed sizes.

\subsection{The linkage effect and economic size}
\label{oa:linkage-size}

Section~\ref{subsec:power_size_base} of the paper measures the base effect and the linkage effect of each of the $3{,}160$ pairs, and we now ask how the linkage effect varies with the size ratio of a pair. Between economies within a factor of two of one another in size, the linkage effect is on average as large as the base effect, whichever of the two economies it favors. Between economies that differ in size by an order of magnitude or more, the linkage effect is on average less than one seventh as large as the base effect.\footnote{We measure both effects in logs, the linkage effect by its absolute value, and average each over the pairs. The linkage effect averages $0.35$ and the base effect $0.35$ over the $746$ pairs whose size ratio is below two. They average $0.49$ and $3.67$ over the $1{,}072$ pairs whose size ratio is ten or more.} This drop is merely a consequence of the fact that the linkage effect varies little across the trading pairs, while the base effect varies by several orders of magnitude. Table~\ref{tab:linkage_by_size}, which sorts the pairs by their size ratio, captures this difference in variation. The third column of the table shows that the linkage effect is much the same size in every row, lying between about $\nicefrac{1}{2}$ and $2$ on nine pairs in ten where the size ratio is below two, and between about $\nicefrac{2}{5}$ and $2 \frac{1}{2}$ where the size ratio exceeds five. What differs from row to row is the base effect, which is the size ratio itself. The last two columns make this point precise. They show that, among the $746$ pairs whose size ratio is below two, the linkage effect is at least as large as the base effect on nearly half of them, as one expects of an effect that is on average as large as the other. Among the $1{,}603$ pairs whose size ratio exceeds five, the linkage effect is at least as large as the base effect on only two.

\begin{table}[!htbp]
\centering
\footnotesize
\begin{adjustbox}{max width=\textwidth}
\begin{tabular}{lrrrrr}
\toprule
 & & Linkage effect between & Larger economy & \multicolumn{2}{c}{Linkage effect as large as} \\
 & & $\nicefrac{1}{f}$ and $f$ on nine & loses more & \multicolumn{2}{c}{the base effect, in favor of the} \\
\cmidrule(lr){5-6}
Size ratio & Pairs & pairs in ten, $f$ & dollars (\%) & larger economy & smaller economy \\
\midrule
Below $2$ & $746$ & $2.1$ & $53$ & $163$ & $175$ \\
$2$ to $5$ & $811$ & $2.0$ & $63$ & $10$ & $18$ \\
Above $5$ & $1{,}603$ & $2.6$ & $76$ & $0$ & $2$ \\
\midrule
All pairs & $3{,}160$ & $2.4$ & $67$ & $173$ & $195$ \\
\bottomrule
\end{tabular}
\end{adjustbox}
\caption{The linkage effect by the size ratio of the pair, each pair oriented with the smaller economy first. The third column gives the factor $f$ for which the linkage effect lies between $\nicefrac{1}{f}$ and $f$ on nine pairs in ten, and the fourth the percentage of pairs on which the larger economy loses more dollars than the smaller one. The last two columns count the pairs on which the two dollar losses differ by at least the size ratio. Where the larger economy loses the more dollars by that margin, the linkage effect favors the smaller economy and makes it the more powerful side. Baseline calibration $\tau=0.30$, $\rho=-1$, $\varrho=-1$, $\delta=0.10$.}
\label{tab:linkage_by_size}
\end{table}

On average the linkage effect runs against the larger economy, and the more so the further apart the two economies are in size. The larger economy loses more dollars than its partner two times in three, and the fourth column of Table~\ref{tab:linkage_by_size} gives the frequency by size ratio. Across the $3{,}160$ pairs the loss ratio rises almost one for one with the size ratio, at a slope of $0.90$ in logs through the origin, and the log of the linkage effect falls with the log size ratio at a slope of $-0.10$, as the two solid lines of Figure~\ref{fig:size_decomposition} of the paper show. Since the logs of the two factors of \eqref{eq:m-size-split} of the paper add up to the log loss ratio, the two slopes are a single finding, the base effect entering with its coefficient of one and the linkage effect offsetting a tenth of it.\footnote{Weighting the pairs by their gross trade gives slopes of $0.88$ and $-0.12$, and the $1{,}309$ pairs trading at least a billion dollars give $0.90$ and $-0.10$. Over both orientations of every pair the variance of the log loss ratio splits into $117\%$ for the log size ratio, $5\%$ for the log of the linkage effect, and $-22\%$ for twice their covariance, negative because the linkage effect runs against the larger economy. The size ratio alone correlates with the loss ratio at $0.98$ in logs.} On the fitted lines, an economy ten times the size of its partner loses about a quarter more dollars than the partner and about one eighth as large a fraction of its own activity.

The reason why the linkage effect runs against the larger economy has to do with how the loss of its exports and the loss of its imports affect its production at home. More specifically, the larger economy has the higher export multiplier on $71\%$ of the pairs and the higher import multiplier on $65\%$. And when these multipliers are higher, they are higher by a sizeable amount, with the mean difference being about a third of a dollar of activity per lost dollar of sales, at $1.8$ against $1.5$ dollars, and about one dollar per lost dollar of purchases, at $2.3$ against $1.2$. The gap in each multiplier also widens with the size ratio, the log ratio of the smaller economy's export multiplier to the larger economy's falling with the log size ratio at a slope of $-0.05$ through the origin, and that of their import multipliers at $-0.13$, on either side of the $-0.10$ of the linkage effect.

The export multiplier of the larger economy is higher because of its greater dependence on production at home, compared with the smaller economy. More specifically, the export sectors of the larger economies tend to source, directly and indirectly, a greater share of their inputs from domestic sectors. This home sourcing means that a contraction of the export sectors, after the trade severance, generates larger upstream effects in the domestic economy. Since the content of a country runs higher in its own producers than in foreign producers of the same sector, an order moved home raises that content, by part (i) of Lemma~\ref{lem:m-homechain} of the paper.\footnote{The domestic share of intermediate purchases, the home share $\varphi_c$ of Section~\ref{subsec:power_size} of the paper, has a median of $0.75$ across the eighty economies, and together with the leakage it accounts for nine tenths of the variation of the export multiplier across them. Part (ii) of Lemma~\ref{lem:m-homechain} of the paper leads one to expect the fit. Its formula tracks the export multiplier that the solved equilibria return, averaged over each country's partners in proportion to its severed sales to them. Across the eighty economies the formula and the export multiplier correlate at $0.95$, the export multiplier exceeds the formula by a median of two percent, and the export multiplier rises with the formula at an elasticity of $1.2$. The condition of part (i) holds as well, at the benchmark orders of the 2022 trade table and full delivery, for every one of the $3{,}949$ active country-sectors against every foreign producer of the same sector, and by a factor of at least two.} The domestic share of intermediate purchases correlates with log size at $0.39$. Irish producers buy $36\%$ of their inputs at home, the lowest share among the eighty economies, and Chinese producers buy $93\%$, the highest. A lost dollar of sales costs Ireland somewhat more than $1$ dollar of activity on average over its partners, and it costs China somewhat less than $3$.

The import multiplier of the larger economy is higher for a very different reason. Here the problem is one of missing inputs, since a lost purchase leaves the producers that use the input short of it, and these producers then produce less, order less from their suppliers at home, and deliver less to their customers at home. A lost purchase of a final good, by contrast, idles no production at home. Per dollar of intermediate purchases, the import multiplier exceeds the export multiplier for all but five of the trading partners with intermediate purchases, at a median of $2.7$ against $1.6$ dollars of activity, as the ordering of Lemma~\ref{lem:m-ordering} of the paper, which holds producer by producer, leads one to expect.\footnote{Over the $6{,}320$ trading partners of the $3{,}160$ pairs the export multiplier has a median of $1.63$ dollars of activity per lost dollar and the import multiplier a median of $1.61$. The two medians lie so close only because the final goods among the severed purchases dilute the import multiplier.} More of what the larger economy buys from its partner is inputs, at a median intermediate share of $0.63$ of its severed purchases against $0.58$ for the smaller economy. And the producers left short of inputs in the larger economy have more of their suppliers and customers at home, so that more of the shortfall stays within the country (Section~\ref{oa:rates-content}).

Note that the higher export multiplier and the higher import multiplier of the larger economies are an empirical phenomenon, which emerges out of the particular patterns of trade in the 2022 trade table. Our model does not in principle sign the relation between the size of an economy and its export multiplier or its import multiplier, since the sizes of the economies leave the pattern of their trade open (Lemma~\ref{lem:m-fiber} of the paper). It works just as well for a world in which the large economies do not source much from home and buy mostly final goods from their partners, thereby having the smaller multipliers. Two statistics of the benchmark table alone, namely the domestic content of each trading partner's severed sales and the domestic content of the shortfalls that its severed purchases induce, absorb nearly all of the association of the two multipliers with size (Section~\ref{oa:rates-content}).

The trade relation between the United States and Mexico illustrates the role of the export multiplier and the import multiplier in generating the linkage effect. A lost dollar of Mexico's sales to the United States carries $1.5$ dollars of Mexican activity, namely the dollar of output itself and $50$ cents of inputs that Mexican producers supply along the chain of orders at home, since much of what Mexico sells to the United States is assembled from parts bought abroad. A lost dollar of American sales to Mexico carries $2.0$ dollars of American activity. The dollar of output and the American parts behind it make up $1.9$ dollars of it. The other ten cents come back from Mexico, where the assembly lines running short of American parts withhold orders from their other American suppliers and deliveries from their American customers. On the purchase side Mexico's import multiplier exceeds that of the United States, as the smaller economy's does on a third of the pairs. The reason is that the United States buys from Mexico mostly final goods, which idle no American production, and Mexico buys mostly parts, which idle its assembly lines and their Mexican suppliers. A lost dollar of purchases therefore costs the United States $1.3$ dollars of activity and costs Mexico $1.6$ dollars. The United States loses more per dollar on the sales side and Mexico on the purchase side, and the two differences nearly offset, so that over the two flows Mexico loses about as many dollars as the United States. The linkage effect of the pair is close to one, and what separates the two countries is the base effect, which is to say how large the severed trade is next to each economy. Mexico's severed sales and severed purchases are $15$ and $10$ percent of Mexican activity, and those of the United States are $0.6$ and $0.9$ percent of American activity. The export elasticity and the import elasticity of Proposition~\ref{prop:m-exposure} of the paper are therefore about $0.2$ each for Mexico against about $0.01$ each for the United States, and the severance costs Mexico $10.9\%$ of its activity and costs the United States $0.6\%$.

\subsection{The export multiplier, the import multiplier, and domestic content}
\label{oa:rates-content}

Two statistics of the benchmark table account for the finding of Section~\ref{oa:linkage-size} that the larger economy of a pair usually loses more on each lost dollar. Write $\mathbf m^{\top}_c:=\mathbbm 1_c[\mathbf I-\mathbf A(\mathbf I-\boldsymbol\beta)]^{-1}$ for the content row \eqref{eq:m-content} at the benchmark orders and full delivery. For a pair $\{c,c'\}$, write $X_i:=\sum_{b\in c'}Z_{ib}$ for the severed sales of a seller $i$ of country $c$ and $M_j:=\sum_{i\in c'}Z_{ij}$ for the severed purchases of a producer $j$ of country $c$. The domestic content of the severed sales is the content of country $c$ in a dollar of them,
\begin{equation}
\bar m^{X}_c:=\frac{\sum_{i\in c}m_{c,i}X_i}{X_{cc'}}.
\label{eq:m-content-sales}
\end{equation}
On the purchase side an unrecovered dollar of intermediate purchases idles at least a dollar of the buyer sector's output (Lemma~\ref{lem:m-shortfall} of the paper). The buyer sector then withholds deliveries from its domestic customers in proportion to the shares of its activity that each of them takes, and the customers idle output in turn. The induced shortfalls $M^{\mathrm{ind}}_j=M_j+\sum_{k\in c}M^{\mathrm{ind}}_k\,Z_{kj}/s_k$ collect that chain over the producers of country $c$. The recursion runs over the producers of country $c$ alone, so it omits the shortfalls induced abroad and the paths that leave the country and return to it, which can matter at first order. The domestic content of the induced shortfalls is the content of country $c$ in them per severed dollar of purchases,
\begin{equation}
\bar m^{M}_c:=\frac{\sum_{j\in c}m_{c,j}M^{\mathrm{ind}}_j}{M_{cc'}}.
\label{eq:m-content-purchases}
\end{equation}
Both statistics are built from the sourcing shares, the shares of each seller's activity that its customers take, the leakages, and the partner's shares of the rows and columns of the table, so neither moves when every flow and every activity is scaled by a common factor. Table~\ref{tab:s6_severed_dollar} regresses the two solved multipliers on the two statistics and on size.

\begin{table}[!htbp]
\centering
\small
\begin{adjustbox}{max width=\textwidth}
\begin{tabular}{lcccccc}
\toprule
 & \multicolumn{3}{c}{Log export multiplier $\ell^{\downarrow}_c$} & \multicolumn{3}{c}{Log import multiplier $\ell^{\uparrow}_c$} \\
\cmidrule(lr){2-4}\cmidrule(lr){5-7}
 & (1) & (2) & (3) & (4) & (5) & (6) \\
\midrule
Log economic size & $0.048$ &  & $0.003$ & $0.147$ &  & $0.001$ \\
 & $(0.012)$ &  & $(0.001)$ & $(0.026)$ &  & $(0.000)$ \\[2pt]
Domestic content of the severed sales (log) &  & $1.007$ & $0.988$ &  &  &  \\
 &  & $(0.012)$ & $(0.012)$ &  &  &  \\[2pt]
Domestic content of the induced shortfalls (log) &  &  &  &  & $1.005$ & $1.003$ \\
 &  &  &  &  & $(0.002)$ & $(0.002)$ \\[2pt]
\midrule
$R^{2}$ & $0.287$ & $0.981$ & $0.982$ & $0.228$ & $0.998$ & $0.998$ \\
Sides of pairs & $6{,}319$ & $6{,}319$ & $6{,}319$ & $6{,}319$ & $6{,}319$ & $6{,}319$ \\
\bottomrule
\end{tabular}

\end{adjustbox}
\caption{Least-squares fits of the log export multiplier and the log import multiplier of \eqref{eq:m-rates} of the paper on log size, on the domestic content of the severed sales \eqref{eq:m-content-sales}, and on the domestic content of the induced shortfalls \eqref{eq:m-content-purchases}. Each observation is one side of a bilateral severance, and standard errors clustered by the country whose export multiplier or import multiplier it is are in parentheses. The export multipliers and the import multipliers are computed at the solved equilibrium of every severance and the two statistics from the benchmark table alone. The one side whose severed purchases are final goods only, and whose purchase statistic is therefore zero, is omitted. Baseline calibration $\tau=0.30$, $\rho=-1$, $\varrho=-1$, $\delta=0.10$.}
\label{tab:s6_severed_dollar}
\end{table}

Across the $6{,}319$ sides, with one observation per side rather than the pair-level ratios of Section~\ref{oa:linkage-size}, the elasticity of the export multiplier with respect to size is $0.05$ and that of the import multiplier $0.15$ (columns 1 and 4).\footnote{The import multiplier is the more closely related to size of the two, against the intuition that a larger home market would offer a lost sale more alternative buyers. Of the elasticity of $0.15$ of the import multiplier $\ell^{\uparrow}_c$, the intermediate share of the severed purchases contributes $0.06$, since the severed purchases of the larger side are intermediate inputs somewhat more often, at a median intermediate share of $0.63$ against $0.58$. The domestic chain behind each intermediate dollar contributes $0.08$, split evenly between the domestic content of the idled buyer sectors' output and the downstream shares $Z_{kj}/s_k$ through which their shortfalls reach their domestic customers. The domestic content of the severed sales accounts for nearly the whole elasticity of $0.05$ of the export multiplier $\ell^{\downarrow}_c$.} The export multiplier follows the domestic content of the severed sales closely. The elasticity is $1.01$ and the $R^{2}$ is $0.98$, and on nineteen sides in twenty the two differ by less than two and a half percent. Where the export multiplier exceeds the content by more, the excess is the part of the export multiplier that the statistic leaves out, the shortfalls that a side's lost sales start at the partner's buyer sectors and that travel back to it through the chain of deliveries. China's export multiplier against Korea, Chinese Taipei, Vietnam, Singapore, and Malaysia exceeds the content of its severed sales by about a tenth, since the producers of these partners build Chinese inputs into their output and fall short when the inputs are cut. Entered beside the content, log size takes a coefficient of $0.003$. The import multiplier follows the domestic content of the induced shortfalls at an elasticity of $1.01$ and an $R^{2}$ of $0.998$. It exceeds the statistic by a median of one percent, and by more than five percent on one side in twenty, where the severed purchases fall on one input sector and the complementarity of inputs adds to the shortfall. Entered beside the statistic, log size falls from $0.15$ to $0.001$.\footnote{The one side that buys only final goods from its partner has no induced shortfall, and Table~\ref{tab:s6_severed_dollar} omits it. At the country level, with each country's export multipliers and import multipliers averaged over its partners in logs, the six fits return the same coefficients, with $R^{2}$ of $0.998$ and $1.000$ in columns 3 and 6.}

The two statistics correlate with log size at $0.56$ and $0.64$ across the economies, and that correlation accounts for the association with size in columns 1 and 4. Once the two statistics are known, the level of activity adds almost nothing to either multiplier on the 2022 trade table. The export multiplier divided by the domestic content of the severed sales lies between $0.999$ and $1.023$ on $90\%$ of the sides, and it exceeds $1.1$ on $26$ sides, $17$ of them sides of pairs trading more than a billion dollars, where the shortfalls the lost sales start abroad come back in force. The import multiplier divided by the domestic content of the induced shortfalls has a median of $1.01$, a ninety-fifth percentile of $1.05$, and a ninety-ninth of $1.10$. The excess can reflect the paths the domestic chain omits, the rerouting of the effective demand, the fulfillment rates, and the curvature of the loss, and it is largest where one input sector accounts for most of a side's severed purchases. Without the downstream part of the chain, the deliveries an idled buyer sector withholds from its domestic customers, the content of the severed purchases falls short of the import multiplier by a median factor of $1.62$.

\subsection{The exact decomposition on the 2022 trade table}
\label{oa:exact-decomposition}

The two exact identities of the paper, the elasticity identity \eqref{eq:m-exact-exposure} of Proposition~\ref{prop:m-exposure} and the decomposition \eqref{eq:m-exact-decomposition} of Proposition~\ref{prop:m-decomposition}, hold to numerical precision on every pair of the 2022 trade table. We solve all $3{,}160$ severances again at the baseline calibration, tightening the tolerance of the rationed exchange from $10^{-9}$ to $10^{-12}$ in the relative row-margin error. At each selected equilibrium we solve the two one-sided fulfillment fixed points of Lemma~\ref{lem:m-onesided} of the paper at the realized effective demand, each by the benchmark path from full delivery with the retention ratios of the spared side's buyer sectors restored by the unrecovered share of their severed sourcing \eqref{eq:m-onesided-theta}. We then record the three parts \eqref{eq:m-three-parts} of each side's loss and the split \eqref{eq:m-shortfall-split} of its shortfall loss into the part its lost purchases cause, the part its lost sales cause, and the interaction. We record as well the two multipliers and the interaction rate, the fulfillment floor $\underline h_c$, the intermediate share $\iota_c$, the two trade shares, and the three ratios. Each part of the loss propagates its own direct loss through the operator $\mathbf R^{*}_\omega$ at the solved fulfillment rates, so the parts and the anatomy of Proposition~\ref{prop:m-anatomy} of the paper are one computation.

On every pair the three log ratios sum to the log ratio of the two solved vulnerabilities to within $1.8\times10^{-15}$, and at the median pair the three parts of each side's loss sum to the solved loss to within a relative $2.9\times10^{-9}$. Against the stored vulnerabilities of Section~\ref{sec:statistics} of the paper, whose rationed exchange was balanced to $10^{-9}$, the vulnerabilities at the tighter tolerance differ by a median relative $3.5\times10^{-6}$ and a ninetieth percentile of $4.4\times10^{-4}$. On $38$ pairs a side differs by more than $1\%$, every one of them a pair whose gross trade is below $7.5$ million dollars, where the balancing residual of the stored run is of the order of the loss itself. The asymmetry moves by at most $0.088$ and by more than $0.01$ on only five pairs, one pair changes sign, and the asymmetry index moves by $4\times10^{-5}$. Every statistic of Sections~\ref{sec:power_size} and~\ref{sec:power_balance} of the paper that uses the export multipliers, the import multipliers, or the three ratios comes from the equilibria solved at the tighter tolerance.

The one-sided fulfillment rates lie above the equilibrium rates on every side up to the tolerance of the fixed-point solver, and the shortfall loss of every side splits into the part its lost purchases cause, the part its lost sales cause, and the interaction to within $10^{-12}$ of itself. The interaction is small, for over the $6{,}320$ sides it is a median $1.2\times10^{-6}$ of the shortfall loss in absolute value and $4.1\times10^{-7}$ of the dollar loss, below one tenth of one percent of the shortfall loss on $98.8\%$ of the sides and below one percent on $99.9\%$. On the $6{,}128$ sides of the pairs that trade at least a million dollars it lies between $-0.003$ and $0.022$ of the shortfall loss and never exceeds $0.3$ percent of the dollar loss. On the $192$ sides of the $96$ pairs that trade less than a million dollars it runs from $-0.55$ to $0.03$ of a shortfall loss of a few thousand dollars. On these pairs the three fixed points carry numerical noise of the same order, as the anatomy identity itself shows, which holds to $0.003$ percent on the larger pairs and only to $8$ percent on these.\footnote{The rationed exchange is balanced to $10^{-12}$ of the largest flow of the trade table, so on a producer whose intermediate sales are a few dollars the capacity bound of the solver can bind by about $10^{-6}$ of its output, which registers as a shortfall of that order in one of the three fixed points and not in the others. The interaction is the one quantity of the decomposition that such a discrepancy reaches at first order, since it is a difference of nearly equal shortfall losses.} The interaction rate $\ell^{\times}_c$ has a median of $6.7\times10^{-7}$, and on the pairs that trade at least a million dollars it is at most $0.08$ percent of the sum $\Sigma_c$ of the side that carries it, so the interaction term of \eqref{eq:m-exact-exposure} of the paper is at most that fraction of the other two on these pairs.

Table~\ref{tab:s6_exact_rates_summary} summarizes the export multipliers, the import multipliers, and the interaction rates over the $6{,}320$ sides. The export multiplier runs from $1.1$ to $3.5$ with a median of $1.6$, above the fulfillment floor $\underline h_c$ of Lemma~\ref{lem:m-rate-bounds} of the paper on every side. The fulfillment floor never falls below $0.83$ and lies above $0.98$ on $95\%$ of the sides, so the export multiplier is close to the content of a country in its export sectors' output. The import multiplier has a median of $1.6$ and a right tail, and it satisfies the bound $\ell^{\uparrow}_c\ge\iota_c$ on every side. It exceeds $5$ on $57$ sides and $10$ on no side.\footnote{Charging every shortfall to the lost purchases, as a simpler convention would, inflates the import multiplier of a side whose purchases are small relative to its sales, since the shortfalls that its own lost sales start at the partner's buyer sectors are then divided by a small denominator. On the 2022 trade table the import multiplier would exceed $10$ on eight sides under that convention and reach $216$, against $8.3$ under the causal split, and the trade-imbalance effect of \eqref{eq:m-exact-decomposition} of the paper would reach $4.5$ log points against $2.2$. On the typical pair the two conventions differ little, for the import premium of a side moves by a median of $0.002$ between them.} The import premia are small, as Section~\ref{subsec:balance_table} of the paper reports, and the sum of a pair's two premia is below $0.05$ in absolute value on $15.7\%$ of the pairs. The trade-imbalance effect of \eqref{eq:m-exact-decomposition} departs from one by $\mathcal B_{cc'}(\mathcal G_c+\mathcal G_{c'})/(1-\mathcal B_{cc'}\mathcal G_{c'})$, so a small sum of the two premia keeps it near one wherever the denominator stays away from zero.

\begin{table}[!htbp]
\centering
\small
\begin{tabular}{lrrrrr}
\toprule
 & Minimum & 5th pct. & Median & 95th pct. & Maximum \\
\midrule
Export multiplier $\ell^{\downarrow}_c$ & $1.06$ & $1.25$ & $1.63$ & $2.08$ & $3.45$ \\
Import multiplier $\ell^{\uparrow}_c$ & $0.00$ & $0.57$ & $1.61$ & $3.26$ & $8.30$ \\
Import premium $\mathcal G_c$ & $-1.00$ & $-0.44$ & $0.00$ & $0.28$ & $0.59$ \\
Fulfillment floor $\underline h_c$ & $0.83$ & $0.98$ & $1.00$ & $1.00$ & $1.00$ \\
Intermediate share $\iota_c$ & $0.00$ & $0.27$ & $0.61$ & $0.89$ & $1.00$ \\
Interaction rate $\ell^{\times}_c$ ($\times10^{3}$, pairs above \$1m) & $-1.78$ & $0.00$ & $0.00$ & $0.06$ & $2.87$ \\
\bottomrule
\end{tabular}

\caption{Quantiles over the $6{,}320$ sides of the $3{,}160$ bilateral severances. The rows are the export multiplier $\ell^{\downarrow}_c$, the import multiplier $\ell^{\uparrow}_c$, and the import premium $\mathcal G_c=(\ell^{\uparrow}_c-\ell^{\downarrow}_c)/\Sigma_c$, with $\Sigma_c=\ell^{\downarrow}_c+\ell^{\uparrow}_c+2\ell^{\times}_c$. The next rows are the lowest fulfillment rate $\underline h_c$ among the producers of country $c$ that sell to the partner and the intermediate share $\iota_c$ of the severed purchases. The last row is the interaction rate $\ell^{\times}_c$ of \eqref{eq:m-rates} of the paper, in thousandths, over the sides of the $3{,}064$ pairs that trade at least a million dollars. Every quantity is computed at the selected equilibrium. Baseline calibration $\tau=0.30$, $\delta=0.10$, and $\rho=\varrho=-1$.}
\label{tab:s6_exact_rates_summary}
\end{table}

Table~\ref{tab:oa-exact-variance} splits the variance of the log loss ratio into the variances and covariances of the three log ratios, an exact accounting since the three sum to the log loss ratio on every pair. Reversing a pair negates all three terms, so the moments are taken over both orientations of every pair, which centers every term at zero. The log loss ratio has a variance of $5.26$. The base effect accounts for $117.3\%$ of it, the linkage effect at balanced trade for $3.7\%$, and the trade-imbalance effect for $0.9\%$. Twice the covariances account for the rest, $-20.7\%$ between the first two ratios, $0.7\%$ between the last two, and $-1.8\%$ between the first and the third. The linkage effect at balanced trade correlates with the base effect at $-0.50$, since the larger economy of a pair usually loses more per lost dollar, and the base effect alone correlates with the log loss ratio at $0.979$.

The trade-imbalance effect has a median absolute value of $0.051$ log points and a ninetieth percentile of $0.29$, against medians of $0.26$ for the linkage effect at balanced trade and $1.63$ for the base effect. Its largest value, $2.2$ log points, is an extreme case in which the split of the linkage effect between its value at balanced trade and the trade-imbalance effect is ill conditioned, while their sum is not. On a pair whose trade runs nearly in one direction and where the country with the trade surplus loses far more per dollar of purchases than per dollar of sales, the denominator $1-\mathcal B_{cc'}\mathcal G_{c'}$ of the trade-imbalance effect is close to zero, which happens on a handful of pairs of the 2022 trade table. The logarithms of the decomposition are finite only on pairs that trade in both directions, so a pair trading in one direction alone would need a separate treatment of its zero flow. The sign of the asymmetry agrees with the sign of the base effect on $93.8\%$ of the pairs and on $98.1\%$ of the $2{,}700$ pairs at $|\mathcal H_{cc'}|>0.2$.

\begin{table}[!htbp]
\centering
\small
\begin{tabular}{lrrrr}
\toprule
 & Base & Linkage effect at & Trade & Log loss ratio \\
 & effect & balanced trade & imbalance & \\
\midrule
Base effect & $+1.173$ & $-0.103$ & $-0.009$ & $+1.060$ \\
Linkage effect at balanced trade & $-0.103$ & $+0.037$ & $+0.003$ & $-0.063$ \\
Trade imbalance & $-0.009$ & $+0.003$ & $+0.009$ & $+0.003$ \\
\bottomrule
\end{tabular}

\caption{Second moments of the three log ratios of \eqref{eq:m-exact-decomposition} of the paper over both orientations of the $3{,}160$ pairs, divided by the second moment of the log loss ratio. The diagonal entries are the variance shares of the three terms, the off-diagonal entries their covariance shares, and the last column the covariance of each term with the log loss ratio, so that the last column sums to one. Baseline calibration $\tau=0.30$, $\delta=0.10$, and $\rho=\varrho=-1$.}
\label{tab:oa-exact-variance}
\end{table}

Grouped as in \eqref{eq:m-size-split} of the paper, the variance of the log loss ratio splits into $117.3\%$ for the base effect, $5.3\%$ for the linkage effect, and $-22.5\%$ for twice their covariance. Through the origin, the log loss ratio rises with the base effect at a least-squares slope of $0.904$, and the linkage effect falls with it at $-0.096$. Of the $-0.096$, the linkage effect at balanced trade accounts for $-0.088$ and the trade-imbalance effect for $-0.008$. The bound \eqref{eq:m-balance-bound} of the paper holds on every pair, with a median of $0.059$ and a ninetieth percentile of $0.258$, and the ratio of the effect of the trade imbalance \eqref{eq:m-balance-transform} of the paper to the bound has a median of $0.51$.

\subsection{The largest bilateral deficits}
\label{oa:largest-deficits}

We now turn to the twenty largest bilateral deficits in dollars of the 2022 trade table. Table~\ref{tab:s6_exact_decomposition} reports the logs of the three ratios of \eqref{eq:m-exact-decomposition} of the paper for each of these pairs, together with each trading partner's import premium. Each pair is oriented so that the country with the trade deficit is country $c$, which makes the trade balance \eqref{eq:direct_trade_asymmetry} of the paper positive on every row. The trade balances of these pairs run from $0.19$ to $0.80$ and lie above the median of $0.35$ across all pairs on thirteen of the twenty, so the trade balance has room to move power here. Among the twenty pairs, the country with the trade deficit is more powerful in fourteen, and in each of the fourteen it is also the larger economy, while in each of the other six it is the smaller economy. With each ratio measured by the absolute value of its log, the base effect is the largest of the three in nineteen of the twenty pairs, the linkage effect at balanced trade is the largest in one, and the trade-imbalance effect is the largest in none. And erasing the trade imbalance (Section~\ref{oa:balanced-trade}) changes which country is the more powerful side in none of the twenty pairs.

Which trading partner the trade imbalance favors is set by the two import premia, and the two premia are set by what each trading partner buys from the other. Where the country with the trade deficit buys mostly inputs, whose loss idles production at home, the two premia sum to a positive number and the trade imbalance works against it, as in the American deficit with Canada and the Japanese deficits with Australia and the United Arab Emirates. Where the country with the trade deficit buys mostly final goods, whose loss idles no production at home, its import premium is negative, and the trade imbalance moves power little or works in its favor, as in the American deficits with China, Mexico, and Vietnam, on which the two premia have opposite signs. Overall, the trade imbalance works against the country with the trade deficit in sixteen of the twenty pairs, and in its favor only in the American deficits with Mexico, Vietnam, Switzerland, and Italy.

And the effect of the trade imbalance is large only where the trade balance is large and the two premia sum to a large number. On no pair of the table does the trade imbalance move the loss ratio by as much as half a log point. The largest movement, the $0.47$ log points of the United Kingdom and Norway, whose trade runs nine to one in Norway's favor, rests mostly on the British import premium. Britain buys from Norway mostly oil and gas, both of them inputs, so its import premium is $0.54$, while Norway's is $0.13$. By \eqref{eq:m-balanced-loss} of the paper the trade imbalance raises what a dollar of lost trade costs Britain by $43\%$ and lowers what it costs Norway by $10\%$.\footnote{The effect of the trade imbalance on this pair, measured by the normalized difference \eqref{eq:m-balance-transform} of the paper, is $0.23$, which is about half of the $0.43$ that the bound \eqref{eq:m-balance-bound} allows, since Norway's import premium is far smaller than Britain's.} But Britain's economy is about six times the size of Norway's, and Britain remains the more powerful side of the pair. The severance costs Britain about $2.2\%$ of its activity and Norway about $3.5\%$ of its.

\begin{table}[!htbp]
\centering
\small
\begin{adjustbox}{max width=\textwidth}
\begin{tabular}{llrrrrrrrr}
\toprule
Deficit country & Surplus country & Deficit (\$bn) & $\mathcal B_{cc'}$ & $\mathcal G_c$ & $\mathcal G_{c'}$ & Base & Linkage effect & Trade & Advantage \\
 & & & & & & effect & at balanced & imbalance & \\
 & & & & & & (log) & trade (log) & (log) & \\
\midrule
United States & China & 321 & $+0.45$ & $-0.24$ & $+0.26$ & $0.00$ & $-0.91$ & $+0.01$ & $+0.42$ \\
United States & Canada & 151 & $+0.21$ & $+0.13$ & $+0.09$ & $-2.48$ & $+0.22$ & $+0.04$ & $+0.80$ \\
United States & Mexico & 123 & $+0.19$ & $-0.22$ & $+0.04$ & $-2.86$ & $+0.03$ & $-0.03$ & $+0.89$ \\
China & Chinese Taipei & 114 & $+0.45$ & $+0.23$ & $+0.09$ & $-3.33$ & $+0.71$ & $+0.14$ & $+0.84$ \\
Mexico & China & 105 & $+0.71$ & $-0.07$ & $+0.28$ & $+2.86$ & $-1.05$ & $+0.17$ & $-0.76$ \\
United States & Vietnam & 86 & $+0.74$ & $-0.40$ & $+0.34$ & $-3.38$ & $-0.87$ & $-0.07$ & $+0.97$ \\
United States & India & 84 & $+0.39$ & $-0.06$ & $+0.15$ & $-1.97$ & $-0.20$ & $+0.04$ & $+0.79$ \\
United Kingdom & China & 69 & $+0.50$ & $-0.14$ & $+0.20$ & $+2.08$ & $-0.78$ & $+0.03$ & $-0.58$ \\
India & China & 66 & $+0.46$ & $+0.08$ & $+0.05$ & $+1.97$ & $-0.41$ & $+0.06$ & $-0.67$ \\
United Kingdom & Norway & 66 & $+0.80$ & $+0.54$ & $+0.13$ & $-1.82$ & $+0.86$ & $+0.47$ & $+0.24$ \\
Japan & Australia & 58 & $+0.63$ & $+0.37$ & $-0.02$ & $-0.93$ & $+0.66$ & $+0.19$ & $+0.04$ \\
France & Germany & 47 & $+0.22$ & $+0.02$ & $+0.03$ & $+0.40$ & $-0.01$ & $+0.01$ & $-0.20$ \\
Japan & United Arab Emirates & 45 & $+0.75$ & $+0.31$ & $-0.16$ & $-2.16$ & $+1.00$ & $+0.10$ & $+0.49$ \\
Canada & China & 42 & $+0.38$ & $-0.08$ & $+0.40$ & $+2.47$ & $-1.09$ & $+0.13$ & $-0.64$ \\
United States & Switzerland & 42 & $+0.43$ & $-0.24$ & $+0.19$ & $-3.18$ & $-0.50$ & $-0.02$ & $+0.95$ \\
China & Australia & 40 & $+0.19$ & $+0.49$ & $+0.05$ & $-2.65$ & $+1.07$ & $+0.10$ & $+0.63$ \\
Vietnam & China & 40 & $+0.19$ & $+0.36$ & $+0.19$ & $+3.37$ & $-0.22$ & $+0.11$ & $-0.93$ \\
China & Singapore & 40 & $+0.25$ & $+0.24$ & $+0.09$ & $-3.65$ & $+0.80$ & $+0.08$ & $+0.88$ \\
United States & Italy & 39 & $+0.40$ & $-0.23$ & $+0.16$ & $-2.38$ & $-0.44$ & $-0.03$ & $+0.89$ \\
China & Saudi Arabia & 38 & $+0.33$ & $+0.50$ & $-0.04$ & $-3.25$ & $+1.48$ & $+0.14$ & $+0.67$ \\
\bottomrule
\end{tabular}

\end{adjustbox}
\caption{The twenty largest bilateral deficits in dollars of the 2022 trade table, each pair oriented so that the country with the trade deficit is country $c$. The trade balance $\mathcal B_{cc'}$ is \eqref{eq:direct_trade_asymmetry} of the paper, and $\mathcal G_c$ and $\mathcal G_{c'}$ are the import premia \eqref{eq:m-gap} of the country with the trade deficit and of the country with the trade surplus. The three columns that follow are the logs of the three ratios of \eqref{eq:m-exact-decomposition} of the paper, the base effect $Q(c')/Q(c)$, the linkage effect at balanced trade $\Sigma_c/\Sigma_{c'}$, and the trade-imbalance effect $(1+\mathcal B_{cc'}\mathcal G_c)/(1-\mathcal B_{cc'}\mathcal G_{c'})$. The three logs sum to the log loss ratio. The last column is the advantage of the country with the trade deficit, which is the negated asymmetry $-\mathcal H_{cc'}$ and is positive when the country with the trade deficit is the more powerful of the two. Every quantity is computed at the solved equilibrium of the trade severance. Baseline calibration $\tau=0.30$, $\rho=-1$, $\varrho=-1$, $\delta=0.10$.}
\label{tab:s6_exact_decomposition}
\end{table}

The largest deficit in dollars is between the United States and China, the one pair of the twenty on which the linkage effect at balanced trade, not the base effect, is the largest. Recall from Section~\ref{subsec:power_size_base} of the paper that the 2022 trade table puts the two countries at near parity in gross activity, at $45.3$ trillion dollars for the United States against $45.1$ for China, so the base effect nearly vanishes and the linkage effect at balanced trade decides which country is more powerful. China loses more than the United States on both sides of the severed trade. A lost dollar of sales costs China $2.7$ dollars of activity and costs the United States $1.8$. And the purchase side separates the two further, for a lost dollar of purchases costs China $4.6$ dollars of activity and costs the United States $1.1$. What China buys from the United States is mostly intermediate inputs, whose loss idles Chinese production and the domestic suppliers that depend on it. What the United States buys from China is mostly final goods, whose loss leaves American production untouched. Which is why the Chinese multipliers ($2.7+4.6$) sum to two and a half times the American ones ($1.8+1.1$).

The deficit offsets almost none of this difference, for the American import premium of $-0.24$ and the Chinese premium of $0.26$ are nearly equal in size and opposite in sign. The trade balance moves power through the sum of the two premia \eqref{eq:m-balance-transform} of the paper, and the sum is only $0.02$, against $0.67$ for the United Kingdom and Norway. The deficit tilts the American losses toward purchases, which cost the United States less per dollar than its sales, and the Chinese losses toward sales, which cost China less per dollar than its purchases. It therefore lowers what a dollar of lost trade costs each of them, the United States by $11\%$ and China by $12\%$, and a trade imbalance that lowers the two costs in nearly the same proportion leaves their ratio where it was. Put simply, the United States would forgo the larger of the two flows, its purchases from China, and would still lose less than half as much, at $0.6\%$ of its activity against China's $1.4\%$. Which is what makes the country with the trade deficit the more powerful of the two partners, at an asymmetry of $0.42$, with China bearing about $70\%$ of the pair's combined proportional loss.

\subsection{Unrecovered shares that differ across the severed links}
\label{oa:unequal-shares}

The rationed exchange of Section~\ref{subsec:model_rerouting} of the paper leaves unrecovered the same share $\tau$ of the trade displaced on every severed link. We now let the unrecovered share differ from one severed link to another. We show that the elasticity identity of Proposition~\ref{prop:m-exposure} and the decomposition of Proposition~\ref{prop:m-decomposition} of the paper hold with the lost sales and the lost purchases of each country in place of the fraction $\tau(1-\delta)$ of its severed sales and of its severed purchases. A difference between the two flows of a pair in how readily they are redirected then enters the loss ratio as a trade imbalance does.

Write $\tau_{ib}\in(0,1]$ for the unrecovered share of the severed link $i\to b$. The budgets \eqref{eq:m-coltarget} and the capacities \eqref{eq:m-rowtarget} of the paper become
\begin{equation}
\begin{aligned}
t^{(\omega)}_{rb}&=\alpha_{rb}-(1-\delta)\!\!\sum_{i\in\mathcal N_r:\,i\to b\text{ severed}}\!\!\tau_{ib}\,a_{ib},\\
g^{(\omega)}_i&=s_i-f^{o}_i-(1-\delta)\!\!\sum_{b:\,i\to b\text{ severed}}\!\!\tau_{ib}\,Z_{ib}.
\end{aligned}
\label{eq:oa-unequal-targets}
\end{equation}
The lost sales \eqref{eq:m-lost-sales} of seller $i$ become $\xi_{\omega,i}=(1-\delta)\sum_{b:\,i\to b\text{ severed}}\tau_{ib}Z_{ib}$, and the rest of the model stays as Section~\ref{sec:model} of the paper defines it. Write
\begin{equation}
\xi^{\downarrow}_c:=\sum_{i\in c}\xi_{\omega,i},
\qquad
\xi^{\uparrow}_c:=\sum_{i\in c'}\xi_{\omega,i}
\label{eq:oa-lost-trade}
\end{equation}
for the lost sales of country $c$ and for its lost purchases, which are the lost sales of the sellers of country $c'$. With a common unrecovered share the lost sales and the lost purchases of country $c$ are $\tau(1-\delta)X_{cc'}$ and $\tau(1-\delta)M_{cc'}$. We measure the export multiplier, the import multiplier, and the interaction rate of \eqref{eq:m-rates} of the paper per dollar of this lost trade,
\begin{equation}
\ell^{\downarrow}_c:=\frac{L^{\downarrow}_c}{\xi^{\downarrow}_c},
\qquad
\ell^{\uparrow}_c:=\frac{L^{\uparrow}_c}{\xi^{\uparrow}_c},
\qquad
\ell^{\times}_c:=\frac{L^{\times}_c}{\xi^{\downarrow}_c+\xi^{\uparrow}_c}.
\label{eq:oa-unequal-rates}
\end{equation}
Here $L^{\downarrow}_c$ and $L^{\uparrow}_c$ are the two parts \eqref{eq:m-two-parts} of the dollar loss of country $c$, and $L^{\times}_c$ is the interaction that remains. The one-sided fulfillment rules behind these parts add back to the retention ratios of the spared buyer sectors the unrecovered fraction of their purchases of each input, as \eqref{eq:m-onesided-theta} of the paper does, and that fraction is now $(\alpha_{rj}-t^{(\omega)}_{rj})/\alpha_{rj}$. The sum $\Sigma_c$ and the import premium $\mathcal G_c$ are built from the multipliers \eqref{eq:oa-unequal-rates} as in Section~\ref{subsec:power_surplus} of the paper. For country $c'$ the two flows exchange roles, so that its lost sales are $\xi^{\uparrow}_c$ and its lost purchases are $\xi^{\downarrow}_c$. We call the normalized difference of the lost purchases and the lost sales of country $c$ the balance of the lost trade,
\begin{equation}
\mathcal B^{\xi}_{cc'}:=\frac{\xi^{\uparrow}_c-\xi^{\downarrow}_c}{\xi^{\uparrow}_c+\xi^{\downarrow}_c}
=\psi\!\left(\frac{\xi^{\uparrow}_c}{\xi^{\downarrow}_c}\right).
\label{eq:oa-lost-balance}
\end{equation}
Write $\bar\tau_X:=\xi^{\downarrow}_c/[(1-\delta)X_{cc'}]$ and $\bar\tau_M:=\xi^{\uparrow}_c/[(1-\delta)M_{cc'}]$ for the mean unrecovered shares of the flow $X_{cc'}$ and of the flow $M_{cc'}$, each mean weighted by the benchmark flows of the links.

\begin{lemma}[Unrecovered shares that differ across the severed links]
\label{lem:oa-unequal-shares}
Suppose that countries $c$ and $c'$ trade in both directions and that the rationed exchange of the trade severance $c'\notnotleftrightarrow c$ leaves unrecovered the share $\tau_{ib}\in(0,1]$ of the trade displaced on each severed link $i\to b$. Lemma~\ref{lem:m-fixedpoint} and Propositions~\ref{prop:m-monotone} and~\ref{prop:m-anatomy} of the paper then hold with the budgets and the capacities \eqref{eq:oa-unequal-targets}, and at the selected equilibrium
\begin{equation}
\begin{aligned}
\gamma^{(c'\notnotleftrightarrow c)}_c
&=\ell^{\downarrow}_c\,\frac{\xi^{\downarrow}_c}{Q(c)}+\ell^{\uparrow}_c\,\frac{\xi^{\uparrow}_c}{Q(c)}+\ell^{\times}_c\,\frac{\xi^{\downarrow}_c+\xi^{\uparrow}_c}{Q(c)},\\
r_{cc'}
&=\frac{Q(c')}{Q(c)}\cdot\frac{\Sigma_c}{\Sigma_{c'}}\cdot\frac{1+\mathcal B^{\xi}_{cc'}\,\mathcal G_c}{1-\mathcal B^{\xi}_{cc'}\,\mathcal G_{c'}}.
\end{aligned}
\label{eq:oa-unequal-decomposition}
\end{equation}
The balance of the lost trade is the bounded sum \eqref{eq:m-oplus} of the trade balance and the normalized difference of the mean unrecovered shares of the two flows,
\begin{equation}
\mathcal B^{\xi}_{cc'}=\mathcal B_{cc'}\oplus\psi\!\left(\frac{\bar\tau_M}{\bar\tau_X}\right).
\label{eq:oa-lost-balance-sum}
\end{equation}
The bound \eqref{eq:m-size-bound} of the paper holds as stated, and the bound \eqref{eq:m-balance-bound} holds with $\mathcal B^{\xi}_{cc'}$ in place of $\mathcal B_{cc'}$.
\end{lemma}

\begin{proof}
We first show that the rationed exchange and the sanction equilibrium are well defined. On every sector block the budgets and the capacities \eqref{eq:oa-unequal-targets} are consistent in the sense of \eqref{eq:m-margincons} of the paper, since each side of that identity is the benchmark flow of the block less the unrecovered dollars $(1-\delta)\tau_{ib}Z_{ib}$ of its severed links. The flow matrix that keeps the fraction $1-\tau_{ib}(1-\delta)\ge\delta$ of the benchmark flow on every severed link and the whole benchmark flow on every other link is positive on the benchmark support, and it meets every budget and every capacity. The proofs of Proposition~\ref{prop:m-balancing} and of part \textnormal{(i)} of Proposition~\ref{prop:m-leak} of the paper use the unrecovered share only through this matrix and these targets. The balanced flow blocks therefore exist, are unique, and take the biproportional form, which gives Lemma~\ref{lem:m-fixedpoint}. In Lemma~\ref{lem:m-kernel} of the paper the sector-block sum \eqref{eq:m-blocksum} becomes the budget $t^{(\omega)}_{rj}\le\alpha_{rj}$, so every column of the effective demand matrix $\mathbf A_\omega$ sums to at most one. The row identity of that lemma, $[\mathbf A_\omega(\mathbf I-\boldsymbol\beta)\mathbf s]_i+f_{\omega,i}=s_i-\xi_{\omega,i}$, holds with the lost sales as redefined. Lemmas~\ref{lem:m-capacity} and~\ref{lem:m-rule} and the proofs of Propositions~\ref{prop:m-monotone} and~\ref{prop:m-anatomy} of the paper use the unrecovered share only through these block sums and this row identity, so the two propositions hold.

We next prove the first line of \eqref{eq:oa-unequal-decomposition}. The retention ratio of a producer $j$ at full delivery is $\theta^{(\omega)}_{rj}(\mathbf 1)=t^{(\omega)}_{rj}/\alpha_{rj}$, the budget \eqref{eq:oa-unequal-targets} over the benchmark share. The second and the third steps of the proof of Lemma~\ref{lem:m-shortfall} of the paper then give $\lambda^{(\omega)}_j(\mathbf 1)\le1-(1-\beta_j)\sum_{r\in\mathcal R_j}(\alpha_{rj}-t^{(\omega)}_{rj})$, and its first and fourth steps give
\begin{equation}
(1-h_{\omega,j})\,s_j\;\ge\;(1-\delta)\!\!\sum_{i:\,i\to j\text{ severed}}\!\!\tau_{ij}\,Z_{ij},
\label{eq:oa-unequal-shortfall}
\end{equation}
the unrecovered part of the severed intermediate purchases of producer $j$. The one-sided fulfillment rule of country $c$ adds $(\alpha_{rj}-t^{(\omega)}_{rj})/\alpha_{rj}$ to the retention ratios of the producers $j$ of country $c'$, which brings each of these retention ratios to one at full delivery, and the rule of country $c'$ does the same for the producers of country $c$. Lemma~\ref{lem:m-onesided} of the paper and its proof then apply, with \eqref{eq:oa-unequal-shortfall} in place of Lemma~\ref{lem:m-shortfall} in its part \textnormal{(iii)}, and so does the split \eqref{eq:m-shortfall-split} of the shortfall loss. A seller outside the pair has no severed link, so the lost sales vanish outside the two countries, and the dollar loss of country $c$ has the three parts \eqref{eq:m-three-parts} of the paper. By the definition \eqref{eq:oa-unequal-rates} these three parts and the split of the shortfall loss sum to $\ell^{\downarrow}_c\xi^{\downarrow}_c+\ell^{\uparrow}_c\xi^{\uparrow}_c+\ell^{\times}_c(\xi^{\downarrow}_c+\xi^{\uparrow}_c)$, and dividing by $Q(c)$ gives the first line. By the same argument the dollar loss of country $c'$ is $\ell^{\downarrow}_{c'}\xi^{\uparrow}_c+\ell^{\uparrow}_{c'}\xi^{\downarrow}_c+\ell^{\times}_{c'}(\xi^{\downarrow}_c+\xi^{\uparrow}_c)$. The weighted-average argument of the proof of Proposition~\ref{prop:m-exposure} of the paper, with the weights $(\ell^{\uparrow}_{c'}+\ell^{\times}_{c'})\xi^{\downarrow}_c$ and $(\ell^{\downarrow}_{c'}+\ell^{\times}_{c'})\xi^{\uparrow}_c$, then gives the bound \eqref{eq:m-size-bound}.

We now prove the second line. By \eqref{eq:oa-lost-balance}, $\xi^{\uparrow}_c=\tfrac12(\xi^{\uparrow}_c+\xi^{\downarrow}_c)(1+\mathcal B^{\xi}_{cc'})$ and $\xi^{\downarrow}_c=\tfrac12(\xi^{\uparrow}_c+\xi^{\downarrow}_c)(1-\mathcal B^{\xi}_{cc'})$. Each country sells to the other and every unrecovered share is positive, so both lost flows are positive and $|\mathcal B^{\xi}_{cc'}|<1$. The proof of Proposition~\ref{prop:m-decomposition} of the paper then applies line by line, with $\xi^{\downarrow}_c$, $\xi^{\uparrow}_c$, and $\mathcal B^{\xi}_{cc'}$ in place of $\tau(1-\delta)X_{cc'}$, $\tau(1-\delta)M_{cc'}$, and $\mathcal B_{cc'}$. That proof and the argument that follows it in Appendix~\ref{app:m-proof-decomposition} of the paper give the second line, the normalized difference \eqref{eq:m-balance-transform}, and the bound \eqref{eq:m-balance-bound}, each with the balance of the lost trade in place of the trade balance.

We last prove \eqref{eq:oa-lost-balance-sum}. By the definition of the two mean unrecovered shares, $\xi^{\uparrow}_c/\xi^{\downarrow}_c=(\bar\tau_M/\bar\tau_X)(M_{cc'}/X_{cc'})$. Part \textnormal{(ii)} of Lemma~\ref{lem:m-psi} of the paper turns the normalized difference of this product into the bounded sum of $\psi(\bar\tau_M/\bar\tau_X)$ and $\psi(M_{cc'}/X_{cc'})$, and the second of the two is the trade balance \eqref{eq:direct_trade_asymmetry}. This proves the lemma.
\end{proof}

Consider first a trade severance in which one flow of the pair is harder to redirect than the other, so that the unrecovered share is $\tau_X$ on every link of the flow $X_{cc'}$ and $\tau_M$ on every link of the flow $M_{cc'}$. The lost sales of country $c$ are then $\tau_X(1-\delta)X_{cc'}$, its lost purchases are $\tau_M(1-\delta)M_{cc'}$, and \eqref{eq:oa-lost-balance-sum} reads
\begin{equation}
\mathcal B^{\xi}_{cc'}
=\mathcal B_{cc'}\oplus\psi\!\left(\frac{\tau_M}{\tau_X}\right)
=\frac{\mathcal B_{cc'}\,(\tau_M+\tau_X)+(\tau_M-\tau_X)}{(\tau_M+\tau_X)+\mathcal B_{cc'}\,(\tau_M-\tau_X)}.
\label{eq:oa-two-flows}
\end{equation}
A flow that is harder to redirect loses more dollars at both of its ends, as sales for the country that sells on the flow and as purchases for the country that buys on it. The difference between the two unrecovered shares therefore tilts the lost dollars of both countries toward the harder flow, as a trade imbalance tilts them toward the larger flow, and the two tilts add in the bounded sum. When the purchases of country $c$ are harder to redirect than its sales ($\tau_M>\tau_X$), the tilt is that of a trade deficit of country $c$. By the second line of \eqref{eq:oa-unequal-decomposition} the tilt reaches the loss ratio through the two import premia. The balance of the lost trade works against the country whose lost purchases exceed its lost sales exactly when the two import premia sum to a positive number, and its effect, measured by the normalized difference as in the paper, is at most $|\mathcal B^{\xi}_{cc'}|\max\{|\mathcal G_c|,|\mathcal G_{c'}|\}$. With a common unrecovered share the normalized difference $\psi(\tau_M/\tau_X)$ is zero, the balance of the lost trade is the trade balance, and \eqref{eq:oa-unequal-decomposition} returns Propositions~\ref{prop:m-exposure} and~\ref{prop:m-decomposition} of the paper.

The unrecovered share can also differ across the products of a flow, some products being harder to redirect than others. The two lines of \eqref{eq:oa-unequal-decomposition} hold as before, with the mean unrecovered shares $\bar\tau_X$ and $\bar\tau_M$ in the balance of the lost trade. The products that are harder to redirect now make up more of the lost dollars of a flow than of its severed dollars. The export multipliers and the import multipliers, which average what a lost dollar costs over the lost dollars of a flow, therefore move with the unrecovered shares, and the shares reach the loss ratio through the linkage effect at balanced trade as well as through the balance of the lost trade. The grouped exercise of Section~\ref{sec:robustness} is of this kind, since it gives fuels a low unrecovered share and export-controlled manufactures a high one, and it gives the two groups different attenuations as well. The lemma holds when the attenuation differs across the severed links, with the factor $(1-\delta_{ib})\tau_{ib}$ of the link in place of $(1-\delta)\tau_{ib}$ in \eqref{eq:oa-unequal-targets} and in the lost sales, since the proof uses the attenuation only through its positivity and through the lost sales. The fractions $\xi^{\downarrow}_c/X_{cc'}$ and $\xi^{\uparrow}_c/M_{cc'}$ of the two flows that are lost then replace the mean unrecovered shares in \eqref{eq:oa-lost-balance-sum}. On the 2022 trade table the grouped exercise moves the asymmetry index $\Psi$ from $0.582$ to $0.577$ and leaves the sign of the asymmetry unchanged on $99\%$ of the pairs (Table~\ref{tab:rob-master}).

We measure what a difference between the unrecovered shares of the two flows does to power by solving each of the $3{,}160$ trade severances of the 2022 trade table again, with an unrecovered share of $0.40$ on one flow of the pair and $0.20$ on the other, in both orders. One flow then leaves unrecovered twice the share of the other, and the mean of the two shares is the baseline $0.30$. Over the $6{,}320$ cases the asymmetries move by a median of $0.009$ in absolute value, against a median absolute asymmetry of $0.63$, and by $0.040$ at the ninetieth percentile. The more powerful side changes in $70$ cases, none of them among the $5{,}400$ cases whose baseline asymmetry exceeds $0.2$ in absolute value, and the asymmetry index $\Psi$ stays at $0.582$ in both orders.

The second line of \eqref{eq:oa-unequal-decomposition} gives the movement of the asymmetry at given multipliers, and the solved severances show how far the multipliers themselves respond to the unrecovered shares. We hold the sizes, the sums $\Sigma_c$ and $\Sigma_{c'}$, and the import premia at the values of the solved equilibria of the baseline, as Section~\ref{subsec:balance_table} of the paper does for the trade balance, and we replace the trade balance by the balance of the lost trade \eqref{eq:oa-two-flows}. The asymmetries then move by a median of $0.009$ and by $0.040$ at the ninetieth percentile, as they do in the solved severances. The two movements differ by less than $0.002$ in $94\%$ of the cases and correlate at $0.998$ across them, and the loss of each side is within half a percent of the loss that the baseline multipliers give on $95\%$ of the sides. A flow that leaves unrecovered twice the share of the other therefore moves power about as much as the trade imbalance of the pair does, since erasing the trade imbalance at the same multipliers moves the asymmetries by a median of $0.011$ (Section~\ref{oa:surplus_counterfactual}). At a ratio of three between the two unrecovered shares the baseline multipliers give a median movement of $0.014$ and a ninetieth percentile of $0.061$. The asymmetry of the United States and China moves from $0.424$ to $0.428$ when the Chinese purchases are the harder flow and to $0.419$ when the American purchases are. The code is \texttt{\_driver\_unequal\_shares.py}, with \texttt{\_numbers\_unequal\_shares.py} for the statistics, and the pair-level output is \texttt{unequal\_shares.csv}.

\subsection{The specification grid of the trade-balance regression}
\label{oa:balance-robustness}

\begin{figure}[!htbp]
\centering
\includegraphics[width=0.62\linewidth]{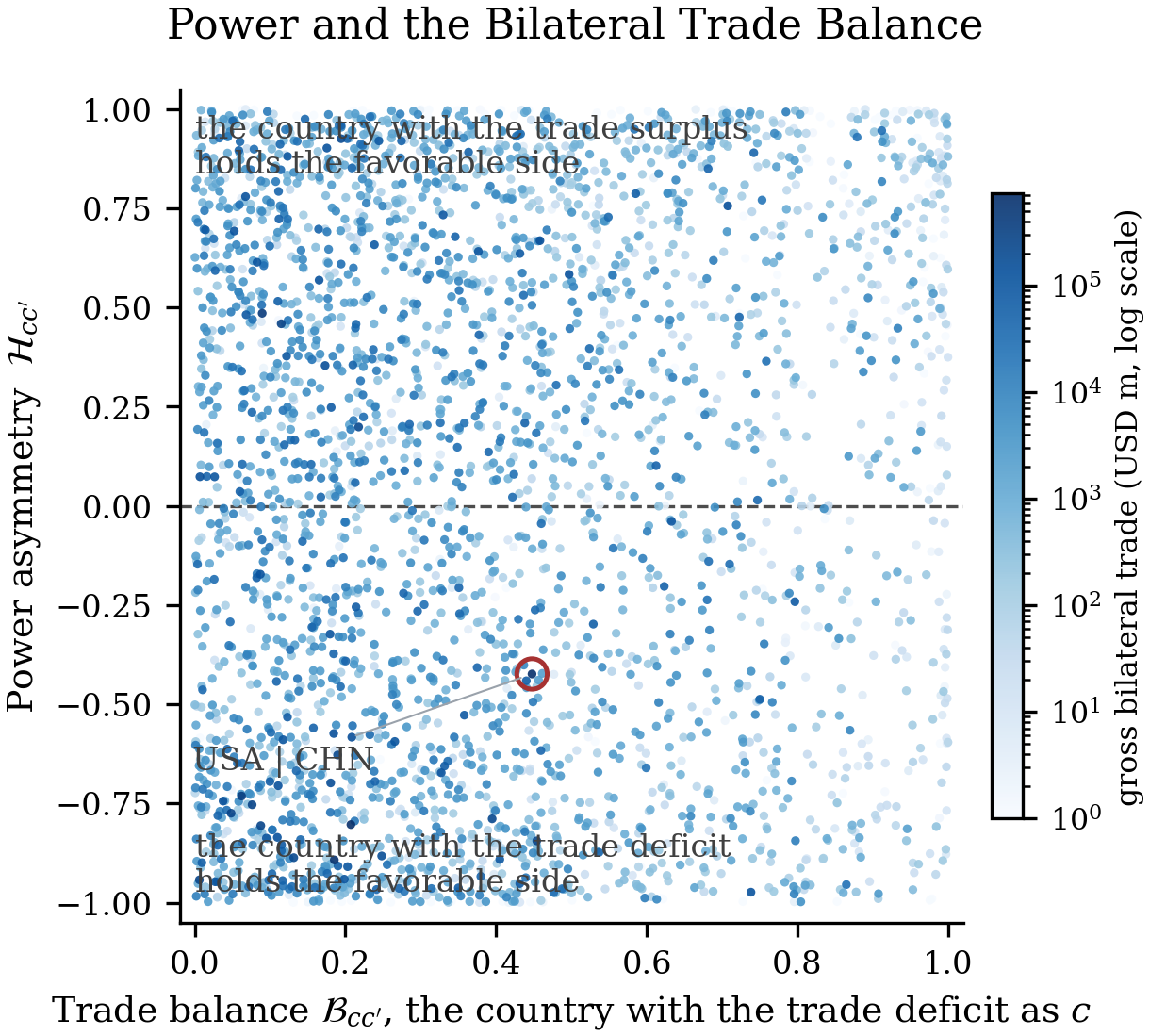}
\caption{The trade balance and power. Each point is one of the $3{,}160$ pairs, with the country with the trade deficit as country $c$, so that a positive asymmetry means the country with the trade surplus holds the favorable side. Color is gross bilateral trade on a log scale, and the circled pair is the United States and China. Baseline calibration $\tau=0.30$, $\delta=0.10$, and $\rho=\varrho=-1$.}
\label{fig:s6_balance_power}
\end{figure}

Figure~\ref{fig:s6_balance_power} plots the asymmetry of every pair against its trade balance, each pair oriented so that the country with the trade deficit is country $c$. At any value of the trade balance the asymmetries range over nearly the whole interval, so which country runs the trade surplus carries little information about which country holds the favorable side. The country with the trade surplus holds the favorable side on $1{,}708$ of the $3{,}160$ pairs, or $54\%$ of them.

The regressions of power on the trade balance that Section~\ref{subsec:balance_table} of the paper cites form a descriptive grid of $270$ fits. The grid crosses five measures of the trade imbalance with three weightings of the pairs and three network-position controls, entered one at a time or all together. The single-control cells run five estimators and the joint-control cells three, which gives $225$ fits with one control and $45$ with all three, and every fit uses two-way dyadic cluster-robust inference. The fits run on the unordered pairs $c<c'$ that carry positive cross-border trade and a positive total bilateral loss, which on the 2022 trade table are all $3{,}160$ pairs. The five measures are the trade balance $\mathcal B_{cc'}$, the raw dollar net flow, its signed logarithm, the sign of the trade imbalance, and the net flow divided by the size of the pair. The three controls are the continuous core score, total cross-border strength, and a nested shell index, which Section~\ref{app:structural_scores_repro} defines.

Across the $270$ fits the trade imbalance is significant at the five percent level in $62$, and in the $90$ nested comparisons it adds a median of one thousandth to the explained variance. Table~\ref{tab:oa-deficit-grid} breaks the grid down by measure and by weighting. The trade balance has the largest median coefficient, $+0.012$, and the median across the $180$ linear single-control fits is $-0.002$. These medians pool different outcome scales, measures, and estimators, which is why they differ from the coefficient of $+0.089$ in the fit that enters the trade balance against the core score alone.

\begin{table}[!htbp]
\centering
\small
\begin{tabular}{lccc}
\toprule
 & Share significant & Median coefficient & Median incremental $R^2$ \\
\midrule
Normalized balance $\mathcal B_{cc'}$ & $0.24$ & $+0.012$ & $0.003$ \\
Raw dollar net flow & $0.41$ & $-0.012$ & $0.001$ \\
Signed-log dollar flow & $0.09$ & $-0.002$ & $0.001$ \\
Sign of the trade imbalance & $0.22$ & $+0.004$ & $0.002$ \\
Net flow divided by pair size & $0.19$ & $+0.005$ & $0.001$ \\
\addlinespace[0.35em]
Equal pair weights & $0.33$ & $+0.02$ & \\
Trade weights & $0.17$ & $-0.01$ & \\
Economic-size weights & $0.19$ & $0.00$ & \\
\bottomrule
\end{tabular}
\caption{The trade imbalance specification grid by measure and by weighting. The first column is the share of the fits of a measure or a weighting, joint-control cells included, in which the trade imbalance is significant at $5\%$ under two-way dyadic cluster-robust inference. The coefficient medians combine standardized level and rank outcomes with the unstandardized binary outcome of the linear-probability fits. They run over the linear single-control fits, $36$ per measure and $60$ per weighting, since the logit coefficients are on the log-odds scale. The incremental-$R^2$ medians use the $18$ nested one-regressor comparisons of each measure.}
\label{tab:oa-deficit-grid}
\end{table}

\subsection{Construction of the network-position scores}
\label{app:structural_scores_repro}

The network-position controls of the specification grid of Section~\ref{oa:balance-robustness} are the continuous core score, total cross-border strength, and the nested shell index. A country-sector score $\mathbf q=(q_1,\dots,q_N)^\top$ built from the benchmark network is aggregated to the country level with benchmark activity weights,
\begin{equation}
\bar q_c
:=
\frac{
\sum_{i:\mathcal C(i)=c}s_i q_i
}{
\sum_{i:\mathcal C(i)=c}s_i
}.
\label{eq:country_structural_score}
\end{equation}
The bilateral gap in such a score is oriented as the asymmetry $\mathcal H_{cc'}$ is,
\begin{equation*}
\Delta\bar q_{cc'}
:=
\bar q_{c'}-\bar q_c.
\end{equation*}
Every score uses the benchmark network alone and no counterfactual loss. Their common input is the transaction matrix $\mathbf Z$ of \eqref{eq:m-flows} of the paper, producer and household columns alike. Summed over the sectors of each country it gives the country flow table, whose off-diagonal entries form the cross-border matrix $\mathbf T^{\mathrm{cb}}$ of \eqref{eq:country_flow_table_crossborder}.

The continuous core score is the leading Perron eigenvector of the symmetric, diagonal-free matrix $\mathbf T^{\mathrm{cb}}+(\mathbf T^{\mathrm{cb}})^{\!\top}$. We compute it by power iteration, and at the spectral radius of that matrix it satisfies
\begin{equation}
\spr\bigl(\mathbf T^{\mathrm{cb}}+(\mathbf T^{\mathrm{cb}})^{\!\top}\bigr)\,\mathrm{core}_c
=\sum_{c'}\bigl(T^{\mathrm{cb}}_{cc'}+T^{\mathrm{cb}}_{c'c}\bigr)\mathrm{core}_{c'}.
\label{eq:oa-corescore}
\end{equation}
A country scores high when it trades heavily with countries that themselves trade heavily, and the computation fits no core--periphery objective. Total cross-border strength is the weighted degree of the same matrix, $\sum_{c'}(T^{\mathrm{cb}}_{cc'}+T^{\mathrm{cb}}_{c'c})$. Its Spearman correlation with the core score is $+0.98$ across the eighty economies and $+0.97$ across the $3{,}160$ pair gaps, so the two are close substitutes rather than independent controls for network position. The alignment of the asymmetry with the core-score gap that Section~\ref{sec:robustness} reports is the cosine of the two dyadic arrays over the unordered pairs. For two signed dyadic arrays $x$ and $y$ on a set $\mathcal E$ of pairs it is
\[
\operatorname{cos}^{\circ}_{\mathcal E}(x,y)
:=
\frac{\sum_{(c,c')\in\mathcal E,\ c<c'}x_{cc'}y_{cc'}}
{\left(\sum_{(c,c')\in\mathcal E,\ c<c'}x_{cc'}^2\right)^{1/2}
\left(\sum_{(c,c')\in\mathcal E,\ c<c'}y_{cc'}^2\right)^{1/2}}.
\]
Reversing the order of a pair changes the signs of both entries, so the alignment does not depend on which country is listed first, and it is defined when both norms in the denominator are positive.

The nested shell index records the round in which an iterated pruning of the country network by strength removes each country. Starting at the smallest strength in the network, we remove every country whose strength within the surviving network is at or below that threshold, recompute the strengths, and repeat until no further country leaves. The threshold then rises to the smallest surviving strength. A country's shell index is the round in which it leaves the network, and the country network of the 2022 trade table has sixty shells.

\section{A Balanced-Trade Counterfactual}
\label{oa:balanced-trade}

Here we ask whether the country with the trade deficit would hold more power if its trade imbalance were erased.\footnote{Section~\ref{subsec:balance_table} of the paper measures the effect of the trade balance on power at the export multipliers and the import multipliers the solved equilibria return, which is as far as the decomposition reaches. But the two multipliers are solved on a trade table that carries the trade imbalance, so the trade imbalance can reach them, through the curvature of the input bundles and through the production shortfalls the two multipliers carry. Erasing the trade imbalance and solving the severance again lets the two multipliers respond.} Naturally, the first decision is how to amend the trade table, since a trade imbalance can be erased in many ways. It helps to think of the trade table as two sets of accounts. The columns record what each buyer spends with each of its suppliers, and the rows record where the output of each seller goes. An amendment that moved a seller's activity would move the sizes and with them the base effect, and one that moved a buyer's spending in a sector would move the proportions in which inputs are bought and therefore the levels of production. Ideally, we would like to measure the effect of the trade balance on power without altering either. To do this, we hold fixed the level of spending of every buyer on every sector, so that inputs are bought in the same proportions, and the activity of every seller, so that the sizes stay where they are.\footnote{Holding the gross volume of the pair's trade fixed as well, as Section~\ref{subsec:power_surplus} of the paper does, isolates the trade balance and loses nothing, since the gross volume does not enter the decomposition of Proposition~\ref{prop:m-decomposition} of the paper.}

To erase the trade imbalance, the buyers of the country with the trade deficit must buy less from its partner, and the buyers of the country with the trade surplus must buy more from the country with the trade deficit. And since a buyer's spending in every sector is held fixed, a buyer that buys less from one supplier must buy as much more from its other suppliers of the same sector, and a buyer that buys more from one supplier must buy as much less from them. These other suppliers can only be producers of the eighty economies, the buyer's own domestic producers and those of third countries, for the rest of the world sector appears in the model only as a buyer. How a buyer spreads the change over its other suppliers we take from the rationed exchange of Section~\ref{subsec:model_rerouting} and Appendix~\ref{app:m-wellposed} of the paper, so that its purchases from each of them rise, or fall, by the same percentage.

The re-sourcing changes what sellers sell to the buyers of the two countries. The export sectors of the country with the trade deficit sell more to the buyers of its partner, and the export sectors of the country with the trade surplus sell less to the buyers of the country with the trade deficit. The other suppliers of the buyers of the country with the trade deficit, its own producers and those of third countries, sell more to them, and the other suppliers of the buyers of the country with the trade surplus sell less. To keep the activity of every seller where it is, each of these changes is matched dollar for dollar by the opposite change in the seller's residual final demand $f^{o}_i$, its sales to the rest of the world sector and to inventories, which no sanction reaches. A seller that sells more to the buyers of the two countries sells as much less to the rest of the world sector, and a seller that sells less to them sells the difference to the rest of the world sector instead. Put differently, the rest of the world sector absorbs the net change in the sales of every seller, but it cannot stand in as a supplier, which is why the amendment needs both the re-sourcing of the buyers and the residual final demand of the sellers.

We now formulate the construction of a trade table on which a chosen pair of countries trades without an imbalance as an optimization problem. Our goal is not merely to keep the primitives that pertain to the two partners, the spending of their buyers by sector and their sizes, but also to disturb as little as possible the trade relations of other nations, since a change in them could in principle move the power between the two partners indirectly. Write $\mathcal J_{cc'}$ for the buyers of countries $c$ and $c'$, their producers and their two households, whose columns alone the amendment changes, so that the purchases of every other nation stay as they are, and write $b\in c$ for the buyers of country $c$. We define the balanced table as the trade table that comes closest to the benchmark while meeting three conditions, with closeness measured by relative entropy as in the RAS method of updating input--output tables \citep{oa-bacharach1970}. The three conditions are that every buyer keeps its spending in every sector, that each direction of the pair's trade carries half its gross volume, and that no seller's residual final demand turns negative. More specifically, the balanced table is
\begin{equation}
\mathbf Z^{\mathrm{bal}}
:=\arg\min_{\mathbf Z'\ge\mathbf 0}\ \sum_{i}\ \sum_{b\in\mathcal J_{cc'}} Z'_{ib}\log\frac{Z'_{ib}}{Z_{ib}}
\label{eq:m-balanced-program}
\end{equation}
subject to
\begin{equation}
\sum_{i\in\mathcal N_r} Z'_{ib}=\alpha_{rb}w_b,
\qquad
\sum_{i\in c}\sum_{b\in c'} Z'_{ib}=\sum_{i\in c'}\sum_{b\in c} Z'_{ib}=\tfrac12\,(X_{cc'}+M_{cc'}),
\qquad
\sum_{b\in\mathcal J_{cc'}}\bigl(Z'_{ib}-Z_{ib}\bigr)\le f^{o}_i,
\label{eq:m-balanced-constraints}
\end{equation}
for every sector $r$, every buyer $b$ of the pair, and every seller $i$, with the convention $0\log0=0$ (so that a flow the benchmark lacks stays absent).\footnote{Since a seller whose residual final demand is exhausted can sell no more to the buyers of the two countries at fixed sizes, some pairs cannot be balanced at half the gross volume. On such a pair we balance the two directions at the nearest feasible level (Section~\ref{app:m-proof-balanced-table}). On the 2022 trade table this happens only for the United Kingdom and Norway, where the British buyers' other suppliers of oil and gas have too little residual final demand to make up the cut in Norway's sales, and the pair balances with its gross trade higher by $29\%$.}

\begin{lemma}[The balanced table]
\label{lem:m-balanced-table}
Suppose that countries $c$ and $c'$ trade in both directions and that the constraints \eqref{eq:m-balanced-constraints} are met by some table that is positive wherever the benchmark table is. Then the program \eqref{eq:m-balanced-program} has exactly one solution, and the solution keeps every size, every producer's input bundle by sector, and every household budget. The solution scales each benchmark flow into the pair's buyers by three factors. The first is $d_c$ on the flows from country $c$ to the buyers of country $c'$, $d_{c'}$ on the flows from country $c'$ to the buyers of country $c$, and one on every other flow. The second is a factor of the seller that lies below one only where its residual final demand is exhausted, and the third is a factor of the buyer's sector cell that restores its spending.
\end{lemma}

\noindent Proof in Section~\ref{app:m-proof-balanced-table}.

By Lemma~\ref{lem:m-balanced-table} each pair has exactly one balanced table, and the three factors of the lemma give a simple way to compute this table. We compute the balanced table of each pair by block-coordinate ascent on the dual of the program, which updates the direction factors, the seller factors, and the cell factors in turn, much as the RAS method updates rows and columns, and we then solve the pair's trade severance on it (Section~\ref{oa:surplus_counterfactual}).\footnote{The factor of the buyer's sector cell carries out the re-sourcing of the buyers, while the direction factor $d_c$ multiplies, in every sector cell of the partner's buyers, the ratio of country $c$'s share to the remaining suppliers' share. The balanced table is therefore the benchmark table at a pair of pseudo-prices, one for each direction of the pair's trade, that erase the trade imbalance. Where no residual final demand binds, the balanced flow from a seller $i$ of country $c$ to a buyer $b$ of country $c'$ is $d_c Z_{ib}/[1+\sigma^{(\omega)}_{rb}(d_c-1)]$, with $\sigma^{(\omega)}_{rb}$ the share of country $c$ in buyer $b$'s sector-$r$ spending (Section~\ref{app:m-shares}). The flow from every other supplier $i$ of the cell is $Z_{ib}/[1+\sigma^{(\omega)}_{rb}(d_c-1)]$.}

\begin{figure}[!htbp]
\centering
\includegraphics[width=0.62\linewidth]{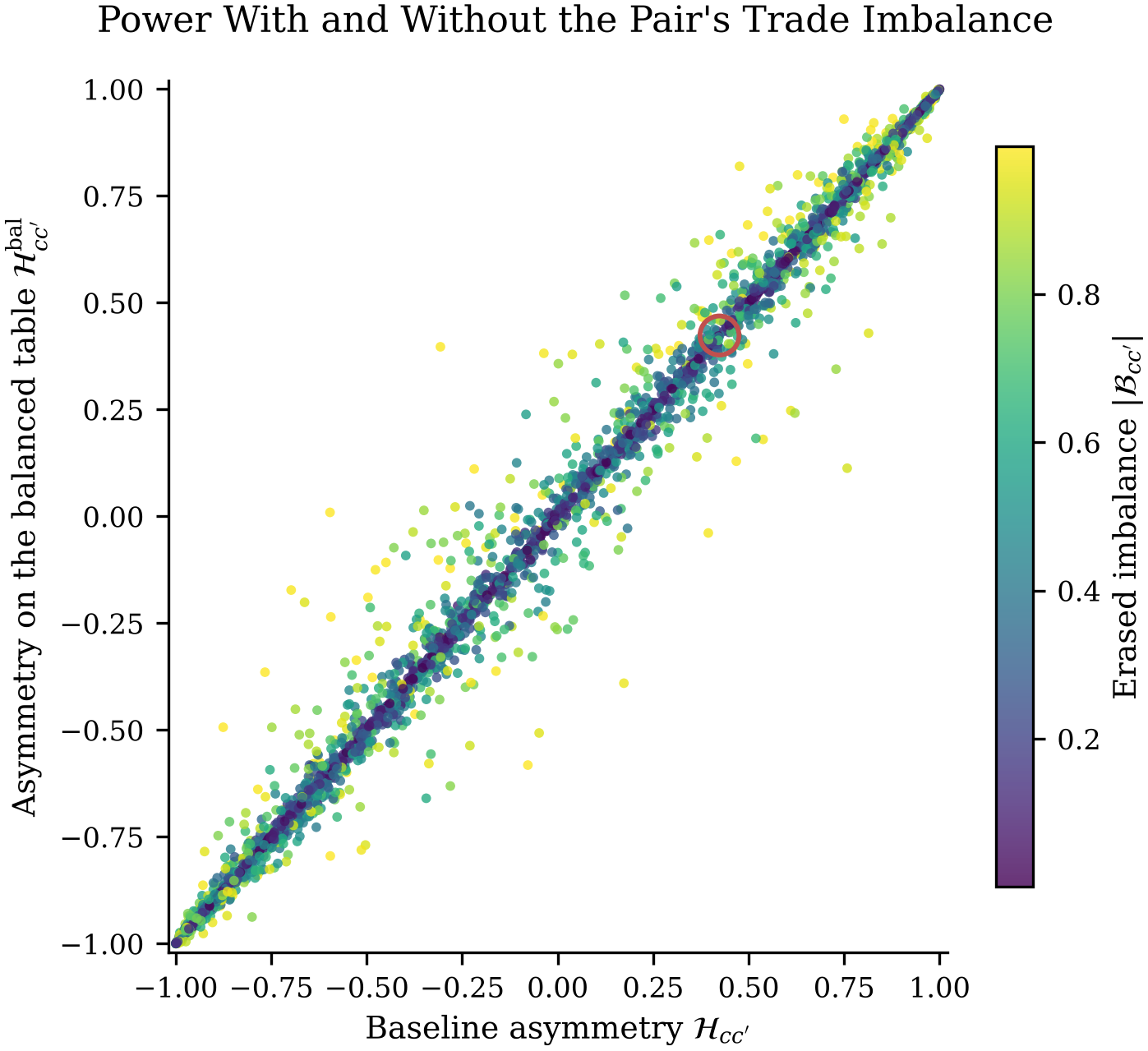}
\caption{The asymmetry $\mathcal H_{cc'}$ of each pair on the 2022 trade table against its asymmetry $\mathcal H^{\mathrm{bal}}_{cc'}$ on its balanced table \eqref{eq:m-balanced-program}. Each point is one of the $3{,}160$ pairs, the dotted line is the identity, color shows the original $|\mathcal B_{cc'}|$, and the circled pair is the United States and China.}
\label{fig:s6_deficit_erasure}
\end{figure}

We find that erasing a pair's trade imbalance barely moves its power. The asymmetries move by a median of $0.010$ and correlate with the baseline asymmetries at $0.995$.
Figure~\ref{fig:s6_deficit_erasure} plots the asymmetry of every pair on the 2022 trade table against its asymmetry on its balanced table. Note that the points lie close to the diagonal. It is not that the trade balance does not matter. It does, but very, very little. The trade balance reverses power on $56$ of the $3{,}160$ pairs and on $8$ of the $2{,}700$ pairs whose asymmetry exceeds $0.2$ in absolute value. The global asymmetry index $\Psi$ moves meagerly from $0.582$ to $0.581$. And in all this, it is the country with the trade surplus that gains power when its surplus is erased, on $1{,}692$ of the $3{,}160$ pairs, so that trade surpluses are, if anything, a minor source of weakness.

The results here echo those of Section~\ref{subsec:balance_table} of the paper, where we measured the effect of the trade balance on power through the decomposition of Proposition~\ref{prop:m-decomposition} of the paper, at the export multipliers and the import multipliers the solved equilibria return. In fact, the correspondence between the two empirical ways of examining the influence of the trade imbalance on power is uncanny. For every pair, the decomposition gives the movement in the asymmetry that erasing the trade imbalance would cause at the solved multipliers, and the balanced table gives the movement it causes when the severance is solved again. The two movements differ by less than $0.002$ on seven pairs in eight, against a median absolute asymmetry of $0.63$, and by less than $0.009$ on more than nine pairs in ten. They have the same sign on $98\%$ of the pairs (Section~\ref{oa:surplus_counterfactual}).

\subsection{Proof of Lemma~\ref{lem:m-balanced-table}}
\label{app:m-proof-balanced-table}

\begin{proof}
Write $\mathcal F$ for the set of tables that meet \eqref{eq:m-balanced-constraints}, that vanish wherever the benchmark table does, and that agree with the benchmark outside the columns of the pair's buyers. The constraints are linear, and the spending constraints bound every free cell by its buyer's spending in the cell's sector, so $\mathcal F$ is a compact convex polytope, and it is nonempty by hypothesis. On $\mathcal F$ the objective of \eqref{eq:m-balanced-program} is continuous, with the convention $0\log0=0$, and strictly convex, being a sum over the free cells of the support of the strictly convex functions $x\mapsto x\log(x/Z_{ib})$. A strictly convex continuous function on a nonempty compact convex set has exactly one minimizer, which proves existence and uniqueness.

We next show that the minimizer $\mathbf Z^{\mathrm{bal}}$ is positive on the support. The hypothesis provides a table $\mathbf Z^{+}\in\mathcal F$ positive on the support. Suppose that $Z^{\mathrm{bal}}_{ib}=0$ on some cell $(i,b)$ of the support. Since $\mathcal F$ is convex, the segment $\mathbf Z^{\mathrm{bal}}+\epsilon(\mathbf Z^{+}-\mathbf Z^{\mathrm{bal}})$ lies in $\mathcal F$ for $\epsilon\in[0,1]$. The derivative of the objective along the segment at $\epsilon=0$ is $-\infty$, because the derivative of $x\mapsto x\log(x/Z_{ib})$ at $x=0$ is $-\infty$ while every other term has a finite derivative there. The objective would fall along the segment, which contradicts the minimality of $\mathbf Z^{\mathrm{bal}}$.

Since the minimizer is positive on the support and the constraints are linear, the Karush--Kuhn--Tucker conditions hold at it. Write $y_{rb}$ for the Lagrange multipliers of the spending constraints, $\log d_c$ and $\log d_{c'}$ for those of the two directions, and $\mu_i\ge0$ for those of the residual final demands, where $\mu_i=0$ on every seller whose residual final demand stays positive. Stationarity in a cell $(i,b)$ of the support reads
\[
\log\frac{Z^{\mathrm{bal}}_{ib}}{Z_{ib}}+1
=y_{rb}+(\log d_c)\,\mathbbm 1[i\in c,\ b\in c']+(\log d_{c'})\,\mathbbm 1[i\in c',\ b\in c]-\mu_i,
\]
that is,
\[
Z^{\mathrm{bal}}_{ib}=Z_{ib}\,d_{ib}\,e^{-\mu_i}\,v_{rb},
\qquad
d_{ib}:=
\begin{cases}
d_c, & i\in c,\ b\in c',\\
d_{c'}, & i\in c',\ b\in c,\\
1, & \text{otherwise},
\end{cases}
\qquad
v_{rb}:=e^{y_{rb}-1}.
\]
This form is the one the lemma states, and the spending constraint of the cell fixes $v_{rb}=\alpha_{rb}w_b\big/\sum_{k\in\mathcal N_r}Z_{kb}d_{kb}e^{-\mu_k}$. The seller factor $e^{-\mu_i}$ equals one unless the residual final demand of seller $i$ is exhausted, by complementary slackness. The amended table keeps the row identity, $\sum_bZ^{\mathrm{bal}}_{ib}+f^{o,\mathrm{bal}}_i=s_i$ with $f^{o,\mathrm{bal}}_i:=f^{o}_i-\sum_{b\in\mathcal J_{cc'}}(Z^{\mathrm{bal}}_{ib}-Z_{ib})\ge0$ by the third constraint, so every activity $s_i$, and with it every size $Q(c)$, is unchanged. The spending constraints keep every producer's spending in every sector, $\alpha_{rj}w_j$, and every household's spending in every sector, $\alpha_{rc}F^{c}$, so every input bundle by sector and every household budget is unchanged. This proves the lemma.
\end{proof}

The levels at which the two directions of a pair's trade can be balanced form an interval. For a level $t>0$ write $\mathcal F(t)$ for the set of tables that meet \eqref{eq:m-balanced-constraints} with $\tfrac12(X_{cc'}+M_{cc'})$ replaced by $t$. The set $\{(\mathbf Z',t):\mathbf Z'\in\mathcal F(t)\}$ is defined by linear equalities and inequalities in $(\mathbf Z',t)$, so it is convex, and its projection on the level $t$, the set of feasible levels, is convex as well, an interval whose endpoints are the values of two linear programs. The construction of Section~\ref{oa:balanced-trade} balances a pair whose interval excludes $\tfrac12(X_{cc'}+M_{cc'})$ at the endpoint of the interval nearer to it, approached from inside the interval so that the hypothesis of the lemma holds.

\subsection{Computation and the comparison with the decomposition}
\label{oa:surplus_counterfactual}

We compute the balanced table of each pair, the solution of the program \eqref{eq:m-balanced-program}, by exact block-coordinate ascent on the dual of the program. Given the seller factors, each direction factor solves a strictly increasing equation in one variable, which we solve by bisection. Given the direction factors and the cell factors, the Lagrange multiplier of every seller's capacity constraint has a closed form, since the dual is separable across sellers once the other factors are fixed. The seller factor is one where the residual final demand covers the sales the seller would make at a factor of one, and it is the ratio of its capacity to those sales where it does not. The cycle of the three updates is a fixed-point iteration in the seller factors, which we accelerate by Anderson mixing with a memory of six iterates and stop when every residual final demand is nonnegative to within $10^{-10}$ of the pair's gross volume and the seller factors move by less than $10^{-9}$. The iteration converges in a median of four cycles and in at most $938$. The construction preserves every size, every input bundle by sector, and every household budget to machine precision, balances the two directions to within $10^{-8}$ of their sum, and keeps the gross volume of $3{,}159$ of the $3{,}160$ pairs to within one part in a million.

The direction factor on the exports of the country with the trade deficit has a median of $1.6$ across the pairs, with quartiles of $1.2$ and $2.6$, and the factor on the exports of the country with the trade surplus a median of $0.7$, with quartiles of $0.6$ and $0.9$. The residual final demand binds on a median of ten sellers per pair, on $133$ sellers for the United States and China, and on no seller on $27$ pairs, where the amendment reduces to the two direction factors and the renormalization of the cells. The relative entropy of the amendment is $0.07$ per dollar of the pair's gross trade at the median. On one pair, the United Kingdom and Norway, no table meets the constraints at half the gross volume. Norwegian sales of crude oil and gas to British buyers amount to $68.9$ billion dollars of the $90.0$ billion the buyers spend in that sector, and the buyers' other suppliers in the sector hold $14.9$ billion of residual final demand between them, so the cut in Norway's sales can be absorbed only in part. The feasible levels of the two directions, which Section~\ref{app:m-proof-balanced-table} shows to form an interval, are the values of the two linear programs that minimize and maximize the common level over the constraint set, and they run from $53.3$ to $610.0$ billion dollars against a half of $41.5$ billion. We balance the pair one tenth of one percent of its gross volume above the lower endpoint, at $53.4$ billion dollars in each direction, which raises its gross trade from $82.9$ to $106.8$ billion dollars.

We compare the movement of the asymmetries on the balanced tables with the movement that setting the trade-imbalance effect of \eqref{eq:m-exact-decomposition} of the paper to one produces at the solved export multipliers, import multipliers, and sizes, with no severance re-solved. The second movement has a median absolute value of $0.011$ and reverses power on $58$ pairs, against $0.010$ and $56$ for the first. The two movements differ by less than $0.002$ on $88\%$ of the pairs and by less than $0.009$ on $97\%$, correlate at $0.99$ across them, and have the same sign on $98\%$. The agreement rests on the causal split of the shortfalls of Section~\ref{subsec:power_size} of the paper. Were every shortfall charged to the lost purchases instead, the two movements would differ by less than $0.002$ on $76\%$ of the pairs and correlate at $0.82$. On the pairs whose trade runs almost entirely one way the movement at the solved multipliers would then miss the re-solved movement by more than one in the units of the asymmetry, since the import multiplier of the side that buys little would carry the shortfalls that its own sales cause. The country with the trade deficit gains power on $1{,}468$ pairs when the severance is re-solved and on $1{,}507$ at the solved export multipliers and import multipliers. At the solved multipliers it gains exactly when the two import premia sum to a positive number, by \eqref{eq:m-balance-transform} of the paper, and that sign predicts the re-solved direction on $98\%$ of the pairs. The code is \texttt{\_cf\_balanced\_table.py}, with \texttt{\_cf\_balanced\_lp.py} for the pairs balanced away from half the gross volume and \texttt{\_assemble\_balanced.py} for the statistics and the figure, and the pair-level output is \texttt{balanced\_table\_pairs.csv}.

\subsection{Checks on the balanced tables}
\label{oa:balanced-benchmark}

The amended tables are checked without solving any severance again. Every producer's column sum and every household's budget match the benchmark to within $10^{-6}$ of the largest spending. The row identity of every seller, its activity as the sum of its deliveries to producers and households and its residual final demand, holds to within $10^{-6}$ of the largest activity. No residual final demand is negative, and the two directions of the pair's trade agree to within $10^{-8}$ of their sum. Every reported asymmetry is rebuilt from the two solved losses. Two independent runs of the construction and the solves, on different machines, cover $533$ pairs in common and agree on every asymmetry to within $3\times10^{-10}$.

The solves on the balanced tables use the baseline calibration and the balancing tolerance of the paper's runs, $10^{-9}$ in the relative row-margin error, and we measure the movements against the baseline solved at the tighter tolerance of Section~\ref{oa:exact-decomposition}. Against the baseline solved at the same tolerance as the balanced tables, the asymmetries move by a median of $0.010$ and correlate with the baseline at $0.995$. They change sign on $57$ of the $3{,}160$ pairs and on $8$ of the $2{,}699$ pairs whose baseline asymmetry exceeds $0.2$ in absolute value, and the asymmetry index moves from $0.582$ to $0.581$. The two baselines differ on two pairs. The asymmetry of Bulgaria and Costa Rica changes sign, from $+0.0005$ on the baseline of the stored run to $-0.0012$ at the tighter tolerance, and that of Bulgaria and Cameroon crosses $0.2$ in absolute value, from $-0.1997$ to $-0.2009$, which is why we count $56$ reversals and $2{,}700$ pairs above $0.2$.

\section{Flexible Prices and the Distribution of Economic Power}
\label{app:flex}

We here extend the model to a flexible-price setting, wherein the prescribed recovery and the fulfillment rule give way to substitution among sources, an endogenous intermediate intensity, and prices that clear every market. We solve the bilateral severances of the paper a second time in this economy, calibrated to the same benchmark network, and compare the distribution of power under the two closures. The fixed-price economy of Section~\ref{sec:model} of the paper rations quantities at posted prices and propagates the fulfillment shortfalls along the effective demand, so the comparison is one between two adjustment closures. The comparison does not isolate price flexibility on its own, and it does not model the passage of time. The two closures agree on the direction of power on most pairs, though reversals occur and bystanders can gain. On the pairs on which both sides lose under both closures, price adjustment relieves the favorable side by more than the weaker side on most of them, so the asymmetry of a relationship widens more often than it narrows. Sections~\ref{subsec:flex-economy} to~\ref{subsec:flex-calibration} set out the economy, the path along which we solve it, its welfare and incidence results, and the calibration of its within-sector curvatures. Section~\ref{subsec:flex-comparison} reports the comparisons, Section~\ref{subsec:flex-incidence-econ} sets out the mechanism behind them, Section~\ref{oa:flex-pivot} shows when a bystander expands, and Section~\ref{app:flex-world} records the computations, the bystander gains, and the severances on which the solver did not converge.

\subsection{The flexible-price economy}
\label{subsec:flex-economy}

A producer facing flexible prices can meet the loss of a supplier by buying more from the other sources of the same input, and the technology must say how substitutable those sources are. We assume the CES form among them, with a curvature $\rho_r\in(0,1)$ for each input sector. The fixed-price economy needed no such curvature, since it treats the sources within a sector as fungible and allocates the effective demand by the rationed exchange. Lemma~\ref{lem:m-curvature} in Appendix~\ref{app:m-coreproofs} of the paper shows that the exchange returns the same allocation under every within-sector CES criterion, each with its own pseudo-prices. That invariance concerns the allocation alone, and it does not make the input services the sources deliver invariant to the technology, which \eqref{eq:flex-technology} now specifies.

We write the technology one level deeper than Section~\ref{subsec:model_ces} of the paper does, so as to separate a producer's substitution among the sources of one input from its substitution across the inputs it uses. Producer $j$ combines its sector bundles with the benchmark shares $\alpha_{rj}$ of Section~\ref{subsec:model_benchmark} of the paper, and each sector bundle combines the sector's source varieties,
\begin{equation}
\eta_j
\;=\;
\Bigl(\sum_{r\in\mathcal R_j}\alpha_{rj}\,\eta_{rj}^{\,\rho}\Bigr)^{\frac{1}{\rho}},
\qquad
\eta_{rj}
\;=\;
\Bigl(\sum_{i\in\mathcal N_r}\Bigl(\frac{\check a_{ij}}{\alpha_{rj}}\Bigr)^{1-\rho_r}x_{ij}^{\,\rho_r}\Bigr)^{\frac{1}{\rho_r}}.
\label{eq:flex-technology}
\end{equation}
Here $q_{ij}$ is the quantity of variety $i$ that producer $j$ uses, $w_j=(1-\beta_j)s_j$ is its benchmark intermediate spending of Section~\ref{subsec:model_benchmark} of the paper, and $x_{ij}:=q_{ij}/(\alpha_{rj}w_j)$ is the quantity in benchmark units, for $i\in\mathcal N_r$ with $\alpha_{rj}>0$. The within-sector weight of source $i$ is $a_{ij}/\alpha_{rj}$ at the benchmark and $\check a_{ij}/\alpha_{rj}$ after the attenuation \eqref{eq:m-disrupted}. Output is $y_j=s_j\,G_j(\eta_j)$. Here $G_j(\kappa):=\bigl[\beta_j+(1-\beta_j)\,\kappa^{\varrho}\bigr]^{1/\varrho}$ is the outer layer of \eqref{eq:m-lambda-rule} with the specific input fixed at $\bar v_j=\beta_js_j$. At the unattenuated benchmark $q_{ij}=w_ja_{ij}$, every composite equals one, and output is $s_j$.

The outer nest, the sectoral shares, and the attenuation of the weights are those of Section~\ref{subsec:model_ces} of the paper, and the source nest with its curvature $\rho_r\in[\underline\rho_r,\overline\rho_r]\subset(0,1)$ is new. The curvature lies on the substitutable side of Cobb--Douglas and short of perfect substitution, as the two nests of the paper lie on the complementary side and short of Leontief. The sources of one input are varieties of the same good, which a buyer trades off against one another, while the inputs of different sectors are complements, which it trades off only at the finite elasticity $1/(1-\rho)$. Section~\ref{subsec:flex-calibration} calibrates the within-sector curvatures from tariff-identified trade elasticities, under a maintained mapping between the two. Since a buyer substitutes along the whole isoquant once prices move, \eqref{eq:flex-technology} is a global assumption on the technology, whereas the nested CES of the paper specifies a fulfillment response and is a local representation (Section~\ref{app:m-local-efficiency}).

Cost minimization directs to source $i$ the fraction of buyer $j$'s sector-$r$ input spending
\begin{equation}
\frac{\check a_{ij}}{\alpha_{rj}}\biggl(\frac{p_i}{P_{rj}}\biggr)^{\frac{\rho_r}{\rho_r-1}}
\qquad\text{with}\qquad
P_{rj}\;:=\;\Bigl(\sum_{i\in\mathcal N_r}\frac{\check a_{ij}}{\alpha_{rj}}\,p_i^{\,\frac{\rho_r}{\rho_r-1}}\Bigr)^{\frac{\rho_r-1}{\rho_r}}.
\label{eq:flex-share}
\end{equation}
The fraction falls in the source's relative price, since the exponent $\rho_r/(\rho_r-1)$ is negative for every $\rho_r\in(0,1)$. The sectoral indices $P_{rj}$ combine into the composite input price index
\begin{equation}
P^{m}_j
\;=\;
\Bigl(\sum_{r\in\mathcal R_j}\alpha_{rj}\,P_{rj}^{\,\frac{\rho}{\rho-1}}\Bigr)^{\frac{\rho-1}{\rho}}.
\label{eq:flex-indices}
\end{equation}
Here the across-sector exponent is the price dual of the curvature $\rho$. The implied elasticities are $1/(1-\rho_r)$ within sectors and $1/(1-\rho)$ across them, one half at the baseline.

Producer $j$ chooses its intermediate intensity $\eta_j$, its composite relative to benchmark, against the composite input price index $P^{m}_j$ and its own output price $p_j$, according to
\begin{equation}
G_j'(\eta_j)=(1-\beta_j)\,\frac{P^{m}_j}{p_j},
\qquad
y_j=s_j\,G_j(\eta_j).
\label{eq:flex-supply}
\end{equation}
The non-intermediate block is fully employed and earns an endogenous rent, the operating surplus per unit of the block, so supply slopes upward rather than standing vertical at the benchmark scale. A producer raises output by using more intermediates with its fixed block, at a marginal cost that rises because the outer curvature makes the intermediate composite and the block poor substitutes. The block itself does not move when prices do, by the specific-factor assumption of Section~\ref{subsec:model_ces} of the paper, and its rent absorbs the adjustment.\footnote{The rent is a quasi-rent in Marshall's sense, since it measures the scarcity of the immobile block and allocates the surplus rather than the factor \citep{oa-marshall1890}. Section~\ref{app:flex-closure} records the rent, the closed-form demands, and the household schedules.}

\begin{definition}[Flexible-price sanction equilibrium]
\label{def:flex-equilibrium}
On the maintained interior domain, a flexible-price sanction equilibrium is a price vector $\mathbf p>\mathbf 0$ and an input allocation satisfying four conditions.
\begin{enumerate}
\item[\textnormal{(i)}] Each producer minimizes input cost at $\mathbf p$ given the attenuated weights and its endowment. Its positive intermediate intensity satisfies \eqref{eq:flex-supply}.
\item[\textnormal{(ii)}] Each producer charges every buyer the same price, equal to marginal cost.
\item[\textnormal{(iii)}] Each household spends its fixed budget according to its benchmark-share demand schedule, which Section~\ref{app:flex-closure} states, with severed-link weights attenuated as for producers. Residual demand $\mathbf f^{o}$ remains fixed in nominal terms.
\item[\textnormal{(iv)}] Every goods market clears.
\end{enumerate}
\end{definition}

Uniform pricing makes a displaced seller's discount a general discount to every buyer it still serves. Buyer-specific discounts, such as those on Russian crude redirected after the 2022 sanctions, are bargaining outcomes outside the model. Pricing at marginal cost is a maintained competitive assumption and not a normalization, since a markup over marginal cost would enter the producer's first-order condition for intermediates rather than relabel rent as profit.

Every quantity in Definition~\ref{def:flex-equilibrium} is a function of prices, so market clearing leaves one equation per good, and a flexible-price sanction equilibrium is a zero of the log excess demands. The households spend fixed nominal budgets (Section~\ref{app:flex-closure}), so the system has no num\'eraire freedom and the world price level is determined. The depth $t\in[0,1]$ of the sanction scales the severed weights by $\delta_t:=1-t(1-\delta)$, the benchmark at $t=0$ and the full severance at $t=1$. We solve the system by starting at the benchmark equilibrium, where every price is one, and raising the depth in small steps to its full value, taking the solution at each step as the starting point of the next. The equilibrium we report is the one this path reaches, as the benchmark path of Section~\ref{subsec:model_counterfactual} of the paper selects the fixed-price equilibrium, and Assumption~\ref{ass:flex-region} in Section~\ref{app:flex-closure} states the condition under which the path reaches full depth and ends at the only equilibrium reachable from the benchmark.

\subsection{The demand system and the continuation path}
\label{app:flex-closure}

The vector $\mathbf f$ of final demand is the benchmark point of a demand schedule, and the $C$ country-households generate its first part in the split \eqref{eq:m-final-split}. Country $c$'s household owns the country's factors, aggregates its final purchasers, and spends the fixed budget $F^{c}=\sum_if^{c}_i$ of Section~\ref{subsec:model_benchmark} of the paper. The budget stays fixed when factor income moves, since saving and the current account absorb the difference, so ownership of the factors does not tie the household's spending to its post-sanction income. Across sectors the household has Cobb--Douglas preferences with the benchmark shares $\alpha_{rc}$. Within each sector it holds the CES aggregator of \eqref{eq:flex-technology} over sources, with the curvature $\rho_r$ and its own weights $a_{ic}/\alpha_{rc}$, attenuated to $\check a_{ic}/\alpha_{rc}$ on severed links as in \eqref{eq:m-disrupted}. The household therefore spends $\alpha_{rc}F^{c}$ on sector $r$ at any prices and directs the share \eqref{eq:flex-share} of that spending to source $i$, built on its own weights and its own sectoral index $P_{rc}$. The residual final demand $f^{o}_i$ of \eqref{eq:m-final-split} is spent at its fixed nominal amount. Dividing spending by price and summing over purchasers gives the outside demand
\begin{equation}
x^{o}_i(\mathbf p)\;=\;\frac1{p_i}\Bigl[\sum_cF^{c}\,\check a_{ic}\Bigl(\frac{p_i}{P_{rc}}\Bigr)^{\frac{\rho_r}{\rho_r-1}}+f^{o}_i\Bigr],
\qquad r=\mathcal S(i).
\label{eq:m-outside}
\end{equation}
At unit prices with no link attenuated the bracket is $\sum_cf^{c}_i+f^{o}_i=f_i$, so the schedule reproduces $\mathbf f$. The attenuation moves the schedule even at unit prices, whereas the fixed-price economy uses the separately balanced effective final demand $\mathbf f_\omega$ in its place. We measure world welfare by the budget-weighted sum of the purchasers' log indirect utilities,
\[
\log W:=\frac1F\Bigl[\sum_cF^{c}\log V^{c}+\sum_if^{o}_i\log\frac1{p_i}\Bigr],
\qquad
\log V^{c}:=\log F^{c}-\sum_r\alpha_{rc}\log P_{rc}.
\]
Here $F:=\sum_cF^{c}+\sum_if^{o}_i=\sum_if_i$ is total outside spending, and the world household is an accounting device and not a primitive. The benchmark table is an equilibrium of this economy. At unit prices, cost-minimizing producers reproduce its input mixes, households reproduce their sourcing shares, outside demand equals $\mathbf f$, every market clears, and world income equals spending, $\sum_j\beta_js_j=\sum_if_i$. The flexible-price economy and the fixed-price economy of Section~\ref{sec:model} of the paper therefore share their benchmark accounts and their disrupted source weights, and they differ in their adjustment closures alone.

The sanction scales the weight of every barred source by $\delta$, for households and producers alike.\footnote{The attenuation $\delta$ stays exogenous in both closures. It describes the sanction, the residual access a barred relationship retains through exemptions, transshipment, and evasion, and the model sets neither the incentives behind that access nor the cost of enforcing against it. Evasion is itself an economic decision, so an endogenous attenuation would move both the losses and their division between the two sides, and the model neither bounds those losses nor signs the change in power.} The disrupted coefficients enter the rationed exchange under fixed prices and the optimizing demands under flexible prices, and in both closures they are weights and not the quantities retained after adjustment. For a household whose benchmark barred share in sector $r$ is $\sigma^{(\omega)}_{rc}$, the demand on a barred source at unit prices is $\delta/[1-(1-\delta)\sigma^{(\omega)}_{rc}]$ times its benchmark value, the denominator being the normalization of the source shares. A producer's demand responds through its sectoral spending and its intermediate intensity as well. Common disrupted coefficients therefore need not imply common retained quantities across the two closures.

On the maintained domain the demand system is closed form. Given producer $j$'s own price $p_j$ and the indices \eqref{eq:flex-share} and \eqref{eq:flex-indices}, the supply condition \eqref{eq:flex-supply} fixes its intermediate intensity $\eta_j$ and its output, and its sector and source demands follow from the CES expenditure shares at \eqref{eq:flex-indices}. At $\varrho=-1$ the marginal-cost condition inverts in closed form. Market clearing for every variety, with the outside demand $x^{o}_i(\mathbf p)$ of \eqref{eq:m-outside}, then closes the system in $\mathbf p$ alone. The rent on producer $j$'s specific endowment is the derived object $\varpi_j:=(p_jy_j-P^{m}_jw_j\eta_j)/\bar v_j$, the operating surplus per unit of the block, positive at every equilibrium, and $w_j\eta_j$ is the intermediate composite in value units.

Eliminating the quantities through the closed-form demands leaves a residual system in the log prices $\log\mathbf p$ on the active set,
\begin{equation}
\mathcal Z(\log\mathbf p)
:=
\log\mathbf D(\log\mathbf p)-\log\mathbf y(\log\mathbf p).
\label{eq:flex-residual}
\end{equation}
Here $\mathbf D(\log\mathbf p)$ collects the total demand for each variety, intermediate and outside, and $\mathbf y(\log\mathbf p)$ the supply from the marginal-cost condition. A flexible-price sanction equilibrium is a zero of $\mathcal Z$, and the benchmark equilibrium is the zero of the unattenuated system. The residual is smooth on the active set, since every index of \eqref{eq:flex-share} and \eqref{eq:flex-indices} is a strictly positive CES aggregate, the marginal-cost inversion is smooth on $\eta>0$, and every denominator stays away from zero on a compact price set with positive intensities. At the benchmark the attenuation is absent and the calibration of the weights gives $\mathcal Z(\mathbf 0)=\mathbf 0$. Write $\mathbf J^{\mathrm p}:=\partial\mathcal Z/\partial\log\mathbf p$ for the Jacobian of the residual in log prices.

We now state the condition under which the continuation path of Section~\ref{subsec:flex-economy} reaches full depth and prove that its endpoint is the only equilibrium reachable from the benchmark. The path could stop short of full depth at a singular Jacobian $\mathbf J^{\mathrm p}$, which removes the local uniqueness that the implicit function theorem supplies and permits the path to fork or to end without requiring either. An unbounded price path would leave every compact set of prices, and an intermediate intensity falling to zero would leave the domain on which the log residual is defined, a corner that the finite marginal product of intermediates at zero, $G_j'(0^{+})=(1-\beta_j)^{1/\varrho}$, does not rule out. The model excludes none of these failures on every network and at every parameter configuration, so we assume that the path meets none of them. The depths to which the benchmark equilibrium continues as a smooth branch of zeros, the continuation depths, form an interval containing $t=0$.

\begin{assumption}[Continuation from the benchmark]
\label{ass:flex-region}
At every continuation depth the equilibrium log prices remain in a fixed compact subset $\mathcal L$ of $\mathbb R^{N}$, and every active producer's intermediate intensity stays above a positive floor, $\eta_j(t)\ge\underline\eta>0$. At every depth the Jacobian $\mathbf J^{\mathrm p}$ is nonsingular at every zero of the system in $\mathcal L$.
\end{assumption}

An equilibrium is reachable by continuation from the benchmark equilibrium if it lies at the end of the path that the benchmark equilibrium traces as the depth of the sanction rises from $t=0$ to $t=1$. The maintained region $\mathcal P$ collects the primitives $(\rho_r,\rho,\varrho,\boldsymbol\beta,\delta)$ and the benchmark networks on which Assumption~\ref{ass:flex-region} holds.

\begin{proposition}[Existence, selection, and regularity]
\label{prop:flex-existence}
Let Assumption~\ref{ass:flex-region} hold.
\begin{enumerate}
\item[\textnormal{(i)}] A flexible-price sanction equilibrium exists.
\item[\textnormal{(ii)}] Exactly one flexible-price sanction equilibrium is reachable by continuation from the benchmark equilibrium. Any other zero of the system at full depth fails to continue back to the benchmark equilibrium as the sanction is lifted.
\item[\textnormal{(iii)}] The equilibrium of part (ii) varies smoothly with the attenuation $\delta$ and the maintained curvatures in a neighborhood where the continuation conditions persist, and it returns to the benchmark equilibrium as the sanction is lifted.
\end{enumerate}
\end{proposition}

\begin{proof}
The equilibrium conditions of Definition~\ref{def:flex-equilibrium} reduce to the residual system \eqref{eq:flex-residual} in the log prices, and the reduction uses the intensity floor of Assumption~\ref{ass:flex-region}, which holds every producer's choice interior. On the active set $\mathcal Z$ is smooth, and the benchmark equilibrium is its zero at depth $t=0$.

We first prove part (i), that the branch reaches full depth. Write $\log\mathbf p_t$ for the branch at the continuation depths. Under Assumption~\ref{ass:flex-region} the Jacobian $\mathbf J^{\mathrm p}$ is nonsingular at every zero along the branch, so the implicit function theorem makes the set of continuation depths open in $[0,1]$ and the branch unique on it. The same assumption confines the branch to the compact set $\mathcal L$, so along any increasing sequence of continuation depths the branch has a limit point in $\mathcal L$, and the continuity of $\mathcal Z$ makes that limit point a zero at the limiting depth $t^{*}$. The limit is unique. The limit set of a continuous path in a compact set is connected, while nonsingularity makes the zeros of $\mathcal Z(\cdot\,;t^{*})$ isolated by the inverse function theorem, and a connected subset of an isolated set is a single point. Nonsingularity at that point continues the branch through it, so the set of continuation depths is closed as well. Being nonempty, open, and closed in the connected interval $[0,1]$, the set is all of $[0,1]$, and the endpoint $\log\mathbf p_1$ is a flexible-price sanction equilibrium. This proves part (i).

We next prove part (ii). Any equilibrium reachable by continuation from the benchmark lies on a branch of zeros through the benchmark zero, and nonsingularity makes that branch unique, so the endpoint of the branch is the only reachable equilibrium. Let $\log\mathbf p'$ be any other zero of $\mathcal Z$ at full depth. Were it to continue back to the benchmark equilibrium it would lie on the same branch, and uniqueness would give $\log\mathbf p'=\log\mathbf p_1$, so no other zero at full depth continues back. This proves part (ii).

We last prove part (iii). The implicit function theorem applied jointly in the state and the parameters gives the smoothness of the endpoint in the attenuation $\delta$ and in the maintained curvatures wherever the three conditions of Assumption~\ref{ass:flex-region} hold on a neighborhood of the branch. Running the continuation backward returns the equilibrium to the benchmark equilibrium as the sanction is lifted. This proves the proposition.
\end{proof}

\subsection{Welfare and incidence along the continuation path}
\label{subsec:flex-welfare}

Two accounting statements separate what price adjustment redistributes from what it cannot undo, and both hold exactly along the continuation path of Proposition~\ref{prop:flex-existence}. The first is an identity for the terms of trade. The second is the envelope property of the cost-minimizing benchmark, stated for the object it governs, the welfare of the world household. Throughout, $\mathbf p(t)$ is the continuation branch in prices, $P^{m}_j(\mathbf p;t)$ is the composite input price index \eqref{eq:flex-indices} at prices $\mathbf p$ and depth $t$, and $\partial_tP^{m}_j$ is its derivative in the depth of the sanction at fixed prices. The log welfare $\log W$ of the world household is the budget-weighted sum of log indirect utilities of Section~\ref{app:flex-closure}. For a buyer $b$, producer or household, and a sector $r$, the severed partner's share $\sigma^{(\omega)}_{rb}$ is the share of the buyer's benchmark sector-$r$ weight that the severed suppliers hold (Appendix~\ref{app:m-content} of the paper).

\begin{lemma}[Zero-sum terms of trade and the welfare loss along the path]
\label{lem:flex-zerosum}
Fix a sanction $\omega\in\Omega$ and an attenuation $\delta\in(0,\bar\delta]$ in the region of Assumption~\ref{ass:flex-region}, and suppose every producer's intermediate intensity is strictly positive along the continuation path. Define country $c$'s cross-border price-revaluation transfer by
\[
\mathcal T_c
:=
\sum_{i:\,\mathcal C(i)=c}\;\sum_{b:\,\mathcal C(b)\ne c}(p_i-1)\,z_{ib}
\;-\;
\sum_{b:\,\mathcal C(b)=c}\;\sum_{i:\,\mathcal C(i)\ne c}(p_i-1)\,z_{ib}.
\]
Here $z_{ib}$ is the equilibrium quantity sold by $i$ to producer or household $b$, and $\mathcal C(b)$ is the buyer's country.
\begin{enumerate}
\item[\textnormal{(i)}] $\sum_c\mathcal T_c=0$ identically.
\item[\textnormal{(ii)}] At every continuation depth $t\in[0,1]$,
\begin{equation}
-\frac{d}{dt}\log W(t)
\;=\;
\frac{1}{F}\Bigl[\sum_j w_j\,\eta_j(t)\,\partial_tP^{m}_j\bigl(\mathbf p(t);t\bigr)
+\sum_c\sum_r\alpha_{rc}F^{c}\,\partial_t\log P_{rc}\bigl(\mathbf p(t);t\bigr)\Bigr]
\;\ge\;0.
\label{eq:flex-welfare-path}
\end{equation}
Here $w_j\eta_j(t)$ is producer $j$'s intermediate composite in value units, and $P_{rc}$ is household $c$'s sectoral price index. The welfare of the world household therefore falls monotonically in the depth of the sanction.
\item[\textnormal{(iii)}] At the benchmark, $-\frac{d}{dt}\log W\big|_{t=0}=\mathcal M_\omega/F$, with
\begin{equation}
\mathcal M_\omega
\;:=\;
\sum_j(1-\beta_j)\,s_j\sum_{r\in\mathcal R_j}\alpha_{rj}\,
\frac{(1-\rho_r)(1-\delta)}{\rho_r}\,\sigma^{(\omega)}_{rj}
\;+\;
\sum_cF^{c}\sum_r\alpha_{rc}\,
\frac{(1-\rho_r)(1-\delta)}{\rho_r}\,\sigma^{(\omega)}_{rc}.
\label{eq:flex-M}
\end{equation}
Here $\mathcal M_\omega$ is the Domar-weighted direct rise in the input and consumption price indices, the direct cost-index loss of the sanction.
\end{enumerate}
\end{lemma}

\begin{proof}
We first prove part (i). Each cross-border flow $z_{ib}$ enters the seller's exports with weight $+(p_i-1)z_{ib}$ and the buyer's imports with weight $-(p_i-1)z_{ib}$, so summing $\mathcal T_c$ over countries cancels every term. The identity holds at any equilibrium and not only to first order. This proves part (i).

We next prove part (ii). Fix a depth $t$ and write $p_j$, $P^{m}_j$, $\eta_j$, $y_j$, and $q_{ij}$ for the producers' equilibrium objects at that depth, and $x^{c}_i$ for household $c$'s purchase of variety $i$. Producer $j$'s operating surplus is the value of its program,
\[
\pi_j
\;=\;
\max_{\eta\ge0}\;\bigl[p_j\,s_j\,G_j(\eta)-P^{m}_j\,w_j\,\eta\bigr].
\]
Its maximizer $\eta_j>0$ satisfies \eqref{eq:flex-supply}. The maximizer is interior by hypothesis and unique by the strict concavity of $G_j$ (Lemma~\ref{lem:m-outer}), so the envelope theorem applies along the path and gives
\[
\dot\pi_j
\;=\;
y_j\,\dot p_j-w_j\eta_j\,\dot P^{m}_j,
\qquad
\dot P^{m}_j
\;=\;
\sum_i\frac{\partial P^{m}_j}{\partial p_i}\,\dot p_i+\partial_tP^{m}_j.
\]
Here a dot is the total derivative along the path, which exists by Proposition~\ref{prop:flex-existence}\textnormal{(iii)}. Shephard's lemma gives the conditional input demand $w_j\eta_j\,\partial P^{m}_j/\partial p_i=q_{ij}$. Substituting and summing over producers yields
\[
\sum_j\dot\pi_j
\;=\;
\sum_j y_j\,\dot p_j-\sum_i\dot p_i\sum_j q_{ij}-\sum_jw_j\eta_j\,\partial_tP^{m}_j
\;=\;
\sum_i\dot p_i\,x^{o}_i-\sum_jw_j\eta_j\,\partial_tP^{m}_j.
\]
The second equality uses market clearing, $y_i=\sum_jq_{ij}+x^{o}_i$ for each good $i$. The left side vanishes identically, because cost minimization gives $\sum_ip_iq_{ij}=P^{m}_jw_j\eta_j$ and the value of market clearing then gives
\[
\sum_j\pi_j
\;=\;
\sum_jp_jy_j-\sum_{i,j}p_iq_{ij}
\;=\;
\sum_ip_i\Bigl(y_i-\sum_jq_{ij}\Bigr)
\;=\;
\sum_ip_ix^{o}_i
\;=\;
\sum_cF^{c}+\sum_if^{o}_i
\;=\;
F,
\]
at every equilibrium, a constant, since the households and the residual final demand spend fixed nominal amounts. Hence $\sum_i\dot p_i\,x^{o}_i=\sum_jw_j\eta_j\,\partial_tP^{m}_j$ at every depth. The outside demand splits into the households' purchases and the residual, $x^{o}_i=\sum_cx^{c}_i+f^{o}_i/p_i$. Household $c$ spends the fixed amount $\alpha_{rc}F^{c}$ on sector $r$ and buys the conditional demands of the unit cost function $P_{rc}$ at the composite quantity $\alpha_{rc}F^{c}/P_{rc}$. Shephard's lemma therefore gives $\sum_{i\in\mathcal N_r}x^{c}_i\,\dot p_i=\alpha_{rc}F^{c}\bigl[\tfrac{d}{dt}\log P_{rc}-\partial_t\log P_{rc}\bigr]$. Summing over households and sectors and adding the residual final demand gives
\[
\sum_i\dot p_i\,x^{o}_i
\;=\;
\sum_c\sum_r\alpha_{rc}F^{c}\,\frac{d}{dt}\log P_{rc}
+\sum_if^{o}_i\,\frac{d}{dt}\log p_i
-\sum_c\sum_r\alpha_{rc}F^{c}\,\partial_t\log P_{rc}.
\]
The first two terms on the right equal $-F\,\frac{d}{dt}\log W$ by the definition of world welfare. Equating the two expressions for $\sum_i\dot p_ix^{o}_i$ proves \eqref{eq:flex-welfare-path}. For its sign, $w_j\eta_j\ge0$ and $\alpha_{rc}F^{c}\ge0$, and raising the depth lowers the severed weights in every within-sector index \eqref{eq:flex-share}, whose exponent $(\rho_r-1)/\rho_r$ is negative, so every $P_{rj}$, every $P_{rc}$, and hence every $P^{m}_j$ rises at fixed prices. Hence $\partial_tP^{m}_j$ and $\partial_t\log P_{rc}$ are nonnegative. This proves part (ii).

We last prove part (iii). At $t=0$ the weights are unattenuated and $\mathbf p=\mathbf 1$, $\eta_j=1$, and $P^{m}_j=1$, so $\partial_tP^{m}_j=\partial_t\log P^{m}_j=\sum_{r\in\mathcal R_j}\alpha_{rj}\,\partial_t\log P_{rj}$. At unit prices $P_{rb}^{\rho_r/(\rho_r-1)}=\sum_{i\in\mathcal N_r}\check a_{ib}(t)/\alpha_{rb}=1-t(1-\delta)\,\sigma^{(\omega)}_{rb}$ for a producer and a household alike, whose logarithmic derivative at $t=0$ times $(\rho_r-1)/\rho_r$ is $\frac{(1-\rho_r)(1-\delta)}{\rho_r}\sigma^{(\omega)}_{rb}$. Substituting into \eqref{eq:flex-welfare-path} gives $\mathcal M_\omega/F$. This proves the lemma.
\end{proof}

What Lemma~\ref{lem:flex-zerosum} establishes is confined to welfare. Price movements transfer welfare between countries and cancel in the aggregate, and the first-order loss of the world household is fixed by the sanction alone, through the severed partner's shares $\sigma^{(\omega)}_{rb}$ it strikes and the curvatures of the sectors it strikes them in. With fixed budgets the revaluation $\mathcal T_c$ does not decompose the change in a country's consumption welfare, and neither statement reaches activity, since the envelope theorem holds a producer's choices fixed when its profit is differentiated and not when its output is.

Gross activity instead moves with intermediate use. Under the hypotheses of Lemma~\ref{lem:flex-zerosum}, differentiating the supply condition at $t=0$ gives $\dot\eta_j=(\dot P^{m}_j-\dot p_j)/[\beta_j(\varrho-1)]$, by $G_j''(1)=\beta_j(1-\beta_j)(\varrho-1)$, and hence
\begin{equation}
\frac{d}{dt}\sum_jy_j\Big|_{t=0}
\;=\;
-\sum_j\frac{(1-\beta_j)\,s_j}{\beta_j(1-\varrho)}\,
\bigl(\dot P^{m}_j-\dot p_j\bigr).
\label{eq:flex-activity-rate}
\end{equation}
Here $\dot P^{m}_j=\sum_i(q_{ij}/w_j)\,\dot p_i+\partial_t\log P^{m}_j$ is the total movement of producer $j$'s input price index. The rate adds up the response of intermediate use to relative prices producer by producer, and it does not reduce to $-\mathcal M_\omega$. Activity, and with it vulnerability, therefore responds to price adjustment at first order in a way welfare does not, and the comparison of the two closures in Section~\ref{subsec:flex-comparison} measures that response together with the other differences between the closures.

The local incidence of a severance differentiates the market-clearing prices and the producers' supply together. Fix the severed set, with indicator $e_{ib}^{\omega}$ on the links to producer and household buyers. The depth path of Section~\ref{subsec:flex-economy} attenuates the weights along $\check a_{ib}(t)=a_{ib}[1-t(1-\delta)e_{ib}^{\omega}]$, so that $t\downarrow0$ is a small increase in depth from the benchmark. At unit prices the direct derivative of a sectoral index is
\begin{equation}
\left.\partial_{t}\log P_{rj}(\mathbf 1;t)\right|_{t=0}
\;=\;
\frac{(1-\delta)(1-\rho_r)}{\rho_r}\;\sigma^{(\omega)}_{rj},
\label{eq:flex-spec}
\end{equation}
as the proof of Lemma~\ref{lem:flex-zerosum}\textnormal{(iii)} computes. The same formula holds for a household's sectoral index $P_{rc}$ through its own share $\sigma^{(\omega)}_{rc}$. The derivative measures the direct effect of input specificity on the cost index, and it is not a derivative of output.

We split the derivative of the residual into a placement source and a specificity source. Write $\mathcal Z_t$ for the partial derivative of the residual \eqref{eq:flex-residual} in the depth at benchmark prices. The placement source differentiates only the attenuated source weights $\check a_{ib}$ that enter demand explicitly, holding prices, price indices, and each buyer's sectoral spending at their benchmark values. Benchmark demand is $s_i$ and the residual is log demand less log supply, so the placement source at seller $i$ is $\varepsilon^{\mathrm p}_{\omega,i}=-(1-\delta)\zeta_{\omega,i}$, with $\zeta_{\omega,i}$ the severed-sales share of Section~\ref{app:m-shares}. The specificity source is the remainder, $\boldsymbol\varepsilon^{\mathrm s}_\omega:=\mathcal Z_{t}-\boldsymbol\varepsilon^{\mathrm p}_\omega$, which collects the direct movement of the indices, the normalization of the source shares, the sectoral spending, and the supply response, and not \eqref{eq:flex-spec} alone. The split is an accounting of one derivative, and it does not define two sanction experiments that could be run separately.

The supply derivatives follow the same split. Write $\mathbf Y(\boldsymbol\ell,t)$ for supply as a function of the log prices $\boldsymbol\ell:=\log\mathbf p$ and the depth, with the partial derivatives $\mathbf Y_{\boldsymbol\ell}$ and $\mathbf Y_t$ evaluated at $(\boldsymbol\ell,t)=(\mathbf0,0)$. The direct supply changes operate through the input indices, so we assign the whole direct supply derivative to the specificity source, $\mathbf Y_{t}^{\mathrm p}=\mathbf0$ and $\mathbf Y_{t}^{\mathrm s}=\mathbf Y_{t}$.

Under flexible prices a country's activity can rise above its benchmark. We therefore write $\gamma^{\mathrm F}_c:=\mathbbm 1_c(\mathbf s-\mathbf y)/Q(c)$ for its flexible-price vulnerability, the counterpart of Definition~\ref{def:country_vulnerability} of the paper. The flexible-price asymmetry puts the sum of the two sides' absolute losses in the denominator of \eqref{eq:bounded-Hirschman} of the paper, $\mathcal H^{\mathrm F}_{cc'}:=(\gamma^{\mathrm F}_c-\gamma^{\mathrm F}_{c'})/(|\gamma^{\mathrm F}_c|+|\gamma^{\mathrm F}_{c'}|)$.

\begin{proposition}[Two-margin incidence at first order]
\label{prop:flex-incidence}
Fix a pair $(c,c')$ and the depth path, with the benchmark in the regular positive-intensity region $\mathcal P$. For $k\in\{\mathrm s,\mathrm p\}$, define
\[
\dot{\boldsymbol\ell}^{\,k}=-(\mathbf J^{\mathrm p})^{-1}\boldsymbol\varepsilon^{k}_\omega,
\qquad
\dot{\mathbf y}^{\,k}=\mathbf Y_{\boldsymbol\ell}\dot{\boldsymbol\ell}^{\,k}+\mathbf Y_{t}^{k},
\qquad
\dot\gamma_c^{k}=-\frac{\mathbbm1_c\dot{\mathbf y}^{\,k}}{Q(c)}.
\]
Each side's flexible-price vulnerability then has the expansion
\[
\gamma^{\mathrm F}_c=t\,\bigl(\dot\gamma^{\mathrm{s}}_c+\dot\gamma^{\mathrm{p}}_c\bigr)+o(t),
\qquad
\gamma^{\mathrm F}_{c'}=t\,\bigl(\dot\gamma^{\mathrm{s}}_{c'}+\dot\gamma^{\mathrm{p}}_{c'}\bigr)+o(t).
\]
Write $\dot\gamma_c:=\dot\gamma^{\mathrm{s}}_c+\dot\gamma^{\mathrm{p}}_c$ and suppose $|\dot\gamma_c|+|\dot\gamma_{c'}|>0$. The normalized components
\[
\Lambda^{\mathrm{s}}_{cc'}:=\frac{\dot\gamma^{\mathrm{s}}_c-\dot\gamma^{\mathrm{s}}_{c'}}{|\dot\gamma_c|+|\dot\gamma_{c'}|},
\qquad
\Lambda^{\mathrm{p}}_{cc'}:=\frac{\dot\gamma^{\mathrm{p}}_c-\dot\gamma^{\mathrm{p}}_{c'}}{|\dot\gamma_c|+|\dot\gamma_{c'}|}
\]
then sum to the limiting flexible-price asymmetry,
\[
\lim_{t\downarrow0}\mathcal H^{\mathrm F}_{cc'}
\;=\;
\Lambda^{\mathrm{s}}_{cc'}+\Lambda^{\mathrm{p}}_{cc'}.
\]
\end{proposition}

\begin{proof}
Apply the implicit function theorem to $\mathcal Z(\boldsymbol\ell(t),t)=\mathbf0$, which gives
\[
\dot{\boldsymbol\ell}=-(\mathbf J^{\mathrm p})^{-1}\mathcal Z_{t}.
\]
Then differentiate supply,
\[
\dot{\mathbf y}=\mathbf Y_{\boldsymbol\ell}\dot{\boldsymbol\ell}+\mathbf Y_{t}.
\]
Both decompositions sum to their respective total derivatives, so linearity gives $\dot{\mathbf y}=\dot{\mathbf y}^{\,\mathrm s}+\dot{\mathbf y}^{\,\mathrm p}$ and the country coefficients. Benchmark output is $\mathbf s$, so vulnerability vanishes at zero depth and has the stated first-order expansion. For $t>0$, divide the numerator and the denominator of $\mathcal H^{\mathrm F}_{cc'}$ by $t$. The limit is $(\dot\gamma_c-\dot\gamma_{c'})/(|\dot\gamma_c|+|\dot\gamma_{c'}|)$, with a positive denominator by hypothesis, and splitting its numerator gives the two normalized components. This proves the proposition.
\end{proof}

The coefficients $\dot\gamma^{\mathrm{s}}_c$ and $\dot\gamma^{\mathrm{p}}_c$ rest on the accounting convention for the direct supply derivative, and the country components combine the propagated price responses with the direct supply responses. Neither propagated component has a general sign, and a positive direct cost-index derivative alone does not determine either country's equilibrium loss. Subtracting the corresponding fixed-price limit gives the difference between the closures whenever the fixed-price losses also have first-order expansions along this path with a nonzero first-order sum of absolute losses. The finite comparisons of Section~\ref{subsec:flex-comparison} measure the total difference between the closures, and they do not estimate the two local components separately, which would require the derivatives to be evaluated.

The sources of the net tilt can rotate along a curvature path. Fix the benchmark network, the attenuation $\delta$, and the severed set, and consider a continuous curvature path in the regular positive-intensity region $\mathcal P$ with $|\dot\gamma_c|+|\dot\gamma_{c'}|>0$ throughout. The direct cost-index factor $(1-\rho_r)/\rho_r$ in \eqref{eq:flex-spec} decreases strictly with $\rho_r$, while the explicit placement source, which depends only on benchmark sales and the attenuation, stays fixed. In the maintained region the residual and supply derivatives and the nonsingular inverse $({\mathbf J}^{\mathrm p})^{-1}$ are continuous, so the derivatives in Proposition~\ref{prop:flex-incidence} are continuous. Dividing by the positive sum of absolute losses preserves the continuity of the normalized components $\Lambda^{\mathrm s}_{cc'}$ and $\Lambda^{\mathrm p}_{cc'}$. If their sum has strictly opposite signs at the endpoints of the path, the intermediate value theorem gives it at least one zero along the path. As $\rho_r\uparrow1$ the direct index factor tends to zero, a limit outside the maintained compact interval of curvatures, and neither the source-share derivatives nor the equilibrium responses need share that limit. The rotation therefore does not make the complete specificity component vanish, and it implies neither a monotone shift toward placement nor a threshold in bilateral dependence at which the order reverses, so those comparisons have to be computed.

\subsection{Calibrating the within-sector curvatures}
\label{subsec:flex-calibration}

The flexible-price economy adds one parameter vector, the within-sector curvatures $\rho_r$. The source weights and the non-intermediate shares $\boldsymbol\beta$ come from the trade table, the attenuation $\delta$ carries over with its evasion evidence, and the curvatures $\rho$ and $\varrho$ keep the paper's baseline, which the intermediate-input substitution literature supports independently. We calibrate the curvatures $\rho_r$ in three steps, a cross-sector profile, an adjustment to the sanction horizon, and a comparison with the paper's own episode evidence.

Table~\ref{tab:flex-sigma} maps the tariff-identified estimates of \citet{oa-caliendoparro2015} for twenty tradable sectors to the fifty ICIO industries. In an Armington source nest the trade-cost exponent of the tariff-inclusive expenditure shares is $\rho_r/(1-\rho_r)$, so for their estimate $\epsilon_r$, their $\theta_r$, we set $\rho_r^{\mathrm{LR}}=\epsilon_r/(1+\epsilon_r)$, an implied within-sector elasticity of $1+\epsilon_r$. Reading these cross-sectional estimates as long-run elasticities is a maintained calibration choice. Of the industries without an estimate of their own, computer, electronic, and optical products take the unweighted mean of the office, communication, and medical estimates, and services and construction take an assumed long-run trade elasticity of $1.5$, which we treat as a robustness parameter.

The adjustment to the horizon draws on \citet{oa-boehmlevchenko2023}, who estimate tariff-exclusive flow elasticities of $0.76$ on impact and roughly $2$ in the long run, with convergence over seven to ten years. Their elasticities are not the tariff-inclusive expenditure-share exponents used here, so we use only their ratio, as a heuristic, setting $\rho_r^{\mathrm{SR}}=0.38\,\epsilon_r/(1+0.38\,\epsilon_r)$, the factor $0.38$ being the ratio of their impact elasticity to their long-run one. The rule preserves the cross-sector relativities, and it neither converts one definition of the elasticity into the other nor estimates a common short-run factor. The resulting baseline elasticities $1/(1-\rho_r)$ run from $1.15$ for other transport equipment to $25.6$ for refined petroleum, with a median of $1.57$, and the long-run profile, whose higher within-sector curvatures lower the direct cost-index factor of \eqref{eq:flex-spec}, serves as the robustness variant.

The third step compares the profile with the paper's episode evidence, used as moments and not as calibration targets. The fixed-price economy prescribes the unrecovered share $\tau$, and the flexible-price economy specifies substitution among sources and optimizing input demand instead, so each solved severance implies the unrecovered share at which the fixed-price economy would reproduce its loss. We define that implied share $\tau^{*}$ by matching the weaker side's flexible-price loss to its fixed-price loss on the paper's $\tau$-grid, by interpolation within a bracket of the grid where one exists. The match is a calibration diagnostic and not an identified fraction of unrecovered trade, and its uniqueness on the unit interval and its monotonicity in substitution are not established.\footnote{Without a bracket on the grid there is no interpolated match. A flexible-price loss above the fixed-price loss at $\tau=1$ does not rule out a match in the interior without monotonicity, and a match extrapolated beyond one lies outside the admissible range of the unrecovered share and is not an equilibrium of the fixed-price economy.} On the paper's sensitivity pairs the implied shares land between $0.13$ and $0.26$, just below the calibrated $0.30$, and their ordering across goods reproduces the ordering of the episode brackets of Section~\ref{app:m-tau}, which run from $0.05$--$0.10$ for fungible commodities to $0.5$--$0.8$ for specific inputs. Those brackets map to within-sector elasticities in the ranges the trade literature estimates for those goods.

\begin{table}[!htbp]
\centering
\small
\begin{tabular}{@{}llrrr@{}}
\toprule
Source estimate & ICIO industries covered & $\epsilon^{\mathrm{LR}}$ & $1+0.38\,\epsilon^{\mathrm{LR}}$ & $1+\epsilon^{\mathrm{LR}}$ \\
\midrule
Agriculture      & A01, A02, A03            & 9.11  & 4.46  & 10.11 \\
Mining           & B05, B06, B07, B08       & 13.53 & 6.14  & 14.53 \\
Food             & C10T12                   & 2.62  & 2.00  & 3.62  \\
Textiles         & C13T15                   & 8.10  & 4.08  & 9.10  \\
Wood             & C16                      & 11.50 & 5.37  & 12.50 \\
Paper            & C17\_18                  & 16.52 & 7.28  & 17.52 \\
Petroleum        & C19                      & 64.85 & 25.64 & 65.85 \\
Chemicals        & C20, C21                 & 3.13  & 2.19  & 4.13  \\
Rubber, plastics & C22                      & 1.67  & 1.63  & 2.67  \\
Minerals         & C23                      & 2.41  & 1.92  & 3.41  \\
Basic metals     & C24A, C24B               & 3.28  & 2.25  & 4.28  \\
Metal products   & C25                      & 6.99  & 3.66  & 7.99  \\
Electronics mean & C26                      & 8.54  & 4.24  & 9.54  \\
Electrical       & C27                      & 12.91 & 5.91  & 13.91 \\
Machinery n.e.c. & C28                      & 1.45  & 1.55  & 2.45  \\
Motor vehicles   & C29                      & 1.84  & 1.70  & 2.84  \\
Other transport  & C301, C302T309           & 0.39  & 1.15  & 1.39  \\
Other mfg.       & C31T33                   & 3.98  & 2.51  & 4.98  \\
Services rule    & B09, D--T                & 1.50  & 1.57  & 2.50  \\
\bottomrule
\end{tabular}
\caption{The within-sector elasticity calibration. The long-run trade elasticities $\epsilon^{\mathrm{LR}}$ use the 99\% sample of \citet{oa-caliendoparro2015}. The implied source elasticities are $1+\epsilon^{\mathrm{LR}}$ in the long run and $1+0.38\,\epsilon^{\mathrm{LR}}$ at the short-run baseline, with the heuristic factor $0.38$ drawn from the tariff-exclusive short-to-long ratio of \citet{oa-boehmlevchenko2023}. The curvatures follow as $\rho_r=1-1/(1+\epsilon)$ at each horizon. Electronics uses the unweighted mean of the office, communication, and medical estimates, and the services rule is an assumption.}
\label{tab:flex-sigma}
\end{table}

\subsection{The distribution of power under fixed and flexible prices}
\label{subsec:flex-comparison}

We now solve the flexible-price economy on the 2022 trade table for every one of its $3{,}160$ country pairs and set each of the $3{,}154$ computed severances against its fixed-price counterpart.\footnote{Six of the $3{,}160$ severances attempted did not converge, and we report no number for them. Section~\ref{app:flex-world} names the six and records their residuals.} The within-sector curvatures $\rho_r$ follow the calibration of Section~\ref{subsec:flex-calibration}. The flexible-price vulnerability $\gamma^{\mathrm F}_c$ and the flexible-price asymmetry $\mathcal H^{\mathrm F}_{cc'}$ are those of Section~\ref{subsec:flex-welfare}, with the asymmetry set to zero when both losses vanish, and $\gamma_c$ remains the fixed-price vulnerability. Since the loss vector $\mathbf s-\mathbf y$ can carry negative entries under flexible prices, the asymmetry divides by the sum of the two absolute losses, and it coincides with that of Section~\ref{subsec:hirschman_matrix} of the paper whenever both losses are positive.

Fix a pair trading in both directions, its severance, the selected fixed-price equilibrium of Proposition~\ref{prop:m-monotone}, and the flexible-price equilibrium of Proposition~\ref{prop:flex-existence}. Write $z_j:=P^{m}_j/p_j$ for the wedge of producer $j$, its composite input price index \eqref{eq:flex-indices} relative to its own price at the flexible-price equilibrium. When the fixed-price loss of country $c$ is positive, write $k_c:=\gamma^{\mathrm F}_c/\gamma_c$ for its loss multiplier, the ratio of its flexible-price loss to its fixed-price loss.

\begin{proposition}[Power when prices adjust]
\label{prop:flex-power}
Maintain Assumption~\ref{ass:flex-region}.
\begin{enumerate}
\item[\textnormal{(i)}] Each producer $j$'s flexible-price output satisfies
\begin{equation}
\frac{y_j}{s_j}
=\Bigl[\frac{1-(1-\beta_j)\,z_j^{\varrho/(\varrho-1)}}{\beta_j}\Bigr]^{-1/\varrho}.
\label{eq:flex-wedge-output}
\end{equation}
The producer contracts if and only if its wedge exceeds one. At $\varrho=-1$, country $c$'s vulnerability is
\begin{equation}
\gamma^{\mathrm F}_c
=\sum_{j\in c}\frac{s_j}{Q(c)}\,\frac{1-\beta_j}{\beta_j}\bigl(\sqrt{z_j}-1\bigr).
\label{eq:flex-vulnerability}
\end{equation}
\item[\textnormal{(ii)}] If both sides lose under both closures, their asymmetries satisfy
\begin{equation}
\mathcal H^{\mathrm F}_{cc'}
=\frac{\mathcal H_{cc'}+\psi(k_c/k_{c'})}{1+\mathcal H_{cc'}\,\psi(k_c/k_{c'})} .
\label{eq:flex-composition}
\end{equation}
The asymmetry is unchanged if and only if $k_c=k_{c'}$. It moves against country $c$ if and only if $k_c>k_{c'}$, and $|\mathcal H^{\mathrm F}_{cc'}-\mathcal H_{cc'}|\le\tfrac12|\log(k_c/k_{c'})|$.
\end{enumerate}
\end{proposition}

\begin{proof}
We first prove \textnormal{(i)}. Fix a producer $j$, write $G_j(\eta)=[\beta_j+(1-\beta_j)\eta^{\varrho}]^{1/\varrho}$ as in Section~\ref{subsec:flex-economy}, and write $q:=G_j(\eta_j)=y_j/s_j$ for the equilibrium output ratio. Differentiating gives
\[
G_j'(\eta)=(1-\beta_j)\,\eta^{\varrho-1}\,G_j(\eta)^{1-\varrho}=(1-\beta_j)\Bigl(\frac{G_j(\eta)}{\eta}\Bigr)^{1-\varrho}.
\]
The supply condition \eqref{eq:flex-supply} therefore reads $(q/\eta_j)^{1-\varrho}=z_j$, which is $\eta_j=q\,z_j^{1/(\varrho-1)}$. Substituting into $q^{\varrho}=\beta_j+(1-\beta_j)\eta_j^{\varrho}$ gives $q^{\varrho}\bigl[1-(1-\beta_j)z_j^{\varrho/(\varrho-1)}\bigr]=\beta_j$. The bracket is positive because $q$ and $\beta_j$ are, and solving for $q$ gives \eqref{eq:flex-wedge-output}. By Lemma~\ref{lem:m-outer}\textnormal{(i)} the aggregator $G_j$ is strictly increasing and strictly concave, so $G_j'$ is strictly decreasing, and $G_j'(1)=1-\beta_j$ by Lemma~\ref{lem:m-outer}\textnormal{(iii)}, since $G_j(1)=1$. The supply condition therefore puts $\eta_j$ below one exactly when $z_j>1$, and the monotonicity of $G_j$ carries the ordering to $q$. At $\varrho=-1$ the exponent $\varrho/(\varrho-1)$ equals one half and $-1/\varrho$ equals one, so $q=(1-(1-\beta_j)\sqrt{z_j})/\beta_j$ and $1-q=\bigl((1-\beta_j)/\beta_j\bigr)(\sqrt{z_j}-1)$. Summing $s_j-y_j=s_j(1-q)$ over the producers of country $c$ and dividing by $Q(c)$ gives \eqref{eq:flex-vulnerability}. This proves \textnormal{(i)}.

We next prove \textnormal{(ii)}. With positive losses on both sides under both closures, $\gamma^{\mathrm F}_c/\gamma^{\mathrm F}_{c'}=(\gamma_c/\gamma_{c'})\,(k_c/k_{c'})$, and on such a pair the flexible-price asymmetry coincides with \eqref{eq:bounded-Hirschman}, so Lemma~\ref{lem:m-rigidity}\textnormal{(i)} gives $\mathcal H^{\mathrm F}_{cc'}=\psi\bigl((\gamma_c/\gamma_{c'})(k_c/k_{c'})\bigr)$ and $\mathcal H_{cc'}=\psi(\gamma_c/\gamma_{c'})$. Lemma~\ref{lem:m-psi}\textnormal{(ii)} turns the product into the bounded sum \eqref{eq:m-oplus}, which is \eqref{eq:flex-composition}. For $u,v\in(-1,1)$ the bounded sum satisfies $u\oplus v-u=v(1-u^{2})/(1+uv)$, whose sign is the sign of $v$, and $v=\psi(k_c/k_{c'})$ vanishes exactly when $k_c=k_{c'}$ and is positive exactly when $k_c>k_{c'}$ by Lemma~\ref{lem:m-psi}\textnormal{(i)}. The asymmetry is therefore unchanged exactly when the multipliers coincide and moves against country $c$ exactly when $k_c>k_{c'}$. The passage from one closure to the other changes the log loss ratio by $\log(k_c/k_{c'})$, so Lemma~\ref{lem:m-rigidity}\textnormal{(iii)} gives the bound. This proves \textnormal{(ii)}.
\end{proof}

Part~(i) is the flexible-price counterpart of the anatomy of Proposition~\ref{prop:m-anatomy}. Under fixed prices a producer's loss is the unfilled fraction of its orders plus its unrecovered lost sales, carried to other producers along the effective demand. Under flexible prices it is a function of one number, the wedge $z_j$ between what the producer pays for its intermediate composite and what it receives for its output, scaled by its intermediate dependence $(1-\beta_j)/\beta_j$, the ratio of its benchmark spending on intermediates to its spending on the fixed block. At the baseline curvature a wedge of one percent costs a producer about half a percent of its activity for every unit of that ratio.

Part~(ii) is Lemma~\ref{lem:m-rigidity} applied to the passage from one closure to the other. Both closures disrupt the same relationships, so only the difference between the two sides' loss multipliers $k_c$ and $k_{c'}$ moves the asymmetry between them, and a relief common to both sides leaves it where it was. Price adjustment widens the asymmetry against the weaker side exactly when it relieves the weaker side by less than the favorable side, and it reverses the direction of power when the favorable side's multiplier exceeds the weaker side's by a factor larger than the fixed-price loss ratio.

The ordering of power changes little, for the United States remains the most powerful economy, with a mean bilateral advantage of $0.96$ over its seventy-eight computed relationships against $0.94$ under fixed prices. China follows at $0.90$ against $0.87$ over its seventy-seven computed relationships, and Germany at $0.74$ against $0.72$ over all seventy-nine. Power stays as concentrated as under fixed prices, since the three most powerful economies hold $17\%$ of the sum of the positive mean advantages across the eighty economies and the single most powerful holds $6\%$, both within a percentage point of their fixed-price shares. The mean bilateral advantages of the eighty economies correlate at $0.98$ and rank-correlate at $0.97$ across the two closures, and the trade-weighted advantages of Table~\ref{tab:s6_power_ranking} rank-correlate at $0.97$. The median economy's mean advantage moves by $0.05$. The largest movements, between $0.16$ and $0.28$, are gains for Cameroon, Bangladesh, Costa Rica, and Argentina and losses for the Democratic Republic of the Congo, Angola, Nigeria, and Vietnam.

\begin{figure}[!htbp]
\centering
\scalebox{0.8}{\includegraphics[width=0.9\linewidth]{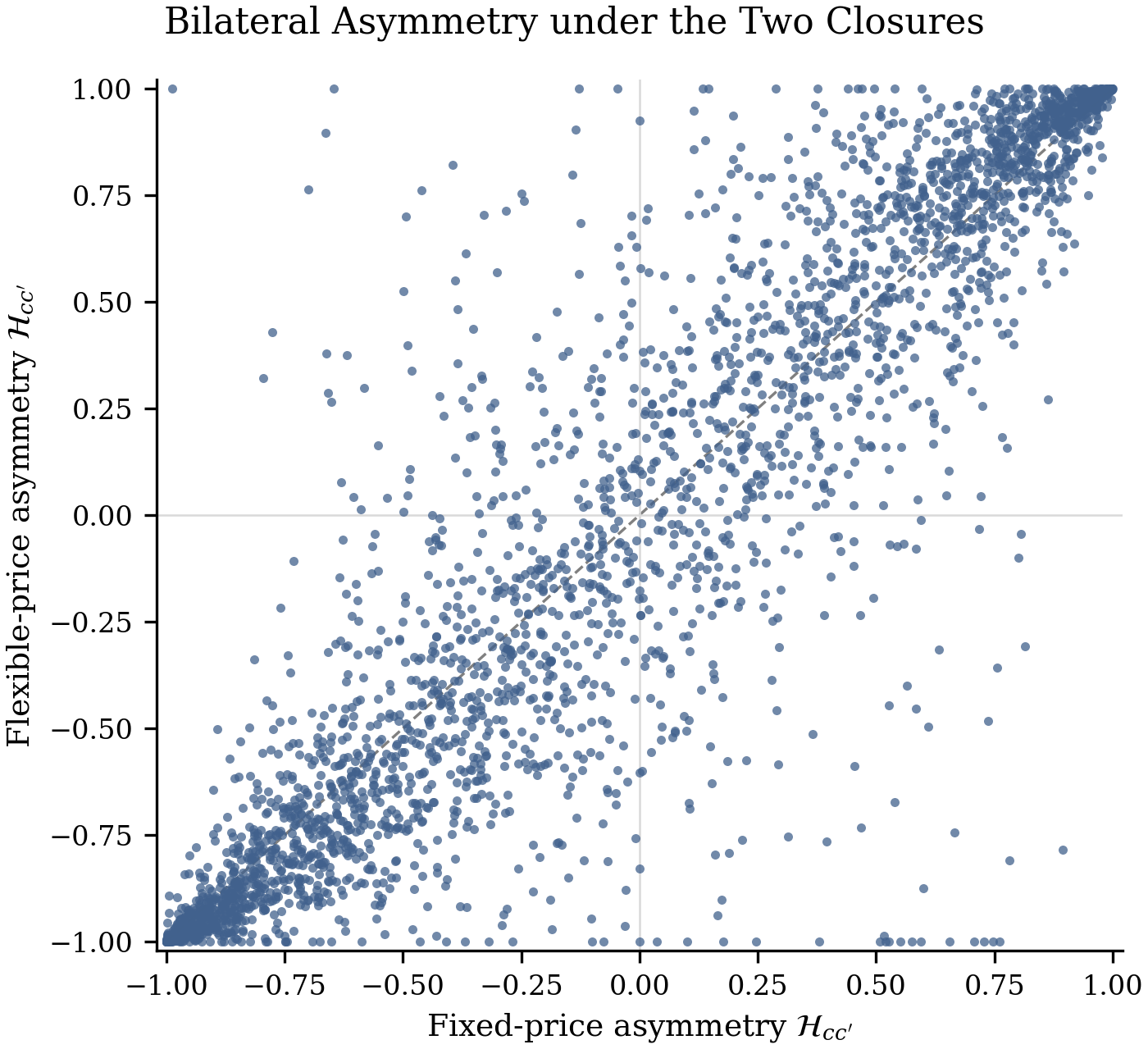}}
\caption{The distribution of power under fixed and flexible prices. Each point is one of the $3{,}154$ computed bilateral severances, with the fixed-price asymmetry of the pair on the horizontal axis and its flexible-price asymmetry on the vertical axis, both signed toward the same country. On the diagonal the asymmetry is the same under the two closures. A pair sits at $\pm1$ on the vertical axis when its two nonzero losses have opposite signs or exactly one of them is zero, strictly between the endpoints when both sides gain, and at zero when both losses vanish. Baseline short-run calibration, $\delta=0.10$.}
\label{fig:s6-flex-scatter}
\end{figure}

Figure~\ref{fig:s6-flex-scatter} plots the fixed-price asymmetry of each pair against its flexible-price asymmetry. The two rank-correlate at $0.91$ across the computed severances, and on the pairs whose baseline asymmetry exceeds $0.2$ in absolute value the two closures agree on the direction of power in $95\%$ of cases.

The levels of the two sides' losses move more than the asymmetry between them, and they move by more on one side of a relationship than on the other. Table~\ref{tab:s6-flex-showcase} reports the severances that the paper discusses by name, the named severances, under the two closures. Across the computed pairs the loss multiplier $k_c$ of the weaker side has a median of $0.89$, with quartiles $0.62$ and $1.23$, so price adjustment relieves the weaker side by about a ninth at the median.\footnote{These relief figures compare two closures of the same severance. The flexible-price closure replaces the unrecovered share $\tau$ with within-sector curvatures calibrated to short-run trade elasticities, so the size of the relief rests on how the two are calibrated together. The comparison of asymmetries rests on that calibration less, since the factor $\tau(1-\delta)$ of Proposition~\ref{prop:m-exposure} is common to the two sides and cancels from the fixed-price loss ratio, though the export multipliers and the import multipliers can still move with the calibration.} The favorable side's multiplier has a median of $0.72$ and falls below the weaker side's on $61\%$ of the computed pairs. The gap between the two, the favorable side's multiplier less the weaker side's, grows with the baseline asymmetry, from $-0.02$ on the pairs whose baseline asymmetry is below $0.2$ in absolute value to $-0.31$ on the pairs above $0.8$. Under the United States--China severance both sides lose less, from $0.6$ and $1.4$ percent of activity to $0.3$ and $0.7$. Kazakhstan against Russia moves the other way, with the weaker side losing more when prices clear than when quantities ration.

\begin{table}[!htbp]
\centering
\small
\begin{tabular}{@{}llrrrrrr@{}}
\toprule
 & & \multicolumn{2}{c}{Loss of $c$ (\%)} & \multicolumn{2}{c}{Loss of $c'$ (\%)} & \multicolumn{2}{c}{$\mathcal H_{cc'}$} \\
\cmidrule(lr){3-4}\cmidrule(lr){5-6}\cmidrule(l){7-8}
$c$ & $c'$ & fixed & flexible & fixed & flexible & fixed & flexible \\
\midrule
USA & CHN & 0.56 & 0.34 & 1.37 & 0.66 & $-0.42$ & $-0.33$ \\
VNM & CHN & 11.49 & 7.66 & 0.44 & 0.10 & $+0.93$ & $+0.97$ \\
CAN & USA & 9.30 & 6.80 & 1.01 & 0.75 & $+0.80$ & $+0.80$ \\
MEX & USA & 10.88 & 9.39 & 0.62 & 0.23 & $+0.89$ & $+0.95$ \\
DEU & RUS & 0.57 & 0.38 & 0.95 & 0.44 & $-0.25$ & $-0.07$ \\
USA & RUS & 0.05 & 0.04 & 0.62 & 0.68 & $-0.84$ & $-0.89$ \\
JPN & CHN & 2.57 & 1.50 & 0.89 & 0.49 & $+0.49$ & $+0.51$ \\
IND & CHN & 1.18 & 0.80 & 0.23 & 0.08 & $+0.67$ & $+0.83$ \\
DEU & CHN & 1.51 & 0.96 & 0.56 & 0.31 & $+0.46$ & $+0.51$ \\
CHN & TWN & 0.63 & 0.26 & 7.43 & 3.27 & $-0.84$ & $-0.85$ \\
KAZ & RUS & 3.46 & 3.92 & 0.54 & 0.20 & $+0.73$ & $+0.90$ \\
AUS & CHN & 3.28 & 1.38 & 0.75 & 0.28 & $+0.63$ & $+0.66$ \\
KOR & CHN & 5.55 & 3.03 & 0.86 & 0.40 & $+0.73$ & $+0.77$ \\
\bottomrule
\end{tabular}
\caption{Named severances under fixed and flexible prices. Losses are own-size-normalized activity losses in percent. The fixed columns are the baseline of Section~\ref{sec:statistics}, and the flexible columns are the calibrated short-run economy of Section~\ref{subsec:flex-calibration}, both with the household columns of the decree. The asymmetry $\mathcal H_{cc'}$ is positive when country $c$ is the weaker side. Baseline calibration $\delta=0.10$, with the short-run $\rho_r$ under flexible prices and $\tau=0.30$ and $\rho=\varrho=-1$ under fixed prices.}
\label{tab:s6-flex-showcase}
\end{table}

By part~(ii) of Proposition~\ref{prop:flex-power}, the asymmetry moves further against the weaker side on exactly the pairs on which that side is relieved by less. The median absolute asymmetry across the computed pairs rises from $0.63$ to $0.72$, and the absolute asymmetry rises on $67\%$ of the pairs and on $65\%$ of the pairs whose baseline asymmetry exceeds $0.2$ in absolute value.\footnote{The $67\%$ exceeds the $61\%$ because a reversal of the direction of power can carry the asymmetry past its fixed-price magnitude, as it does on $166$ of the $301$ reversals. Part~(ii) does not cover the $72$ computed pairs on which a side fails to lose in one of the two closures, and the asymmetry of $71$ of them is plus or minus one. Those $71$ pairs are not the $71$ severances in which a party expands (Section~\ref{app:flex-world}), since a side that fails to lose need not expand.} The direction of power reverses on $301$ of the $3{,}082$ computed pairs on which both sides lose under both closures, $132$ of them pairs whose baseline asymmetry exceeds $0.2$ in absolute value. On the named severances the flexible-price world loss is a median $0.43$ of its fixed-price value (Section~\ref{app:flex-world}), so on those severances prices lower the world loss and raise the asymmetry with which it divides.\footnote{Bystanders can gain when prices adjust, which fixed prices rule out, since every budget and every capacity of the rationed exchange is a benchmark total. A producer's output exceeds its benchmark exactly when its wedge falls below one, by part~(i) of Proposition~\ref{prop:flex-power}, and Lemma~\ref{lem:flex-pivot} in Section~\ref{oa:flex-pivot} gives the same condition for the real rent on its specific factor to exceed its benchmark value. Across the computed severances a third of bystander countries rise above their benchmark, none by as much as half a percent of its activity. Section~\ref{app:flex-world} reports the distributions of bystander activity by country and by industry.}

\subsection{Rationing, prices, and the incidence of a severance}
\label{subsec:flex-incidence-econ}

We now set out the mechanism behind the two findings of Section~\ref{subsec:flex-comparison}, the survival of the ordering of power and the greater relief of the favorable side, neither of which holds on every pair. A severance leaves some sellers without customers and some buyers without suppliers, and the two closures allocate what remains of the world's supply and demand in different ways.

Under fixed prices the rationed exchange reallocates the effective demand on the incumbent support at posted prices, removing the fraction $\tau$ of the displaced benchmark trade and recovering the rest, and demand at new prices plays no part in it. The row and column constraints hold together, so a seller's scaling can move a buyer's relative source shares, and the fulfillment and activity adjustments that follow can lower deliveries further, which is why a recovered target does not guarantee a realized receipt.

Under flexible prices a producer buys inputs up to the point at which the output they add, valued at its own price, covers their cost, so it buys less and makes less when its output fetches a lower price or its input bundle becomes dearer. By part~(i) of Proposition~\ref{prop:flex-power} the output of producer $j$ depends only on its wedge $z_j$. A severance moves the wedge of a seller through the price of its output, the channel that the export elasticity of Section~\ref{subsec:power_size} of the paper measures under fixed prices, and it moves the wedge of a buyer through the price of its inputs, the channel of the import elasticity, and the two price responses can differ.

A seller that loses a customer finds other buyers by lowering its price, and the lower price costs it activity only as far as the price of its input bundle fails to fall with it. Where the seller buys its inputs at home, the fall in its own demand for inputs lowers its domestic suppliers' prices, its bundle becomes cheaper along with its output, and its contraction is limited. Much of the export elasticity then takes the form of a lower price and a lower rent on the seller's specific factor, neither of which enters the measure of activity.\footnote{Fixed budgets are the reading of a transitory sanction, in which households smooth consumption and the current account absorbs the change in factor income. The same reading keeps the final orders autonomous under fixed prices. A budget that moved with the country's rents would add a feedback from income to demand whose equilibrium sign and strength require a re-solved economy, and Section~\ref{app:flex-world} solves that variant on the named severances.} Where the seller buys its inputs abroad, their prices do not follow its own and the discount cuts its output.

A buyer that loses a supplier cannot restore its input bundle in the same way. The remaining sources of the input are imperfect substitutes for the lost one, so the bundle costs more at any prices, by more the larger the share of the bundle the lost source carried and the lower the within-sector elasticity $1/(1-\rho_r)$ of its sector, as \eqref{eq:flex-spec} records. A producer of high intermediate dependence $(1-\beta_j)/\beta_j$ cuts its output steeply when its bundle becomes dearer, by part~(i) of Proposition~\ref{prop:flex-power}. Under fixed prices a displaced dollar of sales and a displaced dollar of purchases go unrecovered in the same fraction $\tau$, and the export multiplier and the import multiplier of most countries are close (Section~\ref{subsec:balance_table} of the paper). These mechanisms can move the burden of a severance from the export elasticity to the import elasticity, as they do in many of the computed comparisons.\footnote{The two effects separate exactly in welfare, which price movements redistribute without adding to the world's loss (Lemma~\ref{lem:flex-zerosum}), and not in activity, which responds to the movements of prices through the wedges.}

The United States--Thailand severance works through both effects. Under fixed prices a Thai producer that bought professional and financial services, chemicals, and electronic components from American firms recovers the fraction $1-\tau$ of them from its remaining suppliers at the old prices. Thailand's rubber-goods and electronics export sectors likewise place the same fraction of their lost American sales with their remaining customers. Under flexible prices these export sectors sell elsewhere what Americans bought only by lowering their prices. A third of the intermediate inputs of Thai producers comes from abroad, at prices that do not fall with Thai prices, and the attenuation of the American source weights raises the input indices of the exposed Thai producers at any prices. Thai producers have an intermediate dependence of about two, more than twice that of American producers, so a rise in their wedges cuts their output steeply. On the American side the sales lost to Thailand are a small fraction of American sales. American producers buy more than nine tenths of their intermediate inputs at home and place those sales elsewhere at a small discount that their domestic suppliers share, and the Thai inputs they lose are a small part of their bundles. Thailand loses more when prices clear than when quantities ration, about three and a half percent of its activity against three, and the United States loses less than half as much.

We read the wedges $z_j$ of every computed severance off its solved prices. The wedges are small, and nearly all of them exceed one. At the median pair the activity-weighted mean wedge is $1.0015$ on the weaker side and $1.0002$ on the favorable side, and producers carrying $98\%$ of the weaker side's activity and $96\%$ of the favorable side's have a wedge above one. Almost every producer on both sides contracts, by a small amount, and the producers that expand return at the median only one percent of the weaker side's loss and three percent of the favorable side's. Section~\ref{app:flex-world} splits the log wedge of every producer on the two sides into the rise in the price of its input bundle that the severed links cause at benchmark prices and the part due to the movement of prices. At the median pair the rise at benchmark prices more than accounts for each side's flexible-price loss, and the movement of prices offsets about a tenth of that rise, because producers pass part of their higher costs on to their customers, and the prices of third countries' goods barely move.

The ordering of power survives price adjustment because prices change what a lost dollar of trade costs each side and leave the trade shares of the two sides where the benchmark put them. The two sides of a pair lose the same flows, so the ratio of their trade shares is the ratio of their sizes, and the size ratios vary far more across the pairs than the costs of a lost dollar do (Section~\ref{subsec:power_size_base} of the paper). On the pairs on which both sides lose under both closures, the log loss ratio correlates with the log size ratio at $0.93$ under flexible prices against $0.98$ under fixed prices. The gain and zero-loss cases lie outside a comparison in logs. The direction of power reverses on pairs of similar size, whose median size ratio is $1.6$ against $6.2$ on the other pairs.

Price adjustment changes the way a loss travels through the economy that suffers it as well as what a lost dollar costs, and on the 2022 trade table the change favors the larger economy. Under fixed prices a loss travels along a supply chain as a loss of output. The larger economy buys more of its inputs at home and more inputs from its partner, so it loses more on each lost dollar, which holds the elasticity of the loss ratio with respect to the size ratio at $0.90$ rather than one (Section~\ref{oa:linkage-size}). Under flexible prices a loss travels along the chain as a movement of prices, and each producer sets its output by its own wedge. The larger economy then no longer loses more on each lost dollar, and the elasticity is $1.01$. The larger economy is the more powerful side on $94\%$ of the pairs, so the change widens the asymmetry of power on most of them, and most on the pairs of greatest asymmetry, where the gap in size is largest.

The shift of the burden from the export elasticity to the import elasticity moves power toward the side of a relationship that sells the other more intermediate inputs than it buys from it. We measure that position by the pair's balance of trade in intermediate inputs, the purchases of the weaker side's producers from the favorable side less the purchases of the favorable side's producers from the weaker side, divided by their sum. On the pairs on which this balance is positive, so that the weaker side buys more intermediate inputs from the favorable side than it sells to it, price adjustment widens the asymmetry on $89\%$. On the other pairs it widens the asymmetry on $30\%$ and moves it in favor of the weaker side on the rest. On nine in ten of the $301$ reversals the fixed-price favorable side buys more intermediate inputs from the weaker side than it sells to it, and on nearly three quarters of them the smaller economy is the more powerful side under flexible prices. The balance of trade in intermediate inputs therefore carries information about power under flexible prices that it does not carry under fixed prices. In a regression of the flexible-price asymmetry on the size asymmetry $\mathcal Q_{cc'}$, adding that balance raises the explained share of the variance from $51\%$ to $70\%$, and under fixed prices it adds nothing.

The United States and China are the largest pair on which the favorable side buys more intermediate inputs from the weaker side than it sells to it, a little under $200$ billion dollars against a little under $150$ billion. Under fixed prices three fifths of the Chinese loss comes from the sales China would lose and three fifths of the American loss from the purchases the United States would lose. Prices absorb much of the Chinese loss on sales and little of the American loss on purchases. The activity-weighted wedge of the United States, $1.0081$, exceeds that of China, $1.0056$, but American producers have an intermediate dependence of $0.9$ against $2.1$ for Chinese producers, so they still lose the smaller share of their activity. The American loss multiplier is the larger, $0.61$ against $0.48$, and price adjustment narrows the asymmetry of the pair from $0.42$ to $0.33$.

Price adjustment lowers the world loss of the named severances and does little to repair the input bundles of the weaker side. Within a sector a displaced seller's price falls against its rivals, and by \eqref{eq:flex-share} it gains a larger share of the spending of every buyer it can still serve. Across sectors substitution runs at the elasticity $1/(1-\rho)>0$ of a finite $\rho<0$, so relative prices can move the sectoral mix, though a sector that delivers no services collapses the CES composite.\footnote{Section~\ref{app:flex-world} measures the reallocation, the part of the loss at benchmark prices that it recovers, and the retention of the severed partner's content under the two closures.}

\subsection{Bystander expansion under flexible prices}
\label{oa:flex-pivot}

Under flexible prices the activity of a country-sector can rise above its benchmark, which fixed prices rule out (Proposition~\ref{prop:m-monotone} of the paper). Every quantity that constrains the placement of displaced orders under fixed prices is a benchmark total, so a bystander that takes up displaced orders in one place must shed orders in another. Under flexible prices a bystander whose output sells for more than its inputs cost deepens its use of intermediates against its fixed block and produces beyond its benchmark. The supply condition \eqref{eq:flex-supply} is the first-order condition of the producer's program $\max_{\eta\ge0}[p_js_jG_j(\eta)-P^{m}_jw_j\eta]$, whose maximizer is unique because $G_j$ is strictly concave (Lemma~\ref{lem:m-outer}\textnormal{(i)} of the paper) and interior under the intensity floor of Assumption~\ref{ass:flex-region}.

\begin{lemma}[Bystander expansion]
\label{lem:flex-pivot}
Fix a sanction $\omega\in\Omega$ and an attenuation $\delta\in(0,\bar\delta]$. At a flexible-price sanction equilibrium, producer $j$ expands, $y_j>s_j$, if and only if $p_j>P^{m}_j$, with the input index given by \eqref{eq:flex-indices}. The equivalence holds at every maintained outer curvature $\varrho<0$. Its specific-factor rent in own-output units, $\varpi_j/p_j$, exceeds the benchmark value one if and only if the same condition holds.
\end{lemma}

\begin{proof}
Fix a flexible-price equilibrium and a producer $j$. By Lemma~\ref{lem:m-outer}\textnormal{(i)} of the paper the outer aggregator $G_j$ is smooth and strictly concave on $(0,\infty)$, so $G_j'$ is continuous and strictly decreasing, and parts \textnormal{(iii)}--\textnormal{(iv)} give $G_j'(1)=e_j(1)=1-\beta_j$. The supply condition \eqref{eq:flex-supply} therefore makes $\eta_j>1$ equivalent to $G_j'(\eta_j)<1-\beta_j$, and hence to $p_j>P^{m}_j$. Since $G_j$ increases strictly and $G_j(1)=1$, the output $y_j=s_jG_j(\eta_j)$ exceeds $s_j$ if and only if $\eta_j>1$.

For the rent, let $\varpi_j$ be the return per unit of the specific endowment $\bar v_j=\beta_js_j$ of Section~\ref{subsec:model_ces} of the paper, so that $\varpi_j\bar v_j=p_jy_j-P^{m}_jw_j\eta_j$ is the operating surplus. Substituting the output $y_j=s_jG_j(\eta_j)$, the input bill $w_j\eta_j$, and the supply condition gives $\varpi_j\bar v_j=p_js_j\,\psi_j(\eta_j)$ for $\psi_j(\eta):=G_j(\eta)-\eta\,G_j'(\eta)$. Strict concavity gives $\psi_j'(\eta)=-\eta\,G_j''(\eta)>0$, and $\psi_j(1)=1-G_j'(1)=\beta_j$. Since $\bar v_j=\beta_js_j$, the real rent is $\varpi_j/p_j=\psi_j(\eta_j)/\beta_j$, which equals one at $\eta_j=1$ and increases strictly in $\eta_j$. The real rent therefore exceeds its benchmark value if and only if $\eta_j>1$. This proves the lemma.
\end{proof}

Under flexible prices a bystander meets $p_j>P^{m}_j$ through a higher output price, a lower input index, or both. Demand reallocated toward it can raise its output price, the inputs that the deprived release can lower its input cost, and neither movement has a general sign in a network. The lemma states the relative-price condition for expansion without requiring every remaining substitute to appreciate or every released input to become cheaper.

At the Leontief limit of the outer aggregator, $\varrho\to-\infty$, the aggregator $G_j(\eta)=[\beta_j+(1-\beta_j)\eta^{\varrho}]^{1/\varrho}$ converges pointwise to $\min\{1,\eta\}$, supply becomes vertical at the benchmark, and no price can induce expansion. The maintained band $[\underline\varrho,\overline\varrho]$ excludes that limit, at which the fixed-price equilibria degenerate as well (Section~\ref{app:m-correction}).

\subsection{The empirical consequences of flexible prices}
\label{app:flex-world}

A purpose-built solver computes the comparisons of Sections~\ref{subsec:flex-comparison} to~\ref{oa:flex-pivot} on the paper's 2022 trade table of eighty economies and fifty industries. Its fixed-price side reimplements the paper's pipeline, the blockwise balancing, the fulfillment fixed point, and the activity solve, and reproduces the paper's country vulnerabilities to six decimal places on the severances checked. Its flexible-price side eliminates quantities through the closed-form demand system of Section~\ref{app:flex-closure} and solves the log-price residual \eqref{eq:flex-residual} by damped iteration with Anderson acceleration and a Newton--Krylov polish. The diagnostics record the convergence of the polish, the residuals, and the range of prices the path visits, and they do not establish membership of the maintained region of Assumption~\ref{ass:flex-region}. A converged polish does not establish that the Jacobian is nonsingular, a full regularity check would add the intermediate-intensity minima and the conditioning of the Jacobian at the solved nodes, and even that check, being finite, would leave the depths between the nodes unverified.

We attempt every country pair at the short-run baseline and ten named pairs additionally at the long-run profile. Of the $3{,}160$ short-run attempts, $3{,}154$ converge to the tolerance of $10^{-6}$ and six do not. The six are China against Saudi Arabia, Belarus against Russia, Iceland against the United States, France against Ireland, Canada against Malta, and China against Cyprus, at residuals of $23$, $8.6$, $2.7$, $2.0$, $1.2$, and $1.0$. We report no number for them and draw no inference from them, since a solver that fails to converge establishes neither that the economy has no equilibrium nor which condition of Assumption~\ref{ass:flex-region} fails. Each of the six strips one side of a partner that supplies a large share of many of its input sectors.

Of the ten named pairs attempted at the long-run elasticities, nine converge, and Belarus against Russia fails at that profile as at the short-run one. On the nine the long-run profile lowers the weaker side's loss below its short-run flexible value in every case, while the asymmetry rises in six, so the change of profile lowers both sides' losses and moves the asymmetry between them in no consistent direction. Across the fixed-price and the flexible-price closures the country losses rank-correlate at $0.97$, a statistic about the levels of the losses rather than about power, and one the paper does not report.

We rebuild the wedges $z_j$ of Proposition~\ref{prop:flex-power} from the solved prices of every computed severance, each producer's composite input price index at the attenuated weights divided by its own price. The output formula of part~(i) reproduces the solved output of every producer of the two parties to within $2\times10^{-6}$, the precision at which the outputs are stored, and the vulnerability formula reproduces each party's solved loss to within $6\times10^{-8}$. The composition of part~(ii) holds to $10^{-15}$ on the $3{,}082$ computed pairs on which both sides lose under both closures. On those pairs the asymmetry moves by a median of $0.10$ against a median bound of $0.25$ from Lemma~\ref{lem:m-rigidity}, and the activity-weighted wedges of Section~\ref{subsec:flex-incidence-econ} have ninetieth percentiles of $1.016$ on the weaker side and $1.002$ on the favorable side.

The split of the wedges in Section~\ref{subsec:flex-incidence-econ} evaluates the composite input price index \eqref{eq:flex-indices} of every producer on the two sides at the attenuated weights and at four price vectors. One vector sets every price at its benchmark value, and each of the other three takes the solved prices of one group of countries, the producer's own country, the partner, or the third countries, with benchmark prices elsewhere. The evaluation at benchmark prices gives the rise in the price of the input bundle that the severed links cause, and the differences of the other three evaluations from it give the parts due to the prices of domestic inputs, of the partner's goods, and of third countries' goods, with a small interaction. The negative of the producer's own log price completes its log wedge. Each producer's term in the loss \eqref{eq:flex-vulnerability} is split in proportion to these parts with the factor $(\sqrt{z_j}-1)/(\tfrac12\log z_j)$, taken as one at $z_j=1$, so the parts sum to the loss the wedges imply to $10^{-15}$. At the median pair the rise at benchmark prices is $1.12$ times the flexible-price loss of the weaker side and $1.09$ times that of the favorable side. The prices of domestic inputs add $0.22$ and $0.25$ of the loss, the producers' own prices subtract $0.40$ and $0.42$, and the prices of the partner's goods and of third countries' goods add less than two hundredths. The balance of trade in intermediate inputs of Section~\ref{subsec:flex-incidence-econ} is positive on $52.6\%$ of the pairs. On the pairs with two positive loss multipliers, a regression of the log ratio of the two multipliers on that balance, on the log size ratio, and on the absolute fixed-price asymmetry gives coefficients of $1.49$, $0.10$, and $-0.06$, with standard errors of $0.02$, $0.01$, and $0.07$. Once the size ratio and this balance are known, the fixed-price asymmetry adds nothing to the difference between the two multipliers.

Retention of the partner's content measures the reallocation of Lemma~\ref{lem:flex-pivot} in embodied value added rather than in gross flows. For each computed severance we rebuild the equilibrium flow matrix from the solved prices and outputs, form the realized input coefficients, and compute Leontief content. The retention of a severance is the value added of the severed partner embodied in the target's deliveries to final demand, at the flexible-price equilibrium relative to the benchmark, and it includes both the residual direct access and the indirect paths through the network, bystanders included. The same reconstruction gives the stand-alone loss $\mathcal E_\omega$, the world activity loss at benchmark prices under the attenuated weights and without market clearing, from the producers' intermediate choices over $\eta_j\ge0$ with zero output at zero intensity. At $\varrho=-1$ that choice is the schedule $y_j^0/s_j=\max\{0,[1-(1-\beta_j)\sqrt{P_j^m(\mathbf1;1)}]/\beta_j\}$. The market recovery $\mathcal R_\omega:=\mathcal E_\omega-\sum_j(s_j-y_j)$ is the part of the stand-alone loss that the flexible-price equilibrium undoes, negative when the equilibrium amplifies it.\footnote{Neither loss is second order in the depth of the sanction, since the stand-alone schedule contains at $t=0$ the same intermediate-use response as \eqref{eq:flex-activity-rate}. The recovery therefore has no definite sign at first order, and both its size and its sign are computed.} Table~\ref{tab:oa-flex-census} reports the census on the named severances.

On the named pairs the median retention of the partner's content is $26.8\%$ under fixed prices and $28.3\%$ under flexible prices, and the flexible retention exceeds the fixed in $17$ of the $26$ directions. Across the $3{,}154$ computed severances the flexible retention has quartiles $32\%$, $43\%$, and $60\%$. Under the United States--China severance the two closures alike retain $35\%$ of Chinese content in American final deliveries and $33\%$ in the reverse direction. Retention this close coexists with different substitution among sources and different input demands, and with the expansion margin of Lemma~\ref{lem:flex-pivot}.

The market recovery is small and mostly positive, for the recovery share $\mathcal R_\omega/\mathcal E_\omega$ is positive on $67\%$ of the computed severances, with quartiles $-5\%$, $8\%$, and $19\%$, and negative, an amplification, on a third. Both tails belong to severances too small for the ratio to carry information, since it divides two world activity losses that fall to $10^{-8}$ of world activity on the smallest pairs. Above a materiality floor of $0.01\%$ of world benchmark activity, which $249$ severances pass, the recovery is positive on $74\%$ with a median of $12\%$, and the deepest material amplification is Kazakhstan against Russia at $-33\%$.

On the named severances the flexible-price world loss runs from $0.3$ to $0.8$ of the fixed-price loss, with a median of $0.4$, which does not make the fixed-price loss a general upper bound. The division of the loss between the two sides in Section~\ref{subsec:flex-comparison} labels each pair's favorable side by its fixed-price outcome and takes the ratio from its flexible-price outcome, so the comparison is not circular, though we do not establish how a label chosen on a noisy baseline moves the measured gap.

Table~\ref{tab:oa-flex-income} replaces the fixed nominal budgets of the households with an income-consistent rule on the named severances. Each budget moves with its country's factor income at the solution, $F^{c}_{\omega}=F^{c}\,\mathrm{VA}_c(\omega)/\mathrm{VA}_c$, so a country whose producers lose rents also spends less. The rule raises the weaker side's loss in every one of the thirteen named severances, by a median of $14\%$ of its fixed-budget value, and widens the asymmetry in nine of them. The median absolute asymmetry across the thirteen is $0.79$ against $0.80$ with fixed budgets, so the rule moves the level of the weaker side's loss and leaves the division of the loss close to where the fixed budgets put it.

\begin{table}[!htbp]
\centering
\small
\begin{tabular}{@{}llrrrrrr@{}}
\toprule
 & & \multicolumn{2}{c}{Loss of $c$ (\%)} & \multicolumn{2}{c}{Loss of $c'$ (\%)} & \multicolumn{2}{c}{$\mathcal H_{cc'}$} \\
\cmidrule(lr){3-4}\cmidrule(lr){5-6}\cmidrule(l){7-8}
$c$ & $c'$ & fixed budgets & income & fixed budgets & income & fixed budgets & income \\
\midrule
USA & CHN & 0.34 & 0.31 & 0.66 & 0.82 & $-0.33$ & $-0.45$ \\
VNM & CHN & 7.66 & 9.11 & 0.10 & 0.11 & $+0.97$ & $+0.98$ \\
CAN & USA & 6.80 & 8.27 & 0.75 & 0.78 & $+0.80$ & $+0.83$ \\
MEX & USA & 9.39 & 10.72 & 0.23 & 0.23 & $+0.95$ & $+0.96$ \\
DEU & RUS & 0.38 & 0.41 & 0.44 & 0.45 & $-0.07$ & $-0.05$ \\
USA & RUS & 0.04 & 0.04 & 0.68 & 0.70 & $-0.89$ & $-0.88$ \\
JPN & CHN & 1.50 & 1.70 & 0.49 & 0.53 & $+0.51$ & $+0.53$ \\
IND & CHN & 0.80 & 0.85 & 0.08 & 0.10 & $+0.83$ & $+0.79$ \\
DEU & CHN & 0.96 & 1.07 & 0.31 & 0.33 & $+0.51$ & $+0.53$ \\
CHN & TWN & 0.26 & 0.27 & 3.27 & 3.82 & $-0.85$ & $-0.87$ \\
KAZ & RUS & 3.92 & 4.00 & 0.20 & 0.22 & $+0.90$ & $+0.90$ \\
AUS & CHN & 1.38 & 1.69 & 0.28 & 0.30 & $+0.66$ & $+0.70$ \\
KOR & CHN & 3.03 & 3.57 & 0.40 & 0.43 & $+0.77$ & $+0.78$ \\
\bottomrule
\end{tabular}
\caption{The flexible-price closure with fixed nominal household budgets, the paper's reading, against the income-consistent variant in which each household's budget moves with its country's factor income at the solution. Own-size-normalized losses in percent and the bilateral asymmetry $\mathcal H_{cc'}$, positive against $c$. Short-run calibration, $\delta=0.10$.}
\label{tab:oa-flex-income}
\end{table}

\begin{figure}[!htbp]
\centering
\includegraphics[width=\linewidth]{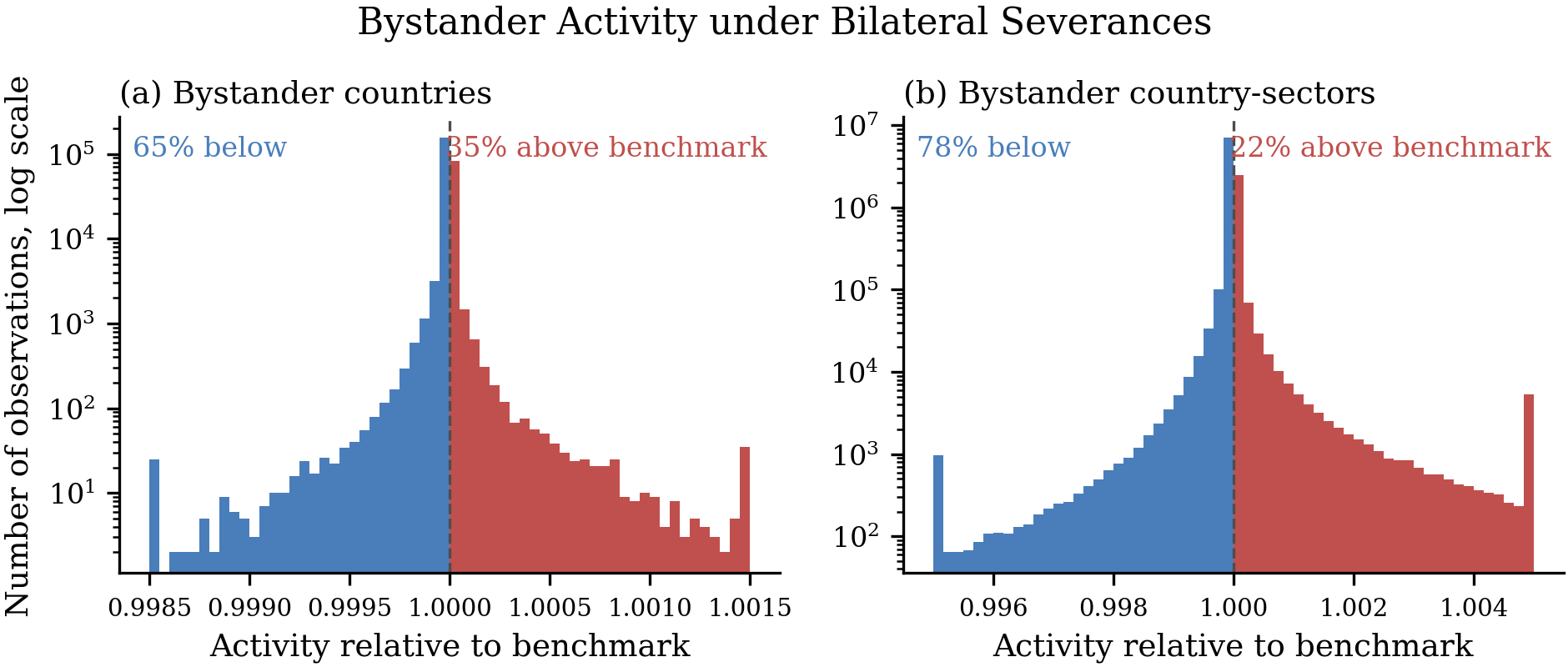}
\caption{Bystander activity under bilateral severances. Panel (a) pools bystander countries across the computed severances, and panel (b) pools bystander country-sectors. Activity is normalized to one at the benchmark. Counts use a log scale, the dashed line marks no change, and the extreme bins collect the observations outside the plotted range. Baseline short-run calibration, $\delta=0.10$.}
\label{fig:s6-flex-bystander}
\end{figure}

Figure~\ref{fig:s6-flex-bystander} pools bystander activity across the computed severances, by country in panel~(a) and by country-sector in panel~(b), with activity normalized so that one is the benchmark. Under fixed prices the mass to the right of one would be empty (Proposition~\ref{prop:m-monotone} of the paper). At the sector level $0.3\%$ of the observations gain more than a tenth of a percent, and the largest single gain is $120\%$.\footnote{Panel~(b) is drawn from activity ratios stored in single precision, so it records a gain only above six parts in a hundred million and puts the share of bystander country-sectors above benchmark at $22\%$, where the solver's double-precision output puts it at $28\%$.}

\begin{figure}[!htbp]
\centering
\includegraphics[width=\linewidth]{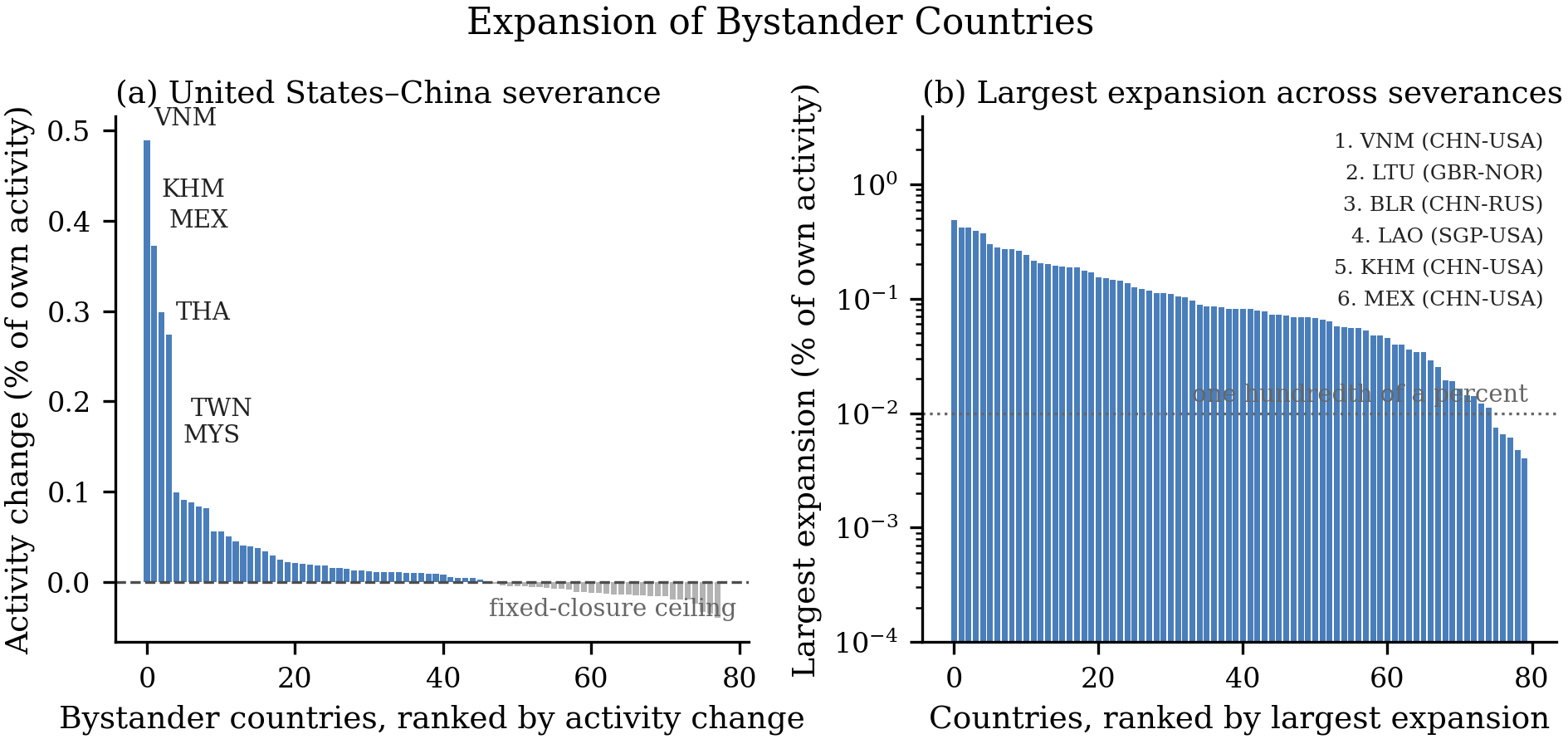}
\caption{Expansion under flexible prices. Panel (a) ranks the bystanders of the United States--China severance by their activity change in percent of own activity, with the gains in blue above the zero line that the fixed-price closure cannot cross (Proposition~\ref{prop:m-monotone} of the paper). Panel (b) ranks the eighty economies by their largest expansion across the computed severances, on a log scale, with the responsible severance named for the leading countries and the dotted line marking one hundredth of a percent of own activity. Baseline short-run calibration, $\delta=0.10$, $3{,}154$ severances.}
\label{fig:s6-flex-expansion}
\end{figure}

Figure~\ref{fig:s6-flex-expansion} reports the gains of each country across the computed severances. A country's largest expansion is the maximum of its activity gains across the computed severances to which it is not a party, in percent of its own benchmark activity. A party expands in seventy-one of the computed severances, all of them small commodity-exporting countries severed from a distant customer and only five gaining as much as a hundredth of a percent of their own activity, so up to that qualification the figure is a census of bystander gains. Fixed prices put a ceiling of zero on every such gain (Proposition~\ref{prop:m-monotone} of the paper).

\begin{table}[!htbp]
\centering
\small
\begin{tabular}{@{}llrrrrrrr@{}}
\toprule
 & & \multicolumn{2}{c}{Retention in $c$} & \multicolumn{2}{c}{Retention in $c'$} & & & \\
\cmidrule(lr){3-4}\cmidrule(lr){5-6}
$c$ & $c'$ & fixed & flexible & fixed & flexible & $\mathcal E_\omega$ (\%) & $\mathcal R_\omega/\mathcal E_\omega$ (\%) & $\dfrac{\text{loss}^{\text{flex}}}{\text{loss}^{\text{fix}}}$ \\
\midrule
USA & CHN & 0.35 & 0.35 & 0.33 & 0.33 & 0.23 & +6 & 0.42 \\
VNM & CHN & 0.22 & 0.28 & 0.22 & 0.18 & 0.11 & +25 & 0.32 \\
CAN & USA & 0.13 & 0.18 & 0.28 & 0.28 & 0.36 & +8 & 0.65 \\
MEX & USA & 0.17 & 0.23 & 0.20 & 0.17 & 0.22 & +21 & 0.46 \\
DEU & RUS & 0.36 & 0.37 & 0.30 & 0.31 & 0.03 & +11 & 0.43 \\
USA & RUS & 0.47 & 0.47 & 0.38 & 0.40 & 0.03 & +4 & 0.81 \\
JPN & CHN & 0.24 & 0.27 & 0.25 & 0.24 & 0.21 & +13 & 0.48 \\
IND & CHN & 0.29 & 0.31 & 0.46 & 0.44 & 0.06 & +27 & 0.39 \\
DEU & CHN & 0.32 & 0.35 & 0.31 & 0.30 & 0.14 & +19 & 0.45 \\
CHN & TWN & 0.27 & 0.28 & 0.23 & 0.26 & 0.12 & +29 & 0.33 \\
KAZ & RUS & 0.18 & 0.25 & 0.14 & 0.13 & 0.01 & -33 & 0.60 \\
AUS & CHN & 0.25 & 0.29 & 0.42 & 0.49 & 0.07 & -28 & 0.32 \\
KOR & CHN & 0.20 & 0.23 & 0.27 & 0.26 & 0.20 & +15 & 0.41 \\
\bottomrule
\end{tabular}
\caption{The conduit-retention census on the named severances. Retention in $c$ is the value added of $c'$ embodied in the deliveries of $c$'s producers to final demand after the severance, relative to benchmark, under each closure, and retention in $c'$ is the reverse direction. $\mathcal E_\omega$ is the stand-alone loss, the world activity loss at benchmark prices under the attenuated weights, in percent of world benchmark activity, $\mathcal R_\omega/\mathcal E_\omega$ the share of it the flexible market recovers, negative where equilibrium interaction amplifies, and the last column the ratio of the flexible to the fixed world loss. Baseline short-run calibration with the household columns, $\delta=0.10$.}
\label{tab:oa-flex-census}
\end{table}

The expansion of the bystanders serves the two parties at the margin, and it is only partly made of the goods they lost.\footnote{In $97.6\%$ of the computed severances the incremental deliveries of the expanding bystanders toward the two parties exceed their entire net expansion, a median of $2.0$ units redirected to the pair per unit of expansion. The rest of what the bystanders send the pair is withdrawn from the customers they keep. The sectoral composition of what bystanders grow overlaps the composition of what the parties lose by $0.5$ at the median, between $0.4$ and $0.7$ over the middle four fifths of severances. About half of the growth is therefore, sector for sector, the goods the pair lost, and the rest is the inputs and margins that go into the new assembly. Overlap is $\sum_r\min(u_r,v_r)$ for $u$ the sector shares of the expansion and $v$ the sector shares of the parties' contraction.} Figure~\ref{fig:s6-flex-expansion} draws the expansion twice. In panel~(a), Vietnam, Cambodia, Mexico, Thailand, Malaysia, and Chinese Taipei stand above a line that fixed prices let no country cross, and forty-six of the seventy-eight bystanders of the United States--China severance stand above it by some margin. In panel~(b), seventy-five of the eighty economies have a severance somewhere in the world worth at least a hundredth of a percent of their activity, and fifty-eight have one worth at least a twentieth. Apart from the seventy-one small expansions of a party, the gains belong to bystanders. The severance that benefits a country most follows from its place in the network. The leaders of panel~(a) expand most when the United States and China sever, since the orders of American households for consumer goods move to them, Belarus when China and Russia do, Laos when the United States and Singapore do, and Luxembourg when the United States and Ireland do. The United States ranks near the bottom, since no other pair's severance is large enough to move it. Power in the paper is the capacity to harm a partner. Figure~\ref{fig:s6-flex-expansion} measures the converse capacity, the capacity to gain from other countries' conflicts, and the same network fixes both.

\section{Our Sanctions Dataset}
\label{oa:datasets}

The empirical test of Section~\ref{sec:validation} of the paper scores decisions rather than countries. Its unit is the decision to restrict goods trade, and its acting party is the group of governments that took the decision. Thirty-one of the thirty-seven economies that sanctioned Russia in 2022 are individually more exposed to Russia than Russia is to them, and none of them acted alone, so scoring each separately against its own bilateral entry would evaluate a decision it did not take. The group severance of that section therefore solves the equilibrium once for each decision. Like the bilateral severance, it bars intermediate and final trade in every sector, and it compares the proportional losses in gross activity that a broad disruption imposes on the acting group and on the target. Without the political objectives and concessions at stake it does not predict the outcome of a bargain, and agreement between the observed initiator and the model's favorable side does not estimate the welfare effect of the legal instrument a decision used.

A decision enters the record if it is a state-imposed or state-directed restriction of goods trade, imposed and not merely threatened between 1995 and 2022, with every sender and every target among the eighty economies of the trade table. General, nondiscriminatory commercial policy is excluded, as are routine tariff schedules, anti-dumping actions, safeguards, and WTO-authorized retaliation that is not used as partner-specific coercion. Discriminatory tariffs and the withdrawal of trade preferences enter when they are directed at a named partner for a coercive purpose, and the decision list marks them. Informal but state-directed restrictions enter as well, sanitary pretexts and customs slowdowns among them. A response counts as a retaliation only if it answers the counterpart's own economic coercion, so a sanction that follows a military or political act is an initiation. The rule yields forty-five decisions, thirty-six initiations and nine retaliations, taken between 1998 and 2022. We screened every candidate in writing and filed each rejected candidate beside the retained decisions with the reason for its rejection. The rejected candidates include measures that reached only arms or dual-use licensing, without a broader goods-trade component.

The decisions come from three sources of unequal documentary quality, and the record reports the source of each. Thirteen form an original core drawn from European, American, and British program lists, twenty-nine come from a later expansion with its own primary sources, and three come from the Global Sanctions Data Base (GSDB). The count of favorable initiations holds within each source. The initiator holds the favorable side in ten of the eleven original-core initiations, in twenty-one of the twenty-four expansion initiations, and in the single initiation that the GSDB supplies. Two decisions are marked as less firmly documented, one of them an initiation, and dropping both leaves thirty-one favorable initiations of thirty-five. Where their coverage overlaps ours, we consult TIES v4 \citep{oa-morganbapatkobayashi2014} and the Chinese Economic Sanctions dataset \citep{oa-zhangshanks2025}. Their units and inclusion rules differ, so the overlap is not treated as independent confirmation.

The count of thirty-two favorable initiations in thirty-six is descriptive (Section~\ref{sec:validation} of the paper), and it does not rest on the harder cases or on the boundary instruments. The initiator holds the favorable side in twenty-one of the twenty-two formal initiations, those announced in a legal instrument, and in eleven of the fourteen informal ones. An informal measure is state-directed but never announced, and it works through a sanitary pretext, a customs slowdown, or an inspection regime. Six initiations and one retaliation work through the tariff schedule or the withdrawal of a trade preference rather than barring trade outright, and dropping those six initiations leaves twenty-six of thirty.

We score each decision on the benchmark table of its own year. The record begins in 1998 and the benchmark series in 1995, so every decision has a trade table of its own year and none needs a fallback vintage, though a trade table of the same year need not precede the decision and can include transactions that follow it.\footnote{The 1995--2010 tables are the small tables of the 2025 edition of the OECD ICIO. The 2011--2015 tables fold that edition's extended tables, which split China and Mexico by intermediate use, back into their parent countries. On a held-out year the conversion reproduces entries of order $10^{6}$ to within $2\times10^{-4}$. The tables of the 2023 edition cannot serve here, since their classification merges coal with crude oil and gas.} Scoring the decisions before 2016 on a single later trade table instead moves one initiation across the line. China against Japan in 2010 has an asymmetry of $+0.23$ on the 2016 trade table and $-0.08$ on the 2010 trade table, on which Japan was the larger economy.

The coverage of the trade table limits what the count supports. Iran, North Korea, Venezuela, Cuba, Syria, Afghanistan, Libya, Sudan, Zimbabwe, Nicaragua, and Mali are absent from it, so many prominent episodes cannot enter. Nor do the two largest senders give the count much identifying variation, since the United States holds the favorable side against all $79$ other economies and China against $78$. China's eight targets of solo initiation have a median advantage of $0.78$ against $0.95$ across all Chinese pairs, and the seven American targets a median of $0.97$, the same as across all American pairs. A Mann--Whitney comparison of the advantage ranks places China's chosen targets lower in its distribution than its other partners ($p=0.01$), and it does not reject equal distributions for the United States ($p=0.12$). Neither sender concentrates its choices where its advantage is greatest. The Hirschman matrix $\boldsymbol{\mathcal H}$ establishes which side of a named relationship holds the advantage, and it does not predict which relationships come to be named. The complete decision list, with the instruments, the trigger, the sources, and the coding fields of every decision and the record of the candidates screened out, is deposited in the Replication Package as \texttt{Data\_CoercionDecisions}.

We use the case-specific GSDB-R4 file rather than its dyadic version \citep{oa-felbermayr2020gsdb,oa-gsdbR4}. The GSDB covers more than our record does, including arms and repression-equipment embargoes that our comparability rule removes when they carry no wider goods-trade component. Its file carries one row per case rather than one per legal instrument, and its cases need not be the decisions of our hand-collected record. We keep the cases with a trade component, restrict their dates to the window of the ICIO tables, and require their senders and targets to be in the trade table. We then expand a multi-target case into one record per target and group the records that share a target and a starting year, with the union of their senders as the acting group. This target-year grouping is a comparison device and not, in general, the acting unit of a documented decision. The grouping can combine distinct senders, instruments, or renewals that share a target and a year, and a single multi-target decision of our record can correspond to several target-specific GSDB episodes.

Matching on target and year, $116$ of the $185$ episodes link to at least one of our forty-five decisions, and twenty-six of the forty-five link to at least one episode. Most of what fails to match is informal measures on our side and arms, dual-use, or renewal cases on theirs. The comparison of the two records is therefore descriptive, and it treats neither as a random sample of coercion decisions. The staged construction, the curation verdicts, the map from cases to episodes, and the surviving episode file are deposited in the Replication Package as \texttt{Data\_GSDB\_Episodes}.

\section{Notation}
\label{app:notation}

This section collects the principal notation of the paper and of this appendix, the orientation of the principal matrices, and the conventions that govern typography and accents. In the country-sector matrices rows are sellers and columns are buyers, so an entry indexed by \(i,j\) is a flow or a share from seller \(i\) to buyer \(j\). The country-level trade matrix \(\mathbf T^{\mathrm{cb}}\) takes the opposite orientation, with buyer countries in its rows and seller countries in its columns. Matrices are uppercase boldface, vectors lowercase boldface, and scalars lowercase. The household matrix \(\mathbb A\) and the indicator vector of a country are double-struck, and level and flow quantities remain plain uppercase even when scalar. Greek letters denote parameters and derived quantities, and calligraphic letters denote sets, mappings, and special objects.

Accents and arrows carry a fixed meaning throughout. For a generic object \(x\), \(\underline{x}\) and \(\overline{x}\) denote a lower and an upper bound, and an overbar also denotes an average where stated. A tilde \(\widetilde{x}\) denotes a no-disruption counterfactual, except on the pseudo-prices \(\tilde p_i\) and pseudo-incomes \(\tilde E_{rb}\) of the rationed exchange, a caron \(\check{x}\) a disrupted object taken before the rationed exchange, and a hat \(\widehat{x}\) an empirical estimate. An up arrow \(x^{\uparrow}\) marks a quantity measured on a country's sourcing side and a down arrow \(x^{\downarrow}\) a quantity measured on its sales side. A subscript or parenthetical superscript \(\omega\), as in \(x_\omega\) or \(x^{(\omega)}\), marks a quantity under the sanction \(\omega\), and superscripts \(0\) and \(1\) mark benchmark and post-severance values.

\renewcommand{\arraystretch}{1.0}
\footnotesize

\begin{longtable}{@{}p{0.25\textwidth}p{0.69\textwidth}@{}}
\caption{Notation used in the main paper and the online appendix.}
\label{tab:oa-notation}\\
\toprule
\textbf{Symbol} & \textbf{Meaning} \\
\midrule
\endfirsthead

\multicolumn{2}{c}{\textit{Table \thetable\ continued}}\\
\toprule
\textbf{Symbol} & \textbf{Meaning} \\
\midrule
\endhead

\midrule
\multicolumn{2}{r}{\textit{continued on next page}}\\
\endfoot

\bottomrule
\endlastfoot

\multicolumn{2}{@{}l}{\textit{Sets, indices, and basic conventions}}\\
\addlinespace[0.35em]

\(C\) & Number of countries. \\

\(S\) & Number of sectors in each country. \\

\(N\le CS\) & Number of active country-sectors after removing zero-activity pairs from the \(CS\) country-sector pairs. \\

\(i,j\) & Country-sector indices over \(\{1,\dots,N\}\), with \(i\) a seller and \(j\) a buyer. \\

\(c,c'\) & Countries, ranging over \(\{1,\dots,C\}\). \\

\(b\) & Buyer index for a producer \(j\) or country-household \(c\). The objects \(a_{ib}\), \(w_b\), \(Z_{ib}\), and \(t^{(\omega)}_{rb}\) apply to both kinds of buyer. \\

\(\circ\) & Entrywise product of two vectors. \\

\(r\) & Sector index, ranging over \(\{1,\dots,S\}\). \\

\(\mathcal C(i)\) & Country of country-sector \(i\) (Definition~\ref{def:m-country-sector}). \\
\(\mathcal S(i)\) & Sector of country-sector \(i\), with \(\mathcal N_r:=\{i:\mathcal S(i)=r\}\) the country-sectors in sector \(r\). \\

\(\mathcal F\) & Sector aggregation operator, \(\mathcal F(\mathbf x)_r=\sum_{i\in\mathcal N_r} x_i\), applied column by column to a matrix. \\

\(\mathbf 1\) & Vector of ones of appropriate dimension. \\

\(\mathbbm{1}_c\) & Indicator vector for country \(c\), acting as a row, so \(\mathbbm{1}_c\mathbf x\) sums over country \(c\). \\

\(\|\cdot\|_1\) & Maximum absolute column-sum for matrices, sum of absolute values for vectors. \\

\(\|\cdot\|_\infty\) & For vectors, the maximum absolute entry. \\

\(\spr(\cdot)\) & Spectral radius. \\

\addlinespace[0.8em]
\multicolumn{2}{@{}l}{\textit{Benchmark network and activity}}\\
\addlinespace[0.35em]

\(\mathbf Z=[Z_{ib}]\) & Benchmark transaction matrix in value units. Rows are sellers, and columns include producer and household buyers. \\

\(w_b\) & Buyer \(b\)'s benchmark spending, \(w_j=\sum_i Z_{ij}=(1-\beta_j)s_j>0\) for a producer and \(w_c=F^{c}\) for a household. \\

\(\mathbf A=[a_{ij}]\) & Buyer-column input-share matrix, \(a_{ij}=Z_{ij}/w_j\), nonnegative with unit column sums. \\

\(\beta_j \) & Leakage parameter of country-sector \(j\), the share of gross activity not spent on retained-network intermediate inputs. The diagonal matrix and bounds are \(\boldsymbol\beta=\diag(\beta_1,\dots,\beta_N)\), \(\bar\beta:=\max_i\beta_i\), and \(\underline\beta:=\min_i\beta_i\). \\

\(\mathbf A(\mathbf I-\boldsymbol\beta)\) & Benchmark propagation matrix, entry \((1-\beta_j)a_{ij}\), with column \(j\) summing to \(1-\beta_j<1\). \\

\(f^{c}_i,\ F^{c},\ \mathbf F\) & Country-household \(c\)'s benchmark purchase of country-sector \(i\)'s output, its budget \(F^{c}=\sum_if^{c}_i\), and the vector of budgets. \\

\(\mathbb A=[a_{ic}]\) & Household matrix, \(a_{ic}=f^{c}_i/F^{c}\), with unit column sums. Its sector shares are \(\alpha_{rc}=\sum_{i\in\mathcal N_r}a_{ic}\). \\

\(\mathbf f^{o}\) & Residual final demand that no sanction reaches: the purchases of the rest of the world sector, inventory changes, and statistical discrepancies. \\

\(\mathbf f\) & Vector of final demand for each country-sector's output, \(\mathbf f=\mathbb A\mathbf F+\mathbf f^{o}\) \eqref{eq:m-final-split}. \\

\(\mathbf s\) & Benchmark activity vector, the solution of
\(\mathbf s=\mathbf f+\mathbf A(\mathbf I-\boldsymbol\beta)\mathbf s\). \\

\(s_i \) & Benchmark activity of country-sector \(i\). \\

\(Q(c)\) & Benchmark size of country \(c\), \(Q(c)=\sum_{i:\mathcal C(i)=c}s_i\). \\

\(\mathbf P_0\) & Column-stochastic completion of the benchmark propagation matrix. Its unique invariant distribution is \(\mathbf s/(\mathbf1^\top\mathbf s)\), as defined in \eqref{eq:m-size-invariant}. \\

\(\sigma^{(\omega)}_{rb}\) & Severed partner's share of buyer \(b\)'s benchmark purchases of input \(r\) (Sections~\ref{oa:diversification} and~\ref{subsec:flex-welfare}). \\

\addlinespace[0.8em]
\multicolumn{2}{@{}l}{\textit{Sanctions and rationed exchange}}\\
\addlinespace[0.35em]

\(\omega\) & A trade sanction: a bilateral country sanction, a group severance, or a sector-level
refinement of either. \\

\(\Omega\) & Set of admissible sanctions: bilateral country
sanctions, group severances, and their sector-level refinements. \\

\(c'\notnotleftrightarrow c\) & Bilateral sanction barring every sector of country \(c'\) from supplying country \(c\)'s producers and household, and every sector of \(c\) from supplying country \(c'\)'s producers and household. \\

\(\tau\) & Unrecovered share of displaced trade \(\tau\in[0,1]\), the fraction of displaced mass removed from the buyer and seller margins. \\

\(\tau_{ib},\ \bar\tau_X,\ \bar\tau_M\) & Unrecovered share of the trade displaced on the severed link \(i\to b\), \(\tau_{ib}\in(0,1]\), and its means over the links of the flows \(X_{cc'}\) and \(M_{cc'}\), weighted by the benchmark flows, when the unrecovered share differs across the severed links (Section~\ref{oa:unequal-shares}). \\

\(\delta\) & Barred-link attenuation, \eqref{eq:m-leak}, the share of a barred link's benchmark weight retained in the disrupted matrix, \(\delta\in(0,\bar\delta]\) with \(0<\bar\delta<1\), so that a higher \(\delta\) is a weaker attenuation. \\

\(\delta^{(\omega)}_{ij}\) & Realized coefficient attenuation on link \((i,j)\) after the rationed exchange, \(\delta^{(\omega)}_{ij}=a^{(\omega)}_{ij}/a_{ij}\) \eqref{eq:m-realized-attenuation}. \\

\(X^{0},\ X^{1},\ \widehat M^{\mathrm{rec}}\) & Benchmark and post-severance flows on a barred relationship, and demand-adjusted recovery with unbarred partners. These flow measures are the inputs to the calibration of the unrecovered share in Section~\ref{app:m-tau}. \\

\(\nu_{rc},\ \widetilde\nu_{rc}\) & Buyer country \(c\)'s total sector-\(r\) intake after a severance and its estimated no-disruption level (Definition~\ref{def:tau-estimator}). \\

\(\check{\mathbf A}=[\check a_{ij}]\) & Disrupted matrix:
\(\check a_{ij}=a_{ij}\) on unsevered links and \(\delta\,a_{ij}\) on severed
links, so
\(\operatorname{supp}(\check{\mathbf A})=\operatorname{supp}(\mathbf A)\). \\

\(\mathbf A_\omega=[a^{(\omega)}_{ij}]\), \(\mathbb A_\omega=[a^{(\omega)}_{ic}]\) & Producer and household standing-order blocks \eqref{eq:m-balanced-coeff}. Their coefficients are the rationed-exchange allocation, satisfying targets \eqref{eq:m-rowtarget} and \eqref{eq:m-coltarget} in each sector block.
\\

\(\mathbf f_\omega\) & Effective final demand, \(\mathbf f_\omega=\mathbb A_\omega\mathbf F+\mathbf f^{o}\), equal to \(\mathbf f\) when nothing is severed. \\

\(\boldsymbol\xi_\omega\) & Lost-sales vector, \(\xi_{\omega,i}=\tau(1-\delta)\sum_{b:\,i\to b\text{ severed}}Z_{ib}=s_i-f^{o}_i-g^{(\omega)}_i\) \eqref{eq:m-lost-sales}. The split \(\boldsymbol\xi^{I}_\omega+\boldsymbol\xi^{F}_\omega\) distinguishes producer and household customers. \\

\(Z_{\omega,ib}\) & Balanced standing-order flow from seller \(i\) to buyer \(b\), \(Z_{\omega,ib}=a^{(\omega)}_{ib}w_b\) \eqref{eq:m-balanced-coeff}. \\

\(Z^{*}_{\omega,ij}\) & Realized post-sanction flow at the selected equilibrium, \(Z^{*}_{\omega,ij}=h_{\omega,i}\,a^{(\omega)}_{ij}(1-\beta_j)s_{\omega,j}\). \\

\(\mathcal I_r^{\omega},\ \mathcal J_r^{\omega}\) & Active sellers and active buyers of the sector-\(r\) block (Definition~\ref{def:oa-rationed-exchange}), \(\mathcal I_r^{\omega}=\{i\in\mathcal N_r:g^{(\omega)}_i>0\}\) and \(\mathcal J_r^{\omega}=\{j:t^{(\omega)}_{rj}w_j>0\}\). \\

\(t^{(\omega)}_{rb},\ g^{(\omega)}_i\) & Buyer \(b\)'s sector-\(r\)
budget \eqref{eq:m-coltarget} and seller \(i\)'s capacity
\eqref{eq:m-rowtarget}. These margins subtract the fraction \(\tau(1-\delta)\) of severed mass from benchmark totals over producer and household buyers. \\

\(u_iv_{rj}\) & Accommodation multiplier of link \((i,j)\) (Section~\ref{subsec:recomposition}), the product of the seller and buyer scalings on affected blocks and \(1\) elsewhere. \\

\(\tilde p_i\) & Pseudo-price of active seller \(i\), \(\tilde p_i=1/u_i\) (Proposition~\ref{prop:m-decentralization}). \\

\(q_r\) & Bound on the contraction modulus of a Sinkhorn half-update on the
sector-\(r\) block,
\(q_r=\tanh(\operatorname{diam}(\check{\mathbf A}^{(r)})/4)\). The full-sweep modulus is at most \(q_r^2\). \\

\(\mathbf B_\omega\) & Sanctioned propagation matrix,
\(\mathbf B_\omega:=\mathbf A_\omega(\mathbf I-\boldsymbol\beta)\), which is nonnegative and column-substochastic and satisfies
\(\|\mathbf B_\omega\|_1\le\|\mathbf A(\mathbf I-\boldsymbol\beta)\|_1<1\). \\

\addlinespace[0.8em]
\multicolumn{2}{@{}l}{\textit{The flexible-price economy}}\\
\addlinespace[0.35em]

\(\rho_r\) & Within-sector CES curvature over source varieties in sector \(r\), maintained in \([\underline\rho_r,\overline\rho_r]\subset(0,1)\) with implied elasticity \(1/(1-\rho_r)\) (Section~\ref{app:flex}). \\

\(\epsilon_r\) & Tariff-identified trade elasticity of sector \(r\) \citep{oa-caliendoparro2015}, from which the within-sector curvature follows as \(\rho_r=\epsilon_r/(1+\epsilon_r)\) in the long run, with \(\epsilon_r\) scaled by \(0.38\) at the short-run baseline (Table~\ref{tab:flex-sigma}). \\

\(\varpi_j,\ \eta_j\) & Respectively, producer \(j\)'s endowment rent, measured as operating surplus per unit of the non-intermediate block, and intermediate intensity relative to benchmark in the flexible-price economy (Section~\ref{subsec:flex-economy}). \\

\(P_{rj},\ P^{m}_j\) & Within-sector and composite input price indices of
buyer \(j\) in the flexible-price economy, displays~\eqref{eq:flex-share} and~\eqref{eq:flex-indices}. \\

\(\mathbf J^{\mathrm p}\) & Jacobian of the flexible-price residual \eqref{eq:flex-residual} in log prices (Section~\ref{app:flex-closure}). \\

\(\mathcal M_\omega\) & Direct cost-index loss of severance \(\omega\), the Domar-weighted direct rise in the input and consumption price indices at the benchmark. Its closed form is \eqref{eq:flex-M}. \\

\(\mathcal E_\omega,\ \mathcal R_\omega\) & Stand-alone loss of severance \(\omega\) at benchmark prices with attenuated weights, and the market recovery \(\mathcal E_\omega-\sum_j(s_j-y_j)\) (Section~\ref{app:flex-world}). \\

\(\mathcal P\) & Maintained regularity region of the flexible-price economy (Assumption~\ref{ass:flex-region}). \\

\(\tau^{*}\) & Implied unrecovered share of a severance, at which the weaker side's fixed-price loss matches its flexible-price loss within a bracket of the paper's \(\tau\)-grid. The implied unrecovered share is a calibration diagnostic (Section~\ref{subsec:flex-calibration}). \\

\addlinespace[0.8em]
\multicolumn{2}{@{}l}{\textit{CES-shaped fulfillment and bundle efficiency}}\\
\addlinespace[0.35em]

\(\alpha_{rj}\) & Buyer \(j\)'s benchmark sector share,
\(\alpha_{rj}=\mathcal F(\mathbf A)_{rj}\), with \(\sum_{r}\alpha_{rj}=1\). \\

\(\mathcal R_j\) & Buyer \(j\)'s benchmark sector support, \(\mathcal R_j=\{r:\alpha_{rj}>0\}\), over which all CES sums run. \\

\(\rho\) & CES exponent of the inner aggregator across sector-level intermediate inputs, maintained on \([\underline\rho,\overline\rho]\subset(-\infty,0)\). \\

\(\varrho\) & CES exponent of the outer aggregator between the intermediate composite and the non-intermediate block, maintained on \([\underline\varrho,\overline\varrho]\subset(-\infty,0)\).
\\

\(G_j(\kappa)\) & Outer aggregator of producer \(j\) \eqref{eq:m-lambda-rule}, \(G_j(\kappa)=[\beta_j+(1-\beta_j)\kappa^{\varrho}]^{1/\varrho}\).
\\

\(e_j(\kappa)\) & Intermediate output elasticity of producer \(j\) \eqref{eq:m-shadow}, \(e_j(\kappa)=(1-\beta_j)\kappa^{\varrho}/[\beta_j+(1-\beta_j)\kappa^{\varrho}]\), with \(\bar e_\omega=\max_i e_i(\underline\kappa_\omega)\) a maintained upper bound.
\\

\(Y_j(\mathbf q_{\cdot j})\) & Buyer \(j\)'s effective intermediate input composite, the normalized CES aggregator \(Y_j(\mathbf q_{\cdot j})=(\sum_{r\in\mathcal R_j}\alpha_{rj}(q_{rj}/\alpha_{rj})^{\rho})^{1/\rho}\). \\

\(q^{(\omega)}_{rj}(x_j,\mathbf h)\), \(q^{0}_{rj}\) & Delivered and benchmark sector-\(r\) input quantities of buyer \(j\) (Section~\ref{app:m-local-efficiency}), \(q^{(\omega)}_{rj}(x_j,\mathbf h)=\alpha_{rj}(1-\beta_j)x_j\,\theta^{(\omega)}_{rj}(\mathbf h)\) and \(q^{0}_{rj}=\alpha_{rj}(1-\beta_j)s_j\). \\

\(\theta^{(\omega)}_{rj}(\mathbf h)\) & Realized sectoral input-retention ratio for sector \(r\) and buyer \(j\) at fulfillment \(\mathbf h\), \(\theta^{(\omega)}_{rj}(\mathbf h)=\sum_{i\in\mathcal N_r}h_i\,a^{(\omega)}_{ij}/\alpha_{rj}\in[0,1]\). \\

\(\kappa^{(\omega)}_j(\mathbf h)\) & Intermediate input efficiency factor of producer \(j\), \(\kappa^{(\omega)}_j(\mathbf h)=(\sum_{r\in\mathcal R_j}\alpha_{rj}(\theta^{(\omega)}_{rj}(\mathbf h))^{\rho})^{1/\rho}\in[0,1]\). Its maintained lower bound is \(\underline\kappa_\omega:=\underline h_\omega\min_i\kappa^{(\omega)}_i(\mathbf 1)\) on \(\mathcal D_\omega\).
\\

\(d_{rj},\ \bar d_j,\ V_j\) & Sectoral retention shortfall \(d_{rj}=1-\theta^{(\omega)}_{rj}\), its \(\boldsymbol\alpha\)-weighted mean, and the composition variance of the local loss expansion (Section~\ref{app:m-local-expansion}). \\

\(\lambda^{(\omega)}_i(\mathbf h)\),
\(\boldsymbol\lambda^{(\omega)}(\mathbf h)\) & Output ratio and output-ratio vector \eqref{eq:m-lambda-rule}, \(\lambda^{(\omega)}_i(\mathbf h)=G_i(\kappa^{(\omega)}_i(\mathbf h))\in[0,1]\). At a sanction equilibrium, the scalar output ratio equals supplier \(i\)'s fulfillment rate.
\\

\(c_\omega\), \eqref{eq:m-cushion-cond} & Interiority floor and interiority condition (Proposition~\ref{prop:m-monotone}\textnormal{(iii)}). The condition gives \(\mathbf h_\omega\ge c_\omega\mathbf 1>\mathbf 0\), with closed-form floor \eqref{eq:m-floor}.
\\

\(\mathcal L^{\mathrm{ces}}_i\), \(\mathcal L^{\mathrm{alt}}_i\),
\(\upsilon_i\) & Maintained CES-shaped fulfillment-response map \eqref{eq:m-form-ces-map}, an alternative response on the same coordinates, and their functional-form deviation \(\upsilon_i=\mathcal L^{\mathrm{alt}}_i-\mathcal L^{\mathrm{ces}}_i\) \eqref{eq:m-form-deviation}.
\\

\(\mathbf J_\omega\), \(\mathbf R_\omega\) & Fulfillment Jacobian \(\mathbf J_\omega=D_{\mathbf h}\boldsymbol\lambda^{(\omega)}(\mathbf h_\omega)\) and fulfillment resolvent \(\mathbf R_\omega=(\mathbf I-\mathbf J_\omega)^{-1}\). Both are nonnegative at the maintained positive CES fulfillment fixed point (Lemma~\ref{lem:m-local-form}).
\\

\(\mathcal X_\omega(\mathbf h)\), \(\mathbf y_{c,\omega}\),
\(\mathcal A_{c,\omega}\) & Respectively, the auxiliary activity map \eqref{eq:m-form-activity-map}, country \(c\)'s activity gradient, and its network amplification factor \eqref{eq:m-form-amplification}. The auxiliary map equals activity at fulfillment fixed points. \\

\addlinespace[0.8em]
\multicolumn{2}{@{}l}{\textit{Fulfillment and the sanction equilibrium}}\\
\addlinespace[0.35em]

\(h_i\) & Fulfillment rate of supplier \(i\), \(h_i\in[0,1]\), the fraction of the effective demand facing supplier \(i\) actually delivered. \\

\(\Phi_\omega\) & Sanction-equilibrium pair map on \(\mathbb R_+^N\times[0,1]^N\), \(\Phi_\omega(\mathbf x,\mathbf h)=(\boldsymbol\lambda^{(\omega)}(\mathbf h)\circ[\mathbf f_\omega+\mathbf B_\omega\mathbf x],\ \boldsymbol\lambda^{(\omega)}(\mathbf h))\). \\

\(\mathcal D_\omega(\underline h_\omega)\) & Maintained bundle-feedback domain \(\mathcal D_\omega(\underline h_\omega):=[\underline h_\omega,1]^N\), abbreviated \(\mathcal D_\omega\). The maintained fulfillment floor is \(\underline h_\omega\in(0,1]\). \\

\((\mathbf s_\omega,\mathbf h_\omega)\) & Benchmark-selected sanction equilibrium under \(\omega\). Its fulfillment vector is the limit of reduced iteration from \(\mathbf h^{(0)}=\mathbf 1\) and the greatest fixed point of \(\boldsymbol\lambda^{(\omega)}\) (Proposition~\ref{prop:m-monotone}).
\\

\(\mathbf R^{*}_\omega\) & Fulfillment-scaled Leontief inverse at the selected equilibrium, \(\mathbf R^{*}_\omega=[\mathbf I-\diag(\mathbf h_\omega)\mathbf B_\omega]^{-1}\). This operator propagates the productivity channel \((\mathbf 1-\mathbf h_\omega)\circ\mathbf s\) and the lost-market channel \(\mathbf h_\omega\circ\boldsymbol\xi_\omega\) in Proposition~\ref{prop:m-anatomy}. \\

\(\mathrm{Lip}_h^\omega,\ \mathrm{Lip}_\lambda^\omega\) & \(1\)-norm Lipschitz constants on \(\mathcal D_\omega\) of the fulfillment rule \(\boldsymbol\lambda^{(\omega)}\) and of the source term \(\boldsymbol\lambda^{(\omega)}(\mathbf h)\circ\mathbf f_\omega\), with \(\|\mathbf f_\omega\|_\infty\) factored out of the source term. Both are bounded by \(\bar e_\omega\,\underline\kappa_\omega^{-\bar\beta}\,\underline\theta_\omega^{\rho-1}\bar a_\omega\) \eqref{eq:m-Lh-bound}. \\

\(\mathcal P_\omega\) & Admissible parameter region for sanction \(\omega\) (Section~\ref{app:m-proofs}). It consists of triples \((\tau,\rho,\varrho)\) for which the fulfillment rule contracts on \(\mathcal D_\omega\) and maps that domain into itself.
\\

\(\bar a_\omega\) & Supplier-side row-sum bound on \(\mathbf A_\omega\), \(\bar a_\omega:=\max_i\sum_j a^{(\omega)}_{ij}\). \\

\(\underline\theta_\omega\) & Lower bound on realized sectoral retention on \(\mathcal D_\omega\), \(\underline\theta_\omega=\underline h_\omega\,\min_{j,r\in\mathcal R_j}\mathcal F(\mathbf A_\omega)_{rj}/\alpha_{rj}>0\).
\\

\(\mathbf s(\mathbf h)\) & Activity vector conditional on the fulfillment vector, \(\mathbf s(\mathbf h)=(\mathbf I-\diag(\mathbf h)\mathbf B_\omega)^{-1}(\boldsymbol\lambda^{(\omega)}(\mathbf h)\circ\mathbf f_\omega)\).
\\

\addlinespace[0.8em]
\multicolumn{2}{@{}l}{\textit{Activity, losses, and country sizes}}\\
\addlinespace[0.35em]

\(\mathbf s_\omega\) & Activity component of the benchmark-selected sanction equilibrium. \\

\(s_{\omega,i}\) & Activity of country-sector \(i\) in the selected sanction
equilibrium under \(\omega\). \\

\(\boldsymbol\Delta\mathbf s_\omega\) & Sanction-induced activity loss,
\(\boldsymbol\Delta\mathbf s_\omega=\mathbf s-\mathbf s_\omega\ge\mathbf 0\). \\

\addlinespace[0.8em]
\multicolumn{2}{@{}l}{\textit{Country-level vulnerability}}\\
\addlinespace[0.35em]

\(Q_\omega(c)\) & Size of country \(c\) in the selected sanction equilibrium
under \(\omega\), \(Q_\omega(c)=\mathbbm{1}_c\mathbf s_\omega\). \\

\(\gamma^{(\omega)}_c\) & Vulnerability of country \(c\) to sanction \(\omega\), its own-size-normalized loss \(\gamma^{(\omega)}_c=(Q(c)-Q_\omega(c))/Q(c)=\mathbbm{1}_c\boldsymbol\Delta\mathbf s_\omega/\mathbbm{1}_c\mathbf s\).
\\

\(\gamma^{(c'\notnotleftrightarrow c)}_c\) & Vulnerability of country
\(c\) under the bilateral sanction \(c'\notnotleftrightarrow c\). \\

\(M_{cc'},\ X_{cc'}\) & Country \(c\)'s benchmark purchases from the severed partner and its sales to it, intermediate and final together, \(M_{cc'}=T^{\mathrm{cb}}_{cc'}\) and \(X_{cc'}=T^{\mathrm{cb}}_{c'c}\) (Section~\ref{subsec:power_size}). \\

\(\boldsymbol\nu_\omega\) & Injury vector of the loss anatomy, \(\boldsymbol\nu_\omega=(\mathbf 1-\mathbf h_\omega)\circ\mathbf s+\mathbf h_\omega\circ\boldsymbol\xi_\omega\), so that \(\mathbf s-\mathbf s_\omega=\mathbf R^{*}_\omega\boldsymbol\nu_\omega\). \\

\(\mathbf m^{*}_c\) & Content row of country \(c\), \(\mathbf m^{*\top}_c=\mathbbm{1}_c\mathbf R^{*}_\omega\) \eqref{eq:m-content}. Entry \(m^{*}_{c,i}\) gives the activity of country \(c\) that a unit of fulfilled direct order at country-sector \(i\) generates (Lemma~\ref{lem:m-content}). \\

\(\mathbf h^{(c)}_\omega,\ \mathbf h^{(c')}_\omega\) & One-sided fulfillment rates of the severance \(c'\notnotleftrightarrow c\): the greatest fixed points of the fulfillment rule at the realized effective demand when only the buyer sectors of country \(c\), or only the buyer sectors of country \(c'\), have lost their inputs \eqref{eq:m-onesided-theta}. They satisfy \(\mathbf h_\omega\le\mathbf h^{(c)}_\omega,\mathbf h^{(c')}_\omega\le\mathbf 1\) (Lemma~\ref{lem:m-onesided}). \\

\(S_c,\ S^{(c)}_c,\ S^{(c')}_c,\ S^{\times}_c\) & Shortfall loss of country \(c\), the part that its lost purchases cause on their own, the part that its lost sales cause on their own, and the interaction of the two causes, \(S^{\times}_c=S_c-S^{(c)}_c-S^{(c')}_c\) \eqref{eq:m-shortfall-split}. \\

\(L^{\downarrow}_c,\ L^{\uparrow}_c,\ L^{\times}_c\) & The parts of country \(c\)'s dollar loss \(L_c=Q(c)-Q_\omega(c)\) that its lost sales and its lost purchases make up \eqref{eq:m-two-parts}, and the interaction of the two that remains. \\

\(\ell^{\downarrow}_c,\ \ell^{\uparrow}_c,\ \ell^{\times}_c\) & Export multiplier and import multiplier of side \(c\) under the severance, the activity it loses per unrecovered dollar of its sales to the partner and per unrecovered dollar of its purchases from it, and the interaction rate, the interaction per unrecovered dollar of the pair's trade, under \(\tau>0\) and positive bilateral flows \eqref{eq:m-rates}. They satisfy \(\gamma_c=\tau(1-\delta)[\ell^{\downarrow}_cX_{cc'}+\ell^{\uparrow}_cM_{cc'}+\ell^{\times}_c(X_{cc'}+M_{cc'})]/Q(c)\) (Proposition~\ref{prop:m-exposure}). \\

\(\Sigma_c,\ \mathcal G_c\) & The sum \(\ell^{\downarrow}_c+\ell^{\uparrow}_c+2\ell^{\times}_c\) of the multipliers and the import premium \((\ell^{\uparrow}_c-\ell^{\downarrow}_c)/\Sigma_c\) of side \(c\) \eqref{eq:m-gap}. These determine how the trade balance enters the loss ratio (Proposition~\ref{prop:m-decomposition}). \\

\(\xi^{\downarrow}_c,\ \xi^{\uparrow}_c\) & Lost sales and lost purchases of country \(c\) under the severance, \(\xi^{\downarrow}_c=\sum_{i\in c}\xi_{\omega,i}\) and \(\xi^{\uparrow}_c=\sum_{i\in c'}\xi_{\omega,i}\) \eqref{eq:oa-lost-trade}, equal to \(\tau(1-\delta)X_{cc'}\) and \(\tau(1-\delta)M_{cc'}\) at a common unrecovered share. \\

\(\underline h_c,\ \iota_c\) & Lowest fulfillment rate among the producers of country \(c\) that sell to the severed partner, and the intermediate share of country \(c\)'s severed purchases. \\

\(\mathbf D\) & Circulation at the activity \(\mathbf s\): a matrix on the support \(E\) with \(\mathbf 1^{\top}\mathbf D=\mathbf 0^{\top}\) and \(\mathbf D(\mathbf I-\boldsymbol\beta)\mathbf s=\mathbf 0\) \eqref{eq:m-circulation}. Along this direction the sizes remain fixed (Lemma~\ref{lem:m-fiber}). \\

\(E,\ K_E\) & Support of the network, the set of links with a positive sourcing share, and the number of connected components of its bipartite seller--buyer graph (Lemma~\ref{lem:m-fiber}). \\

\(\varphi_c\) & Fraction of its intermediate inputs that each producer of country \(c\) buys from producers of country \(c\), in the bound of Lemma~\ref{lem:m-homechain}\textnormal{(ii)}. \\

\(\vartheta\) & Flattening exponent of the flattened networks, whose flow blocks project the kernel \((Z_{ij})^{1-\vartheta}\) onto their own seller and buyer totals (Section~\ref{subsec:power_size_base} of the paper, Section~\ref{oa:fiber}). \\

\addlinespace[0.8em]
\multicolumn{2}{@{}l}{\textit{Hirschman matrix and power}}\\
\addlinespace[0.35em]

\(\boldsymbol{\mathcal H}=[\mathcal H_{cc'}]\) & Hirschman matrix, with \(\mathcal H_{cc'}\in[-1,1]\) the asymmetry between countries \(c\) and \(c'\) defined at \eqref{eq:bounded-Hirschman} and set to zero when its denominator vanishes. The matrix is skew-symmetric. \\

\(r_{cc'}\) & Loss ratio of a pair, \(r_{cc'}=\gamma_c/\gamma_{c'}\) \eqref{eq:m-logratio}, with \(\mathcal H_{cc'}=\psi(r_{cc'})\) (Lemma~\ref{lem:m-rigidity}). \\

\(\psi(r)\) & Normalized difference of a positive ratio, \(\psi(r)=(r-1)/(r+1)\in(-1,1)\) \eqref{eq:m-psi}, with \((x-y)/(x+y)=\psi(x/y)\). The bounded sum is \(u\oplus v=(u+v)/(1+uv)\), with \(\psi(r_1r_2)=\psi(r_1)\oplus\psi(r_2)\) (Lemma~\ref{lem:m-psi}). \\

\(\mathcal Q_{cc'}\) & Size asymmetry of a pair, \(\mathcal Q_{cc'}=\psi(Q(c')/Q(c))\) \eqref{eq:m-size-asymmetry}, positive when country \(c\) is the smaller. \\

\(\mathcal V_c,\ \mu_{cc'}\) & Trade-weighted advantage of country \(c\), \(\mathcal V_c=-\sum_{c'\ne c}\mu_{cc'}\mathcal H_{cc'}\), with \(\mu_{cc'}\) the share of country \(c\)'s cross-border trade that runs with country \(c'\) (Section~\ref{oa:size-advantage}). \\

\(\mathcal V^{Q}_c\) & Size advantage of country \(c\), \(\mathcal V^{Q}_c=-\sum_{c'\ne c}\mu_{cc'}\mathcal Q_{cc'}\) \eqref{eq:m-size-advantage}, the advantage that the sizes alone predict. \\

\(\gamma^{\mathrm F}_c,\ \mathcal H^{\mathrm F}_{cc'}\) & Flexible-price vulnerability of country \(c\), \(\sum_{j\in c}(s_j-y_j)/Q(c)\), and flexible-price asymmetry. The flexible-price asymmetry divides the vulnerability difference by the sum of absolute vulnerabilities (Section~\ref{subsec:flex-welfare}). \\

\(z_j\) & Wedge of producer \(j\) at the flexible-price equilibrium, its composite input price index relative to its own price, \(z_j=P^{m}_j/p_j\), which sets its output through \eqref{eq:flex-wedge-output}. \\

\(k_c\) & Loss multiplier of country \(c\) under a severance, its flexible-price loss over its fixed-price loss, \(k_c=\gamma^{\mathrm F}_c/\gamma_c\). These multipliers determine how asymmetry changes between closures (Proposition~\ref{prop:flex-power}). \\

\(\Psi(\boldsymbol{\mathcal H})\) & Hirschman asymmetry index, the average absolute bilateral asymmetry \(\Psi(\boldsymbol{\mathcal H})=\frac{2}{C(C-1)}\sum_{c<c'}|\mathcal H_{cc'}|\in[0,1]\).
\\

\addlinespace[0.8em]
\multicolumn{2}{@{}l}{\textit{Country-level trade matrices}}\\
\addlinespace[0.35em]

\(\mathbf T^{\mathrm{cb}}=[T^{\mathrm{cb}}_{cc'}]\) & Cross-border flow table \eqref{eq:country_flow_table_crossborder}, with \(T^{\mathrm{cb}}_{cc'}\) the benchmark flow from seller country \(c'\) to buyer country \(c\), intermediate and final together, and zero on the domestic diagonal. \\

\(\mathcal B_{cc'}\) & Trade balance of the pair \eqref{eq:direct_trade_asymmetry}, \(\mathcal B_{cc'}=(T^{\mathrm{cb}}_{cc'}-T^{\mathrm{cb}}_{c'c})/(T^{\mathrm{cb}}_{cc'}+T^{\mathrm{cb}}_{c'c})\), zero when the denominator vanishes. \\

\(\mathcal B^{\xi}_{cc'}\) & Balance of the lost trade, \(\mathcal B^{\xi}_{cc'}=(\xi^{\uparrow}_c-\xi^{\downarrow}_c)/(\xi^{\uparrow}_c+\xi^{\downarrow}_c)\) \eqref{eq:oa-lost-balance}, which takes the place of the trade balance in the loss ratio when the unrecovered share differs across the severed links (Lemma~\ref{lem:oa-unequal-shares}). \\

\(\mathcal J_{cc'},\ \mathbf Z^{\mathrm{bal}},\ d_c,\ d_{c'}\) & The buyers of countries \(c\) and \(c'\), producers and households, the balanced table of the pair, the solution of the program \eqref{eq:m-balanced-program}, and the direction factors on the flows from country \(c\) to the buyers of country \(c'\) and from country \(c'\) to the buyers of country \(c\) (Lemma~\ref{lem:m-balanced-table}). \\

\(\bar m^{X}_c,\ \bar m^{M}_c\) & Domestic content of country \(c\)'s severed sales and of the shortfalls its severed purchases induce, per dollar of severed sales and of severed purchases, built from the benchmark table alone \eqref{eq:m-content-sales}, \eqref{eq:m-content-purchases}. They proxy the solved export multipliers and import multipliers. Differences can reflect omitted global paths, equilibrium changes, and curvature (Table~\ref{tab:s6_severed_dollar}). \\

\addlinespace[0.8em]
\multicolumn{2}{@{}l}{\textit{Structural scores}}\\
\addlinespace[0.35em]

\(\bar q_c\) & Country aggregation of a country-sector structural score
\(\mathbf q\), using the benchmark-size weights in
\eqref{eq:country_structural_score}. \\

\(\Delta\bar q_{cc'}\) & Bilateral structural-score gap,
\(\bar q_{c'}-\bar q_c\), oriented in the same direction as
\(\mathcal H_{cc'}\). \\

\end{longtable}

\normalsize
\renewcommand{\arraystretch}{1}


\end{document}